\documentclass[a4paper,12pt]{mybook} 
\pdfoutput = 1

\usepackage{epsf,epsfig,wrapfig}

\usepackage[dutch,english]{babel}                
\usepackage{fancyhdr}                   
\usepackage{amsmath,dsfont,amssymb,extarrows,multirow,ctable,colortbl,theorem}   
\usepackage[titles]{tocloft}            

\usepackage{makeidx}                    

\usepackage{floatflt,subfig}                   
\ifx\pdfoutput\undefined                
\newcommand*{\includepdf}[2][]{}         
\else                                   

  \usepackage{pdfpages}                  %
\fi

\renewcommand{\chaptermark}[1]          
  {\markboth{\sf\thechapter.\ #1}{}}    
\renewcommand{\sectionmark}[1]          
{\markright{\sf\thesection\ #1}}

\fancypagestyle{plain}{                 
  \fancyhf{}                              
  \fancyfoot[C]{\bfseries\thepage}        
}

\makeatletter                           
\def\cleardoublepage{\clearpage\if@twoside \ifodd\c@page\else
   \hbox{}
   \thispagestyle{empty}                
   \newpage
   \if@twocolumn\hbox{}\newpage\fi\fi\fi}
\makeatother

\newtheorem{theorem}{Theorem}[section]

\newtheorem{corollary}[theorem]{Corollary}

\newenvironment{definition}[1][Definition]{\begin{trivlist}
\item[\hskip \labelsep {\bfseries #1}]}{\end{trivlist}}
\newenvironment{example}[1][Example]{\begin{trivlist}
\item[\hskip \labelsep {\bfseries #1}]}{\end{trivlist}}

\newenvironment{digression}[1][Digression]{\begin{trivlist}
\item[\hskip \labelsep {\bfseries #1}]}{\end{trivlist}}

\newcommand{\qed}{\nobreak \ifvmode \relax \else
      \ifdim\lastskip<1.5em \hskip-\lastskip
      \hskip1.5em plus0em minus0.5em \fi \nobreak
      \vrule height0.75em width0.5em depth0.25em\fi}

\newcommand{\mychapter}[1]{
 \chapter*{#1\markboth{\sf #1}{\sf #1}}
 \addcontentsline{toc}{chapter}{#1}
}

\newcommand{\pp}{{\mathbb P}}
\newcommand{\Z}{{\mathbb Z}}
\newcommand{\zz}{{\mathbb Z}}
\newcommand{\R}{{\mathbb R}}
\newcommand{\rr}{{\mathbb R}}
\newcommand{\C}{{\mathbb C}}

\newcommand{\Q}{{\mathbb Q}}

\newcommand\Tr{{\rm Tr}}

\newcommand{\delbar}{\overline{\partial}}
\newcommand{\zbar}{\bar{z}}
\newcommand{\tbar}{\bar{\tau}}

\newcommand{\ibar}{\bar{\imath}}
\newcommand{\jbar}{\bar{\jmath}}

\newcommand{\bea}{\begin{eqnarray}}
\newcommand{\eea}{\end{eqnarray}}
\newcommand{\ben}{\begin{eqnarray*}}
\newcommand{\een}{\end{eqnarray*}}
\newcommand{\be}{\begin{equation}}
\newcommand{\ee}{\end{equation}}
\newcommand{\bem}{\begin{pmatrix}}
\newcommand{\eem}{\end{pmatrix}}
\newcommand{\bl}{\begin{align}}
\newcommand{\el}{\end{align}}

\def\a{\alpha}
\renewcommand{\b}{{\beta}}
\def\ex{{\bf {e}}}
\def\c{\gamma}
\def\d{\delta}

\def\e{\epsilon}
\def\f{\phi}               

\def\g{\gamma}

\def\im{\mathrm{Im}}
\def\inf{\infty}
\def\j{\psi}
\def\k{\kappa}             
\def\l{\lambda}
\def\m{\mu}
\def\n{\nu}
\def\na{\nabla}
 \def\w{\omega}
\def\p{\pi}
\def\pa{\partial}
\def\del{\partial}
\def\re{\mathrm{Re}}                
\def\r{\rho}                                     
\def\s{\sigma}                                   
\def\t{\tau}
\def\th{\theta}
\def\til{\tilde}

\def\x{\xi}

\def\z{\zeta}

\def\D{\Delta}
\def\F{\Phi}
\def\G{\Gamma}

\def\L{\Lambda}
\def\O{\Omega}
\def\bO{\boldsymbol \Omega}

\def\Q{\Theta}
\def\S{\Sigma}

\def\cn{{\cal N}}

\def \yf {\mathsf{y}}
\def \tf {\mathsf{t}}
\def \Lf {\mathsf{L}}
\def \Qf {\mathsf{Q}}
\def \Hf {\mathsf{H}}
\def \hf {\mathsf{h}}

\def \Sf {\mathsf{S}}
\def \xf {\mathsf{x}}

\def\taubar{\bar{\tau}}
\def\qhat{\hat{q}_0}

\newcounter{secapp}[chapter]            
\newcommand{\sectionappendix}[1]        
{
  \newpage                              
  \renewcommand{\thesection}{\thechapter.\Alph{secapp}}
  \stepcounter{secapp}                  
  \refstepcounter{section}              
  \addcontentsline{toc}{section}{\protect\numberline{\thesection}#1} 
  {\flushleft\Large\sffamily\bfseries\thesection$\;\;$ #1\par}
  \sectionmark{#1}\vspace{\baselineskip}  
}

\begin{document}

%
%

\pagestyle{empty}

\begin{center}
\vspace*{15mm}
{
  \Huge \sffamily \bfseries The Spectra of \\ Supersymmetric States  in \\ String Theory \\[4mm]
}
\vspace*{4.5cm}
{
\large 
PhD Thesis\\
\vspace*{0.2cm}
2008 University of Amsterdam
\\
\vspace*{10cm}
by \\[5mm]
{\sffamily \bfseries Miranda Chih-Ning Cheng} \\[5mm]
}
\end{center}

\cleardoublepage
\pagestyle{fancy}


\selectlanguage{english}
\pagenumbering{roman}\setcounter{page}{1}


\cleardoublepage
\pagestyle{fancy}


\tableofcontents
\cleardoublepage

\newpage
\mychapter{Preface}

The goals of the thesis, apart from for the author to become a doctor,  are the following: 1. To summarise the main results of my research of the past three years. 2. To provide a compact and self-contained survey of the relevant materials for beginning graduate students or researchers in other sub-fields as a shortcut to the frontline of the current research in this area of string theory. 

\vspace{.4cm}
{\it Motivation for the First Goal}
\vspace{.4cm}

My personal motivation to pursue this line of research has two sides. First of all, in order to understand the structure of a theory, it is important to know the spectrum of the theory. Just like the spectrum of  a hydrogen atom holds the key to understanding quantum mechanics, we hope that the same might be true for string theory.
For a very complex theory as string theory is, the supersymmetric part of the spectrum is usually the part which is most accessible to us due to the great simplification supersymmetry offers. Nevertheless, as I hope I will convince the readers in this thesis, it is still a far from trivial task to study this part of the spectrum. In other words, we hope that the study of the spectrum of supersymmetric states of string theory will be a feasible step towards furthering our understanding of string theory. 

In the other direction, it has been a great challenge since the invention of Einstein gravity and quantum mechanics to understand the quantum aspects of gravity. A fundamental question since the work of Bekenstein and Hawking in the 70's, is why black holes have entropy. Only when we can answer this question can we ever claim that we understand the nature of quantum gravity. Conversely, because of the challenging nature of the question, once we can answer this question we have a reason to believe that we are on the right track to the goal of quantising, in one way or the other, Einstein gravity. String theory, at the time of writing, still scores highest in the challenge of explaining the thermodynamical entropy of the black holes, while it is also true that most of the work done along this trajectory still focuses on black holes with supersymmetry, which are unlikely to be directly observable in nature. From this point of view, to study the supersymmetric spectrum of string theory and to use it as information about the black hole entropy, is a part of the effort towards a deeper understanding of the nature of quantum gravity.

\vspace{.4cm}
{\it Motivation for the Second Goal}
\vspace{.4cm}

Now I will move on to explain the motivation to achieve the second goal of the thesis: providing a self-contained material serving as a shortcut to the current research on the topic. 

In the course of development of string theory since its birth in the 70's, it has expanded into an extremely broad and sometimes very complicated field. According Mr. Peter Woit, there might be around 30,000 papers written on the subject so far. While exactly this property makes it, in my opinion, a sufficiently fun field to be working in, it is no good news for beginners. In order to work on a topic in a specific sub-field, she or he is likely to find herself having to go through the labyrinth of a large amount of papers on various totally different but yet somehow inter-connected topics in physics and mathematics, with conflicting notations and conventions.

As it could be fairly time-consuming and frustrating a process, I would like to take the chance of writing my PhD thesis to provide a  service for anyone who might be able to use it, by making an attempt at a relatively compact and self-contained exposition of some of the should-know's for performing research related to the subjects I have worked on in the past three years.

\vspace{.4cm}
{\it Reader's Manual}
\vspace{.4cm}

But the attempt to be self-contained also brings some drawbacks to this thesis, namely that a fair portion of it is probably unnecessary for expert readers only interested in the results of my research. To cure this problem we now give a rough reader's manual so the advanced readers will be less likely finding themselves wasting their time on the introductions.

This thesis is organised as follows. In the first part we introduce some basic concepts of superstring theories in 30 pages, focusing on  the perturbative aspects in the first section and the non-perturbative aspects in the second. The readers who are sufficiently familiar with superstring theory can safely skip this part.

The second part of the thesis is about string theory compactified on Calabi-Yau two- and three-folds. The readers who are not yet familiar with basic concepts of differential and algebraic geometry might want to first read Appendix A before going into this part. Basically all the mathematical background for understanding this thesis, in particular part II and IV, is summarised in  Appendix A with some explanations but without proofs. 

The readers who are familiar with these materials can go directly to chapter two where Calabi-Yau (three-fold) compactification is discussed. This chapter is relevant for understanding the discussion of the BPS spectrum of the \(d=4,\,{\cal N}=2\)  theories. But this chapter can again be skipped if the reader is sufficiently familiar with generalities of Calabi-Yau compactifications. The same goes for chapter three, where K3 compactification is discussed and which is relevant for the understanding the discussion of the BPS spectrum of the \(d=4,\,{\cal N}=4\)  theories.

After chapter four, the reader can go directly to the part III, IV or part V of the thesis, depending on whether she or he is interested in  
\({\cal N}=2\) or \({\cal N}=4\) theories, and which aspects of them. The material presented here are mostly the extended and rewritten version of some of my papers. I wish the readers who have gone through the first half can now enjoy the fruit of reading the hundred pages of ``preliminary knowledge", and the 
expert readers can find this half of the thesis somewhat interesting after having skipped the first half.

\vspace{.4cm}
{ \bf{Acknowledgment}}
\vspace{.4cm}

It is a pleasure to thank Erik Verlinde and Kostas Skenderis for being my thesis advisor, and the committee members B. Craps, J. de Boer, R. Dijkgraaf, B. de Wit, and M. Taylor for reading and commenting on this thesis. Finally I would like to thank M. Esole, L. Hollands and Alessandro Tomasiello for useful discussions and a critical reading of the manuscript. This work is supported by the Netherlands Organisation for Scientific Research (NWO) and Foundation of Fundamental Research on Matter (FOM). 

\cleardoublepage

\ifodd\value{page}
\typeout{Mainmatter starts on odd page. Fine.}
\else
\typeout{Mainmatter starts on even page. Invert odd/even}
\fancyhf{}                              
\fancyhead[LO]                          
{{\bfseries\thepage$\quad$}\leftmark}   
\fancyhead[RE]                          
{\rightmark{$\quad$\bfseries\thepage}}  
\evensidemargin=22mm
\oddsidemargin=18mm
\fi

\pagenumbering{arabic} \setcounter{page}{1}

\part{Superstring Theory }

In this part of the thesis we will give an introduction of various aspects of superstring theory. The readers who are already familiar with general knowledge of string theory can safely skip this part and go directly to other topics of interests.

Facing such a vast subject as string theory is now, it is absolutely not my intention to give a complete account of the subjects. Rather we will try to compactly introduce the key concepts and important results that will be crucial for our study of the spectra of supersymmetric states, in systems resulting from various compactifications of superstring theory. This is done as part of the effort to present a self-contained PhD thesis accessible to beginning graduate students of the field and and should not appeal to all readers.

This part of the thesis is organised as follows. In the first section we will begin with aspects of perturbative superstring theory from a world-sheet viewpoint. The resulting spacetime physics will be introduced in the second section, with a focus on low-energy effective action and the relationship with spacetime coupling constants and the world-sheet fields. In the last section we will turn to the non-perturbative aspects of the ``superstring theory", where fundamental string loses its fundamental status. Topics included in this section are M-theory and S-duality, D-branes and gauge/gravity correspondence.

\chapter{Type IIA and Type IIB Superstring Theory}

\section{ The World-Sheet Action}

Consider the two-dimensional Ramond-Neveu-Schwarz action \index{RNS (Ramond-Neveu-Schwarz)  action}
\be \label{RNS}
S= \frac{1}{4\p} \int d^2z \left(\frac{2}{\a'} \pa X^\m \bar{\pa}  X_\m + \j^\m \bar{\pa}\j_\m + \til{\j}^\m{\pa} \til{\j}_\m
\right)\;,
\ee
where \(\j^\m\) and \(\til{\j}^\m\) are the two components of the super-partner of bosonic fields \(X^\m\). 

The conserved currents corresponding to the symmetry of world-sheet translation and the (\(1,1\)) world-sheet supersymmetry transformation are, respectively,
\bea\notag
T &=& -\frac{1}{2} \left( \pa X^\m \pa X_\m +  \j^\m {\pa}\j_\m\right) \\
\label{current}
T_F &=& i  \j^\m \pa X_\m 
\eea
and similarly for the anti-holomorphic part\footnote{Here and for the rest of this section we have put \(\a'=2\) to simplify the equations.}. Notice that \(T(z)\) and \(\bar{T}(\bar{z})\) are the only non-vanishing component of the energy-momentum tensor of this theory. The vanishing of the trace \(T_{z\zbar}(z,\zbar)\) of the energy-momentum tensor, which is a property special for the above theory in two dimensions, implies a much larger group of symmetry. Indeed one can check that the Noether current corresponding to any infinitesimal world-sheet transformation \(z \rightarrow z + \e(z) \) is conserved, implying the presence of the following conformal symmetry 
\begin{align}\nonumber
&X^\m(z',\bar{z}') = X^\m(z,\bar{z}) \\ \nonumber
&\j^\m (z',\bar{z}') = (\frac{\pa z'}{\pa z})^{-1/2} \j^\m (z,\bar{z})\;,\;
\til{\j}^\m (z',\bar{z}') = (\frac{\pa \bar{z}'}{\pa \bar{z}})^{-1/2} \til{\j}^\m (z,\bar{z})\;,
\end{align}
for any holomorphic function \(z'(z)\) of \(z\).
\index{conformal symmetry}
The world-sheet supercurrent \(T_F\) on the other hand, generates the superconformal transformation \index{superconformal symmetry} 
\bea\nonumber
\d X^\m(z,\bar{z}) &=&- \e \left(f(z) \j^\m(z) + \bar{f}(\bar{z}) \til{\j^\m}(\bar{z}) \right) \\ 
\d \j^\m(z)&=&\e f(z)\pa X^\m(z)\\ \nonumber
\d \til{\j}^(\bar{z})&=&\e \bar{f}(\bar{z}) \bar{\pa} X^\m(z)\;. 
\eea

In order for the above classical symmetries to be realised at the quantum level, one has to make sure that the path integral is well-defined. To work with the gauged-fixed action (\ref{RNS}), the Jacobian factor of the gauge orbits has to be appropriately taken into account for each gauge symmetry of the theory.  This can be done by introducing ghost fields, the so-called Faddeev-Popov ghosts, into the theory. For this particular theory we need ghost fields for both the conformal and the superconformal gauge symmetries. Furthermore, the anomaly associated to the above conformal transformation only vanish when the total central charge of the full conformal field theory, now with the ghosts fields included, vanishes. This condition turns out to imply that the above Ramond-Neveu-Schwarz string theory is only consistent when there are ten of the fields (\(X^\m, \j^\m ,\til{\j}^\m \)). In other words, the critical dimension of the target space is ten for these theories. See for example \cite{Polchinski} or \cite{GSW} for further details of the above argument about the critical dimensions.

\index{ghost field}
\index{Faddeev-Popov ghosts}
\index{critical dimension}

\subsection{ Canonical Quantisation}

Now we would like to quantise this theory in ten-dimensional flat space. We will follow the canonical quantisation formalism. Though arguably not the most elegant way to do it, as opposed to the more systematic way of BRST quantisation, it has the advantage of admitting a simple exposition for our purpose.  Since in this approach we put the ghost sector at its ground state at all stages, we will usually avoid writing down the ghost operators. Interested readers can consult, for example, \cite{Polchinski} or \cite{GSW} for a thorough treatment of the topic.

Let's consider the spectrum of the theory (\ref{RNS}) on a cylinder. Without further identification on the target space, the world-sheet scalars \(X^\m\) have to return to the same value after circling the cylinder once. The fermions on the other hand, can have two different possibilities. The fermions which return to itself after circling once are said to satisfy the Neveu-Schwarz (NS) boundary condition and those which return to minus itself the Ramond (R) boundary condition. 

In other words, upon conformally mapping the cylinder onto a complex plane with coordinate \(z\), the Laurent expansion of the fields can take the form
\begin{align} \notag
\pa X^\m(z) &= -i \sum_{n \in \Z} \frac{a^\m_n}{z^{n+1}}  \\
\notag
\j^\m(z) &=\sum_{r} \frac{\j^\m_r}{z^{r+1/2}} \;,
\end{align}
where one has to choose between the two possible fermion boundary conditions
\be
\begin{cases}
2r =  0\text{  mod   }2 &\text{for R sector} \\ &\\ 
2r = 1 \text{  mod   }2 &\text{for NS sector}\;,
\end{cases}
\ee
and similarly for the right-moving sectors. Notice that all coefficients \(a^\m_\n\) and \(\j^\m_r\) are conserved quantities, due to the presence of the conformal symmetry. Canonical quantisation therefore gives us infinitely many (anti-)commutation relations 
\begin{align}  \notag
[a^\m_m,a^\n_n] &= m \eta^{\m\n} \d_{m+n,0}\\ 
\label{commutation}
\{\j^\m_r, \j^\n_s\} &= \eta^{\m\n} \d_{r+s,0}\;,
\end{align}
where \(\eta^{\m\n} = \text{diag}(-1,1,\dotsi,1)\) is the metric for (9+1)-dimensional flat spacetime, and again the same relations hold also for the right-moving sector. 

Now one can expand the conformal and the superconformal currents in the same way as
\begin{align} \notag
T&= \sum_{n \in \Z}  \frac{L_n}{z^{n+2}}\\
T_F & =\sum_{2r \in \Z^{\pm}} \frac{G_r}{z^{r+3/2}}\;,
\end{align}
and the generators now have the following expansion
\begin{align} \notag
L_m &= \frac{1}{2} \left( \sum_{n \in \Z} :a^\m_{m-n} a_{\m,n}: +  \sum_{2r \in \Z^{\pm}} (r-\frac{m}{2}) :\j^\m_{m-r} \j_{\m,r}: \right)+ L_m^{\text{gh}} + A_{\text{R,NS}} \,\d_{m,0} \\ \label{generator}
G_r& =  \sum_{n \in \Z} a^\m_n \j_{\m,r-n} + G_r^{\text{gh}}\;,
\end{align}
where \(L_m^{\text{gh}}\) and  \(G_r^{\text{gh}}\) denote the contribution from the ghost fields which we won't need, and the \(:...:\) denotes the normal ordering among the operators. From the commutation relations we see that the normal ordering only matters for the generator \(L_0\), and we have introduced the extra constant  ``\(A_{\text{R,NS}}\)" in this case to account for this ordering ambiguity. There are various ways to determine this zero point energy, such as the zeta-function regularization for example. The readers can consult various textbooks, for example, \cite{Polchinski} or \cite{Ginsparg:1988ui} for a careful treatment of this issue. Here we will simple quote the results.
These zero-point energies are
\be
\begin{cases}
A_{\text{R}} = 0 &\text{for R sector} \\ &\\ 
A_{\text NS} = -\frac{1}{2}&\text{for NS sector}\;.
\end{cases}
\ee

Now we are ready to move on to analysing the lowest-lying spectrum of this theory and especially how the tachyonic state can be projected out of the spectrum.

\subsection{ Massless Spectrum}

In the canonical quantisation, extra constraints have to be imposed on a physical state. This is because we expect the scattering amplitudes between two physical states to be invariant under the conformal and superconformal transformation. In other words
\be \label{physical}
 0 = \langle\j_1|T(z)|\j_2\rangle =  \langle\j_1|T_F(z)|\j_2\rangle \ee
for any two physical states \(|\j_1\rangle\) and \(|\j_2\rangle\). 

In the previous two subsections we have seen that the total central charge vanishes, and the currents satisfy the superconformal algebra
\begin{align} \notag
[L_m, L_n ] &= (m-n) L_{m+n}\\ \notag
\{G_r,G_s\} & = 2 L_{r+s} \\ \label{n_1_superconformal_algebra}
[L_m,G_r] & = \frac{m-2r}{2} G_{m+r}\;.
\end{align}
\index{\({\cal N}=1\) superconformal algebra}
It is therefore self-consistent and sufficient for the purpose of ensuring  (\ref{physical}) to impose 
\be\label{constraint_superconformal}
L_m |\j\rangle  = G_r |\j\rangle = 0   \text{    for all     } n,r \geq 0 \;.
\ee
 Especially, since in the canonical quantisation we are putting the ghost sector to its ground state, now we just have to impose the above condition on the matter part of the Hilbert space. In particular, the \(L_0\)-constraint gives the mass-shell condition
 \be\label{mass-shell}
 L_0 |\j\rangle = H |\j\rangle =0\;.
 \ee
In the above equation, \(p^\m\) is the eigenvalue of the center-of-mass mode \(a^\m_0\), \( N=\sum_{n>0} a^\m_{-n}a_{\m,n} + \sum_{r>0}r  \j^\m_{-r} \j_{\m,r}\) is the oscillation number operator, and the ``Hamiltonian operator" is given in terms of them as 
\be \label{level matching}
H = \begin{cases} \frac{1}{2} p^2 + N 
&\text{for R sector} \\ &\\ 
 \frac{1}{2} p^2 + N - \frac{1}{2}  &\text{for NS sector}\;.
\end{cases}\;
\ee

 Apart from the constraints, there are also equivalence relations among the physical states. Namely 
 $$
 |\j\rangle \sim |\j\rangle + |\f\rangle  
 $$
 if \(\langle\j'|\f\rangle=0\) for all physical states \(|\j'\rangle\), since any two states having the same scattering amplitudes with any other physical state must be equivalent. This happens, for example, when 
 \be \label{equivalence}
 |\f\rangle = (\sum_{m>0} \ell_m L_{-m} + \sum_{r>0} \k_r G_{-r})  |\f'\rangle\;\ee 
 for some coefficients \(\ell_m\) and \(\k_r\).
 This equivalence condition together with the constraints will remove for us the two light-cone directions.  To see this we will now study the spectra of R-ground states and NS-ground state and their excited states. 
 
Define \(|0;p\rangle_{\mbox{\tiny R}}\) to be the R-ground state annihilated by all annihilation operators
$$
a^\m_n |0;p \rangle_{\mbox{\tiny R}} = \j^\m_n |0;p\rangle_{\mbox{\tiny R}}  = 0  \quad\text{for all} \quad n > 0 \quad \text{and} \quad a^\m_0 |0;p\rangle_{\mbox{\tiny R}} = p^\m|0;p\rangle_{\mbox{\tiny R}} \;.
$$
\index{Ramond ground state}
From the commutation relations of the fermionic zero modes \(\j^\m_0\), we see that they satisfy the ten-dimensional Clifford algebra. The R-ground states are therefore spacetime fermions. In ten dimensions the smallest representation is the 16-dimensional Weyl-Majorana spinors, which are real and have definite chiralities
$$
\G  |0;p\rangle_{\mbox{\tiny R}}^\pm =\pm |0;p\rangle_{\mbox{\tiny R}}^\pm\;, 
$$
with \(\G\) anti-commuting with all the ten Gamma matrices.
 Now the only non-trivial constraints from (\ref{constraint_superconformal}) are
$$
L_0  |0;p\rangle_{\mbox{\tiny R}}^\pm = G_0 |0;p\rangle_{\mbox{\tiny R}}^\pm =  0\;.
$$
The first condition tells  us that the state is massless and the second now gives the massless Dirac equation.  The massless Dirac equation further reduces the degrees of freedom of the 16-component spinors in half. We therefore conclude that the R-ground states transform in the spinor representation \({\bf 8}\)
 or \({\bf 8'}\) of \(SO(8)\), corresponding to the two possible ten-dimensional chiralities \( \G |0;p\rangle_{\mbox{\tiny R}}^\pm=\pm  |0;p\rangle_{\mbox{\tiny R}}^\pm\). In this case there is no state of this form which can be created by operators of the form as in (\ref{equivalence}) and the equivalence relation does not impose further conditions.

Next we turn to the NS sector. Now there is a unique tachyonic ground state. The first excited states are the ones obtained by acting with the lowest lying fermionic creating operators 
$$
 |v;p\rangle_{\mbox{\tiny NS}} = v^\m \j_{\m,-1/2} |0;p\rangle_{\mbox{\tiny NS}} 
$$
satisfying 
$$
N  |v;p\rangle_{\mbox{\tiny NS}} = \frac{1}{2} |v;p\rangle_{\mbox{\tiny NS}} \;.
$$
 \begin{table}\centering
\begin{tabular}{ccc}
\multicolumn{3}{c}{\bf open string massless spectrum}\\
\rowcolor[gray]{0.8}
sector & SO(8) rep. & G-parity \\ 
R\(+\)& \({\bf 8}\) &1\\
R\(-\)& \({\bf 8'}\)& \(-1\) \\
NS\(+\)&  \({\bf 8_v}\)& \(1\)\\
\midrule
\multicolumn{3}{c}{}\\
\multicolumn{3}{c}{\bf type IIA massless spectrum}\\
\rowcolor[gray]{0.8}
 sector& SO(8) rep. & 10d multiplet\\
 (NS\(+\),NS\(+\))& [0]+[2]+(2) & \(\F\), B, g ,\(\j_{(2/3)}^+, \l^+\)\\
 (R\(+\),NS\(+\))& \({\bf 8'}\)+\({\bf 56}\) &  (Graviton Multiplet )\\
  (NS\(+\),R\(-\))&  \({\bf 8}\)+\({\bf 56'}\) & \(\j_{(2/3)}^-, \l^-\), \(C^{(1)}\),   \(C^{(3)}\)\\
 (R\(+\),R\(-\))&[1]+[3]&  (Gravitini Multiplet)\\
 \midrule
\multicolumn{3}{c}{}\\
\multicolumn{3}{c}{\bf type IIB massless spectrum}\\
\rowcolor[gray]{0.8}
 sector& SO(8) rep. & 10d multiplet\\
(NS\(+\),NS\(+\))& [0]+[2]+(2) & \(\F\), B, g ,\(\j_{(2/3)}^+, \l^+\) \\ 
(R\(+\),NS\(+\))& \({\bf 8'}\)+\({\bf 56}\) &  (Graviton Multiplet )  \\ 
(NS\(+\),R\(+\))&  \({\bf 8'}\)+\({\bf 56}\) & \(C^{(0)}\),   \(C^{(2)}\), 
\(C^{(4)}_+\), \(\j_{(2/3)}^+\), \(\l^+\)  \\ 
(R\(+\),R\(+\))&[0]+[2]+[4]\(_+\)&  (Gravitini Multiplet) \\
       \midrule
\end{tabular}
\caption{\label{massless_spectra}\footnotesize{Summary of the massless spectrum of the open, type IIA, type IIB superstring theories.}}
\end{table}

 The non-trivial constraint \(L_0|v;p\rangle_{\mbox{\tiny NS}} = G_{1/2 } |v;p\rangle_{\mbox{\tiny NS}} =0 \) now gives the mass-shell and the orthogonality condition \(p^2 = p^\m v_\m = 0 \). But here a state of the same form can also be produced by acting on the tachyonic ground state by \(G_{-1/2}\): 
 $$
 G_{-1/2}  |0;p\rangle_{\mbox{\tiny NS}} =  |p ;p\rangle_{\mbox{\tiny NS}} \;,
 $$
 and we are led to  further impose the equivalence condition (\ref{equivalence})  \(v\sim v+ \R p\). Therefore the first excited states of the NS sector are massless and transform in the vector representation \({\bf 8_v}\) of SO(8).  While the degrees of freedom along the light-cone directions for the massless states of the R-sector are forbidden by the constraint which is equivalent to the massless Dirac equation, in the NS sector they are frozen by the constraint together the equivalence relation. 
This is yet another way to see why the zero point energy of the NS sector has to be \(-1/2\): SO(8) is the little group of the massless particles in ten dimensions. For consistency the excited states transforming as a massless photon should indeed be massless \cite{thooft}.


After seeing that the dynamics in light-cone directions are unphysical, we now concentrate on the transversal degrees of freedom and define a parity operator 
\be\label{G_parity}
G = \begin{cases} \,\G\, (-1)^{\sum_{n=1}^\inf \j^i_{-n} \j_{i,n}}&\text{for R sector} \;\;i=1,\dotsi,8\\ &\\ 
(-1)^{\sum_{n=1}^\inf \j^i_{-(n-1/2)} \j_{i,(n-1/2)} +1}  &\text{for NS sector} \end{cases}.
\ee
The massless spectrum of R and NS open string, together with their eigenvalue under the above parity operator, is summarised in Table \ref{massless_spectra}.

Notice that by projecting the NS states onto the positive-G-parity states we eliminate the tachyonic ground state. Another merit of this projection, the so-called GSO (Gliozzi-Scherk-Olive) projection, is that it yields a closed string spectrum with equal number of (spacetime) bosonic and fermionic fields when left- and right-moving copies of the massless fields are combined. In other words, it is a projection which yields a consistent conformal field theory with spacetime supersymmetry. 
It is also possible to derive this projection by various consistency requirement, for example the modular invariance of the loop amplitudes, but we will not do it here. 
 \index{GSO (Gliozzi-Scherk-Olive) projection}
The full GSO projection leaves us with two consistent theories in ten dimensions, corresponding to whether the left- and right-moving R-ground states are combined with the same or the opposite partities. By carefully combing the left- and right-moving components of the NS- and R-massless states discussed above, we obtain the massless spectrum of these two theories as listed in Table \ref{massless_spectra}. The difference between IIA and IIB is that the spin-\(1\!/2\) fields in the type IIB case are of the same chirality and those of IIA are of opposite chiralities. We therefore call type IIB string theory a {\it chiral} theory and IIA a non-chiral one. 
Each theory contains one graviton and one gravitini supermultiplets, and the massless spectrum of type IIA and IIB string theory is therefore identical to that of type IIA and IIB \({\cal N}=2\) supergravity. This is not surprising because the large amount of supersymmetries highly constrains the structure of the fields
\footnote{Strictly speaking, because we begin with the Ramnod-Neveu-Schwarz superstring action (\ref{RNS}) as opposed to the Green-Schwarz one, it is therefore not a priori clear that the theory has spacetime supersymmetry.}. In fact, the supersymmetry does not only fix the field content but also the allowed action; there are two possible supergravity theory with sixteen supercharges in ten dimensions, the type IIA and IIB supergravity, which are are simply the low-energy effective theory of type IIA and IIB string theory. We will see the explicit form of these low-energy effective actions in the following section.  

\subsection{T-Duality}
\label{subsection_T-Duality}

\index{T-duality}
In the previous subsection we introduced two kinds of superstring theories, namely the type IIA and IIB theories. But in fact the two superstring theories discussed above are not so independent from each other as they may seem. They  are related by the so-called T-duality, which involves reflecting the spacetime parity along one (or any odd number in general) spatial  direction on one of the two sides (right- or left-moving) of the world-sheet. 

For concreteness let's choose 
\begin{align} \notag
X^1(\bar{z}) \rightarrow -X^1(\bar{z})\;\;&;\;\; \til{\j}^1(\bar{z}) \rightarrow -\til{\j}^1(\bar{z})\\ \notag
X^1(z) \rightarrow X^1(z) \;\;&;\;\;{\j}^1({z}) \rightarrow {\j}^1({z})\;.
\end{align}

Upon this transformation, the eigenvalues of the parity operator defined in (\ref{G_parity}) flips signs for the right-moving R-sector states and remain unchanged for the NS-sector states. Consulting the table \ref{massless_spectra}, we then see that T-duality, a duality which is stringy by nature, exchanges the chiral (IIB) and the non-chiral (IIA) theories. In this sense type IIA and type IIB string theory are really the same theory. 

For later use we would like to have an explicit map between the bosonic massless degrees of freedom under T-duality. We will now derive it based on the canonical quantisation approach we followed earlier. Let's begin first with the NS-NS sector as its field content is shared by both the IIA and the IIB theory. 
As we saw earlier in Table \ref{massless_spectra}, these are the spacetime fields \(\F, G_{ij}, B_{ij}\) corresponding to the massless states of the form \(\til{\j}^i\j^j |{0}, \til{0}\rangle\)  in the world-sheet theory. Here we remind the readers that the total Hilbert space is the tensor product of left- and right-moving Hilbert spaces, and our notation really means \( |{0}, \til{0}\rangle =  |0\rangle\otimes | \til{0}\rangle\). Now, instead of considering them all in a flat background as in the previous sections, we would like to derive a map between backgrounds. For this purpose it will be more convenient to turn to the vielbein frame, in other words we will consider the spacetime fields \(e^i_{\hat{j}}\) and \(B_{\hat{k}\hat{\ell}} := B_{ij} e^i_{\hat{k}}e^j_{\hat{\ell}}\), and operators \(\j_{\hat{i}}\) etc, where the hatted indices denote the orthonormal indices. Matching the representation under the rotation group we obtain a map between the operators and the perturbations of the spacetime fields under consideration
\ben
\til{\j}_{(\hat{i} } \j_{\hat{j})} &\longrightarrow& \,e^k_{(\hat{i}}\,\d e_{k,\hat{j})} = \frac{1}{2}e^{k}_{\hat{i}}e^{\ell}_{\hat{j}} \,\d G_{k\ell}\\
\til{\j}_{[\hat{i} } \j_{\hat{j}]} &\longrightarrow& e^{k}_{[\hat{i}}e^{\ell}_{\hat{j}]} \,\d B_{k\ell}\\ \notag
\til{\j}_{\hat{i} } \j^{\hat{i}}&\longrightarrow& 2 \d\F\;.
\een
Now consider a T-duality transformation along the 1st direction, with \(e^i_{\hat{1}} \pa_i\) being an isometry (a Killing vector) of the background. For any such background metric we can always choose the vielbein such that \(e_1^{\hat{a}}= e^a_{\hat{1}} =0\), with \(a =2, \dotsi, 8\) being the directions transversal to the light-cone and the T-dual directions. In other words, we can now write the metric in the form
\ben
ds^2 &=& G_{ij} dx^i dx^j = G_{1,1} (dx^1 + A_a dx^a)^2 +{g}_{ab} dx^a dx^b \\
&=& \th^{\hat{1}}\otimes \th^{\hat{1}}+ \th^{\hat{a}}\otimes \th^{\hat{a}}
\een
where \(\th^{\hat{i}} = e^{\hat{i}}_j dx^j\).

In this frame we can rewrite the above relations as follows.

\begin{align}\notag
\til{\j}_{(\hat{1} } \j_{\hat{1})} & \rightarrow \frac{1}{2}\d (\log G_{1,1}) &
\til{\j}_{(\hat{a} } \j_{\hat{b})} & \rightarrow \frac{1}{2} \d g_{ab}\\ \notag
\til{\j}_{(\hat{1} } \j_{\hat{a})} & \rightarrow \frac{1}{2}\, \sqrt{G_{1,1}}\, e^{b}_{\hat{a}}  \d A_{b} &
\til{\j}_{[\hat{1} } \j_{\hat{a}]} & \rightarrow \frac{1}{2} \frac{1}{\sqrt{G_{1,1}}} e^{b}_{\hat{a}}  \d B_{1b}\\ \label{rewrite}
\til{\j}_{[\hat{a} } \j_{\hat{b}]} & \rightarrow e^{c}_{[\hat{a}}e^{d}_{\hat{b}]} \left( \d B_{ab} - 2 A_a \d B_{1b}\right) & \til{\j}_{\hat{i} } \j^{\hat{i}} &\rightarrow 2 \d\F  = \frac{1}{2} g^{ab} \d g_{ab} +  \frac{1}{2}  \d (\log G_{1,1})
\end{align}

When we now reflect the right-moving side along the first direction in the orthonormal frame, namely \(\til{\j}_{\hat{1} } \rightarrow -\til{\j}_{\hat{1} } \), it's easy to see that it's equivalent to a field redefinition as listed in Table \ref{Buscher_rules}. This table rewritten in terms of the full metric \(G_{ij}\) is the usual Buscher rules. Finally we also have to set
\be \label{dilaton_Tdual}
\F \rightarrow\F - \frac{1}{2} \log G_{1,1}\;.
\ee
To understand this map between the dilaton fields under T-duality, first recall that not all the metric fluctuations are physical because of the diffeomorphism invariance \( G_{ij} \sim G_{ij} + \na_{(i}v_{j)}\), with the trace mode here gives rise to the dilaton fluctuation. Of course now the choice of coefficient in  (\ref{dilaton_Tdual}) is just a matter of convention. But it is chosen in such a way that later the world-sheet  action (\ref{world-sheet_action_with_B}) in arbitrary consistent NS background will stay conformally invariant after a T-duality transformation. This choice also renders the nine-dimensional Newton's constant invariant when reduced on the T-circle invariant under T-duality. 

When the 1st direction is a circle, by studying the massive spectrum of the ``compactified" theory before and after the above transformation, one concludes that T-duality also changes the radius of the circle as
\be
R_1 \rightarrow \frac{\a'}{R_1}\;,
\ee
where \(\a'\) is the coupling constant appearing in the world-sheet action (\ref{RNS}). This is indeed consistent with the world-sheet dictionary that \(G_{1,1} \leftrightarrow 1/G_{1,1}\).

We can now do the same analysis for the R-R sector. It is a straightforward exercise involving first inserting \(p\) Gamma matrices to make \(p\)-form fields out out a (spacetime) spinor bilinear, and then flipping the chirality along one spatial direction. Since the R-R gauge fields are different in type IIA and IIB string theory, we expect T-duality to  also exchange the objects charged under these higher-form fields. 

\begin{table}
\centering
\label{T_duality}
\begin{tabular}{c|c}
\toprule
\multicolumn{2}{c}{before \(\longleftrightarrow\) after}\\ 
\midrule
\(G_{1,1} \)& \( \frac{1}{G_{1,1} }\)\\
\(g_{ab} \)& \(g_{ab} \)\\
\(A_a \)& \(B_{1a} \)\\
\(B_{1a} \)&\(A_{a} \)\\
\(B_{ab} - A_{[a}  B_{1b]} \)& \( B_{ab} + A_{[a} B_{1b]} \)\\
\bottomrule
\end{tabular}
\caption{\label{Buscher_rules} \footnotesize{T-duality (Buscher Rules)}}
\end{table}
\index{Buscher Rules}

Furthermore, if there are also open strings in the theory, which have either Neumann or Dirichlet boundary condition at the end points, it is not hard to see that the T-duality transformation exchanges the two boundary conditions. In other words, T-duality together with the presence of open strings  forces  onto us other kinds of objects on which the open strings can end. We will introduce these Dirichlet branes (D-branes) from the point of view of the low-energy effective theory in the following sections. As we will see, these are exactly the objects charged under the R-R fields which get exchanged under T-duality accordingly.
\index{T-duality}
\index{D-branes}

\section{Low Energy Effective Action}
\setcounter{equation}{0}

\subsection{Supergravity Theory in Eleven and Ten Dimensions}
\label{Supergravity Theory in Eleven and Ten Dimensions}
In the previous section we derived the massless spectrum of type IIA and IIB superstring theory. 
In this section we would like to describe the interactions of these massless modes, which is constrained by supersymmetry to be described by the type IIA and IIB supergravity theories in ten dimensions. It will nevertheless turn out to be a rewarding path to begin with the eleven dimensional supergravity theory. Supersymmetry ensures that this theory is unique. Furthermore the IIA ten-dimensional supergravity has to be the dimensional reduction of this higher-dimensional theory,  since the two theories have the same supersymmetry algebras. 

The field content of this theory is rather simple: for the bosons there are just graviton with \(\frac{9\times 10}{2} -1 =44 \) components and a three-form potential with \(\frac{9\times 8\times 7}{3!}=84\) components, in representation of the SO(9) little group of massless particles in eleven dimensions. 
 There is also the gravitino with its \(16 \times 8\) degrees of freedom, in representation of the covering group \(Spin(9)\). This is indeed the same number as the number of massless degrees of freedom of the type II string theory as recorded in Table \ref{massless_spectra}.
\index{eleven-dimensional supergravity}
The bosonic part of the action is
\be\label{S11}
(16 \p G_\text{\tiny N}^{\text{\tiny(11)}}) \, S^{\text{\tiny(11)}}= \int d^{11}x \sqrt{-G} \left(R- \frac{1}{2} |F^{(4)}|^2\right) -\frac{1}{3!} \int A^{(3)}\wedge F^{(4)}\wedge F^{(4)}\;,
\ee
where \(F^{(4)}\) is the field strength of the three-form potential \(F^{(4)} =  dA^{(3)}\) and we use the notation 
$$
|F^{(n)}|^2 = \frac{F^{(n)} \wedge \star F^{(n)}}{\text{volume form}}
$$
for the kinetic term of a (n-1)-form potential. Here and most of the time in this thesis we will avoid writing down the fermionic part of the action. 
This is because we are interested in supersymmetric solutions with zero fermionic fields, which of course have vanishing action for the fermionic part.

To dimensionally reduce it, write the eleven-dimensional metric as
\be\label{reduce_metric}
G_{MN} = e^{-\frac{2}{3}\F} \bem  g_{\m\n}+e^{2\F}A_\m A_\n &  e^{2\F}A_\m \\
e^{2\F}A_\n & e^{2\F}
\eem \;,
\ee
where we use \(M, N, \dotsi = 0,1,\dotsi,10\) to denote the eleven-dimensional and  \(\m, \n, \dotsi = 0,1,\dotsi,9\) the ten-dimensional directions. The above choice of defining the ten-dimensional fields will be justified in subsection \ref{Couplings of String Theory} 
when we make a detailed comparison with the world-sheet theory. 

We also reduce the three-form potential \(A^{(3)}_{MNP}\) as \(A_{\m\n\r}\) when it has no ``leg" in the 11-th direction and as \(A^{(3)}_{MN10}=B_{\m\n}\) and \(H^{(3)} = dB^{(2)} \) when it does.

Under this field redefinition, and truncating all the dependence on the eleventh direction, the action reduces to 

\bea\notag
S^{(\text{\tiny IIA})} &=& S_{\text{NS}} + S_{\text{R}}^{(\text{\tiny IIA})}
 + S_{\text{C-S}}^{(\text{\tiny IIA})}\\ \notag
2\k^2 S_{\text{NS}}  &=& \int d^{10} \sqrt{-g} \, e^{-2\F} \left( R+ 4\pa_\m\F\pa^\m\F -\frac{1}{2} |H^{(3)}|^2  \right)\\ \notag
2\k^2 S_{\text{R}}^{(\text{\tiny IIA})} & =& -\frac{1}{2} \int d^{10}x \left( |F^{(2)}|^2 +  |\til{F}^{(4)}|^2\right) \\ \label{IIA_action}
2\k^2S_{\text{C-S}}^{(\text{\tiny IIA})}&=& -\frac{1}{2!} \int B^{(2)}\wedge F^{(4)}\wedge F^{(4)}\;,
\eea
where \(F^{(2)}\) is the field strength of the Kaluza-Klein gauge field \(A^{(1)}\) and 
\be
\til{F}^{(4)} = dA^{(3)} + A^{(1)} \wedge H^{(3)}
\ee
is the field strength modified by the Chern-Simons term. This is the bosonic action for the type IIA supergravity that we want to construct. The 10d gravitational coupling constant \(\k\) will be discussed in the following subsection.

Type IIB supergravity, on the other hand, cannot be obtained by dimensionally reducing the eleven-dimensional supergravity. Although in principle it is related to the IIA supergravity by T-dualise the IIA string theory and then take the low-energy limit, it is actually not at all a straightforward task to write down a classical action for the field content recorded in Table \ref{massless_spectra}. This is because in d=2 (mod 4) dimensions there is no straightforward way to incorporate in the action the self-duality condition on a middle rank ({\it i.e.}, (\(\frac{d}{2}\))-form) field strength. Recall that in type IIB string theory this is indeed the case at hand, since the R-R field \(C^{(4)}_+\) is constrained to have self-dual field strength. Here we shall write down an action analogous to the IIA version, while the self-duality condition should be imposed as an additional constraint. 

The NS sector bosonic action is identical to the type IIA case, as expected from our notation, while the rest reads

\bea\notag
S^{(\text{\tiny IIB})} &=& S_{\text{NS}} + S_{\text{R}}^{(\text{\tiny IIB})}
 + S_{\text{C-S}}^{(\text{\tiny IIB})}\\ \notag
2\k^2 S_{\text{R}}^{(\text{\tiny IIB})} & =& -\frac{1}{2} \int d^{10}x \left( |F^{(1)}|^2 +  |\til{F}^{(3)|^2}+
\frac{1}{2}|\til{F}^{(5)}|^2\right) \\ \label{IIB_action}
2\k^2S_{\text{C-S}}^{(\text{\tiny IIB})}&=& -\frac{1}{2!} \int C^{(4)}\wedge H^{(3)}\wedge F^{(3)}\;,
\eea
where \(F^{(1)}\) is the field strength of the R-R zero form potential \(C^{(0)}\) and 
\ben
\til{F}^{(3)} &=& dC^{(2)} - C^{(0)} \wedge H^{(3)} \\
\til{F}^{(5)} &=& dC^{(4)}   -\frac{1}{2} C^{(2)} \wedge H^{(3)}+ \frac{1}{2} B^{(2)}\wedge F^{(3)}
\een
are again the field strength with Chern-Simons term, while the self-duality constraint 
\be
\til{F}^{(5)} = \star\til{F}^{(5)} \;,
\ee
must be imposed by hand additionally.
\index{type IIA, IIB supergravity}

\subsection{Couplings of String Theory}
\label{Couplings of String Theory}
In the last section we introduced the type IIA and IIB superstring theory, and in the last subsection the type IIA and IIB supergravity theory. Furthermore we have observed that the massless spectrum of the two sets of theories are the same. We therefore conclude that, for the supergravity theories to be the low-energy effective description of the superstring theories, the dynamics of these massless modes must be the same in both theories. In this subsection we will establish this connection, and furthermore spell out the relation between the coupling constants of the ten- and eleven-dimensional supergravity and the various quantities of string theory .

Let us focus on the NS part of the action (\ref{IIA_action}), which is common for both type IIA and type IIB
supergravity\footnote{Although possible, here we will not attempt to explain the effects caused by a source of the higher-form fields from a world-sheet point of view.}. 
Then the equations of motion obtained from \(S_{\text{NS}} \) is the same as requiring the absence of  the conformal anomaly in the following world-sheet theory 
\be\label{world-sheet_action_with_B}
S_{\text{world-sheet}} = \frac{1}{2\p \a' } \int_\S  d^2\s \sqrt{h} \left( (h^{ab} g_{\m\n} + i \e^{ab} B_{\m\n}) \pa_a X^\m \pa_b X^\n + \a'  \F(X) R(h)\right)\;,
\ee
where \(h^{ab}\) is the world-sheet metric, \(R(h)\) its Ricci scalar, and \(\e^{ab}\) is an anti-symmetric tensor normalised such that the term involving the \(B\)-field simply equal to \(2\p i \int B^{(2)}\). Here and in the rest of the chapter we will not make a difference in notation for the pull-back fields when it is clear from the context. 

It might be surprising that the bulk Einstein equation appears as the requirement of conformal invariance of the world-sheet theory. This indeed requires some justification. Strictly speaking, we have obtained the massless spectrum of type II strings by quantising it in the flat background with B-field and ``dilaton" \(\F\) turned off. How do we know that the same theory is also consistent in other backgrounds, except for the hint from the supergravity theory as a low-energy effective theory? Indeed what we just saw is that a consistent  NS background for the world-sheet theory at the one-loop level is also automatically a solution to the equation of motion of the proposed low-energy effective theory. This connection justifies our choice of frame for dimension reduction (\ref{reduce_metric}). This choice of scaling of the ten-dimensional metric will therefore be called the ``string frame", since in this frame the target space metric is the same one that shows up in the string world-sheet action.  This  ``string frame" is 
different from the usual ``Einstein frame",  in which the curvature term in the action has no pre-factor \(e^{-2\F}\) in front. 

\index{string frame}
\index{Einstein frame}

Now we will comment on the different roles of and relations among various coupling constants in the ten- and eleven-dimensional spacetime and the world-sheet theory. First of all we have \(\a'\) which sets the length scale in the world-sheet action (\ref{world-sheet_action_with_B}). We therefore call the string length
$$
\ell_s \sim \sqrt{\a'}\;.
$$

From the quantisation of the superstring we see that the mass-shell condition gives
$$
m^2 \sim \frac{1}{\a'} N \sim  \frac{1}{\ell_s^2}  N \;,
$$
where \(N\) is the oscillator number above the massless level. From this we see that the low-energy effective action, in which we truncate the fields to keep only the massless ones, is valid when one is only interested in physics at scales much larger than the string length. 

Furthermore, comparing the world-sheet and the supergravity action (\ref{IIA_action}) we conclude that the gravity coupling constant is related to the string scale by
$$
\k^2 \sim \ell_s^8\sim \a'^4\;.
$$

But this is not yet the ten-dimensional Planck length. To discuss that   we should first understand the role played by the ``dilaton" field  \(\F(X)\). From the world-sheet action we see that it controls the scattering between strings. For example, when  \(\F(X)=\F_0\) is constant in spacetime, \(e^{-S_{\text{world-sheet}}}\) has a factor 
$$
e^{-\F \chi} = e^{-\F (2-2g) } \equiv g_s^{2g-2}  \;,
$$
where \(\chi\) is the Euler characteristic and \(g\) the genus of the world-sheet (see (\ref{gauss_bonnet}) for Gauss-Bonnet theorem which relates the two quantities). In other words, the dilaton field \(\F\) controls the genus expansion, the string theory counter-part of the loop expansion in particle physics. We therefore identify as the string coupling's constant
$$
g_s  = e^{\F}\;.
$$
We should emphasize here that this is not just a parameter but really a dynamical field of the theory. 

\index{Planck length (ten-dimensional)}
Now we are ready to identify the ten-dimensional gravitational coupling. By going to the Einstein frame in which the Einstein action takes the standard form, we see that 
\be\label{10d_planck}
G_{\text{\tiny N}}^{(10)} \sim (\ell_{\text{\tiny P}}^{(10)})^8\sim \k^2 e^{2\F_0} \sim \ell_s^8 g_s^2 \sim \a'^4 g_s^2\;,
\ee 
where \(\F_0\) is now the asymptotic value of the dilaton field. Finally we work out the relation between the radius \(R_{\text{\tiny M}}\) of the eleventh-dimensional circle on which we reduce the eleven-dimensional supergravity to obtain the type IIA supergravity, and the eleven-dimensional Planck length. From (\ref{S11}) and ({\ref{reduce_metric}) we get 
$$
\frac{R_{\text{\tiny M}}}{\ell_{\text{\tiny P}}^{(11)}} \sim (e^{4\F/3})^{1/2} \sim g_s^{2/3}
\quad \text{and} \quad \frac{G_{\text{\tiny N}}^{(11)}}{R_{\text{\tiny M}}} \sim  \frac{(\ell_{\text{\tiny P}}^{(11)})^9}{R_{\text{\tiny M}}} \sim G_{\text{\tiny N}}^{(10)} \sim \ell_s^8 g_s^2\;,
$$
in other words
\be
\label{coupling_relation}
 \qquad
\ell_{\text{\tiny P}}^{(11)} \sim \ell_s g_s^{1/3} \qquad \text{and} \qquad R_{\text{\tiny M}} \sim \ell_s g_s\qquad\;.
\ee

Here we see an interesting phenomenon, namely that the radius of the eleventh-dimensional circle in string unit becomes large when the strings are strongly interacting. When the string coupling constant is large, the perturbative string theory we discussed in the last section should not be trusted. 

There is a similar breaking down of the validity of the ten-dimensional theory on the supergravity side. 
In the ``decompactification limit"  in which  the Kaluza-Klein circle becomes larger and larger, the momentum modes along the Kaluza-Klein direction becomes lighter and lighter. As a result, the truncation of  the spectrum to a lower-dimensional one eventually becomes invalid. In other words,

\be\label{decompactify_1}
\text{type IIA supergravity} \xlongleftrightarrow[\text{compactify}]{\text{decompactify}}  \text{11d supergravity}
\ee

\index{decompactification limit}
In subsection \ref{Supergravity Theory in Eleven and Ten Dimensions}  we have used the eleven-dimensional supergravity just as a convenient starting point to write down the ten-dimensional supergravity action. But if we take this reduction a step further, it seems to suggest that the ten-dimensional supergravity is only a valid low-energy description of the full non-perturbative theory at small \(\ell_s\) and small \(g_s\). At large \(g_s\) the eleven-dimensional theory becomes a better description. 
Indeed, later in section \ref{M-theory} we will see that there are dynamical objects other than the fundamental strings which become light at strong coupling and which are captured by the eleven-dimensional supergravity but not by the ten-dimensional one.

\section{Non-Perturbative Aspects}

In the last section we introduced superstring theory as a perturbative theory. But in fact the theory is much richer than that. In particular, for the purpose of studying the supersymmetric spectrum, especially the spectrum which is responsible for the existence of black hole entropy, the non-perturbative aspects of the theory will play a crucial role in our understanding of the problem.

While in general the non-perturbative aspects of string theory is very difficult to study, there are regions in the moduli space that are fortunately accessible to us. The key word here is  ``dual perspective". An example of which  we have seen earlier is the T-duality relating type IIA and IIB string theory, stating that while the two descriptions look different, they offer ``dual perspectives" on the same theory. 

A duality is especially useful if this theory offers a complementary range of computational accessibility. In this section we will introduce a few dualities like this, mapping non-perturbative physics on one side to 
perturbative physics on the other side of the duality. We will also introduce the solitonic objects of the theory, which are generically called ``branes". These objects can often be described from two dual perspectives, a fact that makes black hole counting in string theory possible and motivates an extremely important gauge/gravity duality.

\setcounter{equation}{0}
\subsection{M-theory}
\label{M-theory}

In the last section, we have just saw the interesting possibility that an eleven-dimensional supergravity might be the low-energy effective action for the type IIA string theory at {\it strong} coupling. In this section we will explore this possibility further. 

Historically, eleven-dimensional supergravity theory is interesting because eleven is the highest dimension in which Minkowski signature with Poincar{\'e} and supersymmetry invariance is possible. But one should keep in mind that, just as type IIA and IIB supergravity should only be seen as an effective theory at low energy but not a complete theory because of its non-renormalisability, the eleven-dimensional supergravity can only at best be a low-energy description of a consistent theory. We will refer to this complete theory as ``M-theory", whose non-perturbative description is unfortunately not yet fully developed and out of the scope of the present thesis. 
\index{M-theory}

From the above discussion we see that this ``M-theory", no matter of what nature it actually is, must have the following relationship with type IIA string theory 

$$
{\qquad
\text{IIA string theory} \xrightarrow[]{g_s \gg1}\text{M-theory}\qquad\;
} $$
as suggested by the low-energy relation (\ref{decompactify_1}), where the identification of the compactification radius is given by (\ref{coupling_relation}).
We will later refer to this relation as the ``M-theory lift". 

Without really knowing the non-perturbative definition of M-theory, we will now use its low-energy effective theory as a guide to explore the structure of the theory. As we will see, it will turn out to be a fruitful path towards a simple understanding of many of the non-perturbative features of the type II string theory. 

\subsection{Branes}
\label{subsection_Branes}

Let us begin by finding supersymmetric classical solutions to the eleven dimension supergravity. Since the three-form potential \(A^{(3)}\) is the only bosonic degree of freedom besides the gravitons, from the experience with the usual Maxwell-Einstein theory, we expect to find objects that are charged under these fields. A straightforward generalisation of the Wilson line coupling to one-form potential of a charged point particle is
$$
\int_{\text{world-line}} A^{(1)} \qquad  \longrightarrow \qquad \int_{\text{world-volume}} C^{(n)} \;,
$$
which leads us to expect an object with a \((2+1)\)-dimensional world-volume which plays the role of the ``electron" for \(A^{(3)}\). Indeed there is a \(1/\!2\)-BPS solution (a supersymmetric solution with half of the supersymmetry unbroken) which has a non-zero Noether charge 
\be\label{charge_m2}
Q= \int_{S^7} \star F^{(4)} - \frac{1}{2} A^{(3)}\wedge F^{(4)}
\ee
for the three-form field. Notice that the expression of the Noether charge gets modified in the presence of a Chern-Simons term in the action. The solution reads
\bea \notag
ds_{\text{\tiny M2}}^2 &=& f_{\text{\tiny M2}}^{-2/3} \,ds^2_{3,\text{\tiny L}} +  f_{\text{\tiny M2}}^{1/3} \,ds^2_{8,\text{\tiny E}} \\ \label{M2_metric}
A^{(3)} &=& f_{\text{\tiny M2}}^{-1} \,dV_{3,\text{\tiny L}} \\ \notag
f_{\text{\tiny M2}} &=& 1 + a_2 \,Q\, \bigl(\frac{1}{r}\bigr)^6\;.
\eea

It also has a magnetic cousin which looks like
\bea \notag
ds_{\text{\tiny M5}}^2 &=& f_{\text{\tiny M5}}^{-1/3} \,ds^2_{6,\text{\tiny L}} +  f_{\text{\tiny M5}}^{2/3} \,ds^2_{5,\text{\tiny E}} \\ \label{M5_metric}
A^{(6)} &=& f_{\text{\tiny M5}}^{-1} \,dV_{6,\text{\tiny L}} \quad,\;\; dA^{(3)} = \star dA^{(6)} \\ \notag
f_{\text{\tiny M5}} &=& 1 + a_5 \,P\, \bigl(\frac{1}{r}\bigr)^3\;,
\eea
and satisfies 
\be\label{charge_m5}
P= \int_{S^4} F^{(4)}\;.
\ee 
In the above equations \(ds^2_{n,\text{\tiny L}}\) denotes the usual metric of a flat Lorentzian space with mostly positive signature (1,n-1) and \(dV_{n,\text{\tiny L}}\) its volume form, and 
$$ ds^2_{n,\text{\tiny E}}= dr^2 + r^2 d\O_{(n-1)}^2 $$
is the metric of a flat Euclidean space. The constants \(a_2\), \(a_5\) are chosen such that (\ref{charge_m2}), (\ref{charge_m5}) are satisfied.  

We see that the above solutions carrying electric and magnetic charges have (2+1) and (5+1)  ``tangent" directions respectively. We will therefore call them the M2 and the M5 brane solutions. 
In general, from Hodge duality we see that a \((p+1)\)-dimensional object in \(D\) dimensions must be electric-magnetic dual to another object with \((D-p-4)\) spatial directions, when both objects are required to have a time-span.

Furthermore, as analogous to the Maxwell case, the Dirac quantisation, namely the well-definedness of an electron wave-function in a monopole background, will impose on us\footnote{This Dirac quantisation condition holds when (gravitational) anomaly effects can be neglected. See \cite{Witten:1996md} for a discussion about the correction of the charge quantisation condition in M-theory due to  gravitational effects, and section \ref{Wrapped M-branes and the Near Horizon Limit} of the present thesis for an explicit example in which the charge  quantisation condition is modified.
}  
\be\label{Dirac_quantisation}
QP \in 2 \p \Z\;.
\ee
Notice that this condition cannot be seen from studying the supergravity action alone and is therefore a strictly quantum effect. 

So far we have discovered the two charge-carrying fundamental objects of M-theory, called M2 and M5 branes. In the full theory they should be dynamical objects, but the quantisation of them is not as developed of that of fundamental strings and will not be discussed here. 

Since the only scale of this theory is the eleven-dimensional Planck length, we conclude that their tensions are
\be \label{M_brane_tension}
\t_{\text{\tiny M2}}\sim (\ell_{\text{\tiny P}}^{(11)})^{-3} \quad;\quad
\t_{\text{\tiny M5}}\sim (\ell_{\text{\tiny P}}^{(11)})^{-6}  \quad\Rightarrow \quad
\t_{\text{\tiny M2}}\t_{\text{\tiny M5}} \sim (\ell_{\text{\tiny P}}^{(11)})^{-9} \sim \frac{1}{G_{\text{\tiny N}}^{\text{\tiny(11)}}}\;,
\ee
This can also be checked from an explicit calculation using the gravity solution.

Now we would like to explore what they mean in type IIA string theory when we perform the dimensional reduction to ten dimensions, an operation valid when \(g_s \ll1 \). From the map  of the dimensional reduction (\ref{reduce_metric}) between the field contents, we can deduce a map between charged objects  of the two theories. This is recorded in Table \ref{brane_reduction}.

\begin{table} 
\centering
\begin{tabular}{ccccc}
\toprule
\multicolumn{2}{c}{\multirow{2}*{11d field}} & sources &\multirow{2}*{10d field} & sources \\ 
&& (elec/mag) & & (elec/mag) \\
\midrule[0.07em]
\multirow{4}*{\(G_{MN}\)}  & \(G_{\parallel\parallel}\) & & \(\F\) & \\ 
\cmidrule[0.01 em]{2-5}
& \(G_{\perp \perp}\) & everything& \(g_{\m\n}\) & everything\\ \cmidrule[0.01 em]{2-5}
&\multirow{2}*{\(G_{\parallel \perp}\)} & \(p_{\parallel}\) & \multirow{2}*{\(C^{(1)}\)}& D0\\ 
&&KK-monopole & & D6\\ 
\midrule 
\multirow{4}*{\(A^{(3)}\)} &  \multirow{2}*{\(A^{(3)}_\parallel\)} 
& M2\(_{\parallel}\) &\multirow{2}*{\(B^{(2)}\)} & F1 \\
&& M5\({_\perp}\) && NS5 \\ 
\cmidrule[0.01 em]{2-5}
&  \multirow{2}*{\(A^{(3)}_\perp\)} 
& M2\(_\perp\) &\multirow{2}*{\(C^{(3)}\)} & D2 \\
&& M5\(_\parallel\) && D4 \\
\bottomrule
\end{tabular}
\caption{\label{brane_reduction}\footnotesize Dimensional reduction from M-theory branes to type IIA branes.}
\end{table}

First we will explain the notations in the above table. The subscript ``$\small\parallel$" denotes the field components or the M-theory branes with a leg (legs) or extent in the M-theory circle direction along which we dimensionally reduce the theory. Similar for the transversal direction ``\(\perp\)". Note that 
we leave the sources for the size of the circle direction (\(G_{\parallel,\parallel}\)) empty, since the asymptotic size of the eleventh-direction, or equivalently the ten-dimensional Newton's constant, is a parameter of the theory from the ten-dimensional point of view. 

Now we will explain the objects that appear in this table. 

First we begin with the KK (Kaluza-Klein) monopole. It is the magnetic monopole with respect to the Kaluza-Klein gauge field \(A_\m = G_{\m,\parallel}\), which is a geometry having the structure as the product of a seven-dimensional flat Minkowski space and a four-dimensional Taub-NUT gravitational instanton. The Taub-NUT metric is \index{Taub-NUT space}

\bea \notag
ds^2_{\text{\tiny T-N}} &=& 
V (\vec{x}) d\vec{x}\cdot d\vec{x} + R^2 \,V^{-1}(\vec{x}) (d\f _{10} + \w^0)^2\\ \label{TN}
V(\vec{x})  &=& 1 +  \frac{R}{|\vec{x}|} \\ \notag
d\w^0 &=&  \star_{3}dV \quad;\quad \f_{10} \sim  \f_{10}+ 4\p \;.
\eea
This solution has a self-dual curvature two-form just like the usual Yang-Mills instantons,  and therefore the name ``gravitational instanton"\footnote{Note that this is just an analogy. The term ``gravitational instanton" is also used sometimes to refer to any four-dimensional Cauchy Riemannian manifold that is a solution to the vacuum Einstein's equation, even if it does not satisfy the self-duality condition.}.  The magnetic charge corresponding to this solution is given by
\be
\int_{S^2} dA =-  \int_{S^2}  \star_{3}dV = 1\;.
\ee
The above structure can be easily generalised to obtain multi-instanton solutions. We will now digress to discuss them since they will also be needed in the subsequent parts of the thesis. But the reader can safely skip this part and return at any time. 

\vspace{1.2cm}
\begin{tabular}{p{12cm}}
\hline\\
\end{tabular}

\begin{digression} Some Gravitational Instantons\footnote{See the appendix for some of the background knowledge about classical geometry. An excellent review on the present subject of can be found in \cite{Eguchi:1980jx}.}
\index{gravitational instanton}
\begin{theorem}
Any hyper-K\"ahler four-manifold (four-dimensional Rimannian manifold with \(Sp(1)\sim SU(2)\) holonomy) with a triholomorphic Killing vector, namely any Ricci-flat Riemannian four-manifolds with a Killing vector which preserves all three complex structures, must be of the following Gibbons-Hawking form  \cite{Gibbons:1987sp,Gibbons:1979zt,Gibbons:1979xm}
\index{Gibbons-Hawking metric}
\ben
ds^2_{\text{\tiny G-H}} &=& 
H(\vec{x}) d\vec{x}\cdot d\vec{x} + H^{-1}(\vec{x}) (dx_5 + \w^0)^2\\ 
H(\vec{x})  &=& h + \sum_{a=1}^n \frac{q_a}{|\vec{x}-\vec{x}_a|} \\
\notag
d\w^0 &=&  \star_{3}dH \quad.
\een
It has a self-dual curvature two-form and the anti-self-dual hyper-K\"ahler (three complex) structure
\be
J^{(i)} = (dx_5 + \w^0) \wedge dx^i - \frac{1}{2}\e_{ijk} H dx^j\wedge dx^k\;.
\ee
\end{theorem}

\begin{example}Taub-NUT space

The above Taub-NUT metric (\ref{TN}) can be obtained by taking the special case
\be\label{Taub-NUT as gibbons-hawking}
H(\vec{x}) = \frac{1}{R^2} (1+ \frac{1}{|\vec{x}|})
\ee
and rescale the coordinates appropriately. The coordinate identification comes from requiring the absence of any Dirac-string-like singularity. 

To make the isometry of this space manifest, it will be useful to introduce the SU(2) left- and right-invariant one-forms on the three sphere \(S^3\).

First observe that, parametrising \(\C^2\) using the coordinates
\ben
&z_1 = \r \cos\frac{\th}{2} \,e^{i\frac{\j+\f}{2}}\qquad;\qquad
z_2 = \r \sin\frac{\th}{2} \,e^{i\frac{\j-\f}{2}}\\ \notag
&\p \in [0,\p] , \; \f\in[0,2\p) , \; \j\in [0,4\p)\;,
\een
the flat metric reads
$$
ds^2_{\R^4} = dz_1 \otimes d\zbar_1+ dz_2 \otimes d\zbar_2
= d\r^2 + \frac{\r^2}{4} \left(d\th^2 + d\f^2 + d\j^2 + 2 \cos\th d\f d\j \right).
$$

A general SU(2) rotation takes the matrix form
$$
U(\th,\f,\j) =\frac{1}{\r} \bem z_1 & \zbar_2 \\ -z_2& \bar{z}_1 \eem\;.
$$

Furthermore, SU(2) acts on itself by left- and right- multiplication. We can decompose the left-invariant variation \(U^{-1}dU\) and similarly the right-invariant variation \(dU\,U^{-1}\) in the basis of  Pauli-matrices and get the following right- and left-one forms
\begin{align} \notag
\s_{1,L}&=-\sin\j \,d\th + \cos\j\,\sin\th\,d\f\\ \label{one_form_left}
\s_{2,L}&= \cos\j\,d\th +\sin\j\,\sin\th\,d\f\\ \notag
\s_{3,L}&= d\j+\cos\th\,d\f\\ \notag
\intertext{and} \notag
\s_{1,R}&=\sin\f \,d\th - \cos\f\,\sin\th\,d\j\\ \label{one_form_right}
\s_{2,R}&= \cos\f\,d\th +\sin\f\,\sin\th\,d\j\\ \notag
\s_{3,R}&= d\f+\cos\th\,d\j
\end{align}
satisfying
\ben
d\s_i^L &=& \frac{1}{2} \e_{ijk} \s_j^L \wedge \s_j^L\\
d\s_i^R &=&- \frac{1}{2} \e_{ijk} \s_j^R \wedge \s_j^R\;.
\een

We now see that the above metric of the flat \(\R^4\) can be written as
\bea\notag
ds^2 &=& d\r^2 + \frac{\r^2}{4} \left( \s_{1,R}^2+ \s_{2,R}^2 +\s_{3,R}^2\right)\\
 \label{flat_r4}
& =&d\r^2 + \frac{\r^2}{4} \left( \s_{1,L}^2+ \s_{2,L}^2 +\s_{3,L}^2\right)\;.
\eea
With the above form of the metric, it becomes manifest that \(\R^4\) has a \(SU(2)_L\times SU(2)_R \) symmetry generated by the following dual vectors of the above one-forms
\begin{align} \notag
\xi_{1,L}&=-\cot\th\,\cos\j \,\pa_\j - \sin\j\,\pa_\th + \frac{\cos\j}{\sin\th}\pa_\f\\ \label{left_isometry}
\xi_{2,L}&= -\cot\th\,\sin\j \,\pa_\j + \cos\j\,\pa_\th + \frac{\sin\j}{\sin\th}\pa_\f\\ \notag
\xi_{3,L}&= \pa_\j \\ \notag
\intertext{and} \notag
\xi_{1,R}&=\cot\th\,\cos\f \,\pa_\f + \sin\f\,\pa_\th - \frac{\cos\f }{\sin\th}\pa_\j\\ \label{right_isometry}
\xi_{2,R}&= -\cot\th\,\sin\f \,\pa_\f + \cos\f\,\pa_\th + \frac{\sin\f}{\sin\th}\pa_\j\\ \notag
\xi_{3,R}&= \pa_\f\;.
\end{align}

Let's return to the Taub-NUT space. Now we can rewrite the metric (\ref{TN}) as 
\be
ds^2_{\text{\tiny{T-N}}}
= (1+ \frac{R}{r})\,\left(dr^2 + r^2 (\s_{1,L}^2+\s_{2,L}^2)\right) + (1+ \frac{R}{r})^{-1}\,R^2\, \s_{3,L}^2\;.
\ee
In this form it is manifest that the Taub-NUT space has an \(U(2)=U(1)_L \times SU(2)_R\) symmetry generated by \(\xi_{i,R}\) and \(\xi_{3,L}\). The \(\xi_{3,L}\) isometry has a single fixed point, called a ``nut", at \(|\vec{x}|=0\).
\end{example}
\begin{figure}
\centering
\includegraphics[width=15cm]{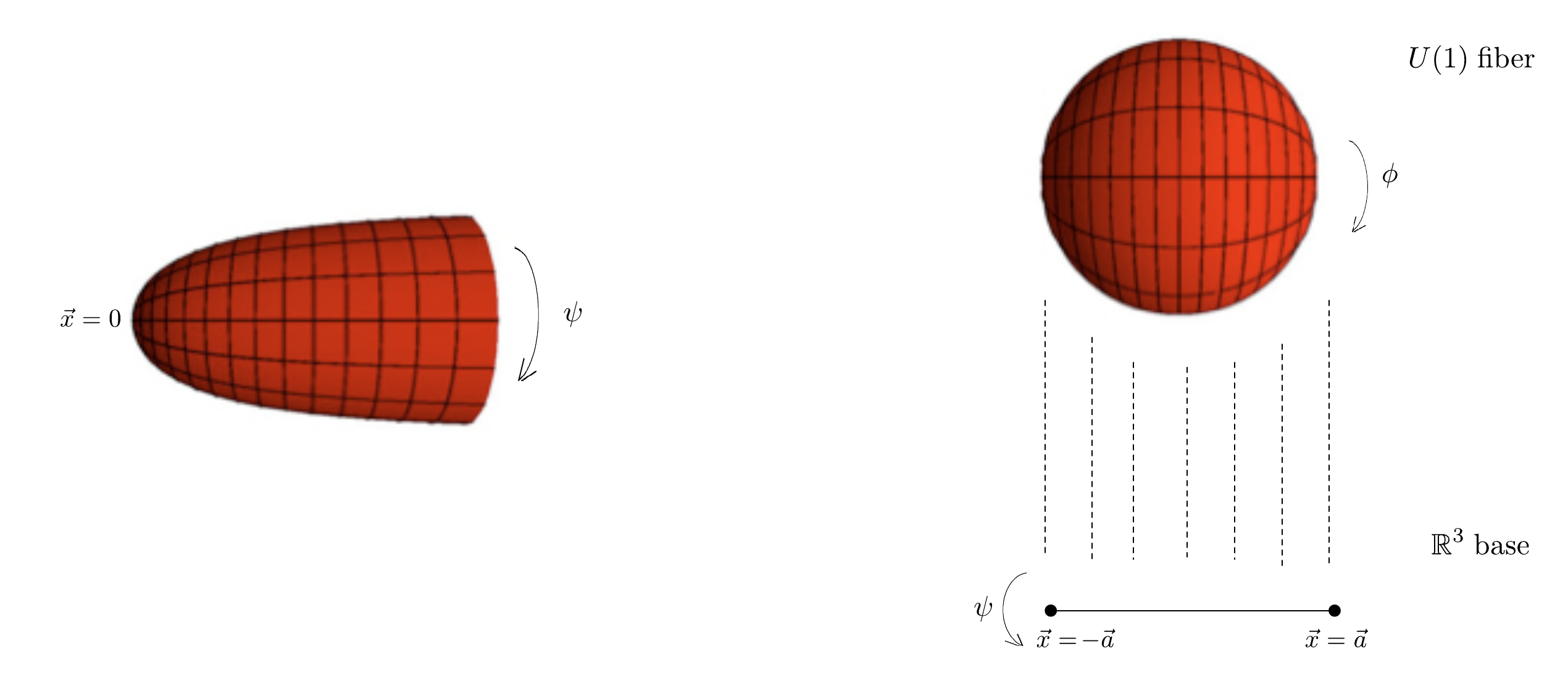} \setlength{\abovecaptionskip}{5pt}
\caption{\footnotesize{(a) The Taub-NUT geometry, with each point representing a two-sphere. (b)The two-cycle (the bolt) given by the fixed point of a U(1) symmetry of the two-centered Gibbons-Hawking space, also known as the Eguchi-Hanson space.}}
\setlength{\belowcaptionskip}{5pt}
\end{figure}

\begin{example} Eguchi-Hanson space

Take the harmonic function in the Gibbons-Hawking Ansatz to be of a two-centered form
\be
H(\vec{x}) = \frac{1}{|\vec{x}-\vec{a}|}+\frac{1}{|\vec{x}
+\vec{a}|}\quad,\quad \vec{a} = a \hat{z}\;. 
\ee
Using the elliptic coordinates
\bea \notag 
x&=& a \sinh\eta\,\sin\th\,\cos\j\\ \label{elliptic_coordinates}
y&=& a \sinh\eta\,\sin\th\,\sin\j \\ \notag
z&=& a \cosh\eta\,\cos\th\;,
\eea
the harmonic function and the metric for the flat \(\R^3\) base becomes
\ben
H &=& 2\, \frac{\cosh\eta}{\cosh^2\eta-\cos^2\th} \\
\notag
a^{-2}\, ds^2_{\R^3}&=& (\cosh^2\eta-\cos^2\th) \, (d\eta^2 + d\th^2) + \sinh^2\eta\,\sin^2\th \, d\j^2\;.
\een
The solution for \(\w^0\) can then be solved to be 
$$
\w^0 = 2 \,\frac{\sinh^2\eta\,\cos\th}{\cosh^2\eta-\cos^2\th}\,d\j\;.
$$
Define now \(R = \sqrt{8a}\) and the new coordinates
\ben
\r &=& R\,\cosh^{1/2}\eta\\ \notag
x_5 &=& 2\f\;,
\een
the metric takes the familiar form
\begin{align}\label{E_H}
&ds^2_{\text{\tiny{E-H}}} =(1- \frac{R^4}{\r^4})^{-1}\,d\r^2 + \frac{\r^2}{4} \,(\s_{1,L}^2 + \s_{2,L}^2) + \frac{\r^2}{4}\,(1- \frac{R^4}{\r^4}) \, \s_{3,L}^2\\ \notag
&R\leq \r < \inf \qquad\th\in[0,\p]\qquad;\;\; \j,\f \in [0,2\p].
\end{align}

It has again a \(U(1)_L \times SU(2)_R\) symmetry generated by \(\xi_{i,R}\) and \(\xi_{3,L}\). Actually the Eguchi-Hanson and the Taub-NUT spaces are the only non-flat, half-flat, asymptotically locally flat spaces with U(2) symmetries. But unlike the Taub-NUT case,  the \(\xi_{3,L}\) isometry has now a \(S^2\) surface of fixed point at \(\r=R\) , called a ``bolt". The unusual identification of the \(\j\) coordinate comes from requiring that the space approaches \(\R^2\times S^2\) without conical singularity near the fixed point \(\r=R\). Comparing the asymptotic form of the metric when \(\r \rightarrow \inf\) with the flat metric (\ref{flat_r4}), we see that it is an asymptotically locally Euclidean (ALE) space, only locally because of the presence of the above-mentioned \(\Z_2\) identification (\(A_1\) in terms of the A-D-E classification). 
\index{Eguchi-Hanson space}
\index{nuts and bolts}
\index{ALE space}

\end{example}
\end{digression}

\vspace{1cm}
\begin{tabular}{p{12cm}}
\hline\\
\end{tabular}
\vspace{1cm}

After discussing the KK monopole, we now turn to the reduction of the M2 brane along the M-theory circle. From the table \ref{brane_reduction} we see that they must be the electric source of the anti-symmetric B-field. Recall that  there is an electric coupling between the anti-symmetric B-field and the string world-sheet (\ref{world-sheet_action_with_B}), we conclude that the M2 brane wrapping the M-theory circle must be reduced to the fundamental string we began with. 

A circle-wrapping M2 brane is electric-magnetic dual to an M5 brane that is transversal to the M-theory circle. From the preservation of the electric-magnetic duality after the dimensional reduction, we know that such an M5 brane must reduce to the magnetic dual of the fundamental string.
In type II string theory, the (5+1)-dimensional object dual to the fundamental string  is called an NS5 brane, the name because it couples to the degrees of freedom coming from quantising the NS-NS sector of the type II strings. We therefore  conclude that an M5 brane transversal to the Kaluza-Klein circle becomes a type IIA NS5 brane upon dimensional reduction.

Finally we turn to the D\(p\)-branes, the official nickname for  the (\(p+1\))-dimensional Dirichlet branes. In the earlier discussion about T-duality we noted that this IIA-IIB duality and the presence of open strings implies the existence of some objects which couple to the Ramond-Ramond fields and on which the open string can end. 
Here we see we indeed get these objects in the spectrum, this time seen from the spacetime point of view. The D\(p\)-branes obtained by dimensionally reducing the M-theory objects are exactly what we need.

In this table we leave out the end-of-the-world M-theory nine branes and the corresponding eight branes in type IIA. Although important for introducing gauge interactions into the theory, we will nowhere need them in the present thesis. The same will be true for the type IIB D7-branes. Albeit fascinating objects, they will play no role in our future discussions. 

From the tension of the M-theory branes (\ref{M_brane_tension}) and the map between ten- and eleven-dimensional units (\ref{coupling_relation}) and by carefully following the reduction procedure, we arrive at the following results for the tension of our newly discovered objects:
\be\label{tension1}
\t_{\text{D}p\text{-brane}} \sim g_s^{-1} \ell_s^{-(1+p)}\qquad;\qquad \t_{\text{F1}} \sim \ell_s^{-2}
\ee
and the rest follows from the relation
\be\label{tension2}
\tau_{\text{\tiny (object)}}\tau_{\text{\tiny  (E-M dual object)}} \sim \frac{1}{G^{(10)}_{\text{\tiny N}}} \sim \ell_s^{-8} g_s^{-2}\;.
\ee

Now we have seen yet another reason why type IIA supergravity is not a good description when strings couplings are large. In this case all other objects are lighter (smaller tension) compared to the fundamental string, and it is therefore also not surprising that the degrees of freedom coming from quantising the fundamental string alone are not sufficient to account for the physics in that regime. 

\subsection{D-brane World-Volume Action}

Besides the closed-string world-sheet action we wrote down earlier (\ref{world-sheet_action_with_B}), for open strings we can add an extra boundary term 
$$
i \int_{\pa \S} A^{(1)}
$$
to the world-sheet action, since the world-sheet has in the open string case a (connected or disconnected) boundary. As we said before, the boundaries of open strings lie on the D-branes that we just introduced, the presence of this boundary coupling suggests that there is an \(U(1)\) gauge field living on the D-branes. We would like to understand how the dynamics of D-branes, including the dynamics of this world-volume \(U(1)\) gauge field, can be described by a world-volume action on the D-brane, in parallel with the way the string dynamics is captured by the string world-sheet action. 

The action for D-branes is a very rich subject and as we won't need too much detail of it later,  it will suffice just to have a pauper's account of the D-brane world-volume action here  (not poor man's because K-theory, the framework needed to discuss the subject properly \cite{Minasian:1997mm,Witten:1998cd} and which we will not introduce here, is a ``poor man's derived category" \cite{Aspinwall:2004jr}).

The basic strategy is to first find the right action for the massless open string modes limited on the D-brane world-volume. 
Let's first begin with the gravitational part. The gravitational coupling of the world-sheet action we used for the strings (\ref{world-sheet_action_with_B}) 
$$
\frac{1}{2\p \a' } \int_\S  d^2\s \sqrt{h}\,h^{ab} g_{\m\n} \pa_a X^\m \pa_b X^\n
$$
seems pretty hard to be generalised to a higher-dimensional world-volume. But in fact, we could have equally well begun with the Nambu-Goto string action \index{Nambu-Goto action} 
\be
\frac{1}{2\p \a'}\int_\S\,d^2\s \sqrt{-\text{det}( g_{\m\n} \,\pa_a X^\m \pa_b X^\n )}\;,
\ee
whose equivalence with the original action can be shown by eliminating the non-dynamical world-sheet metric \(h_{ab}\) using its Euler-Lagrange equation. The Nambu-Goto action, on the other hand, can be generalised easily to higher dimensional object as
\be
-\t_p\,\int d^{p+1}\s \, e^{-(\F-\F_0)} \sqrt{\text{det}G_{ab}}\;,
\ee
where \(\F_0 =\log g_s \) is the asymptotic value of the dilaton , for we have absorbed this factor in the physical string tension \(\t_p\) derived earlier. The quantity \(G_{ab}\) is the pull-back of the metric under the embedding map of the D-brane. This action clearly has the geometric interpretation as the size of the D-brane given a specific embedding. 

Next we turn to the gauge coupling. Consider  the B-field coupling term
\be
\frac{1}{2\p \a'} \int_\S \, B^{(2)}\;.
\ee
in the string world-sheet action (\ref{world-sheet_action_with_B}) , in the presence of world-sheet boundary, the usual gauge transformation
$$
B^{(2)} \rightarrow B^{(2)} + d\e^{(1)}
$$
does not seem to be a symmetry anymore. But this can be repaired by a simultaneous gauge transformation of the \(U(1)\) world-volume field 
$$
A^{(1)} \rightarrow A^{(1)} -\frac{1}{2\p\a'} \e^{(1)}\;.
$$
Now we see that the gauge-invariant field combination is really
\be
{\cal F} =  B + 2\p\a' F
\ee
and is therefore the only combination in which the B-field and the U(1) field strength can appear in the D-brane world-volume action. This leads us to the following so-called Dirac-Born-Infeld action
\be
S_{\text{\tiny{D-B-I}}} = -\t_p\,\int d^{p+1}\s \, e^{-(\F-\F_0)} \sqrt{\text{det}(G_{ab}+ {\cal F}_{ab} )}\;.
\ee
\index{Dirac-Born-Infeld action}
But this is obviously not the whole story. As we mentioned earlier, a higher-dimensional version of the Wilson line coupling
\be
q_p \int C^{(p)}
\ee
should also be included. 

But this is again not the full answer for the R-R coupling. This can be seen from the fact that the world-volume theory has to be anomaly-free. From the fact that D-branes act as sources for the gravitational and the \(p\)-form fields, in general we might expect there to be gauge and gravitational anomalous coupling. By considering two intersecting branes and requiring anomaly cancellation when the open-string zero-modes along the intersection sub-manifold are included, we get the following full ``Chern-Simons" term in the world-volume action \cite{Minasian:1997mm,Green:1996dd}
\be \label{anomalous_brane}
q_p \int  \, C \wedge ch({\cal F}) \wedge \frac{\sqrt{\hat{A}(R_T)}}{\sqrt{\hat{A}(R_N)}}\;,
\ee
where \(C = \sum_{p} C^{(p+1)}\) is the sum of all \(p\)-form fields in the theory, and the Chern character and the A-roof genus are defined and explained in (\ref{chern_character}) and (\ref{A_roof}). \(R_{T,N}\) refers to the tangent and normal bundle of the world-volume manifold respectively. All this is of course only the bosonic part of the action, the supersymmetric action can be built by replacing the above bosonic fields by the appropriate superfields.

One very interesting consequence of this anomalous coupling is, a D\(p\)-brane is not only the source of \(C^{(p+1)}\) but can possibly also source other lower form fields. 


All this is for one single D-brane. In the case of \(N\)-coincident D-branes, things again become more complicated. Now both the gauge potential \(A^{(1)}\) and the transverse coordinate \(X^i\) become \(N\times N\) matrices and it's now not anymore so clear how one should pull-back the bulk fields in this non-commutative geometry. There is a generalisation of both the Dirac-Born-Infeld and the Chern-Simons part of the world-volume action, see for example \cite{Myers:1999ps,Tseytlin:1999dj}. For the purpose of our discussion we will only need the leading in \(\a'\) terms of the non-Abelian Dirac-Born-Infeld action, which reads
$$
-\t_p \frac{(2\p\a')^2}{4}\,\int d^{p+1}\s \, e^{-(\F-\F_0)} \text{Tr} \left(F_{ab}F^{ab}+2 D^aX^i D_a X_i +[X^i,X^j]^2 \right)\;.
$$
From this we conclude that we have \(U(N)\) but not just \(U(1)^N\) world-volume field theory for \(N-\)coincident branes, with gauge coupling
\be \label{coupling_flow}
g_{\text{\tiny{Y-M}}}^2\sim g_s \t_p^{-1} \a'^{-2} \sim g_s \,(\ell_s)^{p-3}\;.
\ee

\subsection{Gauge/Gravity Correspondence}

It is absolutely out of the scope of the present thesis to give a full account of the AdS/CFT correspondence. We will just sketch the ideas we will need later. Please see \cite{Aharony:1999ti,Klebanov:2000me,DHoker:2002aw,Skenderis:2002wp} for reviews of the basic ideas (as opposed to the applications) of the correspondence.

As we mentioned earlier, from dimensional reducing (and taking the T-dual of) the M2- and M5-brane solutions in M-theory (\ref{M2_metric}), (\ref{M5_metric}) we obtain various extremal D\(p\)-brane solutions of type IIA (IIB) string theory.

Let's take the D3-brane solution in type IIB string theory for example. This case is especially simple since the coupling constant of the D-brane world-volume (open string) theory is dimensionless (\ref{coupling_flow}), or, equivalently, that the dilaton of the spacetime solution is constant everywhere. As a result, 
the D3 brane metric reduces to that of \(AdS_{5} \times S^{5}\) 
when we zoom in the region \(r\rightarrow 0\) near the location of the brane.

There are apparently two different ways to describe the physics of this system; one of string theory and one of the D-brane theory. First of all, as we have seen earlier, each of them has its ``low-energy" description, namely the supergravity and the \(U(N)\) gauge theory for \(N\) coincident D-branes respectively. We would like to know when each of them is a valid description. 

Let's begin with the gravitation side. First we note that, because of the infinite redshift factor \(\sqrt{g_{tt}}\) near the horizon, classically the modes near the horizon \(r\rightarrow 0\) never climb up the gravitational potential well and therefore decouple from the rest of the spacetime. From the gravitational point of view it is thus valid to take the ``decoupling limit" and focus on the near horizon geometry \(AdS_5\times S^5\). Secondly, from the relation between the radius of curvature and the string length and the ten-dimensional Planck length
\begin{align} \notag
\frac{R}{\ell_s}&\sim \l^{1/4}\qquad,\quad \l \equiv g_s \,N \sim g_{\text{\tiny{Y-M}}}^2\,N\\ \notag
\frac{R}{\ell_{\text{\tiny P}}^{(10)}} &\sim N^{1/4}\;,
\end{align}
we see that the \(\a'\) corrections and the quantum gravitational effects are controlled by parameters \(\l\) and \(N\) respectively, and that the supergravity is a valid description if
$$
\l \gg 1 \qquad,\qquad N\gg1\;.
$$

On the open string side, for the gauge theory description to be valid we need \(\a' \rightarrow 0\) while keeping the W-boson mass, proportion to \(r/{\a'}\), fixed. This leads us again to the near-horizon limit \(r \rightarrow 0\) where we can consistently truncate the D-brane world-volume theory to \(SU(N)\) gauge theory.\footnote{Notice that we have omitted the \(U(1)\) part of the \(U(N)\), which correspond to the center-of-mass degree of freedom and decouples from the rest of the theory.} Furthermore, it can be shown that the perturbative analysis of the \(SU(N)\) Yang-Mills theory is valid when 
$$
\l \ll 1\;.
$$

Therefore, for large \(N,\) the supergravity theory on the \(AdS_5\times S^5\) background and the \(SU(N)\) super-Yang-Mills theory discussed above are two effective theories describing the system at complementary regimes: the former valid when \(\l\gg1\) and the latter when \(\l\ll1\). This motivates the AdS/CFT conjecture, which in this specific case of D3 branes states that the ten-dimensional type IIB supergravity theory on the \(AdS_5\times S^5\) background is {\it dual} to the \({\cal N}=4\; SU(N)\) super-Yang-Mills theory.

More generally, this conjecture says that the closed string theory on a \(AdS_{p+2}\times K\) background is dual to a conformal field theory living on the conformal boundary \(\pa(AdS_{p+2})\) of \(AdS_{p+2}\). Or, a little bit more concretely, it states 
$$
Z_{\text{\small string}} (\f_0) = \langle e^{\int \f_{0,i} {\cal{O}}^i }\rangle\;,
$$
where \(\f_0\) denotes the boundary condition of the fields on the conformal boundary and \({\cal{O}}^i\) denote the dual operators in the CFT. 

Of course this is an account of the conjecture at the level of caricature. There is first of all the issue of regularisation of the \(AdS\) space on the LHS of the above equation. Secondly there are various interesting and useful generalisations of the above conjecture. This is by itself a vast subject and much more than what we will need later. 

Instead we will simply remark that, first of all, the most remarkable aspect of this conjecture is that it relates a gravitational theory to a theory without gravity. In this sense having a theory of quantum gravity is not too different from solving the field theory. Secondly, this is the first full-blown example of the principle of holography \cite{tHooft:1993gx,Susskind:1994vu}, motivated by the black hole thermodynamics, stating that the degrees of freedom of a \((d+1)\)-dimensional gravitational theory is encoded in the \(d\)-dimensional boundary. In the AdS/CFT setting the extra dimension turns out to be the {\it scale} dimension.

\subsection{S-duality}
\label{S_duality_section}
As some alert readers might have noticed, as we introduced the branes in subsection  \ref{subsection_Branes} by first presenting the two-brane and five-brane solutions in eleven-dimensional supergravity and then dimensionally reducing them, we actually haven't explicitly discussed the extended objects in the type IIB superstring theory. 
In this section we will study the type IIB branes by T-dualising Table \ref{brane_reduction} that appeared when we discussed the reduction of M-theory branes to type IIA objects, and show how a non-perturbative string duality can be revealed in this way.

As explained in subsection \ref{subsection_T-Duality}, the perturbative string T-duality maps type IIA string theory on a circle to type IIB string theory on a dual circle. Furthermore, it exchanges the Neumann with the Dirichlet boundary condition for the open strings, and therefore exchange D-branes of odd and those of even dimensions. Following the world-sheet discussion on the T-duality earlier it's not hard to see that it indeed maps the \(p\)-form field potentials under which the D-branes are charged accordingly. 

We have also learned that M-theory compactified on a small circle is dual to type IIA string theory at weak coupling. Applying subsequently a T-duality, one is led to the conclusion that M-theory compactified on a torus is dual to type IIB theory on a circle. Following the Kaluza-Klein reduction and T-duality rules we can then trace the charged objects of the three different theory in a straightforward way. See Table \ref{M_IIB_chain} for the map under this duality chain. Here we call \(S^1_{(1)}\) the M-theory circle and \(S^1_{(2)}\) the T-duality circle. Branes wrapping at least one of the two circles will be labeled with the number in the parenthesis and ``\(-\)" means they extend only in the directions transversal to  both \(S^1_{(1)}\) and  \(S^1_{(2)}\). The label (i,j)  behind the KK-monopole denotes whether the solution is homogeneous along one of the circle directions (i) and under which Kaluza-Klein gauge field they are charged (j). 

\begin{table}
\centering
\caption{\label{M_IIB_chain}{\footnotesize From M- to type IIB theory. Extended charged objects.}}
\begin{tabular}{ccccc}
\toprule
M-theory &\(\xrightarrow[]{\text{reduce on  }  S^1_{(1)} }\) &IIA &\( \xrightarrow[]{\text{T-dualise along} \; S^1_{(2)} }\) & IIB\\ 
\toprule[0.06em]
M5(1,2) && D4(2) && D3 (\(-\)) \\ 
M2(\(-\)) && D2(\(-\)) && D3(2) \\ 
M2(1,2)&&F1(2) && p(2)\\ 
\midrule
M2(1) && F1 (\(-\)) && F1(\(-\)) \\
M2(2) && D2(2) && D1(\(-\)) \\ 
\midrule
p(2) && p(2) && F1(2) \\
p(1) && D0(\(-\)) &&D1(2) \\ 
\midrule
M5(2) && NS5(2) && NS5(2)\\
M5(1) && D4(\(-\)) && D5 (2)\\ 
\midrule
KK(1,2) && KK(\(-\),2) && NS5(\(-\)) \\
KK(2,1) && D6 (2) &&  D5(\(-\)) \\
\bottomrule
\end{tabular}
\end{table}

Of course, if both circles are small, nothing can stop us from exchanging the two circles \(S^1_{(1)}\) and  \(S^1_{(2)}\), which means we now first reduce along the second and then T-dualise along the first circle. From Table \ref{M_IIB_chain} we see something rather amusing: this simply exchange and fundamental with the D-string, and NS5 and D5 branes, while leaving D3 branes untouched! This exchange is actually a part of a much larger duality group, namely the modular group \(PSL(2,\Z)\) of the torus on which we compactify M-theory on. This means, apart from exchanging the two cycles, we can also consider an arbitrary change of basis. Let's begin with a torus, described as the complex plane \(\C^1\) with the following identification 
\be \label{identification_torus}
z \sim z + v_1 \sim  z + v_2\;. \ee 
A linear change of basis will take the form 
\be\label{SL2Z_1}
\bem v_1 \\ v_2 \eem \rightarrow
\bem a&b\\c&d\eem  \bem v_1 \\ v_2 \eem \;. 
\ee
To preserve the identification (\ref{identification_torus}) we should consider only integral changes of basis, and to keep the volume and orientation invariant we should have \(\vec{v}_1\times  \vec{v}_2  =\im(\bar{v}_1v_2) \)  invariant. These conditions show that the modular group of a torus is 
\be
SL(2,\Z) = \left\{\g=  \bem a&b\\c&d\eem\Bigl\lvert \,ad-bc=1, \;a,b,c,d\in \Z   \right\}\;.
\ee
It takes a point in the upper-half plane \(\t \in {\cal H}_1 = \{z\in\C| \im z> 0 \}\) to another point in the upper-half plane by 
\be
\g(\t)= \frac{a\t+b}{c\t+d}\;,
\ee
where \(\t = \frac{\int_B dz}{\int_A dz} \) encodes the angle between the two one-cycles which we call the \(A\)- and the \(B\)-cycle, or in other words the complex structure of the torus. If we want to be more precise, notice that \(\g  \) and \(-\g=\left( \begin{smallmatrix} -a & -b \\ -c&-d \end{smallmatrix}\right)\) give the same map \({\cal H}_1 \rightarrow {\cal H}_1 \), therefore the modular group of the complex structure of a torus is really \(PSL(2,\Z) = SL(2,\Z)/(\g \sim -\g)\).
\index{upper half-plane}
\begin{figure}
\centering
\includegraphics[width=13.5cm]{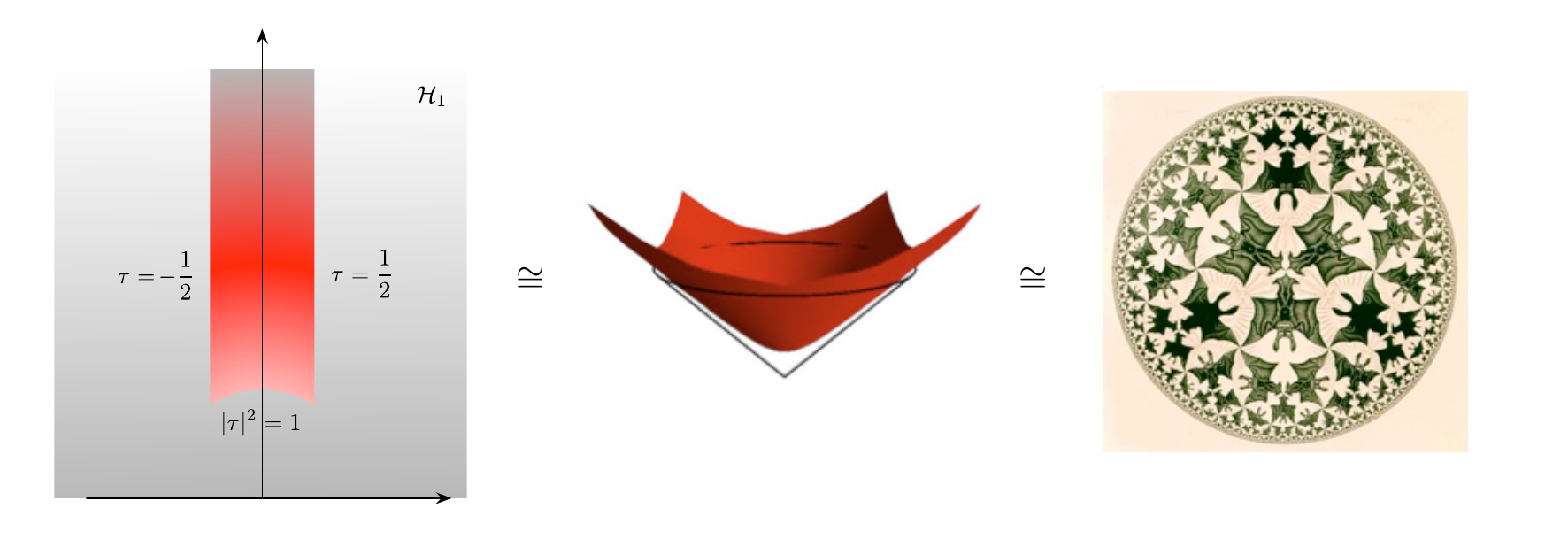} \setlength{\abovecaptionskip}{5pt}
\caption{\label{models_disk_fundamental_domain}\footnotesize{Three different models for the same hyperbolic space: the upper-half plane, the hyperboloid and the Poincar\'e disk. The red region drawn in the upper-half plane is 
a fundamental domain \({\cal H}_1/PSL(2,\Z)\)  under the modular group \(PSL(2,\Z)\). }}
\setlength{\belowcaptionskip}{5pt}
\end{figure}

Considering the mapping of the type IIB fields under the exchange of the two circles \(S^1_{(1)}\) and  \(S^1_{(2)}\), which is 
the so-called S-transformation corresponding to the following \(SL(2,\Z)\) element 
$$ S=\bem
 0&-1\\ 1&0
\eem\;, $$
we are led to the guess that the two-component field has to transform as
$$
B = \bem  C^{(2)} \\B^{(2)}  \eem  \rightarrow \g B 
$$
when the circles are changed as (\ref{SL2Z_1}), while the chiral four-form potential remains invariant. Indeed, the low energy type IIB supergravity has an even larger symmetry which is broken at the quantum level to \(SL(2,\Z)\). To see this, defining also the field combination, the ``axion-dilaton", as
\be
\l = \l_1 + i\l_2 = C_0 + i e^{-\F} \qquad,\qquad {\cal T} =  \frac{1}{\l_2} \bem |\l|^2 & \l_1 \\ \l_1 &1 \eem\;.
\ee
Then it's not hard to check that the IIB supergravity action (\ref{IIB_action}) can be rewritten as 
\bea\notag
2\k^2 S^{\text{\tiny (IIB)}} &=&  \int d^{10}x \sqrt{-g} \left( R - \frac{1}{12} H_{\m\n\r}^T {\cal T}  H^{\m\n\r} + \frac{1}{4} \Tr(\pa^\m {\cal T}  \pa_\m {\cal T}^{-1}) -\frac{1}{4} |\til{F^{(5)}}|^2 \right)\\ \label{IIB_action_SL2Z}
 &&-\frac{1}{4} \int C^{(4)} \wedge B^T \wedge SB 
\eea
where we have gone to the Einsetin frame by rescaling the metric as \(g_{\m\n} \rightarrow e^{-\F/2} g_{\m\n}\) to isolate the \(\F\) dependence in the axion-dilaton combination \(\l\). From 
\be\label{axion-dilaton_matrix_transf1}
 {\cal T} \rightarrow \g  {\cal T} \g^T \qquad \text{when}\qquad \l \rightarrow \frac{a\l+b}{c\l+d} \;,
\ee
and the fact that \(SL(2,\Z) \cong Sp(2,\Z)\), namely \(\g S \g^T = S\), we see that the above action is manifestly invariant under the S-duality transformation
\ben
B \rightarrow \g B\qquad&,&\qquad   \l \rightarrow \frac{a\l+b}{c\l+d} \\
g_{\m\n} \rightarrow g_{\m\n} \qquad&,&\qquad C^{(4)} \rightarrow C^{(4)} \;.
\een

Note that at this level we don't have any reason to require \( \g \in SL(2,\Z)\). Any \( \g \in SL(2,\R)\) is sufficient to ensure the invariance of the above supergravity action. But since we have seen the geometric origin of this symmetry from our M-theory derivation, we conclude that the real symmetry group should be the discreet torus modular group \(PSL(2,\Z)\). Or, another way to see this is the Dirac quantisation condition (\ref{Dirac_quantisation}) which has to be satisfied by branes and strings. We simply cannot map one D5 brane to ``0.32 D5 + 6.7292 NS5" branes without destroying the Dirac quantisation.

Notice that, unlike the T-duality, this ``S-duality" is non-perturbative by nature since we can map from small \(g_s\) to the large coupling regime. Especially, from the brane tensions (\ref{tension1}) and (\ref{tension2}) we see that D1 string becomes the light degrees of freedom instead of the fundamental string after the S-transformation. And similarly for NS5 and D5 branes. This duality suggests that various different objects in string theory should probably be treated at the equal footing, and string theory is really about ``strings" only at a corner of the moduli space of the theory.

\begin{figure}
\centering
\includegraphics[width=11.5cm]{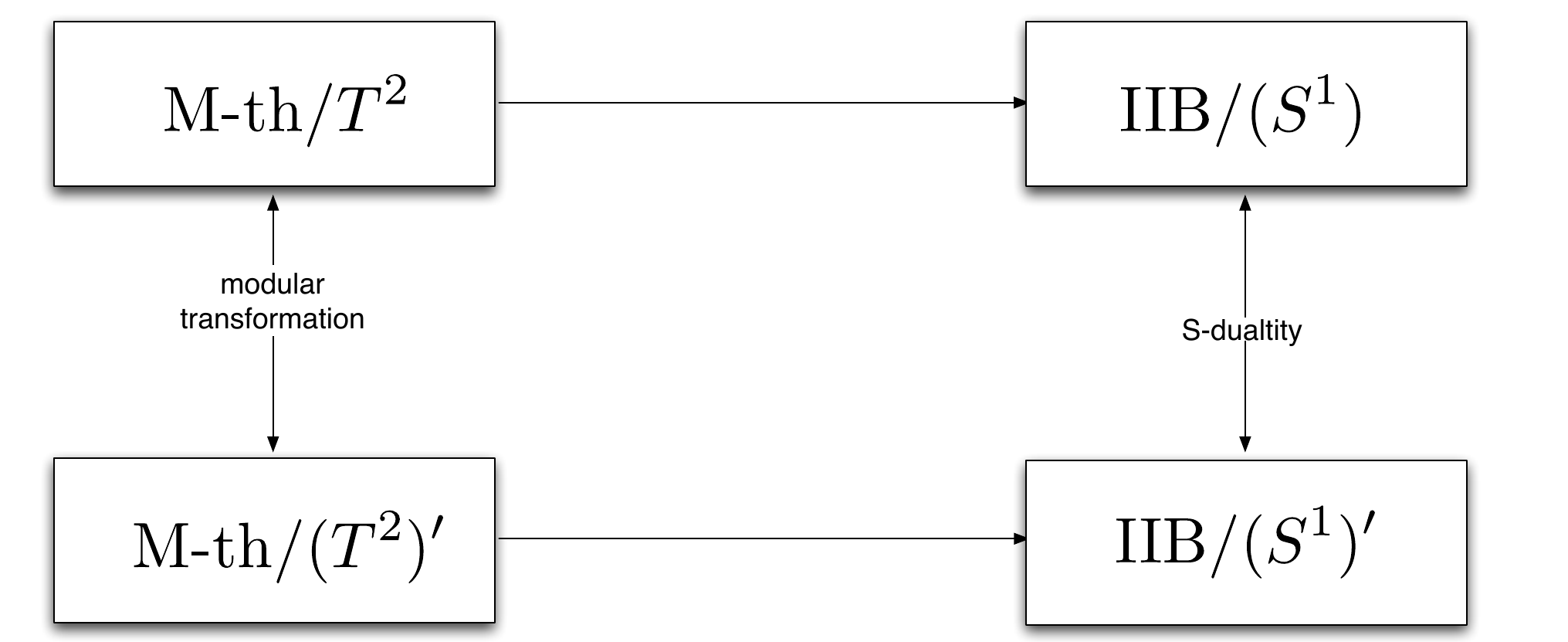} \setlength{\abovecaptionskip}{5pt}
\caption{\footnotesize{S-duality as the modular group of M-theory torus.}}
\setlength{\belowcaptionskip}{5pt}
\end{figure}

\vspace{1.2cm}
\begin{tabular}{p{12cm}}
\hline\\
\end{tabular}\label{upper_half_plane}
\begin{digression}Upper Half-Plane

The upper-half plane is defined in the obvious way as
\be
{\cal H}_1 = \{\t \in \C \lvert\, \im\t>0\}\;.
\ee
As depicted in Figure \ref{models_disk_fundamental_domain}, it is equivalent to the hyperbolic space, namely the Euclidean  \(AdS_2\) space defined as 
\be
-T^2 + X_1^2 + X_2^2 = -1 \;\,,\; T > 0 \;,
\ee
and to the Poincar\'e Disk
\be
\{z\in \C \lvert\, |z|^2 <1\}
\ee
and also to the coset space 
\be
\frac{SL(2,\R)}{U(1)}\;.
\ee

This can be seen using the map
\bea\label{tau_upper_half}
{\cal T} &=&  \frac{1}{\t_2} \bem    |\t|^2 & \t_1 \\ \t_1 &1 \eem\\ \notag
&=& \bem T+X_1 & X_2 \\ X_2 & T-X_1 \eem\\ \notag
&=& 
\frac{2}{\sqrt{3}}\, \frac{1}{1-|z|^2}\bem|z + e^{\frac{5\p i}{6}}|^2 & \sqrt{3}\re z-\frac{1}{2}   |z-i|^2 \\ &\\ \sqrt{3}\re z-\frac{1}{2}   |z-i|^2 &   |z-i|^2\eem \\
&=& \x^T \, \bem1&0\\0&1\eem \x\quad, \;\x  \sim \x_0(\th)\x \in SL(2,\R)\;,
\eea
where 
\be
\x_0(\th) = \bem \cos\th & \sin\th \\ -\sin\th & \cos\th\eem
\ee
is any element of the \(U(1)\) group that stabilses the point \(\t=i\).  Notice that the above map is of course not unique, since one can use the \(SO(2,1)\) symmetry of the space to obtain other equivalent maps. For instance, the complicated-looking map given above between the Poincar\'e disk and the upper half plane is in fact simply the M\"obius transformation
$$
z= i \Big(\frac{\t+e^{-\frac{i\p}{3}}}{\t+e^{\frac{i\p}{3}}}\Big)\;,
$$
but any other M\"obius transformation of the form 
\be
z= e^{i\f}\left( \frac{\t+\tbar_0}{\t+\t_0}\right)\quad ,\quad \t_0 \in {\cal H}_1
\ee
will equally do the job.

All these four different models for the space \({\cal H}_1\) will later make their appearances in different part of the thesis. 
\end{digression}

\vspace{0.2cm}
\begin{tabular}{p{12cm}}
\hline\\
\end{tabular}

\part{String Compactification}

In the previous part of the thesis we have introduced various basic aspects of superstring theory. In this part we will discuss the properties of these theories compactified on the manifolds with special holonomy, in particular the Calabi-Yau three-folds and the K3 manifolds.

There are various motivations to study superstring or M-theory  on some small, compact ``internal" manifolds. First of all, the world we see are not ten- or eleven-dimensional. One way to connect string theory to the ``reality" we see around us is therefore to take some of the ten or eleven dimensions to be extremely small. The other possibility, the so-called brane world scenario, is to assume that we live on a lower dimensional sub-manifold of the total ten- or eleven-dimensional spacetime. At the time of writing, both of them seem to be worth investigating. 

More particularly, there are various reasons to study compactifications with unbroken spacetime supersymmetry. First is the analytic control over the system supersymmetry grants us. It's fair to say that, studying generic non-supersymmetric compactifictions quantitatively is at the time of writing still out of control.

Last but not least, we also want to study these compactifications for purely theoretical purposes. As we have seen in the example of type IIB S-duality and as will see later in this chapter, more string dualities become manifest upon compactification on these special supersymmetry-preserving manifolds. Compactification can therefore be seen as an element of the net connecting different string theories. Furthermore, as we will often see in the discussions about Calabi-Yau three-folds, strings are amazing probes for geometry compared to point particles and compactifying string theory on a particular kind of  manifolds often offers great insights into the nature of these manifolds.

For the purpose of the present thesis, discussing compactification is indispensable since many of the properties of the lower-dimensional theories we will study in details later could be best understood as the properties of the internal manifolds.

This part is organized as follows. We will assume that the readers has some familiarity with the defining properties of Calabi-Yau and K3 manifolds. A self-contained small review of them can be found in Appendix \ref{Mathematical Preliminaries}. In chapter \ref{Calabi-Yau Compactifications} we will discuss the Calabi-Yau compactification and chapter \ref{K3 Compactification} the K3 compactification. In both chapters we will describe the compactification first from a world-sheet and then from a spacetime point of view. These discussions will be quite general, and we will leave the more specific topics, namely the (multi-) black hole solutions of the low-energy supergravity theories and the microscopic counting of the BPS states, for later chapters.

\chapter{Calabi-Yau Compactifications}
\label{Calabi-Yau Compactifications}

In this chapter we will discuss superstring theories compactified on Calabi-Yau three-folds, leading to \({\cal N}=2\) supersymmetry in four dimensions. Some discussions of the basic properties of these manifolds can be found in appendix \ref{Mathematical Preliminaries}. 
We will begin with a world-sheet analysis, meaning studying the (2,2) superconformal theory, describing a string moving in the Calabi-Yau space. Along the way we will introduce various concepts useful for studying the ground states of a supersymmetric conformal theory, which we will often rely on for the studying of supersymmetric spectrum of a black hole system in string theory. 

On the other hand, the geometric intuition will also be indispensable for understanding the compactified string theory. We will therefore switch to a spacetime perspective after basic concepts have been introduced from a world-sheet viewpoint. In particular we will discuss the structure of the geometric moduli space of the Calabi-Yau manifolds in details. 

After that we are ready to introduce the low energy effective actions in lower dimensions, and discuss their range of validity.

\section{(2,2) Superconformal Field Theory}
\setcounter{equation}{0}

In the beginning part of the thesis we have introduced the superstring theory as a two-dimensional conformal theory, considered to have critical central charge equals to \(15\) and thus correspond to a total of ten spacetime dimensions. So-called compactification, can therefore be thought of having a product CFT with a factor with central charge 
\(c=6\) (four spacetime dimensions) and an ``internal" factor with central charge \(c=9\). Furthermore, as we have seen in the superstring example, world-sheet supersymmetries are intimately linked to spacetime supersymmetries. 
Since we would like to end up with a lower-dimensional theory with unbroken spacetime supersymmetry, as we will see shortly it turns out that choosing the internal CFT to have (2,2) world-sheet supersymmetry will serve the purpose. In other words we have a total string theory of \(M_4\times [(2,2), c=9]\) in mind and will now concentrate on the latter ``internal" part. 

\subsection{ ${\cal N}=2$ Superconformal Algebra}

The two-dimensional \({\cal N}=2\) superconformal algebra is rather similar to the \({\cal N}=1\) version we have seen in (\ref{n_1_superconformal_algebra}). The extra supersymmetry means the presence of the second superconformal current \(G\), and an R-current \(J\) under which they are charged. In terms of their Fourier modes the algebra reads

\ben
\lbrack L_m,L_n\rbrack&=& (m-n) L_{m+n} + \frac{c}{12} m (m^2-1) \,\d_{m+n,0}\\ \notag
\lbrack J_m,J_n \rbrack&=&\frac{c}{3} m \,\d_{m+n,0}\\ \notag
\lbrack L_n,J_m \rbrack &=& -m \,J_{m+n}\\ \label{n_2_superconformal_algebra}
\lbrack L_n,G_r^{\pm} \rbrack &=& (\frac{n}{2} - r)\, G_{r+n}^\pm\\ \notag
\lbrack J_n,G_r^{\pm} \rbrack &=&\pm G_{r+n}^{\pm} \\ \notag
\{G^{+}_{r},G^-_{s}\} &=& 2 L_{r+s} + (r-s) J_{r+s} +\frac{c}{3}\, (r^2 -\frac{1}{4})\, \d_{r+s,0}
\een
and as before we have two possible periodic conditions for the fermions
\be
\begin{cases}
2r =  0 \text{  mod   }2 &\text{for R sector} \\ &\\ 
2r = 1 \text{  mod   }2 &\text{for NS sector}\;.
\end{cases} 
\ee

\index{\({\cal N}=2\) Superconformal algebra}

\begin{example}

Consider the {\it non-linear sigma model} with action
\bea\notag 
{\cal S} &=& \frac{1}{2\p \a'}\, \int_\S d^2z \Bigg( \frac{1}{2} g_{\m\n}(X) \pa X^\m \bar{\pa} X^\n 
+ g_{\m\n}(X) (\til{\j}^\m \bar{D} \til{\j}^\n +\j^\m D \j^\n )\\
\label{non_linear_sigma}
&& + \frac{1}{4} R_{\m\n\r\s} \til{\j}^\m\til{\j}^\n \j^\r\j^\s
\Bigg)\;,
\eea
where \(D\) and \(\bar{D}\) is the holomorphic and anti-holomorphic pull-back of the covariant derivative with respect to the metric \(g_{\m\n}\). 

For this action to have (2,2) supersymmetry, the kinetic term and its supersymmetric partner must be able to be written in the superspace form as
\be\label{action_non_linear_superspace}
{\cal S} = -  \,\frac{1}{8\p \a'}\int d^2z\, d^4\th\, K(\Phi^i,\bar{\Phi}^{\ibar})\;,
\ee
where \(\Phi^i\) and \(\bar{\Phi}^{\ibar}\) are the chiral and anti-chiral superfields, satisfying 
$$
{\cal D}_+ \Phi^i = \bar{{\cal D}}_+ \Phi^i = 0 
$$
and the opposite R-charge counterpart for \(\bar{\Phi}^{\ibar}\).  And we have used here the convention 
$$
{\cal D}_\pm = \frac{\pa}{\pa \th^\pm} + \th^\mp \pa\qquad;\qquad
{\cal Q}_\pm = \frac{\pa}{\pa \th^\pm} - \th^\mp \pa
$$
and their holomorphic counterpart. 

In the superspace form it is immediately clear that the presence of  (2,2) supersymmetry imposes the K\"ahlerity condition on the target space. The action (\ref{non_linear_sigma}) is then indeed equivalent to (\ref{action_non_linear_superspace}) with the K\"ahler metric \(g_{i{\jbar}} = \pa_i\pa_{\jbar} K\).

Moreover, as in (\ref{world-sheet_action_with_B}) we can include in the action the topological coupling to the B-field
\be\label{B_term}
2\p i \int B^{(2)}\;.
\ee
We say this coupling is topological because the action only depends on the cohomology classes of \(B^{(2)}\). Furthermore, since an action always appears in the path integral in the form of \(e^{-S}\), we conclude that shifting the B-field by an element of the integral cohomology classes of the target space must be a symmetry of the theory.

Apart from supersymmetry we would like to require conformal symmetry as well. As we have seen in the superstring case, at the leading order of \(\a'\), the vanishing of the beta function imposes that the target space is Ricci flat, at least in the absence of  dilaton gradient or the H-flux, which we will not consider in the present thesis. 
Recalling the fact that a compact manifold admitting a Ricci flat metric must be a Calabi-Yau manifold (\ref{def_calabi_yau}), the conformal symmetry of the superconformal field theory imposes the Calabi-Yau condition on the target space.

We would like to stress that this theory is by far not the only possible ``internal CFT" on one can compactify the superstring on. But it is certainly an obvious candidate and indeed leads to rich structure and analytic control.

\end{example}

\index{non-linear sigma model}

\subsection{Chiral Ring}
\label{Chiral Ring}

To build up a representation for the above algebra, just like in the \({\cal N}=1\) case we are especially interested in the ``highest weight state" annihilated by all the positive modes  
\be
L_m |\phi\rangle = J_n |\phi\rangle = G_r^{\pm} |\phi\rangle= 0
\qquad \text{for all } n,m,r>0\;.
\ee
The reason for this is that  we can build a representation by acting with creation operators on these highest weight states. We say they have conformal weight \(h\) and charge \(q\) if they have eigenvalues \(L_0  |\phi\rangle  = h |\phi\rangle \), \(J_0  |\phi\rangle  = q |\phi\rangle \) under the zero index operators. In the context of state-field correspondence, a highest weight state is said to be created by a ``primary field" \(\phi\), such that \(|\phi\rangle = \phi|0\rangle  \).

\index{highest weight state}
\index{primary field}

Analogous to the case of superstring, more care should be taken for the Ramond sector, because in this case there exist zero index fermionic modes \(G^{\pm}_0\). We will call a state an R-ground state if 
\(G^{\pm}_0 |\phi\rangle =0 \). From the \(\{G^+_0,G^-_0\}\) commutation relation we see that R-ground states always have the weight 
\be \label{R_ground_state}
h(\text{Ramond ground state})=\frac{c}{24}\;.
\ee 
Furthermore, from the 
hermiticity condition \((G_r^{\pm})^{\dag} =G_{-r}^{\mp} \) we see that the above is also a sufficient condition that the state is an R-ground state.
\index{Ramond ground state}
\index{chiral primary field}
For the NS sector, it will turn out to be useful to further refine the concept of primary fields into chiral primary fields. A field is called a chiral primary field if it is a primary field which satisfies the condition 
\be
G^+(z) \phi_c(w) \sim \text{   regular} \;,
\ee
or, using the mode expansion \(G^\pm(z) = \sum_r\,G^\pm_r \,z^{-r-\frac{3}{2}}  \), in the operator language the above equation is equivalent to
\be
G^+_{-1/2} |\phi_c\rangle =0 \;.
\ee

Furthermore, from the \( \{G_{-1/2}^+, G_{1/2}^-\}\) and \(\{G_{-3/2}^+, G_{3/2}^-\}\) commutation relations, we see that the conformal weight and the R-charge of a chiral primary satisfy 
\be \label{bound1}
 0 \leq h_c= \frac{q_c}{2} \leq \frac{c}{6}\;.
\ee

From the OPE between two chiral primaries and the conservation of the R-charge, it's not hard to see that the product also satisfies the chiral primary condition \(h=\frac{q}{2}\).  This suggests that chiral primary fields form a ring, called the ``chiral ring", with the Yukawa coupling given by
\be
\phi_{i,c }\phi_{j,c}  = C^{k}_{\;\;ij} \phi_{k,c} \;.
\ee

Similarly, there is also an anti-chiral ring, with anti-chiral primaries defined as
\be
G_{-1/2}^- |\phi_a\rangle =0 \;.
\ee
and satisfies 
\be \label{bound2}
 0 \leq h_a = -\frac{q_a}{2} \leq \frac{c}{6}\;.
\ee

Combining the left- and right-moving sector, we have the following four rings (c,c), (a,c), (c,a) , (a,a) in a (2,2) superconformal field theory. Their significance for us will be illustrated in the following non-linear sigma model example.

\begin{example}

It is a usual phenomenon that the supersymmetric ground states of a theory are given by the cohomology of a relevant space, at least in the limit in which the string tension is large in the  case  of 2-d CFT. We will now see how this comes about in our non-linear sigma model example. We have seen in the last subsection that the consistency of the CFT requires the target space to be Calabi-Yau, and will therefore assume that this is the case in the following discussion.

As in the case of superstring theory we have seen before, the Ramond ground states are spacetime fermions with definite chirality. In the non-linear sigma model example, using the K\"ahlerity of the target space \(M\), or equivalently \({\cal N}=2\) supersymmetry, we have an extra grading on these spacetime fermions. In other words, writing out the action (\ref{action_non_linear_superspace}) in components and from the supersymmetry transformation we read out the action of the zero modes of the world-sheet current
\be
G_0^+ = \j^iD_i \qquad;\qquad G_0^- = \j^{\ibar}D_{\ibar}
\ee
and similar for the right-moving part. From the chiral and anti-chiral multiplet structure it is easy to see that the fermionic fields carry the following R-charges under (\( J,\til{J}\)):
\be
\begin{array}{lc}
 \j^i  &(1,0) \\
 \j^{\ibar}&(-1,0)\\
 \til{\j}^i   &(0,1)\\
 \til{\j}^{\ibar} &(0,-1)\;.
\end{array}
\ee

From quantising the fermions 
\be \{\j^\m,\j^\n\} = g^{\m\n}\,,\ee
we can choose \(\j^i\) to be the creation and \(\j^{\ibar}\) the annihilation operators on the left-moving side. This choice amounts to a choice of the chirality of the Weyl spinors. Similarly one can now choose the right-moving ground states to have the same or the opposite chirality, namely, apart from  \( \j^{\ibar} |0;0_\pm\rangle = 0 \) we also impose  \( \til{\j}^{\ibar} |0;0_+\rangle = 0 \) or \( \til{\j}^i |0;0_-\rangle = 0 \).

In the first case we see that the ground states correspond to the cohomology class \(H^{n-r,s}(M)\)
\be \label{cc}
\O_{\bar{1}\dotsi \bar{n}}f^{{\jbar}_1\dotsi {\jbar}_r}_{\;\;\;\;\;\;\;\;i_1\dotsi i_s}\til{\j}_{{\jbar}_1}\dotsi\til{\j}_{{\jbar}_r}\j^{i_1}\dotsi \j^{i_s}|0;0_+\rangle\;.
\ee
where we have used the unique harmonic (0,n) form of the Calabi-Yau  (\ref{complex_structure}) to lower the indices. From the index structure one can see that these states have the same sign for the R-charges on the left- and right-moving sides. 

We will see in the next subsection that there is a symmetry of the superconformal algebra which relates R-ground states to NS chiral primary fields. In particular, the R-ground states discussed above 
correspond to (c,c) fields with conformal weights and R-chrages equal to 
\be
(2h,2\til{h}) = (q, \til{q})= (s,r) \;.
\ee

Similarly, there are also ground states of the following form
\be \label{ca}
f_{  {\jbar}_1 \dotsi {\jbar}_r i_1\dotsi i_s}
\til{\j}^{{\jbar}_1}\dotsi\til{\j}^{{\jbar}_r} \j^{i_1}\dotsi \j^{i_s}|0;0_-\rangle\;.
\ee
They correspond to the cohomology class \(H^{r,s}(M)\) and the  corresponding NS-NS fields are (c,a) fields with 
\be
(2h,-2\til{h}) = (q, \til{q})= (s,-r) \;.
\ee

Notice that since \(c=3n\) for \(n\)-complex-dimensional target space, the (anti-)chiral primary condition \(|q| \leq \frac{c}{3}=n \)
(\ref{bound1}), (\ref{bound2}) is indeed in accordance with the correspondence between the CFT chiral ring and the cohomology ring of the target space.

Let's now focus on the case of Calabi-Yau three-folds. We are especially interested in the ring elements with \(h+\til{h}=1\). This is because, by combining them with the appropriate superconformal currents \(G\), we can build from them marginal operators with total conformal weight 2 and which are neutral under R-symmetry. These marginal operators can be then used to deform the superconformal theory. Here we shall remind the readers that marginal operators are the operators that don't become trivial nor dominant when flowing to the IR fixed point, and can therefore be thought of as taking a conformal field theory to another ``nearby" CFT.

From the above analysis we see that in the CY three-fold case, the marginal operators are given by the elements in the (c,c) ring corresponding to elements in \(H^{2,1}(M)\) and those in (c,a) ring corresponding to elements in \(H^{1,1}(M)\). The former ones correspond to deforming the complex structure (the shape) of the Calabi-Yau manifold, and the second one the K\"ahler form (the size) of it. This can be seen from the  expression for their corresponding harmonic forms (\ref{cc}) and (\ref{ca}) as follows.  Roughly speaking, they correspond to the complex structure and the 
K\"ahler part of the metric deformation:
\be
f_{ij} \sim \d g_{ij} \qquad;\qquad f_{i\jbar} \sim \d g_{i\jbar} \;.
\ee

 \begin{table}
 \begin{tabular}{ccc}
 \toprule
 ring&  cohomology & deformation \\
 \midrule
 (c,c) & \(H^{2,1}(M) \)& complex structure\\
 (c,a) & \(H^{1,1}(M) \)& K\"ahler form\\
 \bottomrule
 \end{tabular}
 \caption{\footnotesize{Summary of the relation between chiral rings, marginal deformation and the cohomology class of the target space.}}
 \end{table}
 
It will turn out to be important to have some further knowledge about the structure of the space of all deformations, namely the moduli space of the theory. In principle we can now compute  the moduli space metric from the OPE's of the marginal operators (the Zamolodchikov metric). First of all it's easy to show that the space of the (c,c) and the (c,a) part of the deformation is locally a direct product. Namely
 $$
 {\cal M} =  {\cal M}_{\text{complex}} \times {\cal M}_{\text{K\"ahler}} \qquad\text{(locally)}\;.
 $$
Furthermore, it can be shown that, by employing the \(tt^*\) equation for example \cite{Cecotti:1991me}, the moduli spaces are of the special K\"ahler kind which we defined in (\ref{special_K_def}). But since we will need the geometric picture repeatedly in later analysis, we will postpone the derivation and later derive it in a way that makes its geometric meaning directly manifest.
  
\end{example}

\subsection{Spectral Flow}
\label{section_spectral_flow}

Another important property of the \({\cal N}=2\) superconformal algebra is that it has an inner automorphism, which means that the algebra remains the same under the following redefinition

\begin{align} \notag
L_n &\rightarrow L_n + \eta J_n + \eta^2 \frac{c}{6} \d_{n,0}\\
J_n&\rightarrow J_n + \eta \frac{c}{3} \d_{n,0}\\ \notag
G_r^\pm &\rightarrow G_{r\pm\eta}^\pm\;.
\end{align}

An isomorphism of the algebra implies that of a representation. Namely, when the transformation of the operators is induced by a similarity transformation \(\hat{\cal O} \rightarrow U \hat{\cal O} U^{-1}\), then there is a  corresponding transformation of the representation \(|\phi\rangle \rightarrow U |\phi\rangle\).

Now we will transform the states in the way mentioned above by means of a vertex operator insertion 
\be
U_\eta = e^{-i \eta \sqrt{\frac{c}{3}}H }\;,
\ee
where \(H\) is the free boson from bosonising the R-current
\be
J(z) = i \sqrt{\frac{c}{3}} \pa H\;.
\ee

The net effect is just to shift the \(U(1)\) charge of every state by \(-\eta\frac{c}{3}\), and to change the periodicity of the fermionic current \(G^{\pm}(z)\).

To sum up, beginning with a state \(|\phi\rangle\) with weight \(h\) and R-charge \(q\), the state \(U_\eta|\phi\rangle\) transformed by \(U_\eta: {\cal H} \rightarrow {\cal H}_\eta\) will have
\bea \notag
h_\eta &=& h- \eta q + \frac{c}{6} \eta^2 \\
\label{spectral_flow} 
q_\eta &=& q- \frac{c}{3} \eta \;. 
\eea

This operation is called the ``spectral flow" of \({\cal N}=2\) CFT relating different representations of the algebra. In particular, as promised before, this symmetry relates the R-ground states to the NS sector (anti-)chiral primary states. 
Indeed, from the above relation (\ref{spectral_flow}) it's easy to see that a \(\eta=1/2\) flow takes chiral primary states to Ramond ground states, and another  \(\eta=1/2\) flow takes them again to anti-chiral primary states, and vice versa for the \(\eta=-1/2\) flows. In this sense there is really a unique notion of ``ground states" in \({\cal N}=2\) superconformal field theories. 

\begin{figure}
\centering
\includegraphics[width=15cm]{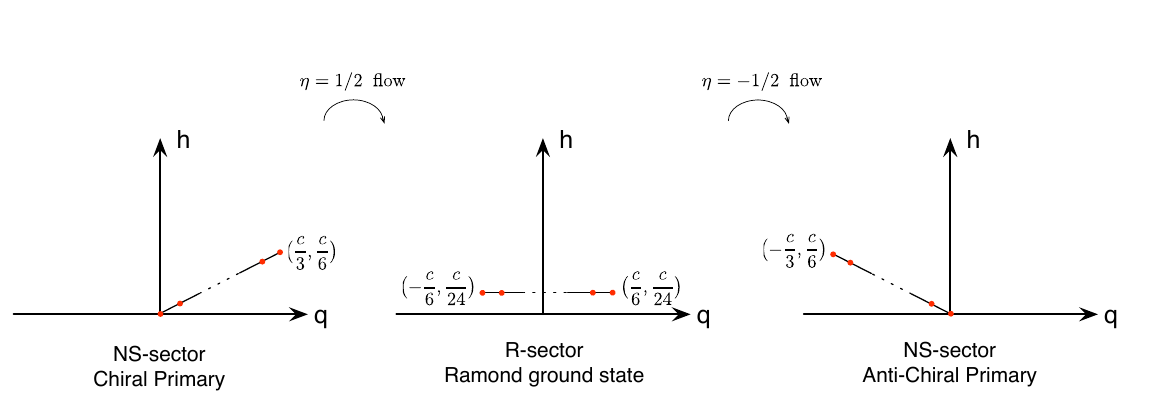} \setlength{\abovecaptionskip}{5pt}
\caption{\label{spectral_flow_pic}\footnotesize{The relationship between the NS- and the R- sector ground states under the spectral flow\cite{Lerche:1989uy}.}}
\setlength{\belowcaptionskip}{5pt}
\end{figure}

\subsection{Topological String Theory}
\label{Topological String Theory}

In this part of the discussion, we will concentrate on our main example, namely the Calabi-Yau sigma model with \(c=9\). We will be very schematic on this subject, since this is not our main topic of interest and also because there exists already a fair amount of excellent  review articles on the topic. See for example 
\cite{Bershadsky:1993cx,Neitzke:2004ni,Vonk:2005yv}.

As we have just discussed, a field \(\f_{(c,a)}\) in the (c,a) chiral-anti-chiral ring satisfies
\be
(G^+ + \til{G}^-) \f_{(c,a)}\sim 0
\ee 
and accounts for the deformation of the K\"ahler moduli in the target Calabi-Yau space. Furthermore the operator which annihilates it satisfies the nilpotency condition \((G^+ + \til{G}^-)^2 \sim 0 \). Similarly a chiral-chiral primary field \(\f_{(c,c)}\) satisfies
\be
(G^+ + \til{G}^+) \f_{(c,c)}\sim 0
\ee 
with \((G^+ + \til{G}^+)^2 \sim 0 \) and accounts for the deformation of the complex structure moduli. 

A question one might ask now is: since we will be mainly interested in this part of the spectrum, why not use the nilpotent operators as BRST operators and focus on the BRST cohomology? But we need one more step before this can be done, since \(G^\pm\) and \(\til{G}^\pm\) have conformal weights \(3/2\) but we need fields of weight \(1\) so that we can integrate them around a loop on the world-sheet to get a conserved charge. This is where the spectral flow property of the theory comes to help. By introducing a coupling term 
\be
\pm\, \frac{1}{2} \int_\S \w \, J  = \pm\, \frac{i\sqrt{3}}{2}\int_\S \w \, \pa H  \;,
\ee
with \(\w\) being the spin connection, we can ``twist" the theory by 
\be
T \rightarrow T  \mp \frac{1}{2} \pa J \;,
\ee 
and especially
\be
L_0 \rightarrow  L_0 \pm\frac{1}{2}  J_0\;.
\ee

This twist shifts one of the two \(G^{\pm}\) to have dimension (1,0) and the other dimension (2,0), depending on the sign of the twist. 
Choosing the opposite (same) sign for the left- and right-movers, we obtain the so-called A- (B-) model topological string theory, with the original (c,a) ((c,c)) ring as now the BRST cohomology. This is summarised in Table \ref{top_str_table}. 

 \begin{table}\label{top_string}
 \begin{tabular}{ccc} 
 \toprule
top. string&  cohomology & deformation \\
 \midrule
B-model & \(H^{2,1}(M) \)& complex structure\\
A-model & \(H^{1,1}(M) \)& K\"ahler form\\
 \bottomrule
 \end{tabular}
 \caption{\label{top_str_table}\footnotesize{Summary of the relation between BRST cohomology of the A- and B-model topological string theory and the cohomology class of the target space.}}
 \end{table}

It is not difficult to check that the energy-momentum tensor is Q-exact, which means the twisted theory is invariant under a continuous change of world-sheet metric and therefore the name ``topological".  But this is not yet the whole story. In order to get an interesting theory we still have to couple it to ``topological gravity" on the world-sheet \cite{Witten:1988xj}, namely to sum over classes of conformally inequivalent metrics. For concreteness let's now focus on a \(L_0 \rightarrow L_0 -  \frac{1}{2}  J_0\) twist. In this case the BRST-charge \(Q = \oint G^+  \) and \(G^-\) satisfies \( T \sim \{ Q, G^-\}\) and \(J G^\pm = \pm G^\pm \), therefore the U(1) current now plays the role of the conserved current for the ghost number and \(G^-\) that of the anti-ghost. Coupling to topological gravity in this case then follows in close analogy with the procedure of computing higher genera amplitudes of bosonic string theory. We refer the reader who needs more background on this topic to, for example,  \cite{Polchinski}.  Recall that the higher-genus vacuum has anomalous ghost number \(-3\chi = 6g-6\) \cite{Witten:1992fb} due to the presence of a non-trivial moduli space for the genus-\(g\) Riemman sueface, which has dim\(_\C {\cal M}_g = 3g-3\). We need therefore \(3g-3\) insertions of anti-ghost on each (left- and right-moving) sector to produce a ghost-neutral amplitude. Or said in another way, to produce the correct measure factor for the moduli space. Therefore we define the genus-\(g\) amplitude for topological strings to be
\be
F_g = \int_{{\cal M}_g}\,\langle \prod_{i=1}^{3g-3}  G^-(\m_i)\,\til{G}^\pm(\m_i)\,\rangle \;,
\ee
where the plus (minus) sign corresponds to the A- (B-)model and \(\m_i\) stands for the Beltrami differential. Now we are ready to define the (perturbative) topological strings partition function as 
\be\label{def_top_pf}
{Z_{\text{\tiny{top}}}} := \exp\,(F_{\text{\tiny{top}}})=  \exp\left(\sum_{g=0}^\inf \, g_{\text{\tiny{top}}}^{2g-2} \, F_g\right)\;,
\ee
where \(g_{top}\) will be referred to as the topological string coupling constant. 

For future use let's also discuss here the expansion of the above partition function. As we have mentioned above, the A-model partition function is, loosely speaking, a function of the K\"ahler moduli \(t^A = B^A+ iJ^A\), in a notation that will be explained in more detail later (\ref{coord_special_kah}).
Around the semi-classical limit \(t\rightarrow \inf\), or the target space large-volume limit, the A-model free energy has the following expansion\cite{Pioline:2006ni,Denef:2007vg}\footnote{Note that we omit all the perturbative terms and MacMahon factors in our definition of \(Z_{\text{\tiny{DT}}}\) and \(Z_{\text{\tiny{GV}}}\).}
\begin{align}\displaybreak[3]
\label{top_free_energy}
F_{\text{\tiny{top}}} &= F_{\text{\tiny{pert}}} +F_{\text{\tiny{GW}}}^{(0)} +F_{\text{\tiny{GW}}} \\ \label{top_pert}
 F_{\text{\tiny{pert}}} &= -i\,\frac{(2\p)^3}{6 g_{\text{\tiny{top}}}^2 }  D_{ABC}\, t^At^Bt^C - \frac{2\p i}{24} c_{2A} t^A \\ \notag
 \label{top_pert}
 F_{\text{\tiny{GW}}}^{(0)}  &= -\frac{1}{2} \chi \, \log [M(\!e^{-g_{\text{\tiny{top}}}}\!) ]\\ \notag
F_{\text{\tiny{GW}}}  &=  \sum_{\b\in H^2(X,\Z)} \sum_{g\geq0}  N_{g,\b}  \,g_{\text{\tiny{top}}}^{2g-2}\, e^{2\p i \b\cdot t} =F_{\text{\tiny{GV}}}
=\log Z_{\text{\tiny{GV}}}\\ \notag
 &= \sum_{\b\in H^2(X,\Z)} \sum_{g\geq 0}  \sum_{m\in \mathbb{N}} \a^{\b}_g \, \frac{1}{m} \, \left( 2i \sinh(\frac{mg_{\text{\tiny{top}}}}{2}) \right)^{2g-2} 
\,e^{2\p i m \b\cdot t} \\ \notag
&=  \log Z_{\text{\tiny{DT}}} \\ \notag
Z_{\text{\tiny{DT}}} &=  \sum_{\b\in H^2(X,\Z)} \sum_{m\in\Z} 
n_{\text{\tiny{DT}}}(\b,m) \, (-e^{-g_{\text{\tiny{top}}}})^m \, e^{2\p i  \b\cdot t}\;,
\end{align}
where 
\be \label{macmahon}
M(q) = \prod_{n\geq1} (1-q^n)^n
\ee 
is the MacMahon function and \(\chi=\chi(X)\) is the Euler characteristic of the target Calabi-Yau space \(X\). \(N_{g,\b}\) is called the Gromov-Witten invariants which are rational numbers counting holomorphic curves, while the \(\a^{\b}_g\) and \(n_{\text{\tiny{DT}}}(\b,m)\), called the Gopakumar-Vafa \cite{GV1,GV2} and the Donaldson-Thomas invariants respectively, have the physical interpretation of counting wrapped M2 branes and D6-D2-D0 bound states respectively. 

Especially, one can show from the above formula that the Gopakumar-Vafa partition function \(Z_{\text{\tiny{GV}}}\) takes the following suggestive product form \cite{GV2,Klemm:2004km}
\begin{align}\notag
Z_{\text{\tiny{GV}}}(g_{\text{\tiny{top}}}, t) =& \prod_{\b\in H^2(X,\Z)}\,\Biggl(\, \prod_{r=1}^\inf\,\Big(1-e^{-rg_{\text{\tiny{top}}}}e^{2\p i \b\cdot t}\Big)^{r\a_0^\b} \\ \label{Z_GV}
& \times \prod_{g=1}^\inf \prod_{\ell=0}^{2g-2}\, 
\Big(1-e^{-(g-\ell-1)g_{\text{\tiny{top}}}}e^{2\p i \b\cdot t}\Big)^{(-1)^{g+\ell} \big(\begin{smallmatrix}2g-2\\ \ell\end{smallmatrix}\big) \a_g^\b}
 \Biggr)\;,
\end{align}
and renders itself intelligible as the partition function of second quantised M2 branes.
Finally we remark that the equivalence of Gromov-Witten partition function with the Donaldson-Thomas partition function is a partially proven conjecture.  See \cite{DVVafa,GWDT1,GWDT2,foam}
for relevant discussions. 
\index{Gopakumar-Vafa invariants}
\index{Donaldson-Thomas invariants}
\index{Gromov-Witten invariants}
\index{topological string partition function}

\subsection{Elliptic Genus and Vector-Valued Modular Forms}
\label{Elliptic Genus and Vector-Valued Modular Forms}

\index{elliptic genus}
Elliptic genus is a useful tool to obtain structured and controllable information about the spectrum of a conformal theory with \((2,2)\) world-sheet supersymmetry\cite{Witten:1993jg,Kawai:1993jk}. It is defined as
\be\label{def_elliptic_genus}
\chi(\t,z) = \Tr_{\text{\tiny{RR}}}\,(-1)^F \, e^{2\p i z J_0}\, e^{2\p i \t (L_0-\frac{c}{24})}\, e^{-2\p i \tbar (\til{L}_0-\frac{c}{24})} \;,
\ee
where \((-1)^F = e^{\p i (J_0 + \til{J_0})}\), and the subscript denotes the fact that the trace should be taken with the R-R boundary condition. It has also the interpretation as a path integral of the theory on the torus with appropriate U(1) coupling to ``label" the left-moving R-charge. This Wilson line coupling is absent on the right-moving side. 
From the \([\til{J}_0,\til{G}^\pm_0]\) and \([\til{L}_0,\til{G}^\pm_0]\) commutation relations we see that only states annihilated by \(\til{G}^\pm_0\), namely states with the right-movers at their R-ground states, contribute to the index. Supersymmetry therefore guarantees its rigidity property which often  renders it computable. A special case of this is the Witten index \(\chi(\t,0)= \Tr_{\text{\tiny{RR}}}\,(-1)^F \). 

Furthermore, the invariance of the spectrum under the spectral flow has interesting implications for the elliptic genus. To shorten the equations we will use in this part of the discussion the symbols
\ben
\hat{c} &=& \frac{c}{3}\\
q &=& e^{2\p i \t}\quad;\quad y = e^{2\p i z} \\
{\ex}[x] &=& e^{2\p i x}\;.
\een
A short manipulation of (\ref{def_elliptic_genus}) using the spectral flow relation (\ref{spectral_flow}) shows that the elliptic genus has the following two properties.
First of all 
\be\label{elliptic_1}
\chi(\t,z+\ell \t + m) = e^{-\p i \hat{c} (\ell^2 \t + 2\ell z)}\, \chi(\t,z)\;,
\ee
and secondly 
\bea\label{theta_decomposition1}
\chi(\t,z) &=& \sum_{\m=-\frac{\hat{c}}{2}}^{\frac{\hat{c}}{2}-1} h_\m(\t)\,\th_\m(\t,z)\\
h_\m(\t) &=& \sum_{n\in \Z_+} c_\m(n)\, q^{-\frac{1}{2\hat{c}}\m^2 + n}
= \sum_{n\in \Z_+} c(n- \frac{\m^2}{2\hat{c}})\, q^{n-\frac{1}{2\hat{c}}\m^2}\\
\label{theta_decomposition3}
\th_\m(\t,z)&=& \sum_{\ell \in \Z} q^{\frac{\hat{c}}{2} (\ell+\frac{\m}{\hat{c}})^2}y^{(\m+\hat{c} \ell)}\;.
\eea
The restriction \(\m \in (-\frac{\hat{c}}{2},\frac{\hat{c}}{2}]\) can be understood as a consequence of the fact that one unit of spectral flow, namely \(\eta = 1\) in  (\ref{spectral_flow}), shifts the U(1) charge by a unit of \(\hat{c}\). 

From the above equation we see that the elliptic genus of a unitary (2,2) CFT has the following form
\be \label{elliptic_2}
\chi(\t,z)  = \sum_{n\in \Z_+, \,\ell+\frac{\hat{c}}{2} \in\Z} c(n,\ell) q^{n}\,y^{\ell} = 
 \sum_{n\in \Z_+, \,\ell+\frac{\hat{c}}{2}  \in\Z} c(n- \frac{\ell^2}{2\hat{c}}) q^{n}\,y^{\ell}\,.
\ee
Notice that \(L_0 - \frac{1}{2\hat{c} }J_0^2 \) is indeed the (up to a multiplicative factor and addition of a constant) unique combination linear in \(L_0\) that is invariant under the spectral flow (\ref{spectral_flow}). Especially, from Figure \ref{spectral_flow_pic} and recalling that \(n=L_0 -\frac{c}{24}\), we see that the coefficients \(c(n- \frac{\ell^2}{2\hat{c}}) =0\) for \(n- \frac{\ell^2}{2\hat{c}} < -\frac{c}{24}\). The functions \(h_{\m=-\frac{\hat{c}}{2}}(\t)\) has therefore \(q\)-expansion beginning from \(q^{-\frac{c}{24}}\). We will call the part of the elliptic genus with negative arguments for \(c(n- \frac{\ell^2}{2\hat{c}})\) the {\it polar part}
\begin{align}
\chi^-(\t,z)  &= \sum_{ n- \frac{\ell^2}{2\hat{c}}< 0} c(n- \frac{\ell^2}{2\hat{c}}) q^n y^\ell\;, 
\intertext{
or equivalently}\label{polar_part_intro}
h^-_\m(\t) &=  \sum_{n- \frac{\ell^2}{2\hat{c}} < 0 } c(n- \frac{\ell^2}{2\hat{c}})\, q^{n-\frac{1}{2\hat{c}}\m^2}\;.
\end{align}
The physical and mathematical significance of this special ``polar part" of the elliptic genus  will be seen later when we discuss the elliptic genus for the black hole CFT in section \ref{The (Modern) Farey Tail Expansion}.

This nice structure brought to us by the spectral flow, or related to it the spacetime supersymmetry, can be made more transparent by thinking about the action of an one-dimensional free Abelian group, namely an one-dimensional lattice, endowed with a \(\Z\)-valued bilinear form
\((\ell,\ell) = \hat{c} \ell^2\) (recall that  \(\hat{c}\) is an integer for a (2,2) non-lieanr sigma model on a Calabi-Yau manifold of any dimension). Let's call this lattice \(\L\). The dual lattice \(\L^*\) is then defined as the lattice of all vectors in \(\L\otimes_\Z \R\) with integral inner products with all vectors in the original lattice \(\L\). In the case at hand, in the integral basis of the one-dimensional lattice \(\L^*\), we see that the lattice \(\L\) is the lattice of the points \(\hat{c}\,\Z\). In this basis in which all points in  \(\L^*\) have integral coefficients, the bilinear form becomes
\be
(\m|\m) = \frac{1}{\hat{c}} \m^2\;.
\ee
We can now rewrite (\ref{theta_decomposition1})-(\ref{theta_decomposition3}) as
\bea\label{(2,2)_elliptic_genus_decomposition}
\chi(\t,z) &=& \sum_{\m\in \L^*/\L} h_\m(\t)\,\th_\m(\t,z)\\
\th_\m(\t,z)&=& \sum_{\l \in \m + \L} q^{\frac{(\l|\l)}{2}} y^{\l}\\& =&
{\ex}[-\frac{\hat{c}}{2}\frac{z^2}{\t}]\,\sum_{\l \in \m + \L} {\ex}[\frac{\t}{2}\big(\l+\frac{z}{\t}\hat{c}\,\big\lvert\l+\frac{z}{\t}\hat{c}\big)]\;.
\eea
From the last expression and using the Poisson resummation formula  we get the modular transformation of the above theta-function
\be\label{modular_tran_theta}
\th_\m(-\frac{1}{\t},\frac{z}{\t}) = {\ex}[\frac{\hat{c}}{2}\frac{z^2}{\t}]\,\sqrt{\frac{1}{\hat{c}}}\,\sqrt{-i\t} \sum_{\m'\in \L^*/\L} {\ex}[(\m|\m')]\,
\th_{\m'} (\t,z)\;.
\ee
We would like to know the modular property of the full elliptic genus as well. The \(z\)-independent part of the transformation must be compensated by the transformation of \(h_\m(\t)\), since the Witten index \(\chi(\t,0)=(-1)^F =\chi(X) \) is clearly modular invariant. Using this fact and the R-charge conjugation symmetry of the CFT, we get the following transformation rule 
\be\label{elliptic_3}
\chi(\frac{a\t+b}{c\t+d},\frac{z}{c\t+d}) = {\ex}[\frac{\hat{c}}{2}\frac{c z^2}{c\t+d}]\, \chi(\t,z)\;.
\ee
This can also be understood as the modular transformation of the path integral with an extra coupling  to the U(1) current.  The property (\ref{elliptic_1}), (\ref{elliptic_2}), (\ref{elliptic_3}) is exactly the definition of a weak Jacobi form of index \(\hat{c}/2\) and weight zero. The elliptic genus of a (2,2) SCFT  therefore enjoys many special properties of a weak Jacobi form. Here we will list one important fact that will be needed later. 

The space \(J_{2*,*}\) of all weak Jacobi forms of even weight and any index is known to be a ring of all polynomials in the following four functions \({\f_{0,1}}(\s,z)\),
\({\f_{-2,1}}(\s,z)\), \(E_4(\t)\),\(E_6(\t)\) \cite{EicZ,Feingold-Frenekel}, where \(E_4, E_6\) are the usual Eisenstein series and \(\f_{0,1},{\f_{-2,1}}\) are the weak Jacobi forms of index one and weight \(0\) and \(-2\) given by
\ben
{\f_{0,1}}(\t,z)& =& \frac{{\f_{12,1}}(\t,z) }{\D(\t)} = y^{-1} +10 + y + {\cal O}(q) \\ \notag
{\f_{-2,1}}(\t,z)& =& \frac{{\f_{10,1}}(\t,z) }{\D(\t)} = y^{-1} - 2 + y + {\cal O}(q) \\ \notag
{\f_{12,1}}(\t,z)&=& \frac{1}{144} \big(E_4^2 E_{4,1} -E_6 E_{6,1} \big)\\ \label{ring_Jacobi_form}
{\f_{10,1}}(\t,z)&=& \frac{1}{144} \big(E_6 E_{4,1} -E_4 E_{6,1} \big)
\een
and the discriminant is 
\be
\D = \eta^{24} = \frac{E_4^3 - E_6^2}{1728}\;.
\ee

This gives great constraint of the form the elliptic genus can take. For example, the space of weak Jacobi forms of weight zero and index one is one-dimensional. Together with the fact that \(\chi(\t,z=0) = \chi(K3)=24 \) for the case of Calabi-Yau two-fold this completely fixes 
the K3 elliptic genus to be 
\be\label{K3_elliptic}
\chi_{K3} (\t,z) = 2\f_{0,1} (\t,z) =\sum_{n\in\Z_+, \ell \in \Z}
c(4n-\ell^2) q^n y^\ell= 2y^{-1} +20 + 2y + {\cal O}(q) \;.
\ee

For the properties of elliptic genus with Calabi-Yau manifolds as the target space, see \cite{Gritsenko1,Kawai:1993jk}. See also \cite{Dijkgraaf:1996xw,Gritsenko1} for a geometric definition of the elliptic genus without reference to a superconformal theory.

\index{polar part}
\index{Poisson resummation}
\index{theta function}

The above lattice formulation can be readily generalized to higher dimensional lattices, with possibly not positive definite bilinear forms. For future use we now digress briefly to discuss them.

For a lattice \(\L\) with a non-degenerate bilinear form of signature \((\s^+,\s^-)\) we can define a modular form with values in  the group ring \(\C[\L^*/\L]\). In other words, in analogy with the concept of a modular form for an one-dimensional lattice, we have vector-valued modular forms in the higher-dimensional cases.  
Here we will only discuss them in terms of their components \(h_\m\) for each vector  \(\m\in\L^*/\L\).

The novel property is that, for a negative-definite lattice, convergence of the series requires that the theta-functions can be written as a sum of \({{\ex}[\frac{\tbar}{2} (x|x)}]\) instead of \({{\ex}[\frac{\t}{2} (x|x)}]\). For a lattice with mixed signature we therefore have both holomorphic and anti-holomorphic  couplings. 

\index{Siegel theta function}
\begin{definition}

For a lattice \(\L\) with a non-degenerate bilinear form \((|)\) and signature \((\s^+,\s^-)\) , given a maximally positive definite subspace in \(\L\otimes \R\), namely, given an element in the Grassmannian \(v\in G(\s^+,\s^-)\), we define the (Siegel or Siegel-Narain) theta-function as
\begin{align}\label{Siegel theta function}
&\th_\m(\t;\a,\b) = \sum_{\l\in\m+\L} {\ex}[\frac{\t}{2}  (\l+\b)_+^2 + \frac{\tbar}{2}  (\l+\b)_-^2  - (\l+\frac{\b}{2}|\a)]\\ \notag
&\text{for}\qquad \m\in \L^*/\L\qquad,\qquad \a,\b\in \L\otimes \R
\;,
\end{align}
where \(x^2:=(x|x)\) and notice that the projection into the positive and negative part depends on \(v\in G(\s^+,\s^-)\). 
\end{definition}

Using the higher-dimensional version of Poisson resummation, one can show that the modular property of the above theta-function is
\be\label{modular_property_Siegel_theta}
\th_\m(-\frac{1}{\t} ;-\b,\a) = \frac{1}{\sqrt{|\L^*/\L|}}\,(\sqrt{-i\t})^{\s^+}
(\sqrt{i\tbar})^{\s^-} \sum_\n {\ex}[-(\m\,|\n)]\, \th_\n(\t;\a,\b)\;,
\ee
where the pre-factor \(\sqrt{|\L^*/\L|} = |\text{Vol}(\L)|\) is the volume of the unit cell of the lattice. For the details of this computation and the generalisation to degenerate lattices, see \cite{Borcherds1}.

\index{weak Jacobi form}
\index{dual lattice}
\index{vector-valued modular forms}

\subsection{Mirror Symmetry and Non-perturbative Effects}

The alert readers might have already noticed that, the differentiation between the (c,c) and the (c,a) ring is rather {\it ad hoc} in our analysis of the chiral ring structure in the Calabi-Yau non-linear sigma model. And in fact also as a warning to the reader, the conventions  do vary in the literature. There is of course nothing to stop us from flipping the sign of the R-current on one (let's say the right-moving) side and exchanging what we call \(G^+\) and \(G^-\) while keeping the left-movers untouched. Although a trivial isomorphism from the world-sheet point of view, it implies something rather drastic on the geometric side. Namely it exchanges what we call the \(H^{n-r,s}\) and \(H^{r,s}\) cohomology classes, where \(n\) is again the complex dimension of the Calabi-Yau manifold. In particular, the Euler character changes its sign for three-folds, as  \(\chi(\text{CY}_3) = 2 ( h^{1,1} - h^{1,2}) \rightarrow -\chi\).

But since the world-sheet theory remains the same, there must be a pair of Calabi-Yau three-folds with exchanging \(H^{3-r,s}\) and \(H^{r,s}\). This symmetry is called the ``mirror symmetry". Geometrically, the easiest way to think about this symmetry is to think of the Calabi-Yau pair as a \(T^3\) fibration over a three-real-dimensional base space, with possibly singular fibre at various points. The mirror symmetry is then implemented by doing three T-dualities along the fibre directions. As we have discussed in the previous chapter, a T-duality exchanges type IIA with type IIB superstring theory. This mirror symmetry must therefore also exchange type IIA and IIB strings living on the Calabi-Yau space. In other words, given a mirror pair \((X,Y)\) of Calabi-Yau three-folds with
$$
h^{2,1}(X) = h^{1,1}(Y)\qquad;\qquad h^{2,1}(Y) = h^{1,1}(X) 
$$
 and a pair of string theories IIA and IIB, there are only two instead of four independent theories one can write down. They are, schematically

\begin{align} \notag
\left(\;\text{IIB }/ X \;\right)\qquad &\simeq \quad\left(\;\text{IIA }/ Y \;\right) 
\intertext{and} \notag
\left(\;\text{IIB }/ Y \;\right)\qquad &\simeq \quad\left(\;\text{IIA }/ X \;\right) \;.
\end{align}

Later we will also see that there is indeed a  corresponding relationship between the spacetime low-energy effective theories.

Combining with different properties of the complex structure and the K\"ahler moduli space, mirror symmetry has been very useful in predicting properties of the theory in different parts of the moduli space, relating classical and quantum geometry. In particular, it predicts that the conformal field theory is well-behaved on the special submanifold in the moduli space where classical geometric intuitions fail, namely when the internal manifold goes through the so-called flop transitions. This is where we again see how strings are superior to point particles as probes for spacetime. But in the following analysis we will concentrate on parts of the moduli space where these non-perturbative effect do not occur. In particular, from now on we will stay well inside the K\"ahler cone and away from conifold points. In other words, we will only consider the part of the moduli space where all the homology cycles are ``large" in the string unit. 

\section{Spacetime Physics}
\setcounter{equation}{0}

\subsection{Moduli Space and Special Geometry}

As we discussed earlier, the (2,2) SCFT's naturally come in families, related to each other by marginal deformations of the theory. We have also seen that these deformations, in the non-linear sigma model case, have the interpretation of deforming the target Calabi-Yau space. Indeed Calabi-Yau manifolds also come in families, meaning we can continuously deform the size and the shape of these internal manifolds without changing the topological properties of the space. As we have noticed in our discussion of the chiral ring, we can separate these deformations into two kinds. First is the shape, or the complex structure deformation, corresponding to \(\d g_{ij}\sim f_{i\bar{k}\bar{\ell}}\,\,\O^{\bar{k}\bar{\ell}}_{\;\;\;j}\), given by the harmonic three-form \(f^{(3)}\in H^{1,2}(X,\C)\) and using the (up to a factor) unique (3,0)-form of the Calabi-Yau \(X\) to contract the indices. Second is the size, or the K\"ahler deformation. Combining it with the NS-NS two-form potential we get \(\d B_{i\jbar} + i \d g_{i\jbar} \sim f_{i\jbar}\), where \(f^{(2)} \in H^{1,1}(X,\C)\). The requirement that the deformation should be given by harmonic forms can be understood as the preservation of the Ricci-flatness of the metric. 
We have also argued that the moduli space is locally a direct product of these two separate moduli spaces. We will therefore study them separately now.

\index{complex structure moduli}
\index{K\"ahler moduli}
\index{special geometry}
\index{special K\"ahler manifold}

\subsubsection{Complex Structure Moduli}

First we look at the complex structure moduli. A change in the complex structure means a different decomposition of the tangent (or equivalently the cotangent) bundle into holomorphic and anti-holomorphic part. Especially, under an infinitesimal change of complex structure, the Calabi-Yau (3,0) form \(\O\) becomes a linear combination of a (3,0) and a (2,1) form. We can therefore think of \(\O\) as a section of a \(H^{3}(X,\C)\) bundle over the moduli space \({\cal M}_{\text{\tiny{complex}}}\). We will call it the Hodge bundle \({\cal E}^{\text{H}}\). This bundle is naturally endowed with the following symplectic structure \(\langle\,,\rangle: H^{3}(X,\C) \times H^{3}(X,\C) \rightarrow \C\), given by
\be
\langle \G_1, \G_2 \rangle  = -\langle \G_2, \G_1 \rangle =\int_X \G_1 \wedge \G_2 \;,
\ee
which has the geometric interpretation as the intersection number of the dual three-cycles of \(\G_1\) and \(\G_2\). 
Furthermore, it defines a natural hermitian metric on \(H^{3}(X,\C)\)
$$
(\G_1,\G_2 ) = i \langle \G_1, \overline{\G}_2 \rangle = (\G_2,\G_1 )^*\;.
$$
But this is not yet the full story. Since the complex structure \(\O\) is only defined up to a constant, or said differently, a rescaling of it will change the section in \({\cal E}^{\text{H}}\) while it really doesn't mean a change of the complex structure of \(X\), we should introduce a line bundle \({\cal L}\) to account for the redundancy. Using the above hermitian metric, is now natural to define the metric on \({\cal L}\) to be \(e^{{\cal K}}\), where
$$
{\cal K} = -\log (\O,\O)= -\log \left( i \langle \O , \bar{\O} \rangle\right)\;.
$$
In other words, due to the extra rescaling symmetry, the complex structure three-form \(\O\) should be thought of a section of  \({\cal E}^{\text{H}} \otimes {\cal L}\).

When the (3,0)-form is rescaled as \(\O \rightarrow e^{f}\,\O \) with a local holomorphic function \(f\), we see that the ``K\"ahler potential" \({\cal K}\)  scales like \({\cal K}\rightarrow {\cal K}  -f - \bar{f}\). In particular, this means that a K\"ahler metric invariant under a local rescaling of the (3,0)-form can be defined on the moduli space \({\cal M}_{\text{\tiny{complex}}}\) using \({\cal K}\) as the K\"ahler potential. See appendix \ref{Mathematical Preliminaries} for properties of K\"ahler manifolds.
The resulting K\"ahler metric is sometimes called the Weil-Peterson metric. \index{Weil-Peterson metric} 

Furthermore, the connection  
\be \label{connection_complex_structure}
\nabla = \pa + \pa {\cal K}\ee 
satisfies the desired property \(\nabla (e^f \O) = e^f \nabla \O\). In particular, it should take value in \(H^{2,1}(M,\C)\). Finally, it follows from the pairing of the cohomology classes that 
\be\label{papo_1}
\langle \pa \O , \O\rangle = 0\;, 
\ee
since \(\pa\O\) has only a \((3,0)\)- and a \((2,1)\)-form part.
These properties mean that  \({\cal M}_{\text{\tiny{complex}}}\) is a (local) special K\"ahler manifold defined in (\ref{special_K_def}).

Since a rescaling of \(\O\) does not have any physical significance, it will be convenient to define a ``unit vector"
\be \label{def_bo}
\bO = \frac{\O}{\sqrt{i\langle \O, \bar{\O}\rangle}} = e^{{\cal K}/2}  \,\O\;,
\ee
satisfying 
$$
(\bO,\bO) =  i\langle \bO, \bar{\bO}\rangle =1\;,
$$
and a ``central charge" function \(Z: H^3(X,\Z) \times {\cal M}_{\text{\tiny{complex}}} \rightarrow \C\), whose name will be justified later, as 
\be\label{central_charge_1}
Z(\G;\O) = \langle \G, {\bO} \rangle\;,
\ee
which in the present case is just \( \int_\G \bO\), where we have used the same symbol \(\G\) for the Poincar\'e dual of the three-form \(\G\) and \(\bO\) for the pull-back of \(\bO\).

From the behaviour of \(\bO\) under a K\"ahler transformation, we see that \(\bO\), and therefore \(Z\), are sections of \({\cal L}^{1/2} \otimes \bar{{\cal L}}^{-1/2}  \) and have therefore the following covariant derivatives
\be\label{cov_normalised}
{\cal D}\bO = (\pa + iQ )\bO \;,
\ee
where the connection is given by
\be\label{connection_normalised}
iQ = \frac{1}{2} \, (\pa {\cal K} - \overline{\pa}{\cal K}) = i \im(\pa {\cal K})\;.
\ee

\subsubsection{In Coordinates}

After this rather abstract derivation of the special K\"ahler properties of \({\cal M}_{\text{\tiny{complex}}}\), to have a better feeling of what is really going on let's now un-package the information by choosing a local coordinates on \({\cal M}_{\text{\tiny{complex}}}\). Of course, writing things out in components in a local coordinate system does not bring new information. The main reason for the following formulation is really to make it easier for the readers to connect to the existing literature.

For this purpose we have to choose a basis for the middle-cohomology \(H^{3}(X,\Z)\) of the Calabi-Yau space \(X\), namely a coordinate system for the fibre of the Hodge bundle \({\cal E}^{\text{\tiny{H}}}\). From the anti-symmetric intersection of the homology \(H_3(X,\Z)\) we see that we can always choose a real basis \((\a_I , \b^J)\) for \(H^{3}(X,\Z)\), and the corresponding basis \((A^I, B_J)\) of  \(H_{3}(X,\Z)\), which is orthonormal in the following sense 
\begin{align} \label{choice_basis}
\langle \a_I , \b^J  \rangle  &= -\langle  \b^J,\a_I   \rangle = \int_X \a_I \wedge \b^J = \d^J_I  = {\#} (A^I\cap B_J)\\ \notag
\int_{A^I} \a_J  &= - \int_{B_J} \b^I = \d^I_J \quad; \quad
\int_{A^I} \b^J  = \int_{B_J} \a_I = 0\\ \notag
& I,J = 0, ... , h^{2,1} \;.
\end{align}

We can then write the (3,0)-form in terms of these coordinates 
as\footnote{Warning: Some authors use the normalized three-form \(\bO = X^I \a_I  - F_I \b^I\) as the definition for \(\big(\begin{smallmatrix} X^I \\ F_I \end{smallmatrix}\big) \), corresponding to the ``gauge choice" \(2\im(X^I\bar{F}_I) = 1\).}
\be \label{def_O_IIB}
\O = - X^I \a_I  + F_I \b^I\;.
\ee
These coordinates then determine the complex structure. Actually it over-determines it. To see this, recall we said earlier that an infinitesimal change of complex structure turns the (3,0)-form into a linear combination of a (3,0)-part and a (2,1)-part, and the dimension of the moduli space is therefore  \(\text{\small dim}_\C{\cal M}_{\text{\tiny{complex}}} = h^{2,1}\) and not \(2h^{2,1}+2\). Together with the fact that the overall scale does not have a physical significance, which is the reason we why introduced the line bundle \({\cal L}\) in the above construction, we can regard \(X^I\)'s as the homogeneous coordinates for the projective space of infinitesimal variation of the complex structure, and \(F_I\) as \(F_I(X)\). Locally, where \(X^0\) does not vanish, we can then use the coordinates, called the ``special coordinates", defined as
$$
t^A = \frac{X^A}{X^0}\qquad;\qquad A = 1, \dotsi, h^{2,1}\;,
$$
as the coordinates of the complex structure moduli space \({\cal M}_{\text{\tiny complex}}\). But as we will see later, it will often be useful to work with the homogeneous coordinates \(X^I\) instead. 

\index{special coordinates}

In these coordinates, the K\"ahler potential reads
$$
e^{-{\cal K} }= 2\im(X^I \bar{F}_I) \;,
$$
and the K\"ahler metric is given by
\be\label{Kahler_metric_1}
g_{I\bar{J}} = \pa_I \delbar_{\bar{J}} {\cal K}  
= -\frac{i\langle \nabla_I \O , \bar{\nabla}_{\bar{J}} \bar{\O} \rangle }{i\langle\O, \bar{\O}\rangle}\,= -i \langle {\cal D}_I \bO, \bar{\cal D}_{\bar{J}} \bar{\bO}\rangle.\ee

The condition (\ref{papo_1}) which follows from the cohomology pairing and which is a part of our definition of special K\"ahler geometry (\ref{special_K_def}), gives in these coordinates 
$$
\langle \pa_I \O,\O \rangle = -F_I + X^J \pa_J F_I = 0 \;,
$$
which implies that \(F_I\) is  homogeneous of degree 1 as a  function of \(X\)'s. We can therefore define
$$
F(X) = \frac{1}{2} X^I F_I \;,
$$
and it's easy to check that it satisfies
\bea\notag
\pa_I F(X) &=& F_I \\ \label{prepo_1}
X^I \pa_IF(X) &=& 2F(X)\;.
\eea
This homogeneous function \(F\) of degree 2  is the so-called ``prepotential". Specifically, for later use we define 
$$
F_{IJ}= \pa_I  \pa_J F(X)\,,
$$
then
\be \label{prepo_2}
F_I = F_{IJ}\,X^J\;.
\ee
\index{prepotential}

Notice that our choice of basis (\ref{choice_basis}) is only fixed up to a symplectic transformation. 
To see this, note that in our ``orthonormal" basis, the symplectic form on \(H^{3}(X)\simeq H_3(X)\) in the matrix representation is just 
$$
\bem 0 & \mathds{1} \\ -\mathds{1} & 0
\eem \;,
$$
where \( \mathds{1} \) is the \((h^{1,2}+1)\times (h^{1,2}+1)\) unit matrix.
Then an integral change of basis  given by \(Sp(2h^{2,1}+2,\Z) \), namely matrices 
$$
 \bem A& B \\C&D  \eem
$$
with \(A,B,C,D\) being \((h^{2,1}+1)\times(h^{2,1}+1)\) matrices satisfying
\ben
A^T C- C^T A &=& 0 \\  \notag
B^T D - D^T B &=& 0 \\ \notag
A^T D -C^T B &=& \mathds{1}\;,
\een
always leaves the above symplectic matrix invariant. 

Under the above change of basis, the coordinates \(X^I\) and \(F_I\) transform as a symplectic vector
$$
V = \bem X^I \\ F_I \eem \rightarrow  \bem A& B \\C&D  \eem \,\bem X^I \\ F_I \eem\;.
$$
From the fact that \(F(X) = \frac{1}{4} V^T \bigl( \begin{smallmatrix} 0 &  \mathds{1} \\  \mathds{1}& 0 \end{smallmatrix}\bigr)V \) we see that the prepotential is not invariant under the symplectic transformation. This is another reason why we chose to define the special geometry without reference to such a prepotential.

\subsubsection{K\"ahler Moduli}

After discussing the complex moduli we now turn to the K\"ahler Moduli of the Calabi-Yau three-fold. As we mentioned earlier in section \ref{Chiral Ring}, the K\"ahler moduli space of a Calabi-Yau \(X\) is given by its cohomology class \(H^{1,1}(X)\). Recall that in the case of complex structure, we double the space \(H^{2,1}(X)\) into \(H^{2,1}(X) \oplus H^{1,2}(X)\) and further enlarge it with \(H^{3,0}(X) \oplus H^{0,3}(X)\) to construct the symplectic bundle. Here we will do a similar thing. To construct the symplectic bundle for the K\"ahler moduli space \({\cal M}_{\text{\tiny{K\"ahler}}}\), we first double \(H^{1,1}(X)\) into \(H^{1,1}(X) \oplus H^{2,2}(X)\) and further enlarge it with \(H^{0,0}(X) \oplus H^{3,3}(X)\). In other words, we consider a \(H^{2*}(X,\C)\) bundle over \({\cal M}_{\text{\tiny{K\"ahler}}}\). 

Again similar to the complex structure case, we want to employ the symplectic pairing of \(H^{2*}(X,\C)\) and thereby see that \({\cal M}_{\text{\tiny{K\"ahler}}}\) is again a special K\"ahler manifold. But this time the geometric meaning of such a pairing will not be as clear. What must stay true is that the pairing should be between the cohomology classes that are Hodge dual to each other. In order for the pairing to be symplectic, we now define a map
$$
\check{\G}  = (-1)^n \G\qquad\text{for}\quad \G \in H^{(n,n)}(X,\C)
$$
and requires it to act component-wise on a general \(\G \in H^{2*}(X,\C)\). The symplectic product \(\langle\,,\rangle : H^{2*}(X,\C) \times H^{2*}(X,\C) \rightarrow \C\) is then given by
\be\label{symplectic_kahler}
\langle\G_1,\G_2\rangle =-\langle\G_2,\G_1\rangle = \int_X \G_1 \wedge \check{\G}_2\;. 
\ee

The next step will be to find a section which gives the K\"ahler potential  (\ref{special_K_def}). Since a deformation in the K\"ahler moduli changes the complexified (with the B-field) K\"ahler form 
\be \label{coord_special_kah}
t = B + i J \in H^{1,1}(X,\C)
\ee 
to a nearby vector in \(H^{1,1}(X,\C)\), a natural way to build such  a section \(\O\) in the \(H^{2*}(X,\C)\) bundle will be
\be\label{IIA_Omega}
\O = - e^{t} = -\left( 1 + t + \frac{1}{2!} t\wedge t +\frac{1}{3!} t \wedge t \wedge t \right) \;.
\ee 
Then the K\"ahler potential
\be\label{kahler_po_IIA}
e^{-{\cal K}} = i \langle \O , \bar{\O}\rangle = \frac{4}{3} \int J\wedge J\wedge J = 8\, \text{vol}(X)
\ee
is given by the Calabi-Yau volume, and the normalised section is 
\be\label{def_O_IIA_1}
\bO =- e^{{\cal K}/2}\,{e^t} =\frac{-e^t}{\sqrt{\frac{4}{3}J^3}}
\ee
Finally, another condition \(\langle\pa\O,{\O}\rangle =0\) for special K\"ahler manifold is now obviously true, from the simple fact that \(\pa (- e^t )= \pa t\,\O\). We therefore conclude that moduli space \({\cal M}_{\text{\tiny{K\"ahler}}}\) is again a special K\"ahler manifold, and the same properties (\ref{connection_complex_structure})-(\ref{connection_normalised}) we discussed for \({\cal M}_{\text{\tiny{complex}}}\) carry straightforwardly to this case. 

As the readers might have noticed, for convenience 
we have actually already used the special coordinates, given by \(t\in H^{1,1}(X,\C)\), in the definition of (\ref{IIA_Omega}) . To formulate the special geometry in homogeneous coordinates as before, we introduce the basis \(\a_A\) for \(H^{1,1}(X,\Z)\) and the dual basis \(\b^A\) for \( H^{2,2}(X,\Z)\) defined by \(\int_X \a_A\wedge \b^B = \d^B_A\), with \(A = 1,\dotsi , h^{1,1}\). For this basis  we then  define the triple-intersection number of the  dual four-cycles as
\be\label{triple-intersection}
\int_X \a_A \wedge \a_B \wedge \a_C = D_{ABC}\;.
\ee
Furthermore we write \(\a_0 =1\) and \(\b^0 =- \frac{J\wedge J\wedge J}{\int_X J\wedge J\wedge J}\) as the basis for \(H^{0,0}(X,\Z) \) and \(H^{3,3}(X,\Z)\). 

In other words, for the special geometry of the K\"ahler moduli we use 
\begin{align} \notag
\a_0& = 1 \\ \notag
\a_A &\in H^{1,1}(X,\Z)\\ \notag
\b^A &\in H^{2,2}(X,\Z)\\ \label{basis_h2*}
\b^0  &= - \frac{J\wedge J\wedge J}{\int_X J\wedge J\wedge J} \\
\intertext{satisfying}\notag
\langle \a_I , \b^J \rangle &= \d_I^J \qquad,\quad I,J = 0,\dotsi,h^{1,1}
\end{align}
as the symplectic basis of \(H^{2*}(X,\Z)\).

Again, for the convenience of the readers we would like to make contact with the convention involving the prepotential used in a large proportion of the existing literature.

Using the projective coordinates \(X^I\), we can 
 introduce the following section of the even cohomology bundle 
 \be\label{def_O_IIA_2}
\O = -X^0 e^{\frac{1}{X^0}X^A \a_A} = - X^I \a_I + F_I {\b}^I \;,
\ee
which reduces to the definition (\ref{IIA_Omega}) when the special coordinates
$$
\frac{X^A}{X^0}=t^A \qquad;\qquad X^0 = 1 
$$
is used.

In these coordinates the K\"ahler potential is again given by
$$
e^{-{\cal K}} = 2 \im(X^I \bar{F}_I)\;,
$$
and one can check that the prepotential reads
\be \label{prepotential_tree_IIA}
F(X) = \frac{1}{2} F^I X_I = \frac{1}{6} D_{ABC} \, \frac{X^A X^B X^C}{X^0}\;,
\ee
which is again a homogeneous function of degree two.

As we will comment more on it later, while the metric of the complex structure moduli space is exact, that of the K\"ahler moduli space receives perturbative and non-perturbative corrections. Here we can see the resemblance of the leading prepotential for the K\"ahler moduli space  (\ref{prepotential_tree_IIA}) and the leading perturbative topological strings amplitude (\ref{top_pert}). As we will see later, the rest of \(F_{\text{\tiny{top}}}\) will also have its role in the prepotential. Specifically, the symmetry of the shift of the B-field 
$$
\re \,t^A \rightarrow \re\, t^A  + \text{  constant}\;,
$$
known as the PQ (Peccei-Quinn) symmetry, constrains the form of the perturbative correction to the prepotential. But non-perturbatively this symmetry is broken and not observed by non-perturbative corrections.

\index{PQ (Peccei-Quinn) symmetry}

This finished our discussion about the moduli space of Calabi-Yau three-folds. For a point of entry into the literature on special geometry, we refer to \cite{Craps:1997gp,Candelas:1990rm,Strominger:1990pd,deWit:1984pk}, or various reviews \cite{Mohaupt:2000mj,Guica:2007wd,deWit:2007dn,Pioline:2006ni}. 

\subsection{Four- and Five-Dimensional Low Energy Supergravity Theory}
\label{Four- and Five-Dimensional Low Energy Supergravity Theory}

When one considers a Calabi-Yau three-fold as a part of the spacetime, from the decomposition of spinors under the rotation group \(Spin(6)\simeq SU(4)\) into representations of the holonomy group \(SU(3)\) as \({\bf 4}  = \, {\bf 3}\oplus {\bf 1}  \), we see that the manifold admits a ``Killing spinor" satisfying \(\nabla_k\eta=0\). In the absence of a dilaton gradient or a background \(H^{(3)}=dB^{(2)}\) flux, the existence of such a Killing spinor means unbroken spacetime supersymmetry. Furthermore, from the above decomposition we see that a quarter of the supersymmetry is preserved. Starting from a ten- or eleven-dimensional theory with 32 supercharges, this leads to eight remaining supercharges after compactification to four- or five-dimensions.

\subsubsection{Four Dimensional Supergravity}

\index{supersymmetry algebra}
Let's first look at the supersymmetry algebra in four dimensions. The minimal supersymmetry algebra reads
$$
\{Q_\a,\overline{Q}_\b\} = -2 P_\m \G^\m_{\a\b}\quad,\quad [P^\m,Q_\a] = 0
\,,
$$
where \(Q\)'s are Majorana fermions with four degrees of freedom, \(\overline{Q} = Q^\dag \G^0 \), \(\G^\m\)'s are gamma matrices furnishing a representation of the Clifford algebra, \(P_\m\) is the spacetime momentum vector and \(\a,\b\) are the spinor indices. By writing out the gamma matrices explicitly, one can easily see that the \(Q\) anti-commutators give one pair of fermionic creation and annihilation operators when \(P_\m\) is massless (lightlike), while the physical fields must be annihilated by the rest two of the \(Q\)'s. On the other hand, when \(P_\m\) is massive (timelike) the \(Q\) anti-commutators give  two pairs of fermionic annihilation and creation operators and the physical fields are annihilated by none of the \(Q\)'s.

\index{(extended) supersymmetry algebra}

For the case of \(4 {\cal N}\) supercharges, the generalisation of the above supersymmetry algebra is 
$$
\{Q_{\a}^A ,\overline{Q}_{\b}^B\} = -2 \d^{AB}\,P_\m \G^\m_{\a\b}\quad,\quad [P^\m,Q^A_\a] = 0\quad A,B = 1,\dotsi, {\cal N}
\,.
$$
Just as before, when \(P_\m\) is massless, \(Q^A\) gives a pair of fermionic creation and annihilation operators for each \(A \in \{1,\dotsi, {\cal N}\}\). As mentioned above, for the Calabi-Yau compactification we have \({\cal N} =2 \) and therefore two pairs of fermionic creation and annihilation operators. As a result, an \({\cal N} =2 \) massless multiplet has  the following helicities 
$$
j, \,j+\frac{1}{2} ,\,j+\frac{1}{2} , j + 1
$$
and accompanied by their CPT conjugate if the multiplet is not conjugate to itself. In particular, as will be seen shortly, there are three massless multiplets, corresponding to \(j = -\frac{1}{2}\,,0\,,1\), which will be relevant for the fields content of the low energy effective theory obtained by compactifying type II string thoery on Calabi-Yau three-folds. These are
\begin{displaymath}
\begin{array}{cc}
\text{hypermultiplet}& (-\frac{1}{2},0^2,\frac{1}{2})+  (-\frac{1}{2},0^2,\frac{1}{2})\\
\text{vector multiplet}& (-1,-\frac{1}{2}^2,0)+  (0,\frac{1}{2}^2,1)\\
\text{supergravity multiplet}& (-2,-\frac{3}{2}^2,-1)+  (1,\frac{3}{2}^2,2)
\end{array}
\end{displaymath}

To obtain the massless spectra of the lower-dimensional theory, we have to compactify the massless spectra of type II superstring listed in Table \ref{massless_spectra}. Using the fact that the Laplacian factorizes into 
$$
\nabla_{\text{\tiny 10d}} = \nabla_{\text{\tiny 4d}}+\nabla_{\text{\tiny C-Y}}\,,
$$
we see that the four-dimensional massless spectrum is given by the cohomology classes of the internal manifold. 
 These massless fields, grouped in terms of the \({\cal N} =2 \) multiplets, is given in Table \ref{massless_spectra_4_5d}.  Note that we have used the self-duality of the \(C^{(4)}_+\) of type IIB theory and the fact that a two-form is dual to a scalar field in four dimensions through \(\star_{\text{\tiny 4}} dC^{(2)} = d\f\) in obtaining this table.

 \begin{table}\centering
\begin{tabular}{ccc}
\multicolumn{3}{c}{\bf IIA/CY massless spectrum }\\
\rowcolor[gray]{0.8}
number& multiplet & bosonic field content\\
 \(1\)& supergravity multiplet & \(G_{\m\n},(C_\m)\)\\
\(h^{1,1}\)& vector  multiplet & \((C_{\m i\jbar}),\,G_{i\jbar},\,B_{i\jbar} \) \\
  \(h^{2,1}\)&  hypermultiplet & \( C_{i\jbar k},\,G_{ij}\)\\
\(1\)& (universal) hypermultiplet&  \(C_{ijk},\Phi,B_{\m\n}  \)\\
 \midrule
\multicolumn{3}{c}{}\\
\multicolumn{3}{c}{\bf IIB/CY massless spectrum}\\
\rowcolor[gray]{0.8}
number& multiplet & bosonic field content\\
 \(1\)& supergravity multiplet & \(G_{\m\n},C_{\m i j k }\)\\
\(h^{2,1}\)& vector  multiplet & \(C_{\m i\jbar k},\,G_{ij} \) \\
  \(h^{1,1}\)&  hypermultiplet & \( C_{\m\n i\jbar },C_{ i\jbar },\,G_{i\jbar},\,B_{i\jbar}\)\\
\(1\)& (universal) hypermultiplet&  \(C,\Phi,B_{\m\n},C_{\m\n}  \)\\
 \midrule
 \multicolumn{3}{c}{}\\
\multicolumn{3}{c}{\bf M-theory/CY massless spectrum}\\
\rowcolor[gray]{0.8}
number& multiplet & bosonic field content\\
 \(1\)& supergravity multiplet & \(G_{\m\n}, (A_{\m i\jbar })\)\\
\(h^{1,1}-1\)& vector  multiplet & \(\f^a,\, (A_{\m i\jbar })\) \\
  \(h^{2,1}\)&  hypermultiplet & \( G_{ij},\,A_{ij\bar{k}}\)\\
\(1\)& (universal) hypermultiplet&  \({\cal V}, A_{ijk} \)\\
 \midrule
\end{tabular}
\caption{ \label{massless_spectra_4_5d}\footnotesize{Summary of the massless spectrum of the type IIA, type IIB superstring theories compactified on Calabi-Yau three-folds. The parenthesis denotes the fact that the gauge field in the supergravity multiplet, the graviphoton field, is actually a linear combination of the (1+\(h^{1,1}\)) gauge fields in the parenthesis. Similarly, in the case of M-theory compactification, the \(h^{1,1}\) vectors \(A_{\m i\jbar}\) split into one supergravity and (\(h^{1,1}-1\)) vector multiplet gauge fields upon dimensional reduction.}}
\end{table}

But we are not done yet with the supersymmetry algebra. For the cases of extended supersymmetry with \({\cal N}>1\), it's possible to have central extensions of the above algebra. Without breaking the Lorentz invariance, namely without incorporating extended sources, the most general form is 
\be\label{superalgebra_w_extension}
\{Q_{\a}^A ,\overline{Q}_{\b}^B\} = -2 \d^{AB}\,P_\m \G^\m_{\a\b}-2 i Z^{AB}\d_{\a\b}\quad,\quad [P^\m,Q^A_\a] =[Z,Q]=[Z,P] =0\quad\,.
\ee
By taking the charge conjugation of the above anti-commutation relation and use the Majorana condition \(\overline{Q} = Q^\dag \G^0  = Q^T C\) we see that the central charge matrix is anti-symmetric,
$$
Z^{AB} = -Z^{BA}\,.
$$ 
In the case that \({\cal N}\) is even, the central charge matrix can therefore be written as a block-diagonal form with the \(i\)-th black being
$$
\bem 0& Z_i \\ -Z_i & 0\eem\qquad,\quad i = 1,\dotsi,\frac{1}{2}{\cal N}\,.
$$
Suppose now that the momentum vector is timelike with mass \(M\), the anti-commutation relation implies that the eigenvalues of the central charge matrix satisfies
\be\label{BPS_bound}
M \geq |Z_i|\qquad,\quad i = 1,\dotsi,\frac{1}{2}{\cal N}\,.
\ee
This is called the BPS (Bogomolny-Prasad-Sommerfield) bound on the  mass respective to the charges. 
\index{BPS (Bogomolny-Prasad-Sommerfield) bound}
\index{central charge matrix}

Analogous to the case without central extensions, the \(\{Q,Q\}\) anti-commutator gives two pairs of fermionic creation and annihilation operators for each \(A \in \{ 1,\dotsi,{\cal N}\} \) if none of the BPS bound is saturated, just as in the massive case in the algebra without central extensions. Now for each \(i\) for which the BPS bound is saturated, two pairs of fermionic creation and annihilation operators are removed. When all of the \(\frac{1}{2}{\cal N}\) BPS bound are saturated there is just one pair of fermionic creation and annihilation operators for each \(A \in \{ 1,\dotsi,{\cal N}\} \), and we have the same representation of this algebra as the massless one in the case with no central charges. To sum up, the relationship between the BPS bound and the unbroken supersymmetry in four dimensions is that the saturation of each BPS bound implies four preserved supersymmetry.

For example, when \({\cal N} = 4\) we have two BPS bounds
\be\label{2_bps_N=4}
M \geq |Z_1|  \geq |Z_2|\;. 
\ee
When only one of the BPS bounds are saturated, we have six fermionic creation operators and the multiplet  therefore contains \(2^6\) states. Especially each state is annihilated by 4 of the total 16 supercharges and is therefore called \(1/\!4\)-BPS. When both of the bounds are saturated, a multiplet contains \(2^4\) states just like in the massless case and the states are said to be  \(1/\!2\)-BPS for obvious reasons.

Back to the \({\cal N} = 2\) case at hand, now there is only one BPS bound
$$
M \geq |Z_1|\,,
$$
and in this case the central charge is given by the graviphoton charge, namely the gauge field in the supergravity multiplet. 

After discussing the massless field content of the type II string theory compactified on a Calabi-Yau, we are now ready to reduce the 10-dimensional low energy supergravity action (\ref{IIA_action}), (\ref{IIB_action}) and obtain the 4-dimensional low energy effective action. The resulting bosonic action for the supergravity and vector multiplets is
\begin{align}\notag
16\p G_{{\text{\tiny N}}}^{(4)} L &= L_{\text{\tiny Eins}}+L_{\text{\tiny scalar}}+
L_{\text{\tiny vector}} &\\ \notag
L_{\text{\tiny Eins}} &= R \star_{\text{\tiny 4}} 1& \\ \notag 
L_{\text{\tiny scalar}} &= -  g_{A\bar{B}} \,dt^A \wedge \star_{\text{\tiny 4}}\,  d\bar{t}^{\bar{B}}\qquad &A= 1,\dotsi,n \\
\label{n2_sugrav_action}
 L_{\text{\tiny vector}} &= -\frac{1}{2} {\cal F}^I \wedge {\cal G}_I\qquad &I= 0,\dotsi,n\;,
\end{align}
where \(\star_{\text{\tiny 4}}\) denotes the Hodge dual in four dimensions and \(n\) is the number of the vector multiplet fields which is given in Table \ref{massless_spectra_4_5d} in terms of the topological data of the internal Calabi-Yau manifolds. Furthermore, \({\cal G}_I\) is given by the requirement that 
\be \label{self_duality}
{\cal F} = {\cal F}^I  \otimes \a_I - {\cal G}_I \otimes \b^I = \star_{\text{\tiny 10}} \,{\cal F}
\ee
when the Hodge dual in the Calabi-Yau space is taken to be
\bea \notag
\star_{\text{\tiny C-Y}} \overline{\O} &=& i \O \\ \label{hodge_dual_CY}
\star_{\text{\tiny C-Y}}{\nabla_I \O} &=& i \overline{\nabla}_{\bar{I}} \overline{\O} \,.
\eea
\footnote{Using this definition, one also has to take the complex conjugate of the coefficient, \(\star a \overline{\O} =  i a^* \O\) for example, in order to have the the bilinear \(\int \G \wedge\star  \G  \) positive definite. }
From the expression of the ten-dimensional gravity coupling constant in terms of the string theory data (\ref {10d_planck}) and following the standard Kaluza-Klein procedure, we conclude that the  four-dimensional gravity coupling constant is given by
$$
G_{{\text{\tiny N}}}^{(4)} \sim (\ell_{{\text{\tiny P}}}^{(4)})^2 
\sim \frac{(\ell_s g_s)^2 }{{\cal V}^{(s)}(\text{CY})}\;,
$$
where
\({{\cal V}^{(s)}(\text{CY})} = {\text{Vol}(\text{CY})}/{(\ell_s)^6}\) is the volume of the internal manifold in string unit.

Not surprisingly, this action is exactly the tree-level action of \({\cal N}=2\), D=4 supergravity action, constructed using the superconformal tensor calculus. See for superconformal supergravity \cite{Bergshoeff:1980is,deWit:1984pk}  and \cite{Ferrara:1988ff} for the dimensional reduction. See also, for example, \cite{Mohaupt:2000mj} and references therein for more details.

Again, for completeness we will now rewrite the above action, written in the form as in \cite{Denef:2000nb}, in a probably more familiar form in terms of the coordinates and prepotential. 

For the scalar part of the vector-multiplet action, using the homogeneous property of the prepotential (\ref{prepo_1})-(\ref{prepo_2}), we can show that the scalar field metric (\ref{Kahler_metric_1}) can be explicitly written as 
\begin{align}\notag
g_{I\bar{J}} & = e^{\cal K}N_{IJ}+ e^{2 {\cal K}}\overline{X}_{\bar{I}}  X_J \intertext{where}\notag
N_{IJ} &\equiv 2 \im F_{IJ} \;,\;X_I \equiv X^J\,N_{IJ}\\
\intertext{satisfies} \notag
X\cdot\bar{X} &\equiv X^I \bar{X}_I= \bar{X}^I X_I = -e^{-{\cal K}}
\;.
\end{align}
Notice that this metric given in terms of the projective coordinates \(X^I\) has one degenerate direction, namely \(g_{I\bar{J}} X^I \overline{X}^{\bar{J}}=0\). The reader should remember that this indeed has to be the case, because the moduli space is really only parametrized by the \(n\)-scalars \(t^A\) and therefore we have to project out the unphysical direction corresponding to rescaling \(\O\rightarrow \l \,\O\). 

Put the above equations together, given a prepotential \(F(X)\), the scalar action is 
$$
L_{\text{\tiny scalar}} = \frac{1}{2}\, \left( \frac{N_{IJ}}{X\cdot \overline{X}} + \frac{\overline{X}_{\bar{I}}  X_J}{(X\cdot \overline{X})^2}  \right) dX^I \wedge\star_{\text{\tiny 4}} d\overline{X}^{\bar{J}}\;.
$$

As for the vector part of the action, define the ``coupling matrix" \({\cal N}_{IJ}\) such that 
\bea \notag
F_I &=& {{\cal N}}_{IJ} \,X^J \\ \label{relation_1}
\overline{\nabla}_{\bar{K}} \overline{F}_{\bar{I}} 
&=&  {{\cal N}}_{IJ} \overline{\nabla}_{\bar{K}} \overline{X}^{\bar{J}}\;,
 \eea
which is solved to be
$$
{\cal N}_{IJ} = \overline{F}_{IJ} + i \frac{X_I X_J}{X\cdot X}
$$
with \(X\cdot X = X^I X_I\).
Using its property (\ref{relation_1}), the Hodge star relation can be written in terms of coupling matrix \({\cal N}_{IJ} \) as
$$
\star_{\text{\tiny C-Y}} \,(\a_I -{\cal N}_{IJ} \b^J ) 
= i 
 \,(\a_I -{\cal N}_{IJ} \b^J ) \;.
 $$
Now it's straightforward to solve the self-duality condition (\ref{self_duality}) and rewrite the vector part of the action as
\ben
L_{\text{\tiny vector}} &=&  \frac{1}{2}\, \left( \re{\cal N}_{IJ} {\cal F}^I \wedge {\cal F}^J +   \im{\cal N}_{IJ} {\cal F}^I \wedge \star_{\text{\tiny 4}} {\cal F}^J \right) \\
&=& \frac{1}{2}\, \left({\cal N} {\cal F}^+ \wedge {\cal F}^+ + \overline{\cal N} {\cal F}^- \wedge {\cal F}^-\right)\;,
\een
where we have split the field strength into the self-fual and the anti-self dual part
$$
\star_{\text{\tiny 4}} {\cal F}^\pm = \pm i {\cal F}^\pm\;,
$$
and the reality of the field strength implies \({\cal F}^- = ({\cal F}^+)^*\). 

We will not discuss the hypermultiplet part of the action since we will not need it later. Let's just remark that it decouples from the supergravity multiplet and vector multiplet part of the action discussed above, in the sense that the hypermultiplet action, including the coupling constants of it, only depends on hypermultiplet fields. Furthermore, the scalar manifold of the hypermultiplet action is not special K\"ahler but the so-called quaternionic K\"ahler manifold. The relationship between the  scalar manifold of the vector- and the hyper-multiplet sectors as predicted by mirror symmetry has to be seen by further compactifying down to three dimensions, using the so-called c-map.

\subsubsection{Five-dimensional Supergravity}
\label{Five-dimensional Supergravity}
We have just discussed the four-dimensional low-energy effective action of type II string theory compactified on Calabi-Yau three-folds, obtained by compactifying the ten-dimensional type IIA and IIB supergravity theories to four dimensions. We can also consider the 
five-dimensional low energy effective action of M-theory compactified on Calabi-Yau manifold. This can be done by Kaluza-Klein reduce the eleven-dimensional supergravity action (\ref{S11}) to five dimensions, since the eleven-dimensional supergravity is supposed to be the low-energy description of M-theory. Not surprisingly, the result is the same as the action of the \({\cal N}=1\), d=5 supergravity \cite{Cadavid:1995bk}. 

First let's look at the massless spectrum of the theory. Again splitting the eleven-dimensional spacetime indices into the internal ones (\(i,j,\ibar,\jbar\)) and the five-dimensional ones (\(\m,\n\)), we get the five-dimensional spectrum as recorded in Table \ref{massless_spectra_4_5d}. Note that the scalar fields \(\f^a\) are now real instead of complex, since there is no B-field in M-theory. More specifically, the \(h^{1,1}\) scalars given by the K\"ahler moduli
$$
J = J^A \a_A \quad,\quad \a_A \in H^{1,1}(X,\Z)
$$
is now divided into the volume factor \({\cal V} = \frac{1}{3!} D_{ABC}J^AJ^BJ^C \) and \((h^{1,1}-1)\) scalars \(\f^a\;,a=1,\dotsi,h^{1,1}-1 \), which are coordinates of the co-dimension one hypersurface inside the  K\"ahler moduli space satisfying \({\cal V}(\f)=1\). The former goes in the (universal) hypermultiplet while the latter make up the \((h^{1,1}-1)\) real scalars of the \((h^{1,1}-1)\) vector multiplets of the theory. In particular, since the hypermultiplet part of the action decouples from the rest, the volume of Calabi-Yau space in eleven-dimensional Planck unit plays no important role in the physical solution.

The vector and supergravity multiplets part of the bosonic action is  \cite{Gunaydin:1984ak}
\begin{align}\notag
16\p G_{{\text{\tiny N}}}^{(5)} L &= L_{\text{\tiny Eins}}+L_{\text{\tiny scalar}}+
L_{\text{\tiny vector}} + L_{\text{\tiny C-S}} &\\ \notag
L_{\text{\tiny Eins}} &= R \star_{\text{\tiny 5}} 1& \\ \notag
L_{\text{\tiny scalar}} &= - h_{ab} \,d\f^a \wedge \star_{\text{\tiny 5}}\,  d\f^b\qquad &a,b= 1,\dotsi,h^{1,1}-1 \\ \notag
L_{\text{\tiny vector}} &= -\frac{1}{2} a_{AB} {F}^A \wedge 
\star_{\text{\tiny 5}} F^B \qquad &A,B= 1,\dotsi,h^{1,1}\\ 
\label{5d_sugrav_action}
L_{\text{\tiny C-S}} &= \frac{1}{3!}\,D_{ABC} A^A\wedge F^B\wedge F^C 
\;,
\end{align}
where \(\star_{\text{\tiny 5}}\) denotes the Hodge dual in five dimensions. To understand the scalar metric \(h_{ab}\) and the gauge coupling \(a_{AB}\), let's consider the natural metric on the \(J^A\)-space
\be\label{5d_gauge_coupling}
g_{AB} = \frac{\pa}{\pa J^A}\frac{\pa}{\pa J^B} {\cal K}=\int_{CY} \a_A \wedge \star \a_B \;,
\ee
where \({\cal K}\) is again given by (\ref{kahler_po_IIA}):
$$
e^{-{\cal K}} = \frac{4}{3} \int J\wedge J\wedge J = 8{\cal V}(J)\;.
$$
Then \(h_{ab}\) and \(a_{AB}\) are given by, up to a convention-dependent coefficient, the induced metric on the hypersurface \({\cal V} =1\) and the restriction of \(g_{AB}\) on the same hypersurface respectively.

\subsubsection{4D-5D Connection}
\label{4D-5D Connection}
\index{4D-5D connection}
As we discussed in the previous chapter, M-theory compactified on a circle is dual to type IIA string theory. Taking the low-energy limit on both sides compactified on a Calabi-Yau manifold, it implies that a solution in the above five-dimensional supergravity theory gives rise to a solution in the four-dimensional supergravity theory when reduced on a circle. Of course, the above statement can also be understood just using the usual Kaluza-Klein reduction of the five-dimensional supergravity theory without reference to string or M-theory. Specifically, a four-dimensional solution can be ``lifted" to a five-dimensional solution with a U(1) symmetry, with the presence of this U(1) isometry assuring the absence of higher Kaluza-Klein modes. To see how it works, let's do some dimensional analysis first. 

Recall the relation between the string length, the eleven-dimensional Planck length and the radius of the M-theory circle (\ref{coupling_relation}), and the usual Kaluza-Klein relation
$$
\ell_{\text{\tiny P}}^{(5)} \sim \frac{\ell_{\text{\tiny P}}^{(11)}}{({\cal V}^{\text{\tiny (M)}})^{1/3}} \qquad,\quad
\ell_{\text{\tiny P}}^{(4)} \sim \frac{\ell_{\text{\tiny P}}^{(10)}}{({\cal V}^{\text{\tiny (s)}})^{1/2}}\;,
$$
and the expression of the ten-dimensional Planck length (\ref{10d_planck}). From the above relations together with the fact that 
\be\label{cy_vol_m_theory}
{\cal V}^{\text{\tiny (s)}} \,(\ell_{\text{\tiny s}})^6 
= {\cal V}^{\text{\tiny (M)}} (\ell_{\text{\tiny P}}^{(11)})^6\;,
\ee
where \({\cal V}^{\text{\tiny (s)}}\) and \({\cal V}^{\text{\tiny (M)}}\) are the Calabi-Yau volume in string and M-theory units respectively, we conclude that 
\be\label{R_M_planck5}
R_{\text{\tiny M}} \sim ({\cal V}^{\text{\tiny (s)}})^{1/3}\,\ell_{\text{\tiny P}}^{(5)} \qquad,\quad \ell_{\text{\tiny P}}^{(5)} \sim ({\cal V}^{\text{\tiny (s)}})^{1/6} \,\ell_{\text{\tiny P}}^{(4)} \;,
\ee
with the volume \({\cal V}\) denotes the volume at the spatial infinity.

Being careful with the coefficients not listed above, this suggests that  a four-dimensional solution \(ds_\text{\tiny 4D}^2\), \(A_\text{\tiny 4D}^A\), , \(A_\text{\tiny 4D}^0\), \(t^A\) gives a five-dimension solution 
\bea \notag
ds_\text{\tiny 5D}^2&=& 2^{2/3} {\cal V}^{2/3}\,(d\j -A_\text{\tiny 4D}^0 )^2 + 2^{-1/3} {\cal V}^{-1/3} ds_\text{\tiny 4D}^2\\ \notag
A^A_\text{\tiny 5D}&=& A^A_\text{\tiny 4D} + \re t^A \,(d\j- A^0_\text{\tiny 4D} )\\ \label{5d_sol}
Y^A &=& \frac{\im t^A}{{\cal V}^{1/3}}
\eea
where the right-hand side of the equations are given by four-dimensional quantities, for example \({\cal V}={\cal V}^{\text{\tiny (s)}}\). This can be checked by a careful comparison of five- and four-dimensional action.

This is the so-called 4D-5D connection reported in \cite{Gaiotto:2005xt,Gaiotto:2005gf}, see also related earlier work \cite{Bena:2004tk,Bena:2005ay}. Very often this connection turns out to be a useful way to generate new BPS solutions in five dimensions, by simply uplifting the known four-dimensional BPS solutions. See for example the discussions in chapter \ref{More Bubbling Solutions}. Nevertheless, it should be stressed that, of course this procedure only gives solutions with at least one U(1) isometries in five dimensions. 

\subsection{Range of Validity and Higher Order Corrections}
\label{Range of Validity and Higher Order Corrections}

It is important to consider when the low-energy effective action discussed above is actually ``effective", namely a good description of the physics occurred. In particular, we would like to know when the classical BPS solutions give a reliable account of the system. For this purpose there are a few scales that are relevant, and we will discuss them beginning with type II compactification. 

First of all, before compactification we want the D-branes not to be too light in the ten-dimensional Planck units. In other words, we have to consider the correction by the creation and annihilation of virtual D-branes unless the D-brane tension  (\ref{tension1}) in ten-dimensional Planck unit (\ref{10d_planck})
\be\label{correction_10d}
\t_{\text{D}p\text{-brane}} \, (\ell_{\text{\tiny P}}^{(10)})^{1+p} \sim g_s^{\frac{p-3}{4}}
\ee
is large. 
Secondly, to suppress the four-dimensional quantum gravitational effect, we need the typical length scale, the radius of curvature near the horizon for example, to be large in the four-dimensional Planck unit. This gives a condition on the charges
$$
S(\G) \gg 1
$$
and thus require that we consider {\it large charges} \(\G\), since the horizon area scales as charge squared. Thirdly, the \(\a'\) stringy correction of the 4d Lagrangian is controlled by the size of the internal manifold in string unit. As a consequence, in order to suppress them we need to stay in the regime where the following is true,
$$
{\cal V}^{\text{\tiny (s)}} \gg1 \;,
$$
namely the {\it large radius} regime. Finally, for the suppression of the stringy effect at the scale in the non-compact spacetime, in particular at the scale of the black hole horizon, we cannot go all the way to the decompactification limit \({\cal V}^{\text{\tiny (s)}} \rightarrow \inf \) either. In other words, we need the horizon to be very large in the string scale. In formulas this means that we require
$$
\frac{S(\G) \,(\ell_{\text{\tiny P}}^{(4)})^{2}}{\ell_{\text{\tiny s}}^2} \sim 
S(\G) \, \frac{g_s^2}{{\cal V}^{\text{\tiny (s)}}} \gg1 \;,
$$
and also because going to the decompactification limit means bad spectrum contamination from the KK-modes.

After discussing the range of validity for various kinds of corrections, let's now take a look at the nature of the corrections. As mentioned before, away from the singularities, local supersymmetry dictates a decoupling between the hypermultiplets Lagrangian from the rest, namely the supergravity and vector multiplets degrees of freedom. Specifically, as we have seen implicitly in the derivation of the BPS solutions, these solutions are specified by the supergravity and vector multiplets degrees of freedom alone and the hypermultiplets can have any vev in these solutions. 

In both type IIA and IIB compactifications, the dilaton field sits in the universal hypermultiplet, as recorded in Table \ref{massless_spectra_4_5d}. Therefore the good news is that we don't have to worry about \(g_s\)-corrections in the supergravity and vector multiplets Lagrangian. In string theories we have to consider \(\a'\)-corrections as well. In the compactification setting, the \(\a'\)-corrections to the lower-dimensional effective theory is controlled by the size of the internal manifold in string unit \({\cal V}^{\text{\tiny (s)}}= \frac{1}{6}D_{ABC} \, J^A J^B J^C\), which is given by the K\"ahler moduli. The conclusion one can draw from this fact is the non-renormalizability property that the effective action of the type IIB compactification is exact in both \(\a'\) and \(g_s\), while the effective action of the type IIA compactification receives both the perturbative world-sheet loop \(\a'\)-corrections and the non-perturbative world-sheet instanton \(\a'\)-corrections. This gives us the possibility to compute the type IIA higher-order-in-\(\a'\) corrections employing the mirror symmetry discussed in the last section.

There is a special family of corrections, the F-term corrections, which is independent of the hypermultiplet fields and is therefore relevant for the correction of black hole entropies. These are studied in 
\cite{LopesCardoso:1998wt,LopesCardoso:1999cv,LopesCardoso:1999ur}. It can be shown that this family of corrections is computed by the topological string theory discussed in the last section \cite{Antoniadis:1993ze,Bershadsky:1993cx}. In particular, as discussed above (\ref{correction_10d}), at the strong string coupling regime of type IIA string theory, which is better described in the language of M-theory, the correction to the four-dimensional F-term has its ten- (or eleven-) dimensional origin as the loop integral of virtual D2-D0 bound states which are light in the large \(g_s\) limit \cite{GV1,GV2}. This identification gives the expression of the topological strings free energy in terms of the so-called ``Gopakumar-Vafa invariants" enumerating D2-D0 bound states as mentioned in (\ref{top_free_energy}).

\index{OSV (Ooguri-Strominger-Vafa) conjecture}
Furthermore, the relation between the F-term correction and the topological strings  leads to a natural conjecture between black hole BPS degeneracies and the topological strings, called the OSV conjecture \cite{Ooguri:2004zv}, which in our convention reads
\begin{align}\label{OSV1}
Z_{BH}(p^I;\f^I):=\sum_{q_I} D(p^I,q_I) e^{-\p \f^I q_I} &= |Z_{\text{\tiny top}}(t^A, g_{\text{\tiny top}})|^2 \\ \notag
t^A = \frac{ip^A + \f^A }{i p^0 + \f^0} \quad&,\quad 
g_{\text{\tiny top}} =\frac{4\p}{i p^0 + \f^0}\;,
\end{align}
where the right-hand side is the topological strings partition function defined in (\ref{def_top_pf}). We will not go into the details about the higher-order derivatives nor OSV conjectures, since there are already many excellent reviews in the literature. See for example, \cite{Mohaupt:2000mj,Guica:2007wd,Pioline:2006ni} and references therein.

\chapter{K3 Compactification}
\label{K3 Compactification}
After discussing the Calabi-Yau compactification of string theories in details, we will be brief in the K3 compactification since many of the basic ideas are fairly similar to the Calabi-Yau case. By discussing the M-theory/type II string theory compactification on K3 manifolds, we will also introduce the toroidally compactified heterotic string theories, which are related to K3 compactified M-theory/type II string theory by dualities.

This chapter is organized as follows. First of all, we assume some basic knowledge about the generic topological properties of \(K3\) manifolds. The readers who are not familiar with them can resort to Appendix \ref{Mathematical Preliminaries}. In the first section we again begin with a world-sheet perspective, introducing the (4,4) superconformal field theory which is relevant for describing the internal CFT with K3 as the target space.  With the knowledge that the marginal deformation of the CFT is given by the moduli space of the target space, in section \ref{Moduli Space of K3} we derive the form of the moduli space using a spacetime viewpoint. In section \ref{Four-Dimensional Theories and Heterotic String Dualities} we dimensionally reduce type II string theory on \(K3 \times T^2\) and study the low-energy effective theory in four dimensions. From the form of the charge lattice and the moduli space we motivate the existence of a toroidally compactified heterotic string theory which is dual to type II superstring on \(K3 \times T^2\), and spell out the correspondence of conserved charges in different frames on the heterotic-IIA-M-IIB chain connected by various dualities. 

\section{(4,4) Superconformal Field Theory}
\label{(4,4) Superconformal Field Theory}
As we mentioned in Appendix A, a Calabi-Yau manifold with \(n\) complex dimensions can be defined as a K\"ahler manifold with \(SU(n)\) holonomy. In particular, a K3 manifold has \(SU(2)\) holonomy and is therefore also hyper-K\"ahler. By decomposing the four dimensional spinor in representations of \(SO(4) = SU(2) \times SU(2)\), we see that the holonomy preserves \(1/2\) of the total thirty-two supersymmetries, as opposed to \(1/4\) in the case of Calabi-Yau three-folds. From our experience with the relationship between spacetime and world-sheet supersymmetry, it is therefore not surprising that 
the relevant superconformal field theory now turns out to have \((4,4)\) instead of  \((2,2)\) world-sheet supersymmetries.  

The action of the non-linear sigma model is again given by (\ref{non_linear_sigma}), supplemented with the coupling to the B-field 
(\ref{B_term}). Instead of the \({\cal N}=2\) superconformal algebra (\ref{n_2_superconformal_algebra}), we have now the following (small) \({\cal N}=4\) superconformal algebra

\index{\({\cal N}=4\) Superconformal algebra}
\bea \notag
\lbrack L_m,L_n\rbrack&=& (m-n) L_{m+n} + \frac{c}{12} m (m^2-1) \,\d_{m+n,0}\\ \notag
\lbrack J_m^i,J_n^j \rbrack&=&-2i\e^{ijk}\,J^k_{m+n}+ \frac{c}{3} m \,\d_{m+n,0}\,\d^{ij}\\ \notag
\lbrack L_n,J_m^i \rbrack &=& -m \,J_{m+n}^i\\ \label{n_4_superconformal_algebra}
\lbrack L_n,G_r^{\pm\pm} \rbrack &=& (\frac{n}{2} - r)\, G_{r+n}^{\pm\pm}\\ \notag
\lbrack J_n^i,G_r^{\a+} \rbrack &=&\s^i_{\a\b}\, G_{r+n}^{\b+} 
\quad,
\lbrack J_n^i,G_r^{\a-} \rbrack = - G_{r+n}^{\b-}\,\s^i_{\b\a}
\\ \notag
\{G^{\a+}_{r},G^{\b-}_{s}\} &=& 2\,\d^{\a\b} L_{r+s} + (r-s) \s^i_{\a\b}\,J_{r+s}^i +\frac{c}{3}\, (r^2 -\frac{1}{4})\, \d_{r+s,0}\,\d^{\a\b}\;,
\eea
where \(\a,\b=\pm\), \(i=1,2,3\), \(\s^i\) are the Pauli matrices and the superscripts ``\(\pm\pm\)" of the fermionic currents \(G^{\pm\pm}\) denote the way they transform under the R-symmetry group \(SU(2)\).
Again we have two possible periodic conditions for the fermions
\be
\begin{cases}
2r =  0 \text{  mod   }2 &\text{for R sector} \\ &\\ 
2r = 1 \text{  mod   }2 &\text{for NS sector}\;.
\end{cases} 
\ee

This \({\cal N}=4\) superconformal algebra shares some important features with the \({\cal N}=2\) superconformal algebra (\ref{n_2_superconformal_algebra}). First of all, there is a natural embedding of the \({\cal N}=2\) algebra into the \({\cal N}=4\) algebra given by
\be
J_m   \to J_m^3 \quad, \quad G_r^+ \to G_r^{++}\quad, \quad G_r^- \to G_r^{+-}\;. 
\ee

As for the representation, a highest weight state is again defined by
\ben
G_r^{\pm\pm} \lvert h,q\rangle &=& J^i_n \lvert h,q \rangle
= L_n \lvert h,q \rangle = 0 \quad\text{for all}\quad r,n > 0\\
L_0  \lvert h,q\rangle &=& h \lvert h,q\rangle\quad,\quad J^3_0 \lvert h,q\rangle =  q\lvert h,q\rangle\;.
\een

As before, a special is played by the ``massless representation", meaning states which are in addition annihilated by 
\be
J_0^+ ,\; G_0^{\pm\pm}
\ee
for states in the R-sector and 
\be
J_0^+ ,\; G_{-1/2}^{-+} ,\; G_{-1/2}^{+-}\quad( \text{or}\;\;
J_0^- ,\; G_{-1/2}^{++} ,\; G_{-1/2}^{--}\;)
\ee
for states in the NS-sector, where we have defined
\(
J^\pm =\frac{1}{\sqrt{2}}\,\big(J^1 \mp i J^2 \big)\;.
\)
These are the counter-part of the R-ground states and the chiral primaries in the \({\cal N}=2\) case respectively, and it can again be seen from various commutation relations that an unitary massless representation satisfies
\be
0\leq h = \frac{c}{24}\;, \quad |q|\leq \frac{c}{6} \quad\text{in the R-sector}
\ee
and 
\be
0\leq h = \frac{|q|}{2} \leq \frac{c}{6} \quad\text{in the R-sector}\;.
\ee

Finally, there is again an automorphism of this \({\cal N}=4\) algebra which generalises the spectral flow of the \({\cal N}=2\) algebra (\ref{spectral_flow}) to

\begin{align} \notag
L_n &\rightarrow L_n + \eta J_n + \eta^2 \frac{c}{6} \d_{n,0}\\
J_n^3&\rightarrow J_n^3 + \eta \frac{c}{3} \d_{n,0}\quad,\quad 
J_n^\pm\rightarrow J_{n\pm2\eta}^\pm\\ \notag
G_r^{\pm+} &\rightarrow G_{r\pm\eta}^{\pm+}\quad,\quad
G_r^{\pm-} \rightarrow G_{r\mp\eta}^{\pm-}
\;.
\end{align}

This in particular implies that the elliptic genus has again the theta-function decomposition as in (\ref{theta_decomposition1}).

There are of course also differences between the \({\cal N}=4\) and \({\cal N}=2\) non-linear sigma models. One important distinction is that, unlike the case for the Calabi-Yau three-folds, the Ricci flat metric is now an exact solution but not just a solution in the leading order of \(\a'\), due to the non-renormalisation theorem brought to us by higher supersymmetries. In the  \({\cal N}=4\) case there is again a notion of mirror symmetry, but since now the complex structure and K\"ahler moduli are in the same cohomology \(H^{1,1}(X,\R)\), the discussion of the mirror symmetry becomes more involved and we will not include it in the present thesis. See \cite{Sevrin:1988ew} for some discussions of \({\cal N}=4\) superconformal algebras and \cite{Eguchi:1988vra,Eguchi:1987wf} for its representations relevant in the present context.

\section{Moduli Space of K3}
\label{Moduli Space of K3}
\setcounter{equation}{0}
Two major differences between the moduli space of Calabi-Yau two-and three-folds are that for the K3 case, first of all there is no clear separation between the complex and K\"ahler moduli space; now both of them are in the same cohomology class \(H^{1,1}(S,\R)\). Secondly, as we have mentioned before, a K3 manifold is not only K\"ahler but also hyper-K\"ahler, which means it has not only one complex structure but a whole \(S^2\) of possible  complex structures, rotated to each other by elements of \(SU(2) \). 

Keeping these facts in mind, a simple counting gives the dimension of the moduli space of the non-linear sigma model:
$$
\text{\small dim}\,{\cal M}_\s = \text{\small dim}\,H^{1,1}(S,\Z) + 2 \text{\small dim}\,H^{1,1}(S,\Z) + \text{\small dim}\,H^{2}(S,\Z) -2 = 80\;,
$$
where the first three terms account for the moduli space for the K\"ahler moduli, the complex structure moduli, and the B-field respectively, and the 2 is subtracted to account for the fact that each metric comes with a sphere of complex structures.

To see the structure of this 80-dimensional moduli space, let's first concentrate on the complex structure and K\"ahler moduli. From 
\begin{align}\notag
\int_S J \wedge J , \;\int_S \O \wedge \overline{\O} &> 0\\ \notag
\int_S \O \wedge \O =\int_S J \wedge \O&=0\;,
\end{align}
where \(J\) is the K\"ahler form and \(\O = \O_1+ i \O_2\) is the complex structure, we see that \(J\), \(\O_1\) and  \(\O_2\) are three vectors that are all mutually perpendicular, with respect to the bilinear (\ref{symmetric_bilinear_four-fold}) on the space \(H^{2*}(S,\R) \cong \R^{4,20}\)  (\ref{K3_lattice_2})
\be
(\a,\b) = \int_S \a \wedge\b
\ee
and that are all spacelike. In other words, \(J\), \(\O_1\) and  \(\O_2\) defines a three-dimensional plane inside \(H^2(S,\R)\cong \R^{3,19}\). Furthermore, a rotation of the three vectors corresponds to a rotation of the \(S^2\) possibilities of complex structures and therefore does not correspond to a change in the 
geometry. In other words, the complex structure and K\"ahler moduli space of K3 is locally a Grassmannian times the positive half of a real line representing the volume \(V\) of the K3. 
Globally, the moduli space is
$$
O(\G^{3,19})\backslash O(3,19,\R)/ \big(O(3,\R)\times O(9,\R) \big)\,\times \R_+\;,
$$
where \(O(\G^{3,19})\) is the automophism group of the lattice \(\G^{3,19}\).

Now we want to incorporate the moduli space for the B-field, which is not considered in the above discussion. Given a choice of B-field two-form and the volume \(V\), define a map \(\x: H^{2}(S,\R)\cong \R^{3,19} \rightarrow H^{2*}(S,\R)\cong \R^{4,20} \) and an additional vector \(\x_4\) by 
\ben
\x(\a) &=& \a - (B,\a) \a^0 \\
\x_4 &=& \a_0 + B + \big(V- \frac{1}{2} (B,B)\big)\, \a^0
\;,
\een
where \(\a_0,\,\a^0\) are the dual basis for \(H^0(S,\Z),\,H^{4}(S,\Z)\) introduced in (\ref{K3_lattice_2}). It can be checked easily that \(\x_4\) is perpendicular to all \(\x(\a)\) with \(\a\in H^2(S,\R)\) and that \(\x(J)\), \(\x(\O_1)\), \(\x(\O_2)\) are again three mutually perpendicular spacelike vectors, but now in the larger space \(\R^{4,20}\). Furthermore, the spacelike four-dimensional plane spanned by \(\x(J)\), \(\x(\O_1)\), \(\x(\O_2)\) and \(\x_4\) contains the same information as the three-dimensional plane spanned by \(J,\,\O_1,\,\O_2\), when a choice of \(B,\,V\) and \(\a_0\) is given. On the other hand, the B-field moduli can be thought of as the moduli of embedding \(\R^{3,19}\cong H^2(S,\R)\) into the larger space \(\R^{4,20}\cong H^{2*}(S,\R)\).

Let's now consider an integral shift of the B-field
$$
B \rightarrow B + \b\quad,\quad \b \in H^2(S,\Z),
$$
which must be a symmetry of the theory. Equivalently, it can be seen as a change in the choice of \(\a_0\)
\begin{align} \notag
\a_0 &\mapsto \a_0 + \b - \frac{(\b,\b)}{2} \,\a^0\\
\intertext{
together with the following shift of the two-form} \notag
\a &\mapsto \a - (\a,\b)\a^0\;.
\end{align}
In other words, when the B-fields are incorporated, the symmetry group involves the whole automorphism group of the larger lattice 
\(H^{2*}(S,\Z) \cong \G^{4,20}\). 
Putting the above together, we then conclude that the moduli space of the \(K3\) non-linear sigma model given by a Grassmannian as
\be
{\cal M}_\s = O(\G^{4,20})\backslash O(4,20,\R)/ \big(O(4,\R)\times O(20,\R) \big)\;.
\ee
See \cite{Aspinwall:1994rg,Aspinwall:1996mn,Nahm:1999ps} and references therein for discussions about the above moduli space.

\section{Four-Dimensional Theories and Heterotic String Dualities}
\label{Four-Dimensional Theories and Heterotic String Dualities}
\setcounter{equation}{0}
As mentioned before, the \(SU(2)\) holonomy of K3 leads to the breaking of half of the supersymmetries.  At the low energy limit, type II string theories compactified on a K3 manifold therefore yield six-dimensional supergravity theories with sixteen supercharges. However, we will be interested in four-dimensional theories instead of six-dimensional ones. 

For concreteness, we will begin with considering type IIA string theory compactified on the internal manifold \(K3\times T^2\) down to four dimensions. The torus has trivial holonomy and thus does not break supersymmetry any further.
The four-dimensional theory has now \({\cal N}=4\) supersymmetry and we anticipate to obtain some  \({\cal N}=4\), d=4 supergravity theory at low energy. We will therefore begin this section by discussing the generalities of these \({\cal N}=4, d=4\) supergravity theories.

\subsection{${\cal N}=4, d=4$ Supergravity}
\label{subsec_N4d4_sugrav}

There are two kinds of supermultiplets relevant in \({\cal N}=4, d=4 \) supergravity theories, namely the supergravity and the matter multiplets. From this we expect the bosonic field content of our low energy effective action to be 
$$
\big(g_{\m\n}, A_\m^{m=1,\dotsi,6}, \l \big) \; \; \text{and}\;\;n \times \big( A_\m , \f^{m=1,\dotsi,6}\big)\;,
$$
where \(\l\) is a complex scalar and \(m\) is an \(SU(4)=SO(6)\) R-symmetry index. Furthermore, the \(2\) and \(6n\) real scalars parametrise the scalar manifold \cite{deRoo:1984gd}
$$
\frac{SL(2)}{U(1)} \times \frac{SO(6,n)}{SO(6)\times SO(n)} \;.
$$

To study the supergravity theory obtained by the IIA/\(K3\times T^2\) compactification, first we would like to determine the number of matter multiplets in the theory. We will do this by counting the number of scalars by dimensionally reducing the massless fields of type IIA string theory (Table \ref{massless_spectra}) using the harmonic forms of the internal manifold. The result is
\begin{displaymath}
\begin{array}{rl}
80& g,B \text{  on  } K3\\
4 & g,B \text{  on  } T^2 \\ 
2& C^{(1)}\\
44&C^{(3)} \\
2 & C^{(3)} \text{ to spatial one-forms and dualize to scalars}\\
2 & \Phi, B_{\m\n} \text{ (axion-dilaton)}\\
&\\
\hline&\\
134&=2 +  6 \times 22\;,
\end{array}
\end{displaymath}
which implies in this case \(n=22\), namely that the massless field content of the four-dimensional theory is one \({\cal N}=4\) supergravity multiplet together with 22 matter multiplets. 

One can easily check that there are 28 vector fields upon compactification, which again decompose into a supergravity multiplet  together with 22 matter multiplets. With respect to these vector fields, there are 28 electric and 28 magnetic conserved charges, forming a charge lattice
\be\label{dyon_charge_lattice_P_Q}
\bem P \\ Q\eem  \in \G^{6,22} \oplus \G^{6,22}\;.
\ee

The Grassmannian part of the scalar manifold 
 $$
 \frac{SO(6,22)}{SO(6)\times SO(22)} 
$$
can be thought of as the moduli space of different ways to separate the above charges into the ``left-moving" and ``right-moving" parts, such that 
\be
P_L^2 - P_R^2 = P^2 
\ee
and similarly for the electric charges, where $P^2$ is the inner product of the vector $P$ with itself using the natural $SO(6,22;\Z)$ invariant metric. Explicitly, we can therefore parametrise this part of the moduli space by a \(28\times 6\) matrix \(\m_{a=1,\dotsi,28}^{m=1,\dotsi,6}\), such that
\be\label{left_moving_charge_dyon}
P_L^m = \m_a^m P^a
\ee
and similarly for the \(Q\)'s. Notice that \(\m\) is only defined up to rotations which leave all  \(P_L^2\) invariant. 

For a very simple example of a moduli space which is a Grassmannian, let's consider string theory compactified on a circle with radius \(R\) and consider the states with winding number \(w\) and momentum \(k\) along the circle. Then the left- and right-charges are 
\ben
P_L &=& k/R + wR \\
 P_R &=& k/R - wR \quad ,\quad P_L^2 - P_R^2 = P^2 = 4kw  \;.
\een
Notice that \( P_L^2 - P_R^2\) does not depend on the radius \(R\) while both \(P_L\) and \(P_R\) do.

The Grassmannian \(SO(1,1,\R)\) is an one-dimensional space parametrised by a real number \(\eta\) as
\be
\bem P_L \\ P_R\eem = \bem\cosh\eta & \sinh\eta \\ \sinh\eta& \cosh\eta\eem \bem k +w \\ k-w \eem\;.
\ee
Then we see that  the modulus  of the compactification circle, in this case the radius \(R\), is related to \(\eta\) by
\be
\cosh\eta = \frac{1}{2}(R+\frac{1}{R}) \quad,\quad
\sinh\eta = -\frac{1}{2}(R-\frac{1}{R}) \;.
\ee

Finally let's turn to the first factor of the scalar manifold
\be
\frac{SL(2)}{U(1)} \cong {\cal H}_1\;.
\ee
As discussed in section \ref{upper_half_plane}, this is nothing but the upper half-plane and we will parametrise it by \(\t \in \C,\, \im \t > 0 \) as in (\ref{tau_upper_half}).

In our present setting of IIA/\(K3\times T^2\) compactification, this complex scalar \(\l\) is the complexified K\"ahler moduli of the torus. As we will see in the following subsection, it becomes the complex structure moduli in the IIB/\(K3\times T^2\) compactification and the axion-dilaton moduli in the heterotic/\(T^6\) compactification, when we apply a chain of dualities. 

 In terms of these scalars \((\l,\m)\) and the conserved charges \((P,Q)\),
 we can now write down the solutions to this supergravity theory. We will leave the details for the Part V of the thesis.

\subsection{Heterotic String Dualities}
\label{Heterotic String Dualities}

\index{heterotic string theory}
In the previous subsection we have seen that the low-energy supergravity theory obtained from compactifying type IIA string theory on \(K3\times T^2\), has a scalar manifold which contains the Grassmannian \(SO(6,22)/SO(6)\times SO(22)\).
This is exactly how the moduli space of a conformal field theory compactified on a \(\G^{6,22}\) lattice looks like locally. 
Notice that there is one unique (up to isomorphism) lattice of this signature (or any \(\G^{\s^+,\s^-}\) with \(\s^--\s^+= 0 \) mod \(8\)) which is even self-dual 
, or sometimes called unimodular. And an even, self-dual lattice is exactly the kind of lattice required for the one-loop modular invariance of the conformal field theory. 
Including four free bosons on both sides corresponding to the four non-compact dimensions, this putative conformal field theory should have \((10,26)\) bosons on the left- and right- moving sector respectively. Notice that they are the critical dimensions, namely the required number of free bosons in order to have total central charge zero with the ghosts included, for \({\cal N}=1\) and \({\cal N}=0\) world-sheet supersymmetry respectively. We therefore conclude that only the left-moving sector of this putative conformal field theory has 
world-sheet supersymmetry. Such conformal field theories are called {\it heterotic string theories}. One way of interpreting such a conformal field theory geometrically is to say that it has ten spacetime dimensions and the rest of the 16 right-moving bosons are always compactified on an internal sixteen dimensional even self-dual lattice. 
In this language, the observation is that the massless fields of type IIA string theory compactified on \(K3\times T^2\) is the same as that of heterotic string compactified on \(T^6\). In more details, one can see from matching the low-energy supergravity theory that the complex scalar \(\l\) in the supergravity multiplet is now the axion-dilaton field of heterotic string, and the 28 vectors are the 16 gauge bosons present in the massless spectrum of the heterotic string theory and the other 12 coming from compactifying the metric and the B-field on the six-torus. 

This motivates the conjecture of the following string duality \cite{Hull:1994ys,Witten:1995ex}
$$
\text{IIA}/K3\times T^2\text{   is dual to   heterotic} /T^6\;.
$$
The U-duality group of the theory is conjectured to be
$$
SL(2,\Z) \times SO(6,22,\Z)\;,
$$
where the presence of the first factor can be seen from the presence of the modular group of  \(T^2\) in the type IIA picture, which then translates into the S-duality (strong-weak-coupling duality) of the heterotic string. This is very reminiscent of the interpretation of the S-duality group of type IIB string theory as the torus modular group in the type M-theory as we saw in section \ref{S_duality_section}. This group acts on the charges and moduli as
\be\label{S-duality_dyons}
\bem P \\ Q \eem \rightarrow \bem a & b \\ c& d \eem \bem P \\ Q \eem
\quad,\quad \l \rightarrow \frac{a\l+b}{c\l+d} \qquad;\quad
\bem a & b \\ c& d \eem \in PSL(2,\Z)
\ee
while leaving the Grassmannian moduli \(\m\) invariant.

The second group, on the other hand, is nothing but the T-duality group of the \(\G^{6,22}\) compactification of the heterotic string, or equivalently the automorphism group of the charge lattice \(\G^{6,22}\). In particular, this group rotates the electric and  magnetic charges separately and does not create a mix between them. 
For convenience we will refer to them in the heterotic language as the S- and the T-duality group respectively in the future. 

Of course, one can combine the dualities between M- and type IIA, IIB string theories discussed earlier in section \ref{M-theory} and \ref{S_duality_section} with the above new IIA-heterotic dualities and thereby construct a new web of dualities: 
\cite{Hull:1994ys,Witten:1995ex}
$$
\text{IIA}/K3\times T^2\sim\text{IIB}/K3\times T^2\sim\text{M-theory}/K3\times T^2 \times S^1\sim\text{heterotic} /T^6\;.
$$
For later reference we will now write down the charged objects giving the charges \(\big(\begin{smallmatrix} P \\ Q\end{smallmatrix}\big) \in \G^{6,22} \oplus \G^{6,22} \) in the above different duality frames. Seperating the charge lattice into four parts
\be\label{het_duality_chain}
\G^{6,22}  \cong \G^{3,19} \oplus  \G^{1,1} \oplus  \G^{1,1}\oplus  \G^{1,1} \cong H^{2}(K3,\Z)  \oplus  \G^{1,1} \oplus  \G^{1,1}\oplus  \G^{1,1}
\ee
with respective bilinear form given by \(C_{AB} = \int_{K3} \a_A\wedge \a_B \;, A,B = 1,\dotsi,22\) and \(U=\big(\begin{smallmatrix} 1&0 \\ 0&1\end{smallmatrix}\big) \) (\ref{lattice_1_1_U}), and using the duality relations between charged objects summarised Table \ref{brane_reduction} and Table \ref{M_IIB_chain}, we obtain the following Table \ref{N=4_brane_dualities} of charged objects of this theory  in its different frames.

Since different duality frames gives different perspectives in counting states, we will use this table extensively when we later derive the microscopic degeneracies of BPS states of this theory.

\begin{table} 
\centering
\begin{tabular}{ccccc}
\multicolumn{5}{c}{\bf magnetic and electric charges (P,Q) \(\boldsymbol{ \in \G^{6,22}\oplus \G^{6,22}}\)}\\
\toprule
&het/&IIA/ &M-th/ & IIB/\\
&
\tiny\( {S_{(2)}^1\!\!\times\!\! S_{(3)}^1\!\!\times\! \!S_{(4)}^1\!\!\times\! \!T^3}\)
&
\tiny\(S_{(2)}^1\!\!\times\!\! S_{(3)}^1\!\!\times\!\! K3\)
&
\tiny\(S_{(1)}^1\!\!\times\!\! S_{(2)}^1\!\!\times\!\! S_{(3)}^1\!\!\times\!\! K3\)
& 
\tiny\(S_{(1)}^1\!\!\times\!\! S_{(3)}^1\!\!\times\!\! K3\)\\
\rowcolor[gray]{0.9}{}
&& &&\\
\multirow{2}*{\(\G^{1,1}\)} 
& p(4) & D0 & p(1) & F1(1) \\
&F1(4) & D4 (K3) & M5(1,K3) & NS5(1,K3)\\
\midrule
\multirow{2}*{\(\G^{1,1}\)} 
& p(2) & p(2) & p(2) & \cellcolor[gray]{0.6}D1(1) \\
&F1(2) & NS5(2,K3) & M5(2,K3) & \cellcolor[gray]{0.6}D5(1,K3)\\
\midrule
\multirow{2}*{\(\G^{1,1}\)} 
& p(3) & p(3) & p(3) &\cellcolor[gray]{0.6}p(3) \\
&F1(3) & NS5 (3,K3) & M5(3,K3) & KKM(\(\hat{1}\))\\
\midrule
\(\G^{3,19}\)& \(q_A\) & D2\((\a^A)\) & M2\((\a^A)\) & D3\((1,\a^A)\) \\
\rowcolor[gray]{0.9}{}
&&&&\\
\multirow{2}*{\(\G^{1,1}\)} 
& NS5(\(\hat{4}\)) & D2(2,3) & M2(2,3)& F1(3) \\
&KKM(\(\hat{4}\)) & D6 (2,3,K3) & TN(2,3,K3) & NS5(3,K3)\\
\midrule
\multirow{2}*{\(\G^{1,1}\)} 
& NS5(\(\hat{2}\)) & F1(3) &  M2(1,3)& D1(3) \\
&KKM(\(\hat{2}\))& KKM(\(\hat{2}\)) & KKM(\(\hat{2}\))& D5(3,K3)\\
\midrule
\multirow{2}*{\(\G^{1,1}\)} 
& NS5(\(\hat{3}\)) & F1(2) & M2(1,2) & \cellcolor[gray]{0.6}p(1) \\
&KKM(\(\hat{3}\)) &KKM(\(\hat{3}\))  & KKM(\(\hat{3}\))& \cellcolor[gray]{0.6}KKM(\(\hat{3}\))\\
\midrule
\(\G^{3,19}\)& \(p^A\) & D4\((2,3,C_{AB}\a^B)\) & M5\((1,2,3,C_{AB}\a^B)\) & D3\((3,C_{AB}\a^B)\)  \\
\bottomrule
\end{tabular}
\caption{\label{N=4_brane_dualities}\footnotesize A chain of dualities relating the charged objects in the different \({\cal N}=4,\,d=4\) string theories, where \(\a_A\)'s are a basis of the twenty-two dimensional lattice \(H^2(K3,\Z) \cong \G^{3,19}\) with the bilinear given by \(C_{AB}= \int_{K3} \a_A \wedge \a_B\).}
\end{table}

\part{Multi-Holes and  Bubbling Solutions}

This part of the thesis contains two chapters. 

In chapter \ref{Black Holes and Multi-Holes} we first discuss various important properties, including the attractor mechanism and the presence of walls of marginal stability for multi-hole solutions, of the supersymmetric solutions of the \({\cal N}=2\), D=4 supergravity theories discussed earlier in section \ref{Four- and Five-Dimensional Low Energy Supergravity Theory}. Later in section \ref{In Coordinates Sugrav} we explicitly work out the details of these solutions using a type IIA compactification language. 

After we have introduced the necessary background, in chapter \ref{More Bubbling Solutions} we move on to discussing properties of the lift of these solutions to five-dimensions using the ideas discussed in section \ref{4D-5D Connection}. In particular we will study closely the case in which the M-theory limit is taken. Finally we focus on specific choice of charges such that the five-dimensional solution is smooth and horizonless. This chapter is based on the results reported in publication \cite{Cheng:2006yq}.

\chapter{Black Holes and Multi-Holes}
\label{Black Holes and Multi-Holes}

In section \ref{Four- and Five-Dimensional Low Energy Supergravity Theory} we have introduced the \(d=4\), \({\cal N}=2\) supergravity theory as the low-energy effective theory of type II compactification, together with the four-dimensional supersymmetric algebra and the concept of BPS states. In this section we will discuss the BPS solutions of this supergravity theory. In particular we will focus on stationary solutions, including the supersymmetric black hole and multi-hole solutions. As we will see, these solutions exhibit very interesting properties, which constitute part of the motivations for the research presented in the present thesis. Among them are the black hole attractor mechanism and the phenomenon of walls of marginal stability for multi-hole solutions. 

This chapter is mostly review by nature and is organised as follows. In the first section we present the generic stationary BPS solutions in a symplectic-invariant formulation. In the second section we discuss the attractor mechanism for solutions with a single black hole. In section three we focus on multi-hole solutions which contain multiple black holes, and summarise their angular momentum, moduli dependence and other existence criteria. In section four, we first unwrap the equations in the previous sections and rewrite them in components given a basis of the symplectic bundle, so that we obtain the set of equations ready to be used in actual calculations. After that, we specialise in 
the type IIA setup and present the explicit solutions in details, which will be needed for the next chapter.

\section{General Stationary Solutions}
\label{General Stationary Solutions}
\setcounter{equation}{0}
Here we are mainly interested in the stationary solutions of the \({\cal N}=2\), d=4 supergravity theories described in section \ref{Four- and Five-Dimensional Low Energy Supergravity Theory}, satisfying the BPS bound and preserving some supersymmetries. Using the stationary and flat base space Ansatz
\be\label{metric_ ansatz}
ds^2 = -\frac{\p}{S(\vec{x})} (dt+\w)^2 +\frac{S(\vec{x})}{\p} dx^i dx^i \;,
\ee
where \(S: \R^3 \rightarrow \R^+\) is a positive-definite function and \(\w = \w_i dx^i\) is a spatial one-form. 
The supergravity and vector multiplet part of the action (\ref{n2_sugrav_action}) can be written in a not manifestly spacetime covariant but manifestly symplectic invariant form \cite{Denef:2000nb}

\be\label{reduced_action}
-16 \p G_{\text{\tiny N}}^{(4)} L  = 
({\cal G}, {\cal G})-4\sqrt{\frac{\p}{S}} \, (Q + d\a + \frac{\p}{2S}\star d\w )\wedge \langle {\cal G}, \im (e^{-i\a} \bO) \rangle 
+ \text{  tot. der.  }\;.
\ee
This action in a very concise form takes some explanation. First of all, 
the \(\star\) without any subscript refers to the Hodge star with respect to the three-dimensional flat metric of the base space, \(d\) refers to the derivative in \(\R^3\), and \( \a: \R^3\rightarrow \R\) is at this point an arbitrary function.
Also recall that  \(\bO\) is the normalised version of the section of the symplectic bundle \({\cal E}\) times the line bundle \({\cal L}\) defined in (\ref{def_bo}) as
\be
\bO = \frac{\O}{\sqrt{i\langle \O, \bar{\O}\rangle}} = e^{{\cal K}/2}  \,\O\;,
\ee
and \(iQ\) is the connection one-form (\ref{connection_normalised}) for it.

Finally, \({\cal G}\) is a spatial two-form taking value in the symplectic bundle \({\cal E}\) over \(\R^3\). Especially it is a spatial two-form times an element of \(H^{3}(Y)\) in the type IIB and a spatial two-form times an element of \(H^{2*}(X)\) in the type IIA setup. It is given by the spatial part \({\boldsymbol {\cal F}}\) of the field strength \({\cal F}\) in ten dimensions (\ref{self_duality}) as
$$
{\cal G} = {\boldsymbol {\cal F}} -2\sqrt{\frac{S}{\p}}\, \im \star D\,(e^{-i\a} \bO) + 2\sqrt{\frac{\p}{S}}\, \re D\,(e^{-i\a} \bO \,\w) \,,
$$
where 
$$
D= d + i \left( Q + d\a + \frac{\p}{2S} \star d\w \right)
$$
and the spatial three-form \( ({\cal G}, {\cal G})\) is defined as
$$
 ({\cal G}, {\cal G}) = \frac{\p}{S}\,\frac{1}{1-\frac{\p^2}{S^2}\,\w^2} 
 \int_{\text{C-Y}} {\cal G} \wedge \star {\cal G} -\frac{\p^2}{S^2} 
( {\cal G} \wedge \w ) \wedge \star ( {\cal G} \wedge \w ) + 
\frac{\p}{S}  {\cal G} \wedge \star (\w \wedge \star  {\cal G})\;.
$$

Finally we remark that, to justify the metric Ansatz (\ref{metric_ ansatz}) one should actually treat the above action (\ref{reduced_action}) as an effective action that has to be supplemented by a constraint \cite{VanProeyen:2007pe}. 

From the form of the effective action (\ref{reduced_action}) we see that a local minimum of the action is given by
\be\label{minimum_action}
Q + d\a + \frac{\p}{2S} \star d\w = {\cal G} = 0 \;,
\ee
which leads to the following solution of the metric, the scalars and the gauge fields in terms of a set of harmonic functions
, namely \(H: \R^3 \rightarrow H^3(X,\R)\) for type IIB and \(H: \R^3 \rightarrow H^{2*}(X,\R)\) for type IIA compactifications:
\begin{align}
\label{eq_1}
2 \sqrt{\frac{S}{\p}}\, \im(e^{-i \a} \bO) &= -H\\
\label{eq_2}
d\w &= \star \langle dH, H \rangle \\
\label{eq_3}
{\cal A} &= 2 \sqrt{\frac{\p}{S}}\,\re(e^{-i \a} \bO) \, (dt+\w) + {\cal A}_d
\\ 
\intertext{where the Dirac part of the gauge field is given by}
\label{eq_4}
d {\cal A}_d &= \star dH\;.
\end{align}
The Dirac part of the gauge field is easy to solve since the equation is linear and we have already seen the solution to the single-monopole case in our construction of the Taub-NUT space (\ref{Taub-NUT as gibbons-hawking}). In what follows we will therefore refer to this part of the gauge fields simply as \({\cal A}_d\) and focus on other more complicated parts of them. 

By imposing the condition that metric approaches that of a flat Minkowski space with the usual normalisation and that \(\int_{S^2_{(i)}} {\cal F} = \G_i\) around a point source of charge \(\G_i\), we see that the harmonic functions are given by the charges and the asymptotic moduli as
\be\label{harmonic}
H = \sum_{i}\,\frac{\G_i}{r_i} + h = \sum_{i}\,\frac{\G_i}{|\vec{x}-\vec{x}_i|} - 2  \im(e^{-i \a} \bO)\lvert_\inf\;,
\ee
where \(\vec{x}_i\) is location of the \(i\)-th center in the flat 3d base and ``\(\inf\)" denotes the fact that the expression should be evaluated at spatial infinity. In other words, the constant term in the harmonic function is determined in terms of the asymptotic moduli and the total charges of the solution.

Recalling that the central charge function is defined as (\ref{central_charge_1})  \be Z(\G;\bO)= \langle \G, \bO \rangle\ee for \(\G\) is a combination of three- (even-) forms  in the type IIB (IIA) language. From the form of (\ref{eq_1}) we now define 
$$
Z(\G) = Z(\G;\bO^*(\G))
$$
with \(\bO^*(\G)\), the so-called attractor moduli, satisfying the equation
$$
2 \sqrt{\frac{S(\G)}{\p}}\, \im(e^{-i \a} \bO^*(\G)) = -\G\;.
$$
Contracting (\ref{eq_1}) with \(\langle,\bO\rangle\)  we then get 
\be\label{central_charge}
Z(H(\vec{x})) = e^{i\a(\vec{x})} \sqrt{\frac{S(H(\vec{x}))}{\p}}\;,
\ee
and from the boundary condition, in particular the correct fall-off of the angular momentum one-form \(\w\), we conclude that
\be\label{bc_alpha}
e^{i\a}\lvert_\inf = \frac{Z(\G;\bO_\inf)}{|Z(\G;\bO_\inf)|}\;.
\ee

Furthermore,  from (\ref{central_charge}) one can see that the solution saturates the BPS bound
$$
M = |Z(\G;\bO)|_\inf
$$
where \(\G= \sum_i \G_i\) is the total charge, and therefore preserves four unbroken supersymmetry.

Our exposition here is similar to that in \cite{Denef:2000nb}. See also  \cite{LopesCardoso:2000qm} for another construction of  these stationary solutions, with \(R^2\) corrections included.

\section{Extremal Black Holes and Attractor Mechanism}
\index{attractor mechanism}
\setcounter{equation}{0}
In this section we will focus on the static black hole solutions. Especially we will discuss an important property of them, namely the attraction mechanism for the scalar fields of the theory, in more details.

Using (\ref{central_charge}) one can now write (\ref{eq_1}) as
\be\label{att_0}
-2 \im (\bar{Z}(H) \bO^*(H)) = H\;.
\ee
It's then obvious that the solution of scalar fields  is invariant under a rescaling of the harmonic functions, namely 
$$
\bO^*(\l H) = \bO^*(H) \quad \forall \;\l \in \R\,.
$$
Especially, considering now a solution with only one center and with arbitrary asymptotic moduli \(\bO\lvert_\inf\)
$$
H = \frac{\G}{r} - 2  \im(e^{-i \a} \bO)\lvert_\inf \;.
$$
In this case we see that the solution is spherically symmetric and static with \(\w=0\).
From the fact that 
$$
|Z(H(\vec{x}))|^2 = \frac{S(H(\vec{x}))}{\p} = -\frac{1}{g_{00}} \rightarrow \inf \quad\text{as} 
\quad r\rightarrow 0
$$
we also conclude that there is an event horizon at the center \(r=0\). Furthermore one can read out the area of the spherical horizon, which is 
\be\label{Bekenstein_Hawking_entropy}
A = 4 S(\G)\;.
\ee
Therefore \(S(\G) = \frac{A}{4} = \p |Z(\G)|^2\) is the macroscopic Bekenstein-Hawking entropy of this extremal black hole, a fact that justifies our choice of notation. For the thermodynamical properties of extremal black holes and the semi-classical analysis of them, see for example \cite{master_thesis} and references therein.

Near the center \(r \rightarrow 0\), the equation for the moduli \(\bO\)  (\ref{att_0}) gives
\be\label{attractor_equation}
-2 \im (\bar{Z}(\G) \bO^*(\G)) = \G\;.
\ee
The magic of the above equation is, no matter what the asymptotic moduli \(\bO\lvert_\inf\) is, near the center of a black hole the moduli is always fixed to be at the ``attractor point" or ``attractor value" given by the above equation \(\bO\lvert_{r\rightarrow 0}= \bO^*(\G) \). This is the so-called attractor mechanism for a single black hole solution and (\ref{attractor_equation}) is called the attractor or the stabilisation equation \cite{Ferrara:1995ih}.  
\index{attractor equation}

To understand the attractor mechanism better, it's illuminating to look at the so-called attractor flow equation, which is obtained by taking the derivative of (\ref{eq_1}). Using the covariant derivative (\ref{connection_normalised}) we get
$$
dH = \sqrt{\frac{S}{\p}}\,\{\,[Q + d\a +\frac{i}{2} d\log(\frac{S}{\p})]\,e^{-i\a} \,{\bO} + i e^{-i\a} \,{\cal D} \bO + \text{ c.c.}\}
$$
Contracting the above equation with \({\cal D}_A \O\), and writing \({\cal D} \O= dz^B \, {\cal D}_B \O\) and using the K\"ahler metric  (\ref{Kahler_metric_1}), we get 
$$
\frac{\pa z^A}{\pa r} = e^{i\a}\,\frac{1}{r^2}\,\sqrt{\frac{\p}{S}}\,g^{A\bar{B}}\,{\bar{\cal D}_{\bar{B}}\bar{Z}}\;.
$$
From this we see that the phase of the central charge is constant along the radial evolution, and the radial evolution of the central charge \(Z(\G;\bO)\) is given by
$$
\frac{\pa |Z|}{\pa r} = \frac{1}{2|Z|} \left( \bar{Z} {\cal D}_r Z +\text{ c.c.}\right) = \frac{1}{r^2} \sqrt{\frac{\p}{S}}\, |{\cal D}Z|^2 \geq 0
$$
where \( |{\cal D}Z|^2 = g^{A\bar{B}} {\cal D}_A Z  \bar{\cal D}_{\bar B} \bar{Z}\). This means that the value of the central charge always increases when one moves further and further away from the black hole center in the spatial directions, or from the attractor point the in the moduli space. See for example Fig \ref{central_charge_fig}.
Furthermore, except for the point at infinity \(r\rightarrow \inf\), the only place where the inequality is saturated is at the attractor point (or at the black hole horizon in the spacetime picture), where \({\cal D}Z=0\). 
Indeed the attractor equation can be derived by requiring that the moduli renders the central charge to be at a local fixed point. To see this, note that \({\cal D}Z=\langle \G, {\cal D}\bO \rangle =0\) requires that the moduli must be such that 
$$
\G = a \bO + a^* \,\bar{\bO}\;
$$
for some complex number \(a\). 
Contracting the above equation with \(\langle\,,\bar{\bO} \rangle\) gives the value of \(a\) and then gives the attractor equation (\ref{attractor_equation}). In particular, for the type IIB setup the above equation has a simple geometric explanation that the moduli adjust themselves as one approaches black hole horizon such that the charge \(\G\in H^3({X,\Z}) \in H^{3,0}(X) \oplus H^{0,3}(X)\).
\begin{figure}
\centering
\includegraphics[width=10cm]{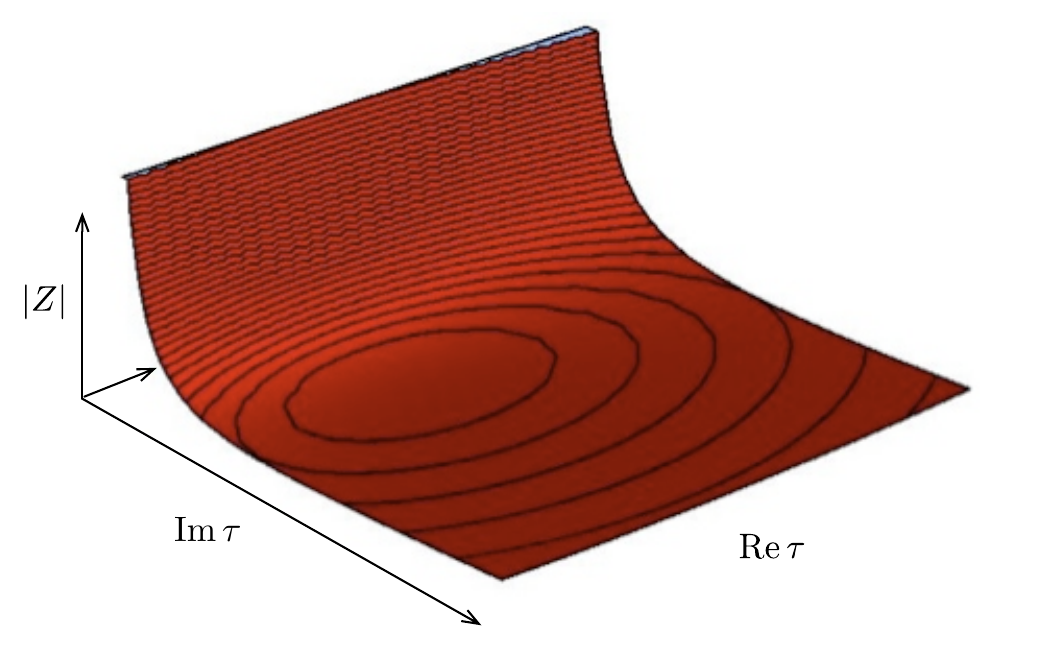} \setlength{\abovecaptionskip}{5pt}
\caption{\label{central_charge_fig}\footnotesize{The magnitude of the central charge \(|Z(\G,\t)|\) in K3\(\times T^2\) compactification as a function of the torus modulus \(\t\), with the rest of the moduli already fixed at the attractor point. The contour lines are the constant \(|Z(\G,\t)|\) line and the function has a unique minimum at the attractor value \(\t^*\).}}
\setlength{\belowcaptionskip}{5pt}
\end{figure}

\section{Properties of Multi-holes}
\label{Properties of Multi-holes}
\setcounter{equation}{0}
\index{multi-hole}
Many qualitative new features emerge when one considers more general solutions with more than one centers \cite{Denef:2000nb,LopesCardoso:2000qm,Bates:2003vx}. First of all, the solution will no longer have the \(SO(3)\) spherical symmetry and will in general no longer be static. In other words, the rotation one-form \(\w\), determined by (\ref{eq_2}) and boundary condition \(\w \rightarrow 0\) as \(|\vec{x}| \rightarrow \inf\), will  generically no longer vanish in the interior of the spacetime. As we will see later, this can be understood as due to the fact that, an electron and a monopole at different locations in the space create an electromagnetic field which induces an ``intrinsic" angular momentum of the spacetime. 
Furthermore,  there is a  integrability condition for \(\w\) (\ref{eq_2}) \cite{Denef:2000nb}
\be \label{integra_1}
d\,d\w = d\star \langle dH,H \rangle =  \langle d\star dH,H \rangle= 0 \;,
\ee
which can be regarded as the condition for the absence of closed timelike curves (CTC's) near the line segments connecting pairs of centers. This condition gives constraints on the possible locations of the centers and in turn constraints on the boundary condition for the scalars for a multi-hole solution with a certain charge distribution to exist. 

In this section we will discuss the above-mentioned properties of these supersymmetric stationary solutions with multiple centers, which will be important in our future discussion. 

\subsection{Walls of Marginal Stability}
\label{Walls of Marginal Stability}
\index{walls of marginal stability}

As mentioned above, solutions with more than one center has to satisfy the integrability condition for the rotation one-form (\ref{integra_1}). 

Plugging in the expression for the harmonic function (\ref{harmonic}), one obtains one equation for each center
\be\label{integrability}
\sum_j \frac{\langle \G_i,\G_j\rangle }{r_{ij}} =  - \langle \G_i,h\rangle 
\ee
where \(r_{ij}=|\vec{x}_i - \vec{x}_j|\) and \(h = -2 \im (e^{-i\a} \bO)\lvert_\inf\), and the sum of the equations for all centers is trivially satisfied
\be
\sum_i\sum_j \frac{\langle \G_i,\G_j\rangle }{r_{ij}} =0=  -\sum_i \langle \G_i,h\rangle\;.
\ee
 In particular, for the case with two centers there is just one integrability condition which gives
\be\label{distance}
r_{12} = |\vec{x}_1-\vec{x}_2| = \frac{\langle \G_1,\G_2\rangle}{2}
\,\frac{M}{\im(Z_1\bar{Z}_2)\lvert_\inf}
\ee
where 
$$
M = |Z(\G;\bO)|_\inf = |Z(\G_1;\bO)+Z(\G_2;\bO)|_\inf 
$$
is the ADM mass of the solution and \(Z_1\lvert_\inf =Z(\G_1;\bO)\lvert_\inf\). 
This shows that a solution with two centers with non-zero intersection (Dirac-Zwanziger) product \(\langle \G_1,\G_2\rangle \neq 0\) is really a bound state of the two centers, since their distance cannot be adjusted freely but instead is constrained to be a certain fixed value.  

Furthermore, an interesting consequence of this is that the solution only exists in some part of the moduli space. More explicitly, from the fact that the distance  must always be a positive number, we see that the moduli space is divided into two parts by the wall
\be\label{wall_marginal_stability}
 \im(Z_1\bar{Z}_2)\lvert_\inf = 0\;,
\ee
and it is only possible to have a solution with centers of these charges at the side of the wall which satisfies
\be\label{stability_condition}
\langle \G_1,\G_2\rangle \, \im(Z_1\bar{Z}_2)\lvert_\inf > 0\;.
\ee
We will therefore call this co-dimenional one wall (\ref{wall_marginal_stability}) ``the wall of marginal stability for charge \(\G_1\) and \(\G_2\)".

The presence of the wall of marginal stability can be understood in the following way. A necessary condition for a bound state of two particles to decay is that the mass of the bound state is the same as the sum of the mass of each individual constituent 
\ben
M &=& M_1 + M_2 \Leftrightarrow\, |Z_1+ Z_2|_\inf = |Z_1|_\inf + |Z_2|_\inf \\ \notag &\Leftrightarrow& \im(Z_1\bar{Z}_2)=0 \quad,\quad\re(Z_1\bar{Z}_2)> 0\;.
\een
From this we conclude that the condition of a physical wall of marginal stability should be further implemented by the requirement that the two central charges are aligned rather than anti-aligned. 

After determining the location of the walls of marginal stability in the moduli space, we still need to determine which side is the stable and which side is the unstable side. We will provide a heuristic derivation of it here. 

Let us consider the scenario that the asymptotic moduli are fixed at the attractor value of the total charge \(\bO\lvert_\inf = \bO^*(\G=\G_1+\G_2)\), then the attractor equation (\ref{attractor_equation}) gives 
$$
\im (Z_1 \bar{Z}_2)\lvert_\inf = -\frac{1}{2} \langle \G_1, \G_2 \rangle 
\quad\text{when}\quad \bO\lvert_\inf = \bO^*(\G)\;.
$$

For this specific choice of the background moduli, we know that there exists a single black hole solution in which the background moduli is constant throughout the whole space. In other words there is no attractor ``flow" in this case. By arguing that this configuration should be more energetically favorable than any other solution with spatial gradient of the scalar fields, including the two-centered solution considered above, we require that the total attractor point must lie on the unstable side of the wall, and derive (\ref{stability_condition}) as a condition for existence of a specific two-centered solution.

\subsection{Angular Momentum of the Spacetime}

Another qualitatively new feature of the multi-hole solution with mutually non-local centers (\(\langle \G_i, \G_j \rangle \neq 0\)) is that the spacetime is no longer static but rather stationary with angular momentum. This can be seen from the expression for \(\w\) (\ref{eq_2}) and can be understood as a consequence of the fact that there are now electrons and monopoles at different points in the space and the electromagnetic fields therefore give contribution to the angular momentum. 

Let's now solve for \(\w\). From (\ref{eq_2}) and using the integrability constraint one obtains
$$
d\w = \sum_{i,j} \frac{\langle\G_i,\G_j\rangle}{2}\,\star\left(
\frac{dr_i^{-1}}{r_j} -\frac{dr_i^{-1}}{r_{ij}} - \;(i\leftrightarrow j) \right)\;.
$$
Recall that the star without subscripts denotes the Hodge star in three-dimensional flat base space. 
It is therefore enough to solve \(\w\) pair-wise as
\begin{align} \notag
\w  &= \sum_{i,j}\,\frac{1}{2}\, \w_{ij} = \sum_{i<j} \w_{ij}\\
\intertext{
where} \notag
d\,\w_{ij} &= \langle\G_i,\G_j\rangle 
\star\left(
\frac{dr_i^{-1}}{r_j} -\frac{dr_i^{-1}}{r_{ij}} - \;(i\leftrightarrow j) \right)\;.
\end{align}

Using again the elliptic coordinates as we have used for Eguchi-Hanson space (\ref{elliptic_coordinates}) for the two centers \(\vec{x}_i\) and \(\vec{x}_j\), with now \(2a = r_{ij}\), then the above equation together with the boundary condition that \(\w \rightarrow 0\) at spatial infinity gives
$$
\w_{ij}  =\frac{\langle\G_i,\G_j\rangle}{2 r_{ij}}\, \frac{\cosh \eta -1}{\cosh^2 \eta-\cos^2\th  } \,\sin^2\th \,d\j\;.
$$

From the asymptotic fall-off
$$
\w_{ij}  \rightarrow \frac{\langle\G_i,\G_j\rangle}{r} \sin^2\th \,d\j 
$$
we conclude that the conserved angular momentum of the spacetime is
$$
\vec{J} = \frac{1}{4}\sum_{i,j}\, \langle\G_i,\G_j\rangle \, \frac{\vec{x}_i-\vec{x}_j}{|\vec{x}_i-\vec{x}_j|}\;.
$$

Furthermore, by solving for \(\w\) without using the integrability condition, one can show that there will in general be closed timelike curve (CTC) around the line segment \(\vec{x}_i-\vec{x}_j\) in the \(\R^3\) base, and the length of the line segment such that this does not happen is exactly the length ordained by the integrability condition. One can therefore also interpret 
the integrability condition (\ref{integrability}) as the physical constraint of the impossibility of time machines.

\subsection{Split Attractor Flow}
\label{Split Attractor Flow}

\index{split attractor flow}
\index{attractor flow tree}
In (\ref{eq_1})-(\ref{eq_4}) we have seen how to solve for the solution given a configuration of centers with particular charges and background moduli. In (\ref{integrability}) we saw that not all arbitrary configurations permit a solution, because the integrability condition does not always permit a solution for any given background moduli. 
In fact the situation is even more subtle than that. 
 Namely, the satisfaction of the integrability condition alone is not sufficient to establish the existence of a solution. 
 This is because there is another condition that the solution must satisfy to qualify as a physical solution, namely that the
 central charge function \(Z(H;\bO)\) does not hit a zero anywhere in the space
\be\label{positive_entropy_function}
\frac{S(H(\vec{x}))}{\p} = |Z(H;\bO)|^2 > 0 \qquad \text{for all   }\vec{x}\in\R^3\;.
\ee

In general, this is of course a very difficult condition to check, because it is a non-local condition in the sense that it has to be satisfied everywhere in the space. In relation to this difficulty we quote here the very useful split attractor flow conjecture, which states that a solution exists if and only if a split attractor flow tree exists in moduli space, which starts at the asymptotic value of the scalars and terminates at the attractor points of each constituent back hole. 
See \cite{Denef:2000nb,Denef:2007vg} for more details.

\section{In Coordinates}
\label{In Coordinates Sugrav}
\index{entropy function}
\setcounter{equation}{0}
In the previous sections in this chapter, we have written down general stationary supersymmetric solutions of \(d=4, \;{\cal N}=2\) supergravity theories, including the static single-hole solutions and the rotating multi-hole solutions. All this was done in a symplectic-invariant formulation. In practice, when we are working with a specific theory we often need to express the solutions in components, in order to be able to directly interpret them.

In this section we will therefore take up the straightforward task of unwrapping the equations in previous sections and rewriting them in terms of components using a specific basis for the symplectic bundle. These rewritten equations are then ready to be used in actual calculations. Specifically, we work out in detail the solutions for the supergravity theory one gets from type IIA compactifications, which will be needed in the next chapter of this part of the thesis.

As discussed in \cite{Bates:2003vx}, a solution can be explicitly written in terms of a single function, called the entropy function
$$
S(\G)= \p |Z(\G)|^2 = \p |Z(\G;\bO^*(\G))|^2 \;,
$$
where the attractor moduli \(\bO^*(\G)\) of  charge \(\G\) are given by the attractor equation (\ref{attractor_equation}). Given such an entropy function, the full solution basically follows from just replacing the charge vector \(\G\) with the harmonic functions \(H\). Let's begin with writing down the moduli, the charge, and the harmonic functions in components. 
Using (\ref{def_bo}), (\ref{def_O_IIB}) and (\ref{def_O_IIA_2})
$$
\bO = e^{{\cal K}/2}\,(-X^I \a_I + F_I \b^I)
$$
and write the charge vector as
\be\label{charge_in_coord}
\G = p^I \a_I + q_I \b^I\;
\ee
and similarly for the harmonic functions
\be\label{harmonics_in_coord}
H(\vec{x})= H^I(\vec{x}) \a_I + H_I(\vec{x}) \b^I\;,
\ee
we obtain the magnetic part of the attractor flow equations in components by contracting (\ref{eq_1}) with \(\langle\, , \b^I\rangle\)
$$
H^I = 2  e^{{\cal K}/2}\, \im (\bar{Z} X^I)\;.
$$
Furthermore, from 
$$
\frac{1}{\p}\frac{\pa}{\pa H_I} S(H) = \frac{\pa}{\pa H_I}  |Z(H)|^2 =
2 e^{{\cal K}/2}\, \re (\bar{Z} X^I)
$$
we get the full solution for the scalar fields
\bea\notag
 2  e^{{\cal K}/2} \bar{Z} X^I (H(\vec{x}))&=& i H^I + \frac{1}{\p}\frac{\pa}{\pa H_I} S\\ \label{scalar_sol}
 \Rightarrow t^A(H(\vec{x})) &=& \frac{X^A}{X^0} = \frac{ i H^A + \frac{1}{\p}\frac{\pa}{\pa H_A} S}{ i H^0 + \frac{1}{\p}\frac{\pa}{\pa H_0} S}\;.
\eea
On the other hand, the vector fields are given by contracting (\ref{eq_3}) with \(\langle \,,-\b^I \rangle\) 
\be\label{vector_general}
A^I = \frac{1}{S}\frac{\pa S}{\pa H_I} (dt+\w) - A_d^I \quad,\quad
dA_d^I =\star dH^I\;.
\ee

As an illustration and for later use, let's work out the entropy function for the type IIA case (\ref{def_O_IIA_1}) with 
$$
\bO = -e^{{\cal K}/2}\,e^{-t}\;.
$$

First of all, the electric part of the attractor equation is obtained by contracting (\ref{attractor_equation}) with \(\langle \,,\a_I \rangle\):
\bea \label{electrc_attractor}
2 e^{{\cal K}/2} \, \im \left(\bar{Z}\frac{(t^2)_A}{2}\right) &=& q_A\\
2 e^{{\cal K}/2} \, \im \left( \bar{Z}\frac{(t^3)}{6}\right)  &=& -q_0\;,
\eea
where we have used the short hand notation \((t^2)_A  \equiv D_{ABC} t^B t^C\) and \(t^3 = D_{ABC} t^A t^B t^C\).
Define the variable \(y^A\) and \(L\) by \cite{Shmakova:1996nz}
\bea\notag
\frac{\pa}{\pa q_0} \big(\frac{S}{\p}\big)^2 &=& 4 (p^0)^2 L\\
\label{IIA_attr_variable}
\frac{\pa}{\pa q_A} \big(\frac{S}{\p}\big)^2 &=& 4 p^0 [-y^A (\frac{y^3}{6}) + p^A L]\;,
\eea
and using 
$$
2  e^{{\cal K}/2} \bar{Z} X^I= i p^I + \frac{1}{\p}\frac{\pa}{\pa q_I} S\quad,\quad t^A = \frac{X^A}{X^0}\;,
$$
the electric half of the attractor equation (\ref{electrc_attractor}) can be written as
\bea\notag 
(y^2)_A &=& -2 q_A + \frac{(p^2)_A}{p^0}\\ \label{attr_IIA_explicit}
L &=& -\frac{q_0}{2} - \frac{p\cdot q}{ 2 p^0} + \frac{p^3}{(p^0)^2}\;.
\eea
Plugging them back into (\ref{IIA_attr_variable}) we finally obtain the expression for the entropy function
\be \label{entropy_IIA}
S(\G) = 2\p \sqrt{p^0 Q^3 - (p^0)^2 \,L^2}\;,   
\ee
where 
\(Q^3 = (\frac{y^3}{6})^2\).

This entropy function, as promised, gives now the full solution including the scalar fields and vector fields
\bea\label{sol_scalar}
t^A(H(\vec{x})) &=& \left(\frac{H^A}{H^0} - \frac{L}{Q^{3/2}}y^A\right) +
i \frac{S}{2\p} \frac{y^A}{H^0 Q^{3/2}}\\ 
A^0(H(\vec{x})) &=&2 \big(\frac{\p}{S}\big)^2 (H^0)^2 L (dt+\w) - A_d^0 \\ \label{sol_vector}
A^A (H(\vec{x}))&=& 2 \big(\frac{\p}{S}\big)^2 H^0\, [-Q^{3/2} y^A + H^A L] (dt+\w)  -A_d^A\;,
\eea
where \(y^A(H)\), \(L(H)\) and \(S(H)\) are given by (\ref{attr_IIA_explicit}) and (\ref{entropy_IIA}) but now with the charges \(p^I, q_I\) replaced by the corresponding harmonic functions \(H^I, H_I\).

But often we will find the above form of the attractor equation clumsy to use. It is easy to see that the equations (\ref{attr_IIA_explicit}) is not ready to be solved when \(p^0=0\). More generally, from the form of the entropy function (\ref{entropy_IIA}), we have to worry about the warp factor of the spacetime \(S(H(\vec{x}))\) vanishes near the region \(H^0 \rightarrow 0\), or from the scalar solution (\ref{sol_scalar}) we have to worry that the Calabi-Yau decompactifies near this region.
But actually these are just consequences of the specific variables (\ref{IIA_attr_variable}) we have used. It is therefore sometimes desirable to use a different set of variables to write our solutions with. 

Instead of defining variables \(y^A\) and \(L\) as in (\ref{IIA_attr_variable}), we now define instead \(\iota^A\) and \(\ell\) by \cite{Cheng:2006yq}
\bea\notag
\frac{\pa}{\pa q_0} \big(\frac{S}{\p}\big)^2 &=& 4 \ell \\
\label{IIA_attr_variable_2}
\frac{\pa}{\pa q_A} \big(\frac{S}{\p}\big)^2 &=& 4  [-{\cal Q}^{3/2} (p^A + p^0 \iota^A) + p^A \frac{\ell}{p^0}]\;,
\eea
where \({\cal Q}^{3/2}\) is given by \(\iota^A\) and \(\ell\) to be
\bea \notag
{\cal Q}^{3/2} &=& \ell + (p^0)^2 \l\\ \label{formulation2_attractor_eq}
\l &=& \frac{q_0}{2} - \frac{\iota\cdot q}{3} - \frac{p \iota^2}{12}\;.
\eea
Again we have used the shorthand notation \(p \iota^2 = D_{ABC} p^A \iota^B \iota^c\).

Now the attractor equations in terms of these new variables \(\iota^A\) and \(\ell\) are
\bea \notag
D_{ABC}p^B\,\iota^C&=& -q_A - \frac{p^0}{2} (\iota^2)_A\\
\label{attractor_IIA_2}
\ell &=& -(p^0)^2 \, \frac{q_0}{2} - p^0 \, \frac{p\cdot q}{2} + \frac{p^3}{6}\;,
\eea
and the entropy function becomes
$$
S(\G) = 2\p \sqrt{2\l \ell + (p^0)^2 \l^2 }\;.
$$

Especially 
\be\label{D4_D2_D0_entropy}
S(\G) = 2\p \sqrt{\frac{p^3}{6}\, \hat{q}_0} = 2\p \sqrt{\frac{p^3}{6}\, (q_0 - \frac{1}{2}D^{AB}q_A q_B )} \quad\text{  when  }p^0 = 0  ,
\ee
where \(D^{AB}\) is the inverse of \(D_{AB} = -D_{ABC}p^C\).

The full solution for the scalar and the vector fields now reads
\bea \notag
t^A (H(\vec{x})) &=& {\cal Q}^{-3/2} \{ (-\ell \iota^A + H^0 H^A \l ) + i 
\frac{S}{2\p}\,(H^A + H^0 \iota^A)\}\\ 
\label{formulation2_sol_vec}
A^0(H(\vec{x})) &=& 2 \big(\frac{\p}{S}\big)^2 \ell (dt+\w) - A_d^0\\ \notag
A^A(H(\vec{x})) &=& 2 \big(\frac{\p}{S}\big)^2[ \, (-{\cal Q}^{3/2} \iota^A - H^0 H^A \l] (dt+\w) -  A_d^A\;,
\eea
where \(\iota^A\), \({\cal Q}^{3/2}\) and \(\ell\) are again solutions to (\ref{attractor_IIA_2}) with the charges \(p^I, q_I\) replaced by the harmonic functions \(H^I,H_I\). From the above form of the solution we see that the region with \(H^0 \rightarrow 0\) is not more prone to singularity than other regions, and verify the claim we made earlier that these are better variables to use in these situations.

\chapter{More Bubbling Solutions}
\label{More Bubbling Solutions}
This chapter of the thesis is based on the result reported in publication \cite{Cheng:2006yq},
in which we construct families of asymptotically flat, smooth, horizonless solutions with a large number of non-trivial two-cycles (bubbles) of \({\cal{N}}=1\) five-dimensional supergravity with an arbitrary number of vector multiplets. They may or may not have the charges of a macroscopic black hole and contain the known bubbling solutions as a sub-family. We do this by lifting various multi-center BPS states of type IIA compactified on Calabi-Yau three-folds, discussed in detail in section \ref{Black Holes and Multi-Holes}, and taking the decompactification (M-theory) limit. We also analyse various properties of these solutions, including the conserved charges, the shape, especially the (absence of) throat region and closed timelike curves, and relate them to the various properties of the four-dimensional BPS states. We finish by briefly commenting on their degeneracies and their possible relations to the fuzzball proposal of Mathur {\it{et al}}.

\label{Introduction}
\section{Introduction}
\setcounter{equation}{0}

The four-dimensional multi-center BPS solutions of type II string theory compactified on a Calabi-Yau three-fold have been derived in  \cite{Behrndt:1997ny,Denef:2000nb,LopesCardoso:2000qm,Denef:2001xn,Bates:2003vx}, and their lift to M-theory was, after the indicative work \cite{Bena:2004tk,Bena:2005ay},  explicitly written down in \cite{Gaiotto:2005xt} (see also \cite{Behrndt:2005he}). Recently, this idea of the 5d lift of 4d multi-center solutions have contributed to the understanding of black ring entropy  \cite{Gaiotto:2005xt,Bena:2005ni,Elvang:2005sa}, the relationship between the Donaldson-Thomas invariants and topological strings \cite{DVVafa}, and the OSV conjecture \cite{Denef:2007vg}. Indeed, with different choices of charges and Calabi-Yau background moduli, one can expect to have a large assortment of BPS solutions to \({\cal N}=1\) (8 supercharges) five-dimensional supergravity with various different properties by simply lifting various multi-center solutions to five dimensions.

On the other hand, Mathur and collaborators have proposed a picture of black holes different from the conventional one. According to this proposal, the black hole could actually be a coarse-grained description of a large number of smooth, horizonless supergravity solutions (``microstates", ``proto-black holes") which have the same charges as that of a ``real black hole". (see \cite{Lunin:2001jy}, \cite{Mathur:2005zp} and references therein). A question one might then ask is, do there exist some solutions in the zoo of the lifted multi-center solutions which possess this property? If yes, how many of them are there? And how to classify them?

To construct a solution like this via the 4d-5d connection, first of all in order to have the right global feature at spatial infinity (that it should approach \(\R_t\times \R^4\) but not \(\R_t\times \R^3 \times S^1\)), one would need to take the decompactification limit in which the M-theory circle is infinitely large at spatial infinity. In this limit the five-dimensional description is also the only valid one. Furthermore, for the smooth and horizonless feature we have to restrict ourselves to D6 or/and anti-D6 branes as the centers in 4D. To obtain non-trivial charges we then turn on the world-volume fluxes on these centers. Finally we lift the solutions with these charges and background to five dimensions. In this way we have indeed obtained a large number of asymptotically flat, smooth and horizonless solutions, to five-dimensional supergravity theories with an arbitrary number of vector multiplets, which may have the total charge of that of a black hole. Actually, if we restrict to the STU Calabi-Yau and make a special Ansatz of the K\"ahler moduli, we retrieve the known bubbling solutions of \cite{Berglund:2005vb,Bena:2005va,Bena:2006is}.\footnote{In \cite{Balasubramanian:2006gi} it has been observed that, if one adds a constant term to one of the harmonic functions in the Bena-Warner {\it{et al}} bubbling solutions, which corresponds to de-decompactify the extra dimension, and then reduce it, one would get a 4D multi-center solution. See also \cite{Saxena:2005uk} for a related discussion. } In a recent paper, through a more explicit study of the above-mentioned solutions, Bena, Wang and Warner \cite{Bena:2006kb} have constructed the first smooth horizonless solutions with charges corresponding to a BPS three-charge black hole with a classical horizon. Indeed, to understand this recent development has been the original motivation of the present work.

To be able to have a solution like this in the case of a general Calabi-Yau compactification further heightens the contrast between the picture of a black hole of Mathur {\it{et al}} and the conventional one . Unlike the torus case, a general Calabi-Yau with its complicated topological data is generically the biggest origin of a large black hole entropy \cite{Maldacena:1997de,Vafa:1997gr}. As we have mentioned, to have a horizonless solution lifted from four dimensions forces us to consider only rigid centers, i.e., those without any (classical) internal degrees of freedom associated to them. To reconcile these two pictures therefore seems to be much more challenging in the case of a general Calabi-Yau compactification. The authors of \cite{Balasubramanian:2006gi} have proposed a following picture: while the system is described by a D-brane bound state at weak  string coupling, it expands into a multi-particle system when we turn on the \(g_s\) and is thus described by a multi-centered supergravity solution, and further grows into a five-dimensional system when the string coupling is increased even further. While this picture has been carefully studied and tested in the case with the total charge {\it{not}} corresponding to that of a classical black hole  \cite{Denef:2002ru}, we don't seem to have much evidence to argue the same for the case with black hole total charges. In other words, {\it{a priori}} we don't see the reason why the D-brane bound state must open up into a multi-center configuration instead of staying together and form a black hole in the conventional sense, as \(g_s\) is slowly turned on. To sum up, how one would be  able to reconcile the two pictures of black holes remains mysterious.

This part of the thesis is organised as follows: in section \ref{Lift_of_Multi_Center_Solutions} we repeat some definitions and and collect the formulas pertaining to the type IIA compactification moduli space, the 4d multi-hole solutions and their lift to five dimensions, as discussed in the previous part of the thesis. In section \ref{Construct the Bubbling Solutions}` we construct our bubbling solutions in 3 steps. First we work out the 4d solution in the M-theory \(\Leftrightarrow\) large IIA Calabi-Yau volume limit, and lift it to five dimensions. Secondly we rescale the five-dimensional coordinates to make it commensurable with the five-dimensional Planck units. Finally we put in the charge vectors of D6 and anti D6 with fluxes and arrive at the final form of the bubbling solutions. 

In section \ref{The Properties of the Solution} we analyse in full details the various properties of these solutions. A large part of the analysis holds also for generic lifted multi-center solutions in the decompactification limit, and some furthermore also holds for generic values of background moduli. Therefore, along the way we have also derived various properties of all the lifted multi-center solutions; or to say, the properties of various configurations of charged objects in type IIA string theory in the very strong coupling limit. Specifically, in \ref{The Conserved Charges} we work out the asymptotic metric, read off the five-dimensional conserved charges, including the electric charges of the M-theory C-field, and the two angular momenta \(J_L\) and \(J_R\), for generic centers. In \ref{The Shape of the Solution} we focus on the metric part and first study the condition for the absence of closed timelike curves (CTC's). Here we find a map between diseases: a CTC pathology in 5D corresponds to an imaginary metric pathology in 4D. We also analyse the possibility of having a throat-like (i.e. AdS-looking) metric in some part of the space. We conclude, also independent of the details of how the charges get distributed, that a multi-center configuration with charges not giving any black hole can never have a region like that, at least in the regime where supergravity is to be trusted.  We also check that, for our specific fluxed D6 and anti-D6 composition, the metric is smooth (at worst with an orbifold singularity when there are stacked D6) and horizonless everywhere, and we do this by establishing that the metric approaches that of a(n) (orbifolded) flat \(\R^4\times\R_t\) in the vicinity of each center. In \ref{Large gauge Transformation} we briefly discuss the role of the large gauge transformation of the M-theory three-form potential in our setting. We end this part of the thesis with discussions about future directions and some more speculative discussions about the degeneracy of ``black holes" or "proto-black holes".

\section{The Lift of Multi-Center Solutions}
\label{Lift_of_Multi_Center_Solutions}
\setcounter{equation}{0}

The lift \cite{Gaiotto:2005xt} to five dimensions, reviewed in section \ref{4D-5D Connection}, of the multi-center solution described in the previous chapter, is the starting point of our construction of the new bubbling solutions. In this section we will collect the relevant definitions and equations regarding the \({\cal N}=2\), D=4 stationary BPS solutions and their lift to five dimensions. In the present part of the thesis we will describe these theories as the low-energy effective theories of type IIA and M-theory compactified on a Calabi-Yau manifolds, although strictly speaking we do not need to know the microscopic origin of these lower-dimensional supergravity theories. 

Our basic strategy is as follows. First we recall that, using the basis (\ref{basis_h2*}) of the second cohomology \(H^{2*}(X,\Z)\)
and the symplectic product \(\langle\,,\rangle\) on them given by (\ref{symplectic_kahler}), in terms of the components (\ref{charge_in_coord}) and (\ref{harmonics_in_coord}), the general multi-hole solutions in four dimensions are given by (\ref{metric_ ansatz}), (\ref{attr_IIA_explicit})-(\ref{sol_vector}), or equivalently (\ref{formulation2_attractor_eq})-(\ref{formulation2_sol_vec}), with the harmonic functions given by the charges and the asymptotic moduli as (\ref{def_O_IIA_1}) and (\ref{harmonic}). Using the dictionary of lifting a four-dimension solution to five dimensions (\ref{5d_sol}), we can then write down the corresponding five-dimensional solution.

Anticipating a rescaling of coordinates later when the M-theory limit is taken, we will begin with writing the four-dimensional quantities in a boldface font and with an explicit  subscripts ``(4)" whenever it is needed. 
Especially, the harmonics functions are written as
\bea \label{harmonics4}
\Hf &=& \Hf^\L \a_\L + \Hf_\L \b^\L = \sum_{i=1}^N \frac{\G_i}{|\vec{\xf}-\vec{\xf}_i|} + \hf \\
\hf& =& \hf^\L \a_\L + \hf_\L \b^\L= - 2 {\mathrm{Im}}\Bigl((e^{-i\a}\O)|_\inf\Bigr)\;,
\eea
where \(\a\lvert_\inf\) is the phase of the total central charge at spatial infinity , \(Z(\G=\sum_i\G_i)\lvert_\inf= \left( e^{i\a}  |Z(\G)| \right)\lvert_\inf\) (\ref{bc_alpha}).
Using the lift dictionary (\ref{5d_sol}), the metric part of the five-dimensional solution is given by
\bea 
\label{lift_metric}
ds_{5d}^2 &=& 2^{2/3} ({\cal{V}}^{(s)})^2 (d\psi -A^0_{4D})^2 + 2^{-1/3}({\cal{V}}^{(s)})^{-1} ds_{4d}^2 \\ \nonumber
&=& 
-  (2^{2/3}\Qf)^{-2} [ \, ( d\tf + \w_{(4)} +2\Lf (d\psi + \w_{(4)}^0) ]^2\\ \label{metric1}
&& + (2^{2/3}\Qf) 
[ \frac{1}{\Hf^0}  (d\psi+\w_{(4)}^0)^2+   \Hf^0 d\xf^a d\xf^a ]\;,
\eea
where the 4d and 5d warp factors \( \Sf(\vec{x}) \), \(\Qf(\vec{x})\) and the 5d rotation parameter \(\Lf(\vec{x})\) appearing here are functions of the \(\rr^3\) coordinates \(\xf^a\) and are given by the above harmonic functions  as 

\bea\notag
\Sf &=&  2\p \sqrt{\Hf^0 \Qf^3 -  (\Hf^0\Lf)^2}\\ \notag
\Lf&=& -\frac{\Hf_{0}}{2} - \frac{\Hf^A \Hf_{A}}{2\Hf^0} + \frac{D_{ABC}\Hf^A \Hf^B \Hf^C}{6(\Hf^0)^2}\\ \notag
\Qf^3 &=& (\frac{1}{6}D_{ABC}\yf^A \yf^B \yf^C)^2\\  \label{aflow}
D_{ABC}\yf^B \yf^C &=& -2\Hf_{A} + \frac{D_{ABC}\Hf^B \Hf^C}{\Hf^0}\;,
\eea
and the cross terms in the 5d metric are determined up to coordinate redefinition by
\ben
d\w_{(4)} &=&\star^3_{(4)}\langle d\Hf,\Hf\rangle\\
d\w^0_{(4)} &= & \star^3_{(4)} d\Hf^0\;,\een
where the \(\star^3_{(4)}\) is the Hodge dual operator w.r.t. the flat \(\R^3\).

Furthermore, as discussed in the previous chapter, for the four-dimensional solution to be physical we have to require the integrability condition (\ref{integrability}) and the positivity of the entropy function (\ref{positive_entropy_function}). As we will show later, in the five-dimensional picture the latter condition manifests itself as the condition of the absence of closed timelike curves.

\section{Construct the Bubbling Solutions}
\label{Construct the Bubbling Solutions}
\setcounter{equation}{0}

After reviewing the formulae we need, now we can construct the bubbling solutions in three steps: first taking the limit, second rescaling the solution, and finally specifying the centers.

\subsection{M-theory Limit}
\label{M-theory Limit}

First of all, in order to get an asymptotically flat metric in 5d, it is clear that one should take the decompactification limit in which the M-theory radius \(R_{\text{\tiny M}}\) goes to infinity. From the expression of the radius in the five-dimensional Planck unit (\ref{R_M_planck5}), we see that we should take the type IIA decompactification limit \(J^{\text{\tiny (s)}}\rightarrow \inf\), while keeping 
$$
{J}^{\text{\tiny (M)}} \sim J^{\text{\tiny (s)}} \,\frac{\ell_{\text{\tiny P}}^{(11)} }{R_{\text{\tiny M}}} \sim 
J^{\text{\tiny (s)}} \,\big(\frac{\ell_{s}}{R_{\text{\tiny M}}} \big)^{2/3}
$$
finite.

Therefore, we will now stipulate the background moduli to be 
\ben
B^A|_{\inf} &\equiv& b^A\mbox{    finite} \\ \notag
J^{A(s)}|_{\inf} &\equiv & j^A \rightarrow \inf\;.
\een

In this limit the constant terms \(\hf\) in the harmonic functions take a especially simple form (the general expressions can be found in Appendix \ref{Constant Terms for General Charges and Background}):
\bea \label{hf1}
\hf^0\mbox{ ,  }\hf^A &\rightarrow& 0\\ \label{hf2}
\hf_A &\rightarrow& - \frac{p^0}{|p^0|}\, \frac{(j^2)_A}{\sqrt{\frac{4}{3}\,j^3}}\\ \label{hf3}
\hf_0 &\rightarrow& \frac{1}{|p^0|}\frac{D_{ABC}p^A j^B j^C}{\sqrt{\frac{4}{3}\,j^3}} = -\frac{p^A}{p^0} \hf_A \;.
\eea

\subsection{Rescale the Solution}
\label{Rescale the Solution}

It seems that we are done with the background moduli and all still left to be done is to choose the appropriate charges and fill them in the harmonic functions. But there is a subtlety which is a consequence of the large (IIA) Calabi-Yau volume limit that we are taking. One can see this already from the expression for the constant terms in the harmonic functions (\ref{hf2}), (\ref{hf3}): these remaining constants go to infinity in this limit! Indeed, as a result, the three-dimensional (apart from the time and the 5th dimension) part of the metric goes to \((H^0Q)|_{\inf} d\xf^a d\xf^a \rightarrow \inf \,\frac{d\xf^a d\xf^a}{|\vec{x}|}\) at spatial infinity, while it goes to zero in the timelike direction: \( - g_{tt} = 2^{-4/3} \frac{1}{Q^2} \rightarrow 0\).\footnote{See the next section for detailed asymptotic analysis.} This is a clear signal that we are using a set of coordinates not appropriate for the five-dimensional description.

To find the right coordinates, let's remind ourselves that the four-dimensional metric is measured in the four-dimensional Planck units, while the extra warp factor \({\cal{V}}^{-1}\) rescale the metric to be measured in the five-dimensional Planck length when the the solution gets lifted (see (\ref{lift_metric}) ), whose ratio (\ref{R_M_planck5}) goes to infinity in the present large-IIA-volume limit.
Therefore, in order to obtain a coordinate system natural in five dimensions, we should rescale all the coordinates with a factor \(\L \sim ({\cal V}^{\text{\tiny (s)}})^{1/6} \) and accordingly  the harmonic functions as well. Let's define

\ben
\L&\equiv& \frac{1}{2} \, (\frac{4}{3}j^3)^{1/6} \\ \notag 
x^a&\equiv& \L \xf^a\\
t&\equiv& \frac{1}{2\L} \tf \\ \notag
\{H, L, Q , \w\}&\equiv&\frac{1}{\L} \{\Hf, \Lf, \Qf , \w_{(4)}\}\\ \notag
S &\equiv& \frac{1}{\L^2} \Sf\;.
\een

One can easily check that the lifted five-dimensional metric (\ref{metric1}) can be written in the above rescaled coordinates and functions in exactly the same form:
\bea \nonumber \label{metric2}
2^{-2/3} ds_{5d}^2 &=& - Q^{-2} \, \lbrack dt + \frac{\w}{2} + L (d\psi + \w^0) \rbrack^2 \\ && + Q 
\lbrack \frac{1}{H^0} (d\psi+\w^0)^2+   H^0 dx^adx^a \rbrack \;.
\eea
The only difference the rescaling makes to the metric is that the warp factor \(Q(\vec{x})\) approaches a finite constant (\(= \pm\)1) even in the decompactification limit we are working in. 

Let's now pause and summarise. What we have done so far is to obtain a large number  of BPS solutions of five-dimensional supergravity with an arbitrary number of vector multiplets, by lifting the four-dimensional solutions in the limit that the extra direction is infinitely large. These solutions might have singularities or/and horizons, depending on the charges of each center and their respective locations. For later use, we will now spell out explicitly the five-dimensional solutions. 
 
The metric part of the solution is given by (\ref{metric2}) and 
(\ref{attr_IIA_explicit})-(\ref{entropy_IIA}), (\ref{eq_2}) and (\ref{vector_general})
\be
d\w^0 = dA^0_d = \star dH^0\;,
\ee
where \(\star\) is again the Hodge star with respect to the flat \(\R^3\) base given by \(x^a\).

The harmonic functions are given by, in their most explicit form:

\bea \notag
H^0(\vec{x})&=&\sum_i \frac{p^0_i}{r_{i}}\\ \notag
H^A(\vec{x})&=& \sum_i \frac{p^A_i}{r_i}\\  \label{H_A} 
H_A(\vec{x})&=&\sum_i \frac{q_{A,i}}{r_{i}}  +h_A\;\;;\;\;h_A=- \frac{|p^0|}{p^0}\frac{2\,(j^2)_A}{(\frac{4}{3}j^3)^{2/3}}\\  \notag
H_0(\vec{x})&=& \sum_i \frac{q_{0,i}}{r_{i}}  
+h_0\;\;;\;\;h_0=-\frac{p^A}{p^0}h_A = \frac{2}{|p^0|} \frac{D_{ABC}p^A j^B j^C}{(\frac{4}{3}j^3)^{2/3}}
\\ \nonumber 
\mbox{where } &r_{i}& = |\vec{x}-\vec{x}_i| .
\eea
Notice that now the remaining constant terms \(h_A\), \(h_0\) are insensitive to the rescaling of \(j\). We can therefore as well interpret the \(j\) to be the M-theory asymptotic K\"ahler moduli \(j^A=\lim_{|\vec{x}|\rightarrow \inf}J^{A(M)}(\vec{x})\), which we keep as finite.

Since the integrability condition (\ref{integrability}) is going to play an important role in the analysis in the following section, we also rewrite it as
\be \label{integrability2}
\langle\G_i,H_i\rangle = 0 \Leftrightarrow \sum_{j}\frac{\langle\G_i,\G_j\rangle}{r_{ij}} = -h_A \til{p}^A_i \;,
\ee
where
\bea
H_{i} &\equiv& (H-\frac{\G_i}{r_i})|_{\vec{x}=\vec{x}_i}\\
\til{p}^A_i &\equiv& p^A_i - p^0_i \frac{p^A}{p^0} \;\;;\;\; 
r_{ij} = |\vec{x}_i-\vec{x}_j|\;.
\eea
Notice that the right hand side of (\ref{integrability2}) would in general have a much more complicated dependence on the charges of the centers, if we hadn't taken the M-theory limit.

Now we turn to the vector multiplets. Using the 4d solution (\ref{sol_scalar}) and (\ref{sol_vector}), the 4d-5d dictionary (\ref{5d_sol}) now gives the lifted solution
\begin{align}
Y^A &= \frac{y^A}{Q^{1/2}}\\
A^A_{5D} &=
-\frac{y^A}{Q^{3/2}} (dt +\frac{\w}{2}) +(\frac{H^A}{H^0} -\frac{L}{Q^{3/2}} y^A )(d\psi+\w^0) - {A}_d^A \\
\intertext{where \(A_d^A\) again denotes the Dirac monopole part of the gauge field}
d {A}_d^A &= \star dH^A
\;. 
\end{align}

In a form more familiar in the five-dimensional supergravity literature, these solutions can be equivalently written as
\bea
2^{-2/3} ds_{5D}^2 &=&  -Q^{-2}\,e^0 \otimes e^0 + Q ds_{base}^2\\ \label{field_strength_5d}
F^A_{5D} &=& d A^A_{5D} =  - d(Q^{-1}Y^A e^0 )  + \Theta^A \;,
\eea
where
\bea
ds_{base}^2 &=& H^0 dx^a dx^a + \frac{1}{H^0} (d\psi + \w^0)^2 \\
e^0 &=&  dt + \frac{\w}{2}  + L (d\psi + \w^0) \\
\Theta^A &=& \star_{base} \Theta^A= d\lbrack \frac{H^A}{H^0} (d\psi + \w^0) \rbrack - \star_3 dH^A \;.
\eea

For example, taking one D6 charge the base metric becomes that of the Taub-NUT space (\ref{TN}). Taking two D6 charges at different points the base metric is that of the Eguchi-Hanson gravitational instanton (\ref{E_H}).

\subsection{Specify the 4D Charges}
\label{Specify the 4D Charges}
Now we would like to know what kind of 4d charges for the centers we should take, in order to obtain an asymptotically flat, smooth, horizonless solution when lifted to five dimensions. We now argue that the only possibility is the multi-center configurations composed of D6 and anti-D6 branes with world-volume fluxes turned on, and with the constraint that the total D6 brane charge equals to \(\pm1\).\footnote{Furthermore, each center must have D6 charge \(\pm 1\), if one also wants to exclude orbifold singularities at the center. But we will keep the formulae as general as possible and do not specify the D6 charges of each center.}
This can be understood as the following: if we take D2 or D4 branes or their bound states with other branes, the uplift to M-theory will have also M2, M5 brane sources and thus won't have the desired smooth and horizonless virtue. In other words, the uplifted metric near a D2 or D4 center will not be flat. One might also wonder about the possibility of adding D0 branes into the picture. First of all, in contrast to the usual scenario \cite{Witten:2000mf},  a D0-D6 bound state doesn't exist in the large volume \(J^{(s)}|_\inf \rightarrow \inf\) limit we are taking, irrespective of the (finite) value of the background B-field. But one could still imagine a multi-center KK monopole-electron-antimonopole-positron juxtaposition living in the large coupling limit. But this time the metric near the D0 centers is not smooth; more specifically, the metric in the 5th direction blows up while remaining flat in the \(\rr^3\) direction. In summary, in order to get a smooth and horizonless solution, we have to restrict our attention to D6 and anti-D6 branes with world-volume fluxes.

From the part of the D6 world-volume action coupling to the RR-potential \cite{Douglas:1995bn,Li:1995pq}
\be\label{wv_action}
\int_{\S_7} e^{B+F} \wedge C \;\;\;;\;\; C\in H^{2*}(X,\rr)\;,
\ee
one sees that the world-volume flux induces a D4-D2-D0 charge. Specifically, neglecting the B-field which can always be gauged into world-volume fluxes locally on the six brane, the charge vector of a center of \(p^0_i\) D6 and with world-volume two-form flux \(\frac{f_i}{p^0_i} =\frac{f_i^A}{p^0_i}  \a_A\) turned on is
\be \label{chargei}
\G_i = p^0_i e^{\frac{f_i }{ p^0_i }} = p^0_i  + f_i + \frac{1}{2} \frac{f_i^2}{p^0_i} +\frac{1}{6} \frac{f_i^3}{(p^0_i)^2}\;.
\ee

Thus the total charge vector is\footnote{In the case of stacked D6 branes, we only turn on the Abelian  fluxes. The reason for this restriction is that for non-Abelian \(F\), the induced D4-D2-D0 charges are proportional to \({\rm Tr} F\), \({\rm Tr} F\wedge F\) and \({\rm Tr} F\wedge F\wedge F\) respectively. In this case one can easily see that the corresponding solution will in general develop a singularity or a horizon. }
\bea \nonumber
\G&=&p^0 + p^A \a_A + q_A \b^A + q_0 \b^0 \\
\label{totalcharges}
&=& \sum_{i=1}^N \G_i =\sum_{i=1}^N p^0_i  + \sum_{i=1}^N f_i +\sum_{i=1}^N \frac{1}{2} \frac{f_i^2}{p^0_i} +\sum_{i=1}^N\frac{1}{6} \frac{f_i^3}{(p^0_i)^2}\;.
\eea
As mentioned earlier, we are especially interested in the case \(p^0 = \pm  1\), since this condition ensures asymptotic flatness. More specifically, only for the case \(p^0 = \pm  1\) the metric approaches that of \(\rr_t\times \rr^4\) in spatial infinity without identification.

Simply filling these charges into the harmonic functions in the last subsection gives us, as we will verify later, a metric that is asymptotically flat, smooth and horizonless everywhere, and may or may not have the conserved charges of those of a classical black hole.

\section{The Properties of the Solution}
\label{The Properties of the Solution}
\setcounter{equation}{0}

\subsection{The Conserved Charges}
\label{The Conserved Charges}

\subsubsection{4D and 5D Charges}
\label{4d and 5d Charges}
When lifting a four-dimensional solution to five dimensions, the charged objects in IIA get mapped into charged objects in M-theory. The Kaluza-Klein monopoles and  electrons, namely the D6 and D0 charges, show themselves as Taub-NUT centers and the angular momentum in the five-dimensional solution. Especially we expect \(q_0 \sim -J_L\). The (induced) D4 charges, as can be seen in (\ref{field_strength_5d}), parametrize the magnitude of the part of the field strength that is self-dual in the Gibbons-Hawking base. In the type IIA language, in the case with non-zero D4 charges, one also has non-zero internal B-field in various regions in space. When lifted to M-theory they give a new contribution to the vector potential and we expect those to modify the definition of the electric charges. Therefore, as suggested in \cite{Gaiotto:2005gf}, \(q_{A,(5D)}\) and \(J_L\) will get extra contributions involving \(p^A\) through the Chern-Simons coupling and the Poynting vectors of the gauge field. An inspection of the five-dimensional attractor equation for a 5d black hole
\bea
S_{5D}&=& 2\p \sqrt{Q^3 - J_L^2} \\
Q^3 &=& (\frac{y_{(5D)}^3}{6})^2 \;\;;\;\;D_{ABC} y_{(5D)}^B y_{(5D)}^C = -2 q_{A,(5D)}\;,
\eea
and comparing it to the four-dimensional ones (\ref{aflow})  with \(p^0=1\) suggests that, when \(p^A\) becomes non-zero, \(q_{A,(5D)}\) and \(J_L\) must get an extra contribution as
\bea
-2 q_{A,(5D)} &\rightarrow& -2 q_{A,(5D)}  + \frac{(p^2)_A}{p^0}
\\
J_L &\rightarrow&J_L- \frac{p^A q_A}{2p^0 } +\frac{p^3}{6\,(p^0)^2}\;.
\eea

We will now verify this through explicit asymptotic analysis, while more discussion related to the role of \(p^A\) charges can be found in section \ref{Construct the Bubbling Solutions}.

\subsubsection{The Asymptotic Analysis}
\label{The Asymptotic Analysis}

Now we would like to work out the asymptotic form of the solution. We are interested in it for the following two reasons. First of all we would like to verify that our metric is indeed asymptotically flat; secondly we would like to read off all the conserved charges of these solutions.
The following asymptotic analysis applies to all the solutions in the form of that presented in the end of the last section, i.e., to {\it{all}} the solutions of the \({\cal N}=1\) five-dimensional supergravity obtained by lifting four-dimensional solutions in the decompactification limit. \footnote{Apart from the fact that we are assuming in this subsection that the sign of the total D6 charge is positive, to avoid messy phase factors everywhere. The adaptation to the case in which \(p^0 < 0 \) is straightforward. }

Let's first look at the metric part. In the limit \( r = |\vec{x}| \rightarrow \inf\) we have the various quantities in the metric approaching\footnote{One has to be a bit careful with the order of taking the two limits \( r \rightarrow \inf\) and \(j^A\rightarrow \inf\). Here we restrict ourselves to the range \( 1 \ll r \ll \frac{R_M}{\ell^{(5)}_P} \rightarrow \inf\), in other words, where the spacetimes remains appearing to be five-dimensional. In this range one can indeed ignore the extra constant terms \(h^0\), \(h^A\) (see Appendix \ref{Constant Terms for General Charges and Background}).  }
\ben
Q &=& 1 + {\cal O}(r^{-1})\\ \notag
H^0 &=& \frac{p^0}{r} +  {\cal O}(r^{-2})\\ \notag
\w^0  &=& p^0\cos\th d\f + {\cal O}(r^{-1}) \\ \nonumber
L &=& \frac{1}{r}\, \lbrack\,  (-\frac{q_0}{2} - \frac{p^A q_A}{2p^0} + \frac{D_{ABC}p^A p^B p^C}{6(p^0)^2})  + \frac{\hat{r}}{p^0} \cdot (\sum_{i,j=1}^N \frac{ \langle\G_i,\G_j\rangle}{4} \frac{(\vec{x}_i-\vec{x}_j) }{|\vec{x}_i-\vec{x}_j|} )  \rbrack\\ \notag&& + {\cal O}(r^{-2})\;,
\een
where the second term in the last equation is derived from the dipole term in the expansion and we have used the integrability condition (\ref{integrability2}) to put it in this form. 

We have now a natural choice of coordinates of the \(\rr^3\) factor of the metric. This is because the dipole term picks out a unique direction in the spatial infinity. Let's now choose the spherical coordinate in such a way that the vector
\be
\vec{J}_R = \sum_{i,j} \vec{J}_{ij}  =\sum_{i,j}  \frac{ \langle\G_i,\G_j\rangle}{4} \frac{\vec{x}_i-\vec{x}_j}{|\vec{x}_i-\vec{x}_j|}
\ee
points at the north pole. The second term in \(L\) can then be written as \(\frac{1}{p^0}\vec{J}_R \cdot \hat{r} = \frac{1}{p^0} J_R \cos\th\).

Finally, solving the \(\w\) equation asymptotically gives us
\be
\frac{1}{2}\w =  \frac{1}{r}\,J_R\,\sin^2\th d\f +  {\cal O}(r^{-2})\;,
\ee
up to trivial coordinate transformations. 

After a change of coordinate \(r= \r^2/4\), the metric at infinity now reads
\bea \nonumber  2^{-2/3} ds_{5D}^2 &=&
-\lbrace dt + \frac{4}{\r^2} \lbrack p^0  J_L (\frac{1}{p^0}d\psi + \cos\th d\f) + J_R (d\f+ \frac{1}{p^0}\cos\th d\psi )  \rbrack + {\cal O}(\r^{-4}) \rbrace^2 
\\  \label{metric_asymp}
 &+& p^0 \lbrace d\r^2 + \frac{\r^2}{4} \lbrack d\th^2 + \sin^2\th d\f^2 + 
(\frac{1}{p^0}d\psi + \cos\th d\f)^2 \rbrack+ {\cal O}(\r^{-2})\rbrace\;,
\eea
with
\bea \label{J_L}
J_L & = &-\frac{q_0}{2} - \frac{p^A q_A}{2p^0} + \frac{D_{ABC}p^A p^B p^C}{6(p^0)^2}\\
 \label{J_R}
J_R &=& |\sum_{i<j} \frac{ \langle\G_i,\G_j\rangle}{2} \frac{\vec{x}_i-\vec{x}_j}{|\vec{x}_i-\vec{x}_j|} | \;
\eea
being the two angular momenta, corresponding to the \(U(1)_L\) exact isometry and the  \( U(1)_R\) asymptotic isometry, generated by \(\xi^3_L = \pa_{\psi}\) and \(\xi^3_R =\pa_\f\) respectively, as the unbroken part of the \(SU(2)_R \times SU(2)_L\) isometries (\ref{left_isometry})-(\ref{right_isometry}) .

Indeed we see that, the metric approaches that of a flat space without identification when \(|p^0| =1\). In that case it can be more compactly written as
 \bea \nonumber
 2^{-2/3} ds_{5D}^2 &=& -\lbrack dt + \frac{4}{\r^2} (J_L \s_{3,L} + J_R \s_{3,R} ) \rbrack^2  \\
&+& ( d\r^2 + \frac{\r^2}{4}  (\s_{1,L}^2+ \s_{2,L}^2 + \s_{3,L}^2)  ) + ...\\ \nonumber
&=& -\lbrack dt + \frac{4}{\r^2} (J_L \s_{3,L} + J_R \s_{3,R} ) \rbrack^2 + ( d\r^2 + \frac{\r^2}{4}  (\s_{1,R}^2\\&+& \s_{2,R}^2 + \s_{3,R}^2)  ) + ...
 \eea
 where the \(\s\)'s are the usual \(SU(2)_L\) and \(SU(2)_R\) invariant one-forms of \(S^3\) (\ref{one_form_left})-(\ref{one_form_right}).

After working out the angular momenta we now turn to the electric charges of the 5d solutions. From the gauge field part of the action of \({\cal N}=1\) 5d supergravity (\ref{5d_sugrav_action}), we see that the conserved electric charges are then given by the Noether charge
\ben
q_{A(5D)} &=& -\frac{16\p G^{(5)}_{\text{\tiny N}}}{V_{S^3} } \int_{S^3_\inf} \frac{\pa L}{\pa F^A} \\
&= &\frac{1}{V_{S^3} }\,\int_{S^3_\inf} a_{AB} \star_5 F^B - \frac{1}{3} D_{ABC}\,F^B\wedge A^C\;,
\een
where the gauge coupling \(a_{AB}\) is given by the scalar solution by (\ref{5d_gauge_coupling}) and \(V_{S^3}\) denotes the volume of a unit 3-sphere.

We need to know the asymptotic behaviour of the vector potential and the field strength in order to compute the charges. They are given by
\bea\label{A_5D}
A^A_{5D} &=& \frac{p^A}{p^0}d\psi -\frac{j^A}{(\frac{1}{6}j^3)^{1/3}} dt + {\cal O}(\r^{-2})\; \; (
+ \mbox{gauge transf.})\\
F^A_{5D} &=& -d(\frac{y^A}{\frac{1}{6}y^3}) \wedge dt + {\cal O}(\r^{-2})d\s  + {\cal O}(\r^{-3}) d\r\wedge \s\;.
\eea

From these equations it is clear that the Chern-Simons term does not contribute to the charges, and from
\bea \nonumber
a_{AB} F^B_{5D} &=& -\frac{1}{2} d\Bigl( \frac{y^B}{y^3/6} \Bigr) \left\{ \frac{\pa}{\pa y^B}\Bigl(\frac{(y^2)_A}{y^3/6}\Bigr)\, \right\}|_{\frac{y^3}{6}=1} \wedge dt +... \\  \nonumber
&=&  -\frac{1}{2} d(y^2)_A \wedge dt + ... \\
&=& (q_A - \frac{(p^2)_A}{2p^0})\,(\frac{\r}{2})^{-3} dt\wedge d\r ...\;.
\eea
we get after integration
\be
\label{q_5d}
q_{A(5D)} = q_A - \frac{(p^2)_A}{2p^0}\;.
\ee

This finishes our analysis of the conserved charges of our solutions. As mentioned earlier, the expressions for the charges and for the the asymptotic metric (\ref{metric_asymp}), (\ref{J_L}), (\ref{J_R}) and (\ref{q_5d}) apply to {\it{all}}  solutions lifted from four dimensions in the infinite radius limit, i.e., all the solutions presented in section \ref{Rescale the Solution}. For the specific case we consider in the last section (let's focus on the case \(p^0 = +1\)), they are given simply by the D6 charge and the flux of each center as
\bea
q_{A(5D)} &=& q_A - \frac{(p^2)_A}{2p^0} = \sum_i \frac{(\til{f}_i^2)_A}{2 p^0_i}\\
J_L &=& \sum_i \frac{\til{f}_i^3}{6 (p^0_i)^2}\\ 
J_R &=&  |\frac{1}{4} \,\sum_{i,j=1}^N \,p^0_ip^0_j \frac{f_{ij}^3}{6} \frac{\vec{x}_i-\vec{x}_j}{|\vec{x}_i-\vec{x}_j|}\,|\,
\eea
where 
\bea \label{til_f}
\til{f}_i^A  &\equiv& f^A_i - p^0_i (\sum_j f^A_j) \\ \label{f_ij}
f_{ij}^A & \equiv& \frac{f_i^A}{p^0_i}- \frac{f_j^A}{p^0_j} =\frac{\til{f}_i^A}{p^0_i}- \frac{\til{f}_j^A}{p^0_j} \;.
\eea

As we will see later, \(\til{f}_i^A\) has the physical interpretation as the quantity invariant under the gauge transformation, and \(p^0_i p^0_j f_{ij}^A \) has the interpretation as the fluxes going through the \(ij\)-th ``bubble".

\subsection{The Shape of the Solution}
\label{The Shape of the Solution}

After analysing the solution at infinity, now we would like to know more about the metric part, i.e. the shape, of these solutions. First of all we would like to spell out the criterion that the metric is free of pathological closed timelike curves. Having black hole physics in mind, we would also like to see if the solution exhibits a throat (AdS-looking) behaviour in some region. These two parts of the analysis, unless otherwise stated, apply to general solutions presented in section \ref{Rescale the Solution}. 

There is another region of special interest here. Namely, we would like to explicitly verify our claim that the metric, provided that the CTC-free condition is satisfied, is smooth and horizonless near each center. As discussed in section \ref{Specify the 4D Charges}, this property only pertains to the special charges (D6 or anti-D6 with fluxes) that we have chosen.

\subsubsection{Closed Timelike Curves}
\label{Closed Timelike Curves}

Before jumping into the equations, let's first make a detour and look at the four-dimensional metric (\ref{metric_ ansatz}) we started with. Apart from the integrability condition (\ref{integrability2}), it's apparent that we 
also need to impose the condition 
\be \label{CTC}
(\frac{S(\vec{x})}{2\p})^2 = H^0 Q^3 - (H^0)^2 L^2 > 0\;,
\ee
in order to have an everywhere real metric in four dimensions. Indeed, in the case this is not satisfied, the volume of the internal Calabi-Yau goes through a zero and things stop making sense in all ten dimensions.  

A look at the 5d metric:  
\be
2^{-2/3} g_{\psi\psi} = (\frac{S(\vec{x})}{2\p})^2 (\frac{1}{H^0 Q})^2\;,
\ee
makes it clear that as long as the 4D metric is real everywhere, the lifted metric has its 5th direction always spacelike. Furthermore, from 
\be
(\frac{S(\vec{x})}{2\p})^2 = H^0 Q^3 - (H^0)^2 L^2 > 0 \Rightarrow H^0 Q> 0\;,
\ee
it also ensures that the warp factor in front of the \(\rr^3\) part of the metric is always positive, and therefore another danger for CTC is also automatically eliminated. In more details, this is because the harmonic functions are real by default, and it's really the \(Q\), or rather the \(y^A\), attractor flow equations that are not a priori endowed with a real solution. 

Now we can worry about the more subtle \(-Q^{-2} (\frac{\w}{2})^2\) part of the metric. Looking at the equation for \(\w\)
\be
d\w = \star_3 \langle dH,H\rangle\;, 
\ee 
one sees that the danger zone is the region very close to a center, since it's the only place where \(dH\) and \(H\) blow up. But as we will see later, the integrability condition always guarantees that \(\w\) actually approaches zero at least as fast as the distance to the center under inspection. We can therefore believe that this term poses no threat. To sum up, what we find is
\be \label{CTC2}
\mbox{4d metric real  } \Leftrightarrow \mbox{  5d metric no CTC}\;.
\ee

Of course, mapping one problem to the other does not really solve anything. Indeed, at the moment the author does not know of any systematic way of checking this condition. Especially, the integrability condition, while often ensures the real (4d) metric condition (\ref{CTC}) to be satisfied near a center, is in general not sufficient to guarantee that it is satisfied everywhere.\footnote{In the four-dimensional context, a conjecture about the equivalence between the existence of a solution with an everywhere well-defined metric with given background and charges, and the existence of a split attractor flow connecting the asymptotic moduli and the attractor points of all the centers, has been proposed and studied in \cite{Denef:2001xn}, \cite{Denef:2000ar}, and \cite{Denef:2007vg}. See subsection \ref{Split Attractor Flow}. If this conjecture is indeed true, it provides us a more systematic way of studying the existence of multi-centered solutions.} On the other hand, this is how it should be, since: given \(N\) centers, the naive moduli space of their locations grows like \( (\rr^3)^N \), the number of distances between them grows like \(N^2\), but the number of integrability condition grows only like \(N\). Given the possibility that one can always {\it{a priori}} add one more pair of centers with opposite charges while still keeping the total charge unaltered, it seems extremely unlikely to be able to obtain a reasonable moduli space for BPS states with a given total charge, if there are no rules of the game other than the integrability condition. 

We finish this subsection by noting that our discussion here about the closed timelike curves, especially the conclusion (\ref{CTC2}), applies to all 4D-5D lift solutions irrespective of the background moduli. That is, it applies even without taking the decompactification limit.

\subsubsection{The Throat Region}
\label{The Throat Region}

In section \ref{The Conserved Charges} we have seen that, when we look at the asymptotic region: \be  h \gg \frac{1}{r} \gg \frac{r_{ij}}{r^2}\;,\ee the harmonic function can be expanded, in the order of decreasing magnitude, as
\be
H= h + \frac{\G}{r} + \mbox{dipole terms} +\mbox{quadrupole terms}  + ... \;,
\ee
where the non-vanishing constant terms \(h\) are of order one in our renormalization (see section \ref{Rescale the Solution}
).

If the (coordinate) distances \(r_{ij}\) of each pair of centers are all much smaller than one, namely \(r_{ij} \ll 1\;\forall\; i, j\), one can consider another region in which 
\be \label{throat_region}
\frac{1}{r} \gg h \;\; , \mbox{      } \frac{1}{r}  \gg \frac{r_{ij}}{r^2}\;.
\ee
 In other words, when the centers are very close to each other, one can zoom in a bit more from the asymptotic region so that  the constant terms become subdominant, while still not getting substantially closer to any of the centers than the others, and can still see the conglomeration of centers (the blob) as an entity without seeing the structure of distinct centers. 

In this region, the harmonic functions are expanded, again with descending importance, as
\be
H=  \frac{\G}{r} + \Bigl( h + \mbox{dipole terms} \Bigr) +\mbox{quadrupole terms}  + ... \;,
\ee
and attractor flow equation is given by
\be
D_{ABC}y^B y^C = \frac{1}{r} \,(-2q_A + \frac{(p^2)_A}{p^0} ) + ... \;.
\ee
Define \(y^A_{bh}\) to be the solution to the equation \((y_{bh}^2)_A =-2q_A + \frac{(p^2)_A}{p^0} \) and \(Q_{bh}^3 = (\frac{y_{bh}^3}{6})^2\), one arrives at
\be
Q = \frac{Q_{bh}}{r} + ...  .\ee

At the same time, 
\be
L  = \frac{1}{r}\,J_L+ ... = \frac{1}{r}\,\left(-\frac{q_0}{2}-\frac{p\cdot q}{ 2 p^0} + \frac{p^3}{6 (p^0)^2 }\right) + .... \,.
\ee
Notice that, unlike in the asymptotic region, the dipole contribution to \(L\) is sub-leading because now 
\( \frac{1}{r} \gg h\). Again using the integrability condition to relate the dipole contribution of \(L\) to the magnitude of \(\w\), one sees that \(\w\) as well is of minor importance in this region.

Now the 5th dimension part of  the metric reads
\be \label{CTC_throat}
g_{\psi\psi}  = 2^{2/3} \, (\frac{Q}{H^0} - \frac{L^2}{Q^2}) = \frac{1}{(p^0)^2 Q_{bh}^2
} (\frac{S_{bh}}{2\pi})^2 + ...\;,
\ee
where \be S_{bh}  = 2\p \sqrt{p^0 Q_h^3 - (p^0)^2 J_L^2}  \ee is a constant equal to the (classical) black entropy with the charges corresponding to that of the total charges of our multi-center configuration. 

Putting everything together, we find that the metric in the region (\ref{throat_region}) looks like\footnote{For the readability we have imposed in the this equation that the total monopole charge \(p^0=1\). It's trivial to put back all the \(p^0\) factors, and the metric one obtains in the case of \(|p^0|\neq 1\) is that of an orbifolded BMPV near horizon geometry.}
\bea \notag
2^{-2/3} ds_{5D}^2 &=&  -(\frac{r}{r_{bh}})^2 dt_{bh}^2 + (\frac{r_{bh}}{r})^2 dr^2 + 2 r \,(\frac{J_L}{r_{bh}^3})\, dt_{bh}\,\s_{3,L} 
\\  \label{near_horizon_1} &+& r_{bh}^2 \Bigl( \s_{1,L}^2 +\s_{2,L}^2+ \s_{3,L}^2 -(\frac{J_L^2}{r_{bh}^3})^2 \s_{3,L}^2 ) \Bigr)\;,
\eea
where \(r_{bh}\equiv \sqrt{Q_{bh}}\) and we have rescaled the time coordinate \(t_{bh} = \frac{t}{\sqrt{Q_{bh}}}\). 

One can now readily recognise this metric as the \(AdS_2\times S^3\) near horizon metric of a BMPV black hole\footnote{Or, more precisely, an identification of \(AdS_3\times S^3\) which leaves a cross term \(dt \,\s_{3,L}\) behind \cite{AlonsoAlberca:2002wr}. Also the \(S^3\) is squashed in such a way that its area again gives the black hole entropy.}  \cite{Kallosh:1996vy}. Therefore we can identify the region (\ref{throat_region}) as a sort of near horizon region of the multi-center BPS solution.  

So far it all seems very satisfactory: the 5D solutions obtained from lifting multi-center 4D solutions have a throat region which looks like the near horizon limit of a classical black hole with charge given by the total charge of the 4D centers via the prescription we give in section \ref{The Conserved Charges}. But we should not forget that the analysis here depends on the existence of the region (\ref{throat_region}). Indeed, it's obvious that this region cannot exist for all choices of charges: when the total charge does not give a classical black hole, namely when \(S_{bh}^2 < 0\), the existence of this region together with (\ref{CTC_throat}) would imply the presence of a CTC, or equivalently, an imaginary metric in 4D, in this region. One thus conclude that the region (\ref{throat_region}) can only exist when the total charge of all the centers together corresponds to that of a black hole. This also justifies our notation \(y_{bh},Q_{bh},t_{bh},r_{bh}\). 

In other words, when the total charge doesn't give a black hole, at least one pair of the centers must be far away from each other:
$$
\exists\;i, j \mbox{    s.t.    } r_{ij} \sim h \mbox{   or   }  r_{ij} > h  \mbox{      if    }  S_{bh}^2 < 0\;.
$$
 
This argument applies actually not only to multi-center solutions in the large volume limit with arbitrary charges, but also to those with {\it{arbitrary}} background moduli \(j, b\), with the only difference being that we have to include in general much more complicated constant terms in the harmonic functions (see Appendix \ref{Constant Terms for General Charges and Background}) to estimate the lower bound on the distances between the centers. Therefore we conclude that, for a choice of charges such that the total charge doesn't give a black hole, the centers cannot get arbitrarily close to each other, at least as long as we stay in the regime where the supergravity description is to be trusted $$ \frac{R_M}{\ell^{(11)}_P } \gg 1\,, \,J^{(M)} \gg 1 \Leftrightarrow g_s \gg 1\,,\,J^{(s)} \gg 1\,,$$ apart from other conditions discussed in section \ref{Range of Validity and Higher Order Corrections}. What happens to these multi-center configurations with total charge of no black holes, when \(\frac{R_M}{\ell^{(11)}_P} = g_s^{2/3}\) is lowered beyond the supergravity regime is described in terms of microscopic D-brane quiver theory and the higgsing thereof in \cite{Denef:2002ru}. From the five dimensional point of view, it would be interesting to refine the result of \cite{DVVafa} in a similar spirit. 
 
We finish our throat examination with two remarks. First of all, the reverse of what we just said is not always true:  when the total charge does correspond to that of a classical black hole, the centers don't have to sit very close to each other. We can also imagine them to be far apart  and still have a well-defined metric. For example, the centers can split themselves up into two blobs far away form each other, with each blob having its throat region and can therefore be coarse-grained as an AdS-fragmentation kind of scenario \cite{Dijkgraaf:2005bp},\cite{Maldacena:1998uz}. 
Furthermore, it should be clear that our analysis given above does not exclude the presence of any kind of throat other than the ``common throat" encompassing all the centers as we discussed here. Especially, when the total charge of a subset of the centers corresponds to the charge of a black hole, one might also expect the presence of a ``sub-throat" encompassing just the subset in question, given that the other centers are sufficiently far away. The most well-known example of this phenomenon is that of the black ring geometry, which can be seen as the uplift of a D6 and a D4-D2-D0 center in the M-theory limit\cite{Elvang:2005sa,Gaiotto:2005xt,Bena:2005ni}. In the case that the total charge corresponds to that of a D6-D4-D2-D0 black hole (the case of small D0 charge), one has indeed a common throat of the BMPV type we discussed above. But apart from that, if one zooms in further near the D4-D2-D0 center there is another \(AdS_3\times S^2\) ``sub-throat" region, which is locally the same as the uplift of the D4-D2-D0 near horizon geometry and which gives the Bekenstein-Hawking entropy of the black ring\footnote{which is the same as the entropy of the D4-D2-D0 blak hole.}. For the special case of \(T^6\) compactification, a related issue is discussed in the dual D5-D1-P language in \cite{Elvang:2004ds,Bena:2004tk}.

Finally, the presence of a throat region opens the possibility to learn more about the CFT states these solutions correspond to: by treating the throat region as an asymptotically AdS spacetime, we can employ the AdS/CFT dictionary to read off the relevant vevs of these proto-black holes, see for example \cite{Skenderis:2006ah}. It will be interesting to see what kind of CFT states our bubbling solutions (including the known ones of Bena-Warner {\it{et al}}) correspond to.

\subsubsection{Near a Center}
\label{Near a Center}

While much of the discussion above applies generally to all the lifted solutions in the large radius limit and depend only on the total charges, the solution near a center is of course strongly dependent on how the charges are allocated. Indeed, as we discussed in section \ref{Specify the 4D Charges}, we've chosen the specific D6 and anti-D6 with Abelian world-volume fluxes as our centers because we'd like the metric to be free from horizons and singularities. Now we will explicitly verify this by analysing the metric near a center. Therefore, unlike most of the equations in the previous subsections, our discussion here applies only to the charges we described in section \ref{Specify the 4D Charges}:
\be
\G=
\sum_{i=1}^N \G_i =1 + \sum_{i=1}^N f_i +\sum_{i=1}^N \frac{1}{2} \frac{f_i^2}{p^0_i} +\sum_{i=1}^N\frac{1}{6} \frac{f_i^3}{(p^0_i)^2}\;.
\ee

In the region very close to the {\it{i}}th center, where
$$
\frac{1}{r_i} \gg \frac{1}{r_{ij}}, \;h_0,\;h_A\;,
$$
we can expand the harmonic functions as
\be
H = \frac{\G_i}{r_i } + H_i +{\cal O}{(\frac{r_i}{r_{ij}^2})}\;,
\ee
with \(H_i\) defined below (\ref{integrability2}).

If we plug this into the attractor flow equation, and notice that the possible \(\frac{1}{r_i}\) term cancels because our choice of charges has the virtue
\be
-2q_{A,i} + \frac{(p_i)_A^2}{ p^0_i} = 0 \;,
\ee
we get
\be
D_{ABC}y^B y^C = -2 c_{A,i} + {\cal O}(\frac{r_i}{r_{ij}})\;,
\ee
where
\ben
c_{A,i} &=& H_{A,i}+\frac{1}{p^0_i}H^0_i \,q_{A,i} - \frac{1}{p^0_i} D_{ABC} p^B_i H^C_i \\
&=& h_A + \sum_j \frac{p^0_j}{r_{ij}}\frac{(f_{ij}^2)_A}{2}
\een
is a constant. 

The condition that the \(\rr^3\) part of the base metric is positive \( QH^0 > 0 \) can be satisfied if \be 
\label{near_center_condition} p^0_i c_{A,i} < 0\,.\ee

Assuming that our choice of locations and fluxes satisfies this condition, we have a solution
\ben
y^A &=& y^A_i +  {\cal O}(\frac{r_i}{r_{ij}}) \mbox{   where}\\ \notag
\frac{(y_i^2)_A}{2}  &=& -c_{A,i}\\
\Rightarrow Q^3 &=& Q^3_i + {\cal O}(\frac{r_i}{r_{ij}}) = (\frac{y_i^3}{6})^2 + {\cal O}(\frac{r_i}{r_{ij}})\;.
\een
 
With a similar expansion and exploit the integrability condition (\ref{integrability2}) at the \(i\)th center and the explicit expression of the charges (\ref{chargei}), we get  
 \ben
 L &=& {\cal O}(\frac{r_i}{r_{ij}}) \\ \notag
  \w^0 &=& p^0_i \cos\th d\f + {\cal O}(\frac{r_i}{r_{ij}})\\ \notag
d\w &=& \star_3\langle dH,H\rangle = \star_3 dr_i \, {\cal O}(\frac{1}{r_{i}})\\
 \Rightarrow  \w &=& {\cal O}(r_i)\;.
 \een

Notice here that the first equation guarantees that (\ref{near_center_condition}) is enough to ensure that there is no closed timelike curve near this center. 

With everything put together, we obtain the metric near the \(i\)th center:
$$
2^{-2/3} ds_{5D}^2 = -dt'^2 + d\r^2 +\frac{\r^2}{4} \lbrack d\th^2 + \sin^2\th d\f^2 + (\frac{1}{p^0_i}d\psi+\cos\th d\f)^2 \rbrack + + {\cal O}(\frac{r_i}{r_{ij}})\;,
$$
where we have rescaled the coordinates as \(t' = \frac{t}{Q_i}\), \(\r^2 = 4p^0_i Q_i r_i\). Therefore we conclude that metric approaches that of a \(\C^2/ \zz_{p^0_i}\) orbifold, and has nothing more singular than a usual orbifold singularity. Specifically, the solutions with only \(p^0_i = \pm 1\) for all the centers will be completely smooth everywhere.

Furthermore, one sees that the \(U(1)_L\) isometry generated by \(\xi_L^3 = \pa_\psi\) has a fixed point at the center. Thus a non-trivial two-cycle which is topologically a sphere (the bubbles) is formed between any two centers and therefore the name ``bubbling solutions" (or rather the ``sausage network" solutions).  These two-cycles can support fluxes and indeed, the fluxes going through the \(ij\)th bubble is \(p^0_i p^0_j {f_{ij}^A}\), with \(f_{ij}^A\) defined as (\ref{f_ij}) \cite{Berglund:2005vb}. Furthermore, the amount of fluxes going through the bubbles constrains the distance between them through the integrability condition (\ref{integrability2}), which in this case reads
\be
\sum_j \frac{1}{r_{ij}} \,p^0_i p^0_j \frac{f_{ij}^3}{6} = -h_A \til{p}^A_i = -h_A \til{f}^A_i \;. 
\ee

\subsection{Large gauge Transformation}
\label{Large gauge Transformation}

It is well known that there is a redundancy of description, namely a gauge symmetry, in type IIA string theory or equivalently M-theory, which is related to the large gauge transformation of the B-field and the three-form potential \(C^{(3)}\) respectively. Physically, this large gauge transformation can be incurred by the nucleation of a virtual M5-anti-M5 pair and thus the formation of a Dirac surface in five dimensions \cite{deBoer:2006vg}. See Fig \ref{fig:nucleation}. This shift of \(C^{(3)}\) also shifts the definition of the charges, but leaves all the physical properties of the solution intact. 

While this is a generic feature for all choices of charge vectors and all background moduli one might begin with, what we are going to do here is just to check this gauge symmetry explicitly for our bubbling solutions. 

Indeed, in our case, the transformation
\be \label{gauge_sym}
f_i^A \rightarrow f_i^A + p^0_i a^A\;\;\;;\;\;\; a^A \in \zz^{b_2(X)}
\ee
will in general change the charges (\ref{chargei}) of the configuration, especially the total D4 charge will transform like
\be
p^A \rightarrow p^A + a^A
\ee
in the case \(p^0 =1\). 
Especially, one can always exploit this symmetry to put \(p^A=0\).
It's trivial to check that the quantities \(Q, L, \w, \w^0\) in the metric are also invariant under this transformation, since all the combinations of harmonic functions involved can equally be written in terms of the ``invariant flux parameters" \(\til{f}_i\) and \(f_{ij}\) defined in (\ref{til_f}) and (\ref{f_ij}). Especially, all the conserved charges are invariant under the transformation. On top of that, we see that the right hand side of the integrability condition (\ref{integrability2}) is also invariant.\footnote{In general, in the four-dimensional language, this also implies that the existence of a BPS bound state of given, fixed charges such that \(\til{p}^A_i= p^A_i -\frac{p^A}{p^0} p^0_i \neq 0\) for every center, is insensitive to the shift of B-field in the large volume limit.} We can therefore conclude that the metric part of the solution has a symmetry (\ref{gauge_sym}). 

Furthermore, a look at the gauge field (\ref{A_5D}) tells us that this transformation indeed corresponds to a large gauge transformation of the \(A_{5D}^A\); equivalently, in the full eleven and ten dimensions, it corresponds to 
\be
C^{(3)} \rightarrow C^{(3)}  + a^A d\psi \wedge \a_A \mbox{  (M-theory)}\;\;\;;\;\;\;B\rightarrow B + a^A \a_A \mbox{  (IIA)}\;. 
\ee

Indeed, a look at the D6 brane world-volume action (\ref{wv_action}) makes it clear that the transformation (\ref{gauge_sym}) can be seen as turning on an extra integral B-field. This explains the origin of this extra symmetry.

\section{Conclusions and Discussion}
\label{Conclusions and Discussion}
\setcounter{equation}{0}

What we have done in this part of the thesis is to motivate and present a large number of asymptotically flat, smooth, and horizonless solutions to the five-dimensional supergravity obtained from the Calabi-Yau compactification of M-theory. We also analysed their various properties and along the way described various properties of  generic five-dimensional solutions obtained from lifting the multi-center four-dimensional solutions. 

A natural question to ask is the degeneracies of such solutions. From our analysis it is obvious that these bubbling solutions we describe have the same degeneracies as their four-dimensional counterparts. Especially, these are charged particles without internal degrees of freedom; their degeneracies have to come from the non-compact spacetime. 

Relatively little is known about the degeneracies of such states, though. The core of this supergravity problem is really that, although we have the integrability condition (\ref{integrability2}) to constrain the type of the solutions we can have, generically it is not enough. Indeed, while in many cases this condition alone can exclude the existence of a bound state of given charges and background moduli, generically the fact that it can be satisfied does not mean that the solution has to exist.  
Another criterion a valid solution has to conform to is the real metric condition (\ref{CTC}), which gets translated in five dimensions as the no CTC condition. Though the integrability condition helps  to exclude the presence of an imaginary metric near a center, in general it does not guarantee anything. For the purpose of counting bubbling solutions and also for the greater ambition of counting multi-center degeneracies in general, it would be extremely useful to have a systematic way to see when the integrability is enough and when we have to impose additional conditions, and of what kind. Please see section \ref{Black Holes and Multi-Holes} for a conjecture (the split attractor flow conjecture) pertaining to this issue.

For the case that is of special interest, that is the case in which the total charge is that of a black hole, the problem is also of special difficulty. The situation is described in \cite{Denef:2002ru} as the following: if we tune down the string coupling, at certain point the distances between the centers will be of the string length (recall that \(\ell^{(4)}_P \sim \frac{l_s g_s}{\sqrt{(\frac{(J^{(s)})^3}{6})}}\)) and the open string tachyons will force us to end up in a Higgs branch of the D-brane quiver theory and thus a wrapped D-brane at one point in the non-compact dimensions. But in the other direction, for the case with a black hole total charge at least, things are much more complicated. As one increases the \(g_s\), {\it{a priori}} the state doesn't necessarily have to open up, but rather it can just collapse into a single-centered black hole, or any other kind of possible charge splittings. Therefore, seen from this cartoon picture, the D-brane degeneracy really has to be the sum of degeneracies of all of the allowed charge splittings. While at the same time, if the total charge doesn't give a black hole, from the real metric condition (\ref{CTC}) we see that the system has to split up when \(g_s\) is tuned up, since these charges only have multi-centered configurations as supergravity embodiments. 

Now let's come back to the quest of smooth, horizonless solutions with black hole charges. We have argued that the bubbling solutions we presented seem to be the only kind of solutions which can be lifted from four dimensions with these virtues. In any case it would be interesting to find explicit BPS solutions to the 5D supergravity of M-theory on Calabi-Yau {\it{without}} any exact \(U(1)\) isometry. For example, some wiggly ring structure or other things our imagination permits. These can of course never be obtained by lifting 4D solutions. 

\newpage

\section{Appendix 1: Reproduce the old Bubbling Solutions}
\label{Reproduce the old Bubbling Solutions}
\setcounter{equation}{0}

The known bubbling solutions are given by (See \cite{Berglund:2005vb,Bena:2005va,Bena:2006is,Bena:2006kb})

\bea \nonumber
ds_{5d(b)}^2 &=& - (\frac{1}{Z_1 Z_2 Z_3})^{\frac{2}{3}}\,(dt+ k )^2 \\
&+& (Z_1 Z_2 Z_3)^{\frac{1}{3}} \lbrace \frac{1}{V} (d\psi + \O^0)^2 + V dx^a dx^a \rbrace 
\eea

where

\bea \notag
V&=& \sum_{i=1}^{N} \frac{p^0_i}{r_i}\;\;;\;\; r_i = |\vec{x} - \vec{x}_i |\;\;;\;\sum_{i=1}^N p^0_i = 1\\ \notag
L_A &=& 1- \frac{1}{2} D_{ABC} \sum_i \frac{1}{r_i } \frac{f_i^B f_i^C }{ p^0_i}\\ \notag
K^A &=& \sum_i \frac{f_i^A}{r_i } \\ \nonumber
M & =& -\frac{1}{2} \sum_i \sum_A  f_i^A \,+ \frac{1}{12} \sum_i  \frac{1}{r_i }  \frac{f_i^3 }{ (p^0_i)^2}\\ \notag
d\O^0  &=& \star^3 dV\\ \notag
k &=& \m (d\psi + \O^0) + \O\\ \notag
Z_A 
&=& L_A + \frac{1}{2V} D_{ABC} K^J K^K  \;\;\;;\;\;D_{ABC} = |\e_{ABC}| \\ \notag
\m &=& M + \frac{1}{2V} K^A L_A + \frac{1}{6V^2} K^3\\ \label{tilo}
\nabla \times \O &=& V \na M - M \na V + \frac{1}{2} (K^A \na L_A - L_A \na K^A)
\eea

Let's now see how our solutions contain these as a special case. 

Firstly, apply the formulae to the special 3-charge (STU) case
$$
D_{ABC} = |\e_{ABC}| \;\;\;\;A, B, C = 1,2,3\;.
$$

In general, the attractor flow equation (\ref{aflow}) is difficult to solve, but not in this case:

\bea \notag
Q^3 &=&  (\frac{1}{6}D_{ABC}y^A y^B y^C)^2 =  (y^1 y^2 y^3)^2\\
\notag
y^2 y^3 &=& -H_1 + \frac{H^2 H^3}{H^0}\mbox{      and permutations}\\ \label{Q1}
\Rightarrow Q ^3 &=&  (-H_1 + \frac{H^2 H^3}{H^0})
(-H_2 + \frac{H^1 H^3}{H^0})(-H_3 + \frac{H^1 H^2}{H^0})\;.
\eea

Secondly we take the special Ansatz that the K\"ahler form is the same in the asymptotics for all the three directions:
\be J^1|_\inf =J^2|_\inf =J^3|_\inf = j \rightarrow \inf \;,
 \ee
and that the background B-field is finite
\be B^A|_\inf = b^A \ll j  \;.
\ee

In this case we have
\ben H_A & = &\frac{1}{2} \sum_i \frac{1}{r_{i}} \frac{(f_i)^2_A}{p^0_i} - 1\;\;\;\;A=1,2,3 \\ \notag
H_0&=& -\frac{1}{2} \sum_i \frac{1}{r_{i}} \frac{(f_i)^3}{(p^0_i)^2} + \sum_i (f_i^1+f_i^2+f_i^3)\;.
\een

Now, if we rename the coordinates and quantities appearing in our solution as 
\ben
V&=& H^0\\ \notag
L_A &=& -H_A\\ \notag
K^A &=& H^A\\ \notag
M &=&- \frac{H_0}{2} \\ \notag
\O &=& \frac{1}{2}\w\\ \notag
\O^0 &=& \w^0\\ \notag
\m &=& L \\ \notag
\Rightarrow Q^3 &=&  Z_1 Z_2 Z_3\;,
\een
one can easily check that our solution (\ref{metric2}) reduces to 
$$ ds_{5d}^2 = 2^{2/3} ds_{5d(b)}^2 \;,$$
 and the equations for and relations between quantities defined in our solutions correctly reproduce those appearing in the known bubbling solutions.

\section{Appendix 2: Constant Terms for General Charges and Background}
\label{Constant Terms for General Charges and Background}
\setcounter{equation}{0}
\ben
Z&=&<\G,\bO> \\ \notag
&=&  \frac{1}{\sqrt{\frac{4}{3}J^3}}\,
\left(p^0\,\frac{(B+iJ)^3}{6}  - \frac{p\cdot(B+iJ)^2}{2}   + q\cdot(B+iJ) -q_0  \right)\\ \nonumber
\hf &=& -2 \im \Bigl( (e^{-i\th} \O )|_\inf \Bigr) \\ \nonumber
&=&\frac{2}{\sqrt{\frac{4}{3}j^3}}\,\frac{1}{|p^0\,\frac{(b+ij)^3}{6}  - \frac{p\cdot(b+ij)^2}{2}   + q\cdot(b+ij) -q_0|} \; \im\lbrace \\ \nonumber
&&\lbrack p^0\,\frac{(b-ij)^3}{6}  - \frac{p\cdot(b-ij)^2}{2}   + q\cdot(b-ij) -q_0  \rbrack \\ 
&&\cdot\lbrack \frac{(b+ij)^3}{6} + \frac{(b+ij)^2}{2} + (b+ij) + 1 \rbrack
\rbrace
\een

\begin{align} \nonumber
\hf^0&= \frac{2}{\sqrt{\frac{4}{3}j^3}}\,\frac{1}{|p^0\,\frac{(b+ij)^3}{6}  - \frac{p\cdot(b+ij)^2}{2}   + q\cdot(b+ij) -q_0|}\\ & \lbrace \frac{p^0}{6} (j^3-3jb^2) + pjb - qj \rbrace 
\end{align}

\begin{align} \notag
\hf^A &= \frac{2}{\sqrt{\frac{4}{3}j^3}}\,\frac{1}{|p^0\,\frac{(b+ij)^3}{6}  - \frac{p\cdot(b+ij)^2}{2}   + q\cdot(b+ij) -q_0|}\\ 
\nonumber
& \lbrace b^A\, \lbrack \frac{p^0}{6} (j^3-3jb^2) + pjb - qj  \rbrack \\ 
&+ j^A \,\lbrack  \frac{p^0}{6} (b^3-3j^2b) -\frac{p(b^2-j^2)}{2} + qb - q_0 \rbrack \rbrace
\end{align}

\begin{align} \notag
\hf_A& = \frac{2}{\sqrt{\frac{4}{3}j^3}}\,\frac{1}{|p^0\,\frac{(b+ij)^3}{6}  - \frac{p\cdot(b+ij)^2}{2}   + q\cdot(b+ij) -q_0|}\\ \nonumber & \lbrace \frac{(b^2-j^2)_A}{2}  \,\lbrack \frac{p^0}{6} (j^3-3jb^2) + pjb - qj  \rbrack \\
&+ (jb)_A \,\lbrack  \frac{p^0}{6} (b^3-3j^2b) -\frac{p(b^2-j^2)}{2} + qb - q_0 \rbrack \rbrace
\end{align}

\begin{align} \notag
\hf_0&= \frac{-2}{\sqrt{\frac{4}{3}j^3}}\,\frac{1}{|p^0\,\frac{(b+ij)^3}{6}  - \frac{p\cdot(b+ij)^2}{2}   + q\cdot(b+ij) -q_0|}\\ \notag& \lbrace 
\frac{b^3-3j^2b}{6}\,\left(pjb-qj\right) \\
& -\frac{j^3-3jb^2}{6}\,\left(  -\frac{p(b^2-j^2)}{2} + qb -q_0 \right)
\rbrace
\end{align}

\part{A Farey Tail for Attractor Black Holes}

This part of the thesis is based on the result reported in publication \cite{deBoer:2006vg}. Other publications in the same period on closely related topics are \cite{Kraus:2006nb,Gaiotto:2006wm,Denef:2007vg}. An important follow-up publication which refines our results is \cite{Manschot:2007ha}.

The goal of the present part of the thesis is to gain a better understanding of the microstates of D4-D2-D0 black holes in IIA string theory
compactified on a Calabi-Yau manifold. 
These microstates  are counted by a
(generalized) elliptic genus of a (0,4) conformal field theory. By
exploiting a spectral flow that relates states with different
charges, and using a generalised Rademacher formula, we find that the elliptic genus has an exact asymptotic expansion in terms of
semi-classical saddle-points of the dual supergravity theory. This
generalizes the known "Black Hole Farey Tail" of
\cite{Dijkgraaf:2000fq} to the case of \({\cal N}=2\), \(d=4\) black holes in string theory.

\chapter{A Farey Tail for Attractor Black Holes}

\section{Introduction}
\label{Introduction_farey}

One of the main successes of string theory has been the
microscopic explanation of black hole entropy. The microstates
for extremal BPS black holes are well understood in
theories with 16 or more supercharges. This includes the
D1-D5-P system in type IIB theory on \(K3\times S^1\) for
which the microstates are represented by the elliptic genus of a
(4,4) CFT with target space given by a symmetric product of $K3$ \cite{Strominger:1996sh, Dijkgraaf:1996xw}.
The elliptic genus for this target space can be explicitly
computed, leading to a concrete and exact expression for the
number of BPS-states.

The D1-D5-P system has a well understood dual
description in terms of type IIB theory on $K3\times AdS_3\times
S^3$. A rather remarkable result  is that the elliptic genus has an exact asymptotic
expansion, which has a natural interpretation as a sum over
semi-classical contributions of saddle-point configurations of the
dual supergravity theory.  This exact asymptotic expansion, together with its
semi-classical interpretation, has been coined
 the Black Hole Farey Tail \cite{Dijkgraaf:2000fq,Manschot:2007ha} \footnote{In the paper \cite{deBoer:2006vg} we based our analysis on the Farey Tail expansion developed in \cite{Dijkgraaf:2000fq}, which was later improved in \cite{Manschot:2007ha}. In this part of the thesis I will use the ``Modern Farey Tail" of \cite{Manschot:2007ha} without any derivation and refer the reader to the paper and the PhD thesis of fellow student Jan Manschot for further details.}.
Although the Farey tail was first introduced in the context of the D1-D5 system,
it applies to any system that has a microscopic
description in terms of a (decoupled) 2d conformal field theory
and has a dual description as a string/supergravity theory on a
spacetime that contains an asymptotically $AdS_3$.

The aim of this part of the thesis is to apply the generalised Rademacher formula \cite{Manschot:2007ha} to black
holes in theories with eight supercharges and in this way extend
the Farey Tail to \({\cal N}=2\) (or ``attractor" in a slight abuse of language) black holes.  Specifically, we
consider M-theory compactified on a Calabi-Yau three-fold $X$ and
study the supersymmetric bound states of wrapped M5-branes with
M2-branes. These states correspond to extremal four dimensional black holes
after further reduction on a circle. For this situation a
microscopic description was proposed quite a
while ago by Maldacena, Strominger and Witten (MSW)
\cite{Maldacena:1997de}, who showed that the black hole microstates are
represented by the supersymmetric ground states of a (0,4) conformal field theory.
These states are counted by an appropriately defined elliptic genus of the (0,4) CFT.

The interest in attractor black holes has been revived in recent
years due to the connection with topological string theory
discovered in \cite{LopesCardoso:1998wt, Ooguri:2004zv} and
subsequently studied by many different authors.
As was discussed in (\ref{OSV1}), it was conjectured by Ooguri, Strominger and Vafa
(OSV) in \cite{Ooguri:2004zv} that the mixed partition function of
4d BPS black holes is given by the absolute value squared of the
topological string partition function. Earlier, in a separate
development, a different connection between BPS states and
topological strings was discovered by Gopakumar and Vafa (GV)
\cite{GV2}, who showed that topological string theory
computes the number of five-dimensional BPS-invariants of wrapped
M2 branes in M-theory on a Calabi-Yau. The GV-result differs from
the OSV-conjecture (\ref{OSV1}) in the sense that the topological string
coupling constant appears in an S-dual way. Recently, this aspect of 
 the OSV conjecture have been considerably clarified in the
work of Gaiotto, Strominger, and Yin \cite{Gaiotto:2006ns}. These
authors used the CFT approach of MSW to show that the elliptic
genus of the (0,4) CFT has a low temperature expansion which
(approximately) looks like the square of the GV-partition
function. The OSV conjecture then follows from the modular
invariance of the elliptic genus, which at the same time naturally
explains the different appearances of the coupling constant.

In this part of the thesis we will show that elliptic genus of the (0,4) SCFT
can be written as a generalised Rademacher series (or the ``Modern Farey tail" expansion)
similar to that of the previously studied (4,4)
case.\footnote{Some related results were obtained independently in
\cite{Denef:2007vg,Gaiotto:2006wm,Kraus:2006nb}. }  An important
property of the SCFT is the presence of a spectral flow that
relates states with different charges, and implies that the
elliptic genus can be expanded in terms of  theta functions. The presence of these
theta functions signal the presence of a set of chiral scalars in
the SCFT, while from a spacetime point of view their appearance
naturally follows from the Chern-Simons term in the effective
action.  We find that the 
(modern) Farey tail expansion contains subleading
contributions to each saddle point that can be interpreted as
being due to a virtual cloud of BPS-particles (actually, wrapped
M2-branes) that are ``light" enough such that they do
not form a black hole by themselves. The degeneracies of these particles are, in
the large central charge limit, given in terms of the
Gopakumar-Vafa invariants. In this way we see that the results of
\cite{Gaiotto:2006ns} naturally fit in and to some extent
follow from our Attractor Farey Tail.

The outline of this part of the thesis is as follows: in section \ref{Wrapped M-branes and the Near Horizon Limit} we review the
bound states of wrapped M5 and M2 branes in M-theory on a
Calabi-Yau three-fold and explain the emergence of the spectral
flow. We then discuss the decoupling limit and the near horizon
geometry and describe the dimensionally reduced effective action
on $AdS_3$. In section \ref{The (0,4) SCFT} we turn to the M5 brane world-volume
theory and its reduction to the (0,4) SCFT. Here we also define
the generalised elliptic genus which counts the graded black
hole degeneracies. 
In section \ref{Spacetime Interpretation of the Attractor Farey Tail} we interpret our
result from the dual supergravity perspective and discuss its
relation to the OSV conjecture. Finally in section \ref{Summary and Conclusion} we conclude by summarising this part of the thesis and raise some open questions.

\section{Wrapped M-branes and the Near Horizon Limit}
\label{Wrapped M-branes and the Near Horizon Limit}

To establish the notation, in this section we describe the BPS
bound states of wrapped M5 and M2 branes in M-theory on a
Calabi-Yau from a spacetime point of view. We will derive a
spectral flow symmetry relating states with different M2 and M5
brane charges, first from an eleven-dimensional perspective and
subsequently in terms of the effective three dimensional
supergravity that appears in the near horizon limit.

\subsection{Wrapped Branes on Calabi-Yau and the Spectral Flow}
\label{Wrapped branes on Calabi-Yau and the spectral flow}

Consider M-theory on a Calabi-Yau threefold \(X\) and a circle \(\R^{3,1}\times X \times S^1\), 
and an M5-brane wrapping a
4-cycle \({\cal P}\) with  $[{\cal{P}}]=p^A S_A$ in the Calabi-Yau three-fold $X$. Here $\{S_{A=1,\dotsi,h^{1,1}}\}$ is a
basis of integral 4-cycles $H_4(X,\Z)$ in \(X\), where \(h^{1,1}(X)\) is the second Betti number of the Calabi-Yau manifold. In order for this five-brane to be
supersymmetric, the 4-cycle ${\cal{P}}$ has to be realized as a positive
divisor. Therefore we will assume that ${\cal{P}}$ is a smooth ample divisor so that classical geometry is a valid tool for our analysis. There is a line bundle \({\cal L}\) with
 \be c_1({\cal L})= [{\cal P}]= p^A \a_A := P
 \ee
 associated to this divisor, where $\alpha_A \in H^2(X,\Z)$ is a basis of harmonic 2-forms Poincar\'e dual to the four-cycles $S_A$, such that the position of the four-cycle ${\cal{P}}$ can be thought of as the zero locus of a section of this line bundle.

The wrapped M5-brane reduces to a string in the
remaining five dimensions. In addition there are five-dimensional particles
corresponding to M2-branes wrapping a two-cycle $[\Sigma] =q_A 
\Sigma^A$, where \(\Sigma^A\) is a basis of $H_2(X,\Z)$ dual to $\{S_A\}$, i.e.
\(\Sigma^A \cap S_B = \delta_B^A \).
These particles carry charges $q_A$ under the $U(1)$
gauge fields $A^A$ which arise from the dimensional reduction of the M-theory
3-form $A^{(3)}$, as summarised in Table \ref{massless_spectra_4_5d}, 
\be
\label{CA} A^{(3)} = \sum_A A^A \wedge \alpha_A\;.
\ee

 Such an ensemble of
strings and particles can form a BPS bound state which leaves four
of the eight supersymmetries unbroken.

Eventually we are interested in the
BPS states of the 4d black hole that is obtained by further
compactifying the string along an $S^1$. These states carry an
additional quantum number $q_0$ related to the Kaluza-Klein momentum
along the string. From the four dimensional perspective, the quantum
numbers $(p^A,q_A,q_0)$ are the D4, D2
and D0 brane charges in the type IIA compactification on $X$. In this
part of the thesis we will be switching back and forth between a spacetime perspective from eleven (M-theory), ten (type IIA), 
five (M-th/CY), four (IIA/CY), or even three dimensions \((AdS_3)\), and a world-volume perspective of
the M5-brane or its reduction to a world-sheet.

Before going to the world-volume description of the M5-brane and
its reduction to a string, let us describe the spectral flow symmetry of  BPS states from the spacetime perspective. 

First recall that, as discussed in section \ref{4D-5D Connection} and chapter \ref{Black Holes and Multi-Holes}, a supergravity solution with \(U(1)\) isometry of the low-energy effective action obtained from compactifying M-theory on a Calabi-Yau space \(X\) can be thought of as the ``lift" of a four-dimensional solution of the \({\cal N}=2,\,d=4\) supergravity theory and  is specified by \(2h^{1,1}+2\) harmonic functions \(H^I,H_I: \R^3 \rightarrow \R\), \(I= 0,\dotsi,h^{1,1}\).
There is a symmetry between different solutions whose corresponding set of harmonic functions are related by \cite{Cheng:2006yq,deBoer:2008fk}
\begin{align} \notag
H^0 & \rightarrow H^0\\ \notag
H^A & \rightarrow H^A - H^0 k^A\\ \notag
H_A & \rightarrow H_A - D_{ABC}H^B k^C + \frac{H^0}{2} D_{ABC}k^B k^C \\ \label{harmonic_shift}
H_0 & \rightarrow H_0 + k^A H_A - \frac{1}{2} D_{ABC}H^A k^B k^C+\frac{H^0}{6} D_{ABC}k^A k^B k^C\;,
\end{align}
where \(D_{ABC}\) is the triple-intersection number for the basis \(\a_A\) introduced in (\ref{triple-intersection}). From (\ref{formulation2_sol_vec}) or equivalently (\ref{sol_scalar}), and (\ref{5d_sol}), we see that the above transformation in particular leaves the geometry part of the solution invariant and induces a shift in the five-dimensional vector field by
\be\label{spectral_flow_U1}
A^A \rightarrow A^A - k^A  d\j\;,
\ee
where \(\j \in [0,1)\) is the coordinate of the M-theory circle \(S^1\). 
In particular, for the case at hand we have \(H^0 = 0\) and the above transformation corresponds to the following shift of charges 
\begin{align}\notag
p^A & \rightarrow p^A\\ \label{qflow}
q_A& \rightarrow q_A- D_{ABC}k^Bp^C \\ \label{q0flow}
q_0& \rightarrow q_0+k^A q_A- {1\over 2}D_{ABC}k^A k^Bp^C\;,
\end{align}
which leaves the geometry part of the solution invariant. 

From M-theory point of view, the above symmetry among solutions of the five-dimensional supergravity can be understood in the following way.  The low energy action of
M-theory (\ref{S11}) contains the Chern-Simons coupling 
\be
S_{CS}=-\frac{1}{3!}\,\int A^{(3)}\wedge F^{(4)}\wedge F^{(4)}\;,
\ee
where $F^{(4)}=dA^{(3)} = F^A\wedge\alpha_A$ is the four-form field strength.
As a result, the M2-brane charge is defined as (here we work in
11D planck units)
\be
q_A=\int_{S^2\times S^1\times S_A }  \left( * F+C\wedge F\right)\;.
\ee
The charge thus contains a Chern-Simons type contribution
depending explicitly on the $A^{(3)}$-field. This term can be written as
a volume integral of $ F \wedge F$ and hence is invariant under
small gauge transformations that vanish at infinity. However, it
can still change under large gauge transformations corresponding
to shifts in $A^{(3)}$ by a closed and integral three-form
\be
A^{(3)} \rightarrow A^{(3)} - \sum_{A=1,\dotsi,h^{1,1}} k^A  d\j\wedge \a_A\;, 
\ee
which gives exactly the shift of the lower-dimensional gauge field (\ref{spectral_flow_U1}) upon dimensional reduction. This
transformation should be an exact symmetry of M-theory. The value
of the charge $q_A$, though, is not invariant but instead receives
an extra contribution proportional to the M5-brane charge $p^A$.
Namely, using
\be
\int_{S_A\times S^2\times S^1} d\j \wedge \alpha_B \wedge F
=D_{ABC}\int_{S^2} F^C = D_{ABC} p^C\;,
\ee
one finds that
\be
q_A\to q_A- D_{ABC}k^Bp^C
\ee
under a large gauge transformation of the \(A^{(3)}\) field.

Alternatively, this shift of the conserved charges can also be understood from a type IIA point of view. From Table \ref{massless_spectra_4_5d}
we see that the integral shift of the three-form fields now translates into a shift of internal Neveu-Schwarz B-field 
\be B^{(2)} \to B^{(2)}  - \sum_{A=1,\dotsi,h^{1,1}} k^A  \a_A\;, \ee
which should be an exact symmetry of string theory. 
From the Wess-Zumino part of the D-brane action (\ref{anomalous_brane}) of a D4 brane wrapping the four-cycle \({\cal P}\) with $[{\cal{P}}]=p^A S_A$, we see that the presence of the factor \(ch({\cal F}) = e^{F+B}\) causes the shift of the induced D2, D0 brane charges (\ref{qflow},\ref{q0flow}) when the B-field is shifted\footnote{Notice that our convention defines the D0 brane to be the objects that couples to 
\(-C^{(1)}\)}.

It will turn out to be convenient to introduce the symmetric
bilinear form
\be\label{bilinear_lattice_def}
 D_{AB} = -D_{ABC} p^C= -\int_P \a_A \wedge \a_B, \ee
where $\a_A$ should be understood as the pullback of the harmonic 2-forms from the ambient Calabi-Yau \(X\) to the 4-cycle ${\cal{P}}$. Notice that the extra minus sign in the definition implies that the anti-self-dual directions now have positive signatures. 

By the Lefschetz hyperplane theorem, which states that the inclusion map \(H^{2}({\cal P}) \rightarrow H^{2}(X) \) is surjective and \(H^{1}(X)\) and \(H^{1}({\cal P})\) are isomorphic, this form is non-degenerate.
In fact, according to the Hodge index theorem it has signature $(h^{1,1}-1,1)$. Thus, for every
positive divisor ${\cal{P}}$ we obtain a natural metric $D_{AB}$
on $H^2(X,\Z)$ which turns it in to a Lorentzian lattice
$\Lambda= \L^{h^{1,1}-1,1}$. 

Generically this lattice is not self-dual (or equivalently, unimodular), namely that the inverse metric \(D^{AB}\) is not integral. We will call the dual lattice, defined as the set of all vectors  \( v^A \in \R^{h^{1,1}-1,1}\) such that the inner product with all vectors in \(\L\) take integral values, \(\L^*\). 

This dual lattice can be naturally identified with the lattice \(H^4(X,\Z)\)
 and the Dirac quantization condition suggests that 
M2-brane charges takes value in $\Lambda^*$, whose bilinear form
$D^{AB}$ is given by the inverse of $D_{AB}$. However, due to the
presence of the Freed-Witten anomaly the M2 brane charge gets shifted and the above statement is no longer true. 

The shift of the charges can be understood as the fact that when the divisor is not spin, namely when the second Whitney class is non-vanishing or equivalently when \([{\cal P}]=c_1({\cal L})\) is odd,
the U(1) gauge field on it has to have a non-trivial holonomy in order for it to have a \(Spin^c\) structure, and as a result the M2-brane charge does not satisfy the
usual Dirac quantization condition, but rather 
\cite{Freed:1999vc,Minasian:1997mm}
\be\label{qshift}
q_A \in {1\over 2} D_{ABC}p^Bp^C \oplus \Lambda^*\;,\quad\text{or just}\quad q\in \L^* + \frac{p}{2}.
\ee
In terms of the bilinear form \(D_{AB}\) the flow equations
(\ref{qflow}) and (\ref{q0flow}) read
\bea
 \label{qflow2}q_A &\to& q_A+ D_{AB}k^B ,\\
\label{q0flow2} q_0 &\to& q_0+k^A q_A+ {1\over 2}D_{AB}k^A k^B\;.
\eea
In this form one can see explicitly that the following combination of charges
\be\label{qhat}
\hat{q}_0=q_0-{1\over 2} D^{AB}q_A q_B
\ee
is the unique combination (up to additive constant and multiplication factors) that is invariant under the combined spectral flow of $q_A$ and $q_0$. This phenomenon is familiar from our study of the spectral flow symmetry of the \({\cal N}=2\) superconformal algebra in section \ref{section_spectral_flow} and \ref{Elliptic Genus and Vector-Valued Modular Forms}, where the flow-invariant combination \(L_0 - \frac{1}{2\hat{c}}J_0^2\) plays a similar role. 

From (\ref{qflow}) we see that the spectral flow
transformation amounts to shifting the vector $q$ by an element \(k \in \L \).
Note that due to the integrality of the symmetric bilinear $D_{AB}$
one has $\Lambda\subset\Lambda^*$. In general, $\Lambda$ is a
proper subset of $\Lambda^*$, which means that not all charge
configurations (\(\Lambda^*\)) are related to each other by
spectral flow (\(\Lambda\)). Explicitly, any given M2 charge vector \(q\in \L^* + p/2\) has a unique decomposition
\be\label{decomposition_m2_charge}
q= \m + k +\frac{p}{2}\quad,\quad \m \in \L^*/\L \quad,\quad k \in \L
\ee
where \(\m \neq 0\) for generic charges. 

From the above argument we conclude that the combined spectral
flow transformations (\ref{qflow2}) and (\ref{q0flow2}) constitute
a symmetry of M-theory/string theory. This gives a
non-trivial prediction on the BPS degeneracies that the number of
BPS states $d_{{\cal P}} (q_A, q_0)$ should be invariant under these
transformations.

We can now compare this microscopic prediction with the macroscopic result and see that it indeed passes the consistency check. The leading macroscopic entropy of the 4d black hole with charges $p^A$,
$q_A$ and $q_0$ is given by (\ref{D4_D2_D0_entropy})
\be\label{M5_entropy}
S=2 \pi\sqrt{{\hat{q}_0 D}}
\ee
where
\be
6D\equiv  \int_X P \wedge P \wedge P= \,D_{ABC} \, p^A p^B p^C.
\ee
is the norm
of the vector \(p \in \L\). From this we see that the entropy formula is indeed consistent with our
prediction that the entropy must be invariant under the spectral
flow.

Finally, we would like to point out that the spectral flow
(\ref{qflow2}) can be induced spontaneously by the nucleation of a
M5/anti-M5 brane pair with magnetic charge $k^A$, where the M5
loops through the original (circular) M5 brane before annihilating
again with the anti-M5 brane. We will make use of this comment in
the next section where this same process is translated to the near
horizon geometry.
\newline

\begin{figure}[htb!]
\centering%
\includegraphics[height=3cm]{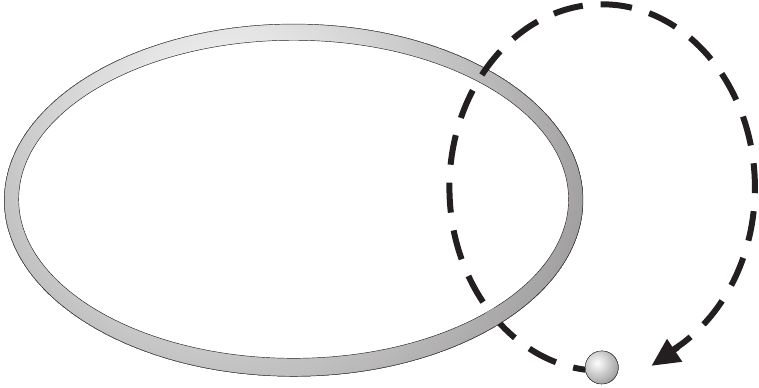}
\caption{\small{An M5 brane loops through the original (circular) M5 brane and then annihilates again with an anti-M5 brane.}}
\label{fig:Loop}
\end{figure}

\subsection{The Near-Horizon Geometry and Reduction to Three
Dimensions}

In the decoupling near-horizon limit, the spacetime physics can be entirely captured by the
world-volume theory of the brane . In this limit the
11-dimensional geometry becomes
\be
X \times AdS_3 \times S^2\;,
\ee
with the K\"ahler moduli $J = J^A \a_A
\in H^{1,1}(X)$ of the Calabi-Yau fixed by the
attractor mechanism to be proportional to the charge vector $P = p^A\a_A$.

More explicitly, the attractor equation reads (\ref{5d_sol}), (\ref{formulation2_sol_vec})
$$\label{attractor_M5}
{J^A \over {\cal V}^{1/3}} = { p^A \over D^{1/3}},
$$
where $ {\cal V}$ denotes the volume of the Calabi-Yau in the eleven-dimensional Planck unit
\be
 {\cal V}= {1\over 6} \int_X J  \wedge J \wedge J = {1\over 6} D_{ABC} J^AJ^B J^C\;.
\ee
As explained in section \ref{Five-dimensional Supergravity}, the volume $ {\cal V}$ sits in the universal hypermultiplet and  is not fixed by the attractor equation.
Furthermore, from the relation between the eleven- and five-dimensional Newton's constant and the expression for the five-dimensional solution (\ref{5d_sol}), (\ref{formulation2_sol_vec})
\be
\ell^{(5)}_{\text{\footnotesize p}} \sim \ell^{(11)}_{\text{\footnotesize p}} \, {\cal V}^{-1/3}\quad,\quad \ell \sim \ell^{(5)}_{\text{\footnotesize p}} D^{1/3}\;,
\ee
we see  the ratio $ {\cal V}/D$ turns out to be related to the curvature radius
$\ell$ of the $AdS_3$ and $S^2$ as
\be
\frac{\ell}{\ell^{(11)}_{\text{\footnotesize p}}} \sim \left(\frac{D}{{\cal V}}\right)^{1/3}\;. 
\ee

Therefore, for the five-dimensional supergravity to be a valid description we need the universal hypermultiplet scalar to satisfy 
\(D \gg {\cal V} \gg 1\).  For our purpose it will be
useful to consider a further reduction along the compact $S^2$  to
a three dimensional theory on the non-compact $AdS_3$ spacetime.
In the low energy limit, this theory contains the metric and the
$U(1)$ gauge fields $A^A$ as the massless bosonic fields.

From the five-dimensional perspective, the five-brane flux of
 M-theory background gets translated into a magnetic flux $F^A= dA^A$ of the \(U(1)\) gauge fields through the $S^2$:
$$
\int_{S^2} F^A =  p^A.
$$
The eleven-dimensional Chern-Simons term of the $A^{(3)}$-field can therefore
be reduced in two steps. First to five dimensions, where it takes
the form (\ref{5d_sugrav_action})
$$
16 \p G_{\text{\footnotesize N}}^{(5)}S_{CS} = \frac{1}{3!}\int_{AdS^3 \times S^2} D_{ABC} A^A \wedge F^B \wedge
F^C,
$$
and subsequently, by integrating over the $S^2$, to three
dimensions, where it turns into the usual (Abelian)
Chern-Simons action for the gauge fields on $AdS_3$. In combination
with the standard kinetic terms, we get
$$
16 \p G_{\text{\footnotesize N}}^{(3)} S =  \int d^3x \sqrt{g}\left(R - \frac{2}{\ell^{2}}\right) -\frac{a_{AB}}{2}
 \int  F^A \wedge \star F^B +D_{AB} \int  A^A
\wedge dA^B
$$
as the terms in the bosonic action relevant for our discussion,
where  $a_{AB}=\int_X \a_A \wedge \ast\a_B$ is given in (\ref{5d_gauge_coupling}).
The 3d Newton constant is given by
\be
{1\over G_{\text{\footnotesize N}}^{(3)}} \sim \frac{\ell^2}{(\ell^{(5)}_{\text{\footnotesize p}})^3}\sim \frac{D}{\ell}\;.
\ee

We will end this section by some discussions about the spectral
flow in the setting of the attractor geometry. First we note that
the spectral flow argument in the previous section can be carried
to the three-dimensional setting by dimensional reduction . The
M2-brane charge $q_A$ is defined now as an integral over a circle
at spatial infinity of the $AdS_3$ as
$$
q_A = \int_{S^1} \Bigl(- a_{AB}\star F^B+ D_{AB}A^B\Bigr)\;.
$$
Again one easily verifies that it changes as in (\ref{qflow2}) as
a result of a large gauge transformation $A^A\to A^A - k^A
d\j$ in three dimensions. The charge $q_0$ is related to the
angular momentum in $AdS_3$. To understand the shift in $q_0$
under spectral flow, one has to determine the contribution to the
three-dimensional stress energy tensor due to the gauge field.

As mentioned above, the spectral flow has a nice physical
interpretation in terms of the nucleation of an M5/anti-M5 brane
pair. Let us now describe this process in the near horizon
geometry. The following argument is most easily visualized by
suppressing the (Euclidean) time direction and focusing on a
spatial section of $AdS_3$, which can be thought of as a copy of
Euclidean $AdS_2$ and hence is topologically a disk (\ref{upper_half_plane}). Together with
the $S^2$ it forms a four dimensional space. First, recall that a
wrapped M5 brane appears as a string-like object in this four
dimensional space. Since an M5-brane is magnetically charged under
the five-dimensional gauge fields $A^A$, it creates a "Dirac
surface" of $A^A$.  Of course, the location of the Dirac surface
is unphysical and can be moved by a gauge transformation. Now
suppose at a certain time an M5/anti-M5-brane pair nucleates in
the center of $AdS_2$ in a way that the M5 and the anti-M5 branes
both circle the equator of the $S^2$.  Subsequently, the M5 and
the anti-M5 branes move in opposite directions on the $S^2$, say
the M5 brane to the north pole and anti-M5 to the south pole.  In
this way the M5 and anti-M5 brane pair creates a Dirac surface
that stretches between them. Eventually both branes slip off and
self-annihilate on the poles of the $S^2$. What they leave behind
now is a Dirac surface that wraps the whole $S^2$ and still sits
at the origin of the $AdS_2$. To remove it one literally has to
move it from the center and take it to the spatial infinity. Once
it crosses the boundary circle, its effect is to perform a large
gauge transformation that is determined by the charge $k^A$ of the
M5 brane of the nucleated pair. We conclude that spectral flow can
thus be induced by the nucleation of pairs of M5 and anti-M5
branes.
\newline

\begin{figure}[htb!]
\centering%
\includegraphics[height=5.5cm]{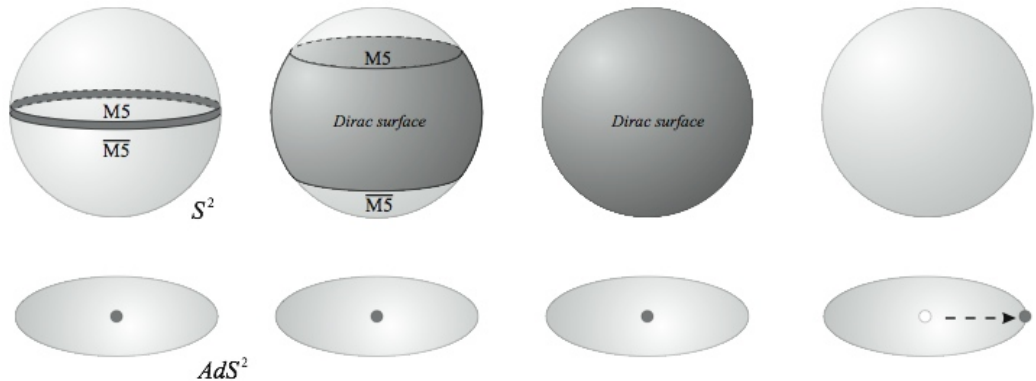}
\caption{\label{fig:nucleation}\small{A large gauge transformation: (i) An M5-anti-M5
pair wrapping the equator of the \(S^2\)  nucleates at the center
of the \(AdS_2\). (ii) The M5 and anti-M5 begin to move in the
opposite directions in the \(S^2\), while still stay at the center
of the \(AdS_2\). (iii) A Dirac surface wrapping the whole \(S^2\)
is formed. (iv) Finally one moves the Dirac surface from the bulk
of the \(AdS_2\) towards the spatial infinity across the
boundary.}} 
\end{figure}

\section{The (0,4) Superconformal Field Theory}
\label{The (0,4) SCFT}
\setcounter{equation}{0}

The existence of the bound states of M2-branes to the M5-brane can
be seen in an elegant way from the point of the view of the
five-brane world-volume theory. This world-volume theory is
a six-dimensional (0,2) superconformal field theory whose field content
are five scalars, two Weyl fermions, and a tensor field $b^{(2)}$
with self-dual 3-form field strength $H$ \cite{Callan:1991ky}. The spacetime $A^{(3)}$-field
couples to $H$ through the term \be \label{CH} \int_W A^{(3)} \wedge H
\ee where $W = {\cal{P}} \times S^1 \times \R_t$ denotes the world-volume of
the five-brane.

In a bound state the M2-brane charges are
dissolved into fluxes of $H$ in the following way: the self-dual tensor $H$ that carries the
charges $q_A$ has spatial component
\be
\label{HA} H_\j =- D^{AB} \,q_A \
\a_B \wedge d\j,
\ee
with \(\j \) being again the coordinate of the $S^1$ and 
the two-forms \(\a_A\) should be understood as their pull-back from the Calabi-Yau \(X\) to the divisor \({\cal P}\).  The timelike components follow from the self-duality condition.
Combining the formulas (\ref{CA}) and (\ref{HA}), one sees that this
produces the right coupling
$$
\int_W A^{(3)} \wedge H = q_A \int_{\R_t} A^A
$$
of the $U(1)$ gauge fields \(A^A\) to the charges $q_A$.

When we take the scale of the Calabi-Yau to be much smaller than
the radius of the M-theory circle, the M5 world-volume theory
naturally gets reduced along the
4-cycle ${\cal{P}}$ to a two-dimensional conformal field theory with
$(0,4)$ supersymmetry.
As usual the superconformal symmetries are identified
with the supersymmetric isometries of the $AdS_3 \times S^2$
manifold. In this case the right-moving superconformal algebra contains ``small" \({\cal N}=4\) superconformal algebra (\ref{n_4_superconformal_algebra}) as a sub-algebra. In particular the $SU(2)$ R-symmetry
corresponds to the rotations of the $S^2$ factor.

\subsection{Counting the Degrees of Freedom}

Let us now find the degrees of freedom of the CFT by dimensionally reducing the massless fields of the five-brane theory on the divisor 
\({\cal P}\) in the Calabi-Yau which the M5 brane wraps. Our treatment here follows closely that of \cite{Maldacena:1997de,Minasian:1999qn}. Here we assume some familiarity with basic algebraic geometry. Especially we will use some formulas which we haven't introduced before, including the adjunction formula, Kodaira vanishing theorem, and Lefschetz hyperplane theorem, which can be found in the chapter (I), section two of \cite{GH}.

First let's begin with the five scalar fields, which in the original five-brane theory correspond to the locations of the five-brane in the remaining five spatial dimensions normal to the world-volume. Upon dimension reduction, three of them correspond to the center of mass location of the string in the three non-compact directions transversal to it, and simply reduce to three scalar fields of the CFT which have both left- and right-moving components. The remaining two scalars, let's call them \(X^1\) and \(X^2\), now correspond to the position of the cycle  \({\cal P}\) inside the Calabi-Yau. In other words, they correspond to the deformations of \({\cal P}\) in the Calabi-Yau while keeping the homology class invariant. In order to reduce the complex scalar \(X^1 + i X^2\) we have to know the space of deformations of \({\cal P}\). 

Since the four-cycle can be thought of as the zero locus of a section of the line bundle \({\cal L}\), this space is 
\(\pp H^0(X,{\cal L} )\). From the Hirzebruch-Riemann-Roch theorem
(\ref{Hirzebruch_Riemann_Roch}), which gives
\ben
w&=& \sum_{k} (-1)^k \text{\small dim}H^k(X,{\cal L})  = 
\int_X ch({\cal L}) \,Td(X)  
\\ &=& \int_X  e^{P} \, (1+\frac{1}{12} c_2(X))
= D + \frac{1}{12} c_2\cdot P\;,
\een
and the fact that \(\text{\small dim}H^k(X,{\cal L}) = 0 \) for \(k>0\) from the Kodaira's vanishing theorem (recall that we have assumed \({\cal P}\) to be a positive divisor in section \ref{Wrapped branes on Calabi-Yau and the spectral flow}), we obtain the complex dimension of the space of infinitesimal deformations of \({\cal P}\) 
\be
\text{\small dim}_\C\pp H^0(X,{\cal L})  = D + \frac{1}{12} c_2\cdot P-1:= N\;,
\ee
and notice that the space is projective because sections related by a complex multiplication have the same zero locus. We therefore conclude that the complex scalar \(X^1 + i X^2 \) reduces to \(N\) complex scalars which are both left- and right-moving.

As the next step we dimensionally reduce the chiral two-form on the five-brane to the string world-sheet. Rewriting the two-form field  using the basis \(\{w_I \}\) of \(H^{2}({\cal P})\)
\be\label{b2}
b^{(2)} =  \f^{I}w_I \quad,\quad I = 1,\dotsi, b^2({\cal P})\;, 
\ee
then the self-duality condition on \(db^{(2)} = H\) implies that the scalars 
\( \f^{I}\) is left-moving when \(w_I\) is anti-self-dual and right-moving if \(w_I\) is self-dual.

To compute the dimension of the self-dual and anti-self-dual part of \(H^{2}({\cal P})\) we have to first collect a few facts about the topology of the divisor \({\cal P}\). First of all, from the adjunction formula we have 
\be
T{\cal P} =\frac{TX\lvert_{{\cal P}}}{{\cal L}\lvert_{{\cal P}}}\;,
\ee
which combined with the composition rule for Chern classes gives the Chern classes for the divisor \({\cal P}\)
\be
c( T{\cal P}) = \frac{1+c_2(X)}{1+P} = 1- P + c_2( X)  + P^2 \;,
\ee
where we have used the fact that \(X\) is a Calabi-Yau and therefore has vanishing first Chern class. Explicitly, this gives
\be \label{chern_classes_divisor}
c_1( {\cal P}) = - c_1( {\cal L}) = -P \quad,\quad 
c_2( {\cal P}) = c_2( X)  + P^2\;. 
\ee
On the other hand, the Lefschetz hyperplane theorem tells us that \(b_1({\cal P})= b_1(X) = 0\). Combining these facts with the Gauss-Bonnet theorem (\ref{gauss_bonnet}) and the signature index theorem
(\ref{signature_index_theorem}), we conclude that the numbers of right- and left-moving bosons obtained by reducing the chiral two-form to the two dimensional world-sheet, which are the same as the dimensions of the self- and anti-self-dual part of \(H^2({\cal P})\), are
\ben
b_2^R &=& \frac{1}{2} (\chi + \s) -1 = 2D +\frac{1}{6}c_2\cdot P -1 = 2N+1 \\
b_2^L &=& \frac{1}{2} (\chi - \s) -1 = 4D +\frac{5}{6}c_2\cdot P -1 \;.
\een

Especially, the direction proportional to the K\"ahler class of the divisor \({\cal P}\) is the only direction in \(H^{1,1}({\cal P})\) that is self-dual, a statement that can be checked by using the Hirzebruch-Riemann-Roch theorem (\ref{Hirzebruch_Riemann_Roch}) with the bundle \(V\) taken to be the bundle of \((q,0)\) forms on  \({\cal P}\), which gives
\bea \notag
h^{2,0}({\cal P})&=& N=D + \frac{1}{12} c_2\cdot P-1 \\
\label{anti_self_dual_11}
h^{1,1}({\cal P})&=& b_2^L+1 =4D +\frac{5}{6}c_2\cdot P\;.
\eea

Finally we will reduce the two Weyl spinors. By decomposing the spin bundle on the ambient Calabi-Yau space \(X\) into the product of the spin bundles on \({\cal P}\) and its normal bundle \cite{Minasian:1999qn}, we see that the fermionic zero modes are all right-moving and given by the zero-forms and holomorphic two-forms on \({\cal P}\). Namely that there are 
\be\label{fermions_reduction}
4 \left( h^{2,0}({\cal P}) + 1  \right) = 4 (N+1)
\ee
real right-moving fermionic degress of freedom and no left-moving ones. 
 
This ends our derivation of the massless fields of the two-dimensional conformal field theory. In particular, putting all the bosons and fermions together we get the following counting of the left- and right-moving central charges of the CFT
\ben
c_R &=& 6D + \frac{1}{2} c_2\cdot P\\
c_L &=& 6D +  c_2\cdot P= \chi({\cal P})\;,
\een
where the equality between the left-moving central charge and the Euler characteristic follows from the expression for the Chern classes of the divisor \({\cal P}\) (\ref{chern_classes_divisor}) and the Gauss-Bonnet theorem (\ref{gauss_bonnet}).

\subsection{The Universal Sigma Model}
\label{The Universal Sigma Model}

For the discussion of the BPS states of the CFT it will be useful to separate the CFT into two factors of heterotic sigma models. The first factor, which we call the universal sigma model following \cite{Minasian:1999qn}, is the heterotic sigma model obtained by reducing the five-brane theory on the part of cohomology classes of \({\cal P}\) which are images of the injective map \( H^{2}(X) \rightarrow H^{2}({\cal P})\) induced by the inclusion map \({\cal P} \rightarrow X\). This separation is useful because the universal factor is the one containing the information about conserved charges. One way to understand this is the following. Although the two-form field \(b^{(2)}\) reduces also on the two-cycles in \({\cal P}\) which do not correspond to any element in \(H_2(X)\) (\ref{b2}), the corresponding charges are not conserved due to the existence of the membrane instantons which wraps the three-ball in \(X\) that have the two-cycle in \({\cal P}\) as boundary. Notice that as \(\text{\small dim}H^{2}({\cal P}) \) is generically much larger than \(\text{\small dim}H^{2}(X)\), the universal part of the CFT actually accounts for only a small portion of the central charges.

The five five-brane world-volume scalars give rise to three left- and right-moving bosons in the universal sector, corresponding to the position of the brane in the three non-compact transversal directions. On the chiral two-form side, from the fact that the K\"ahler form direction, which is given by \(J \sim p^A \a_A \) at the attractor point (\ref{attractor_M5}), is the only self-dual direction inside \(H^{1,1}({\cal P})\), we conclude that the chiral two-form field \(b^{(2)}\) on the world-volum reduces to a single right-moving boson and \(h^{1,1}(X)-1\) left-moving bosons. Explicitly, following (\ref{HA}), the M2 brane charges \(q_A\) can be thought of as the H-fluxes through the two-cycles dual to
\bea\notag
-q_A D^{AB} \a_B &=& \frac{p\cdot q}{6D}\,(p^B\a_B) 
+ (-q_AD^{AB}- \frac{p\cdot q}{6D} p^B) \a_B \\
\label{m2_charge_left_right}
&=& q_- \a_- + q_+^a \a_{+,a} \;,\; a = 1,\dotsi, h^{1,1}(X)-1\\ \notag
q^2 &=& D^{AB}q_A q_B= q_+^2 - q_-^2\;.
\eea
Notice that we have separated the charges, which can now be thought of the charges under the \(U(1)\) gauge fields on the world-sheet, into the right-moving (\(q_-\)) and left-moving (\(q_+\)) parts. The way to separate them, namely the choice of a vector inside the Grassmannian \(O(h^{1,1}-1,1)/O(1)\times O(h^{1,1})\), is independent of the moduli of the divisor \({\cal P}\). In general it will depend on the K\"ahler moduli of the Calabi-Yau but for convenience we have fixed it to be at the attractor value given by the total M5 charges. 

Finally we will look at the fermions. From the decomposition of the spin bundle (\ref{fermions_reduction}), we see that the spinor which is (projectively) covariantly constant along \({\cal P}\) is independent of the moduli of \({\cal P}\) and therefore belongs to the universal factor of the sigma model.

Putting everything together, we conclude that the field content of the universal sigma model we are interested in includes three left- and right-moving scalars from the non-compact direction,
(\(h^{1,1}-1,1\)) compact scalars from reducing the chiral two-form fields, and four right-moving fermionic zero modes. Especially the right-moving degrees of freedom form a \({\cal N}=4\) scalar multiplet, which we will sometimes refer to as the ``universal" hypermultiplet. We can therefore interpret this universal part of the CFT as a \((0,4)\) heterotic sigma model, with the target space  being \(\R^3\times S^1\) and with the left-moving Narain model of gauge group \(\big(U(1)\big)^{h^{1,1}-2}\).

The presence of the four fermion zero modes, which can be thought of as the Goldstinos of the broken supersymmetry, makes the criterion for a state to be supersymmetric somewhat more involved. To see this we have to know more about the superconformal algebra of the right-moving (supersymmetric) side of the CFT. Now with the presence of the universal scalar multiplet, the superconformal algebra is not just the ``small" \({\cal N}=4\) SCA written down in (\ref{n_4_superconformal_algebra}) but is enlarged into the \({\cal A}_{\k,\inf}\)  SCA \cite{Maldacena:1999bp,Gaiotto:2006wm,Sevrin:1988ew,Hasiewicz:1989vp}, where the $SU(2)$ level $\k$ is given by the central charge by $\k= c_R/6$. Especially, in the zero-momenta (\(\vec{\p}=0\)) sector  for the center of mass degrees of freedom in \(\R^3\), we have on top of (\ref{n_4_superconformal_algebra}) an extra piece of superconformal algabra which includes
\bea\notag
\lbrack J_n^i,Q_r^{\a+} \rbrack &=&\s^i_{\a\b}\, Q_{r+n}^{\b+} 
\quad,
\lbrack J_n^i,Q_r^{\a-} \rbrack = - Q_{r+n}^{\b-}\,\s^i_{\b\a}\\
\notag
\lbrack U_n,U_m \rbrack &=& n \d_{n+m,0}\\
\notag
\lbrack U_n, G_r^{\pm\pm} \rbrack &=& n Q_{n+r}^{\pm\pm}\\
\notag
\{Q^{\a+}_r, Q^{\b-}_s\} &=& \d^{\a\b}\d_{r+s,0}\quad\text{and other  }
\{Q,Q\}=0 \\ \label{middle_SCA}
\{Q^{\a+}_r, G^{\b-}_s\} &=& \d^{\a\b}U_{r+s}\quad\text{and other  }
\{Q,G\}=0\;,
\eea
where \(Q^{\pm\pm}(\zbar)\) are now the fermionic currents and \(U(\zbar)\) is the \(U(1)\) current corresponding to the right-moving part of the M2 charge (\ref{m2_charge_left_right}).  Notice that, here and in the following discussion we write all the operators without the tilde's for the readability of the formulas , despite of the fact that  they are the right-movers. We hope that the fact that only the right-moving sector is supersymmetric will prevent any possible confusion.

In the sector in which all states have not only \(\vec{\p}=0\) but also \(U_0|0\rangle =  q_- |0\rangle =0\), the \(G\)'s and the \(Q\)'s decouple and the states that preserve unbroken supersymmetries are those annihilated by \(G^{\pm\pm}_0\) as well as by all positive modes. This is in direct analogy with the Ramond ground states we discussed in (\ref{R_ground_state}), except for now the four Goldstino's \(Q^{\a\b}\) act non-trivially as two pairs of creation and annihilation operators and produce a short multiplet with four BPS states. It's easy to check that this short \({\cal N}=4\) multiplet does not contribute to the \({\cal N}=2\) elliptic genus defined in (\ref{def_elliptic_genus}) but does contribute to the modified version of it when we insert the factor \(F^2=({J}_0^3)^2\) in the trace, where \({J}^3\) is the \(U(1)\) current of the R-symmetry group \(SU(2)_R\). We will therefore insert this factor when we later define the generalised elliptic genus for the present theory. 

Now we turn to the more interesting cases with non-vanishing right-moving M2 charge \(q_-\). If we ``bosonize" the \(U(1)\) current and write \(U(\zbar) = i \bar{\pa} \varphi\), then such a state can be thought of as being created from a state with  \(U_0|0\rangle =0\) by adding a vertex operator
\be
 |q_-\rangle = e^{iq_{-} \varphi}|0\rangle\;. 
\ee
Then using the \(\lbrack U,G\rbrack\) commutator we see that the supersymmetry condition gets modified into
\be
0=e^{iq_{-} \varphi} G_0^{\pm\pm}|0\rangle
= \big( G_0^{\pm\pm}- q_-\,Q_0^{\pm\pm}\big)  |q_-\rangle \;,
\ee
while the other four combination of \(G_0^{\pm\pm}\) and \(Q_0^{\pm\pm}\) generate the four-component short multiplet which contributes to the modified elliptic genus. 

Another consequence of the modification of the supersymmetry condition by the presence of the right-moving charges is a change of the value of the conformal weight of the BPS states. Instead of (\ref{R_ground_state})
\be
\frac{1}{4} \d^{\a\b} \{G_0^{\a+},G_0^{\b-}\} |0\rangle = (\til{L}_0 - \frac{c_R}{24})  |0\rangle 
= 0\;,
\ee
we now have 
\be\label{L0_R}
\frac{1}{4} \d^{\a\b} \{G_0^{\a+}-qQ_0^{\a+},G_0^{\b-}-qQ_0^{\b-}\} |q_-\rangle = (\til{L}_0 - \frac{c_R}{24} - \frac{1}{2} q_-^2)  |q_-\rangle 
= 0\;,
\ee
where 
\be
q_-^2 = \int_{\cal P} q_-\a_- \wedge q_-\a_-  = \frac{(p\cdot q)^2}{6D}\;.
\ee
The above relation between the conformal weight  and the right-moving charges will be important when we discuss the modular properties of the modified elliptic genus later.

In the general cases in which we also have charges \(\vec{\p}\neq 0\) in the non-compact \(\R^3\) directions, there are terms other than those listed in (\ref{middle_SCA}) in the \({\cal A}_{\k,\inf}\) superconformal algebra which will play a role. They can be found in, for example, \cite{Hasiewicz:1989vp,Sevrin:1988ew}. Incorporating these extra charges and repeating exactly the same analysis as above, we conclude that for a BPS state with right-moving charge \(q_-, \vec{\p}\)  the following relation is satisfied
\be
(\til{L}_0 - \frac{c_R}{24} - \frac{1}{2} q_-^2 - \frac{1}{2} \vec{\p}^2)  |q_-,\vec{\p}\rangle_R=0 \;.
\ee
Similarly there is also a \(\frac{1}{2} \vec{\p}^2\) contribution to the \(L_0\) eigenvalue. 

The spectral flow relations (\ref{qflow2}) (\ref{q0flow2}) are
implemented in the (0,4) CFT as a symmetry of the superconformal
algebra. It is given by
\begin{align} \notag\label{spectral_flow_SCA_L}
\til{L}_n &\rightarrow \til{L}_n + k_- U_n + \frac{1}{2} k_-^2\,\d_{n,0} \\
U_n &\rightarrow U_n + k_- \d_{n,0}
\\
\intertext{for the bosonic part of the right-moving side, and }\notag
{L}_n &\rightarrow L_n + k_+^a A_{n,a} + \frac{1}{2} k_+^2 \,\d_{n,0}\\
\label{spectral_flow_SCA_R}
A_{n,a} &\rightarrow A_{n,a} + k_+^b \,D_{ab}\,\d_{n,0}\quad,\;a,b=1,\dotsi,h^{1,1}(X)-1
\end{align}
for the left-moving side.
In the above formulae, the projection of the large gauge transformation parameter \(k\)  into the left- and right-moving component  is again given by  (\ref{m2_charge_left_right}), with \(q_A\) now replaced by \(D_{AB}k^B\), and the metric 
\be
\int_{\cal P} \a_{+,a} \wedge \star \a_{+,b} = -  
\int_{\cal P} \a_{+,a} \wedge \a_{+,b}  = D_{ab}
\ee
is given by the restriction of \(D_{AB}\) onto the hypersurface in 
\(H^2(X)\) orthogonal to the K\"ahler form. 
Note that these transformations leave
$L_0-\frac{1}{2}A_0^2$ and $\til{L}_0-\frac{1}{2}U_0^2$
invariant.

\section{A Generalised Elliptic Genus}

The generating function of BPS index for a fixed M5-brane
charge $p^A$ can be identified with a generalised elliptic genus
of the CFT (see also
\cite{Denef:2007vg,Gaiotto:2006wm,Kraus:2006nb}). More precisely, we
want to compute the partition function
\be \label{part}
Z'_{{\cal P}}(\t,y)= \Tr_R\left[ F^2 (-1)^F e^{\pi ip\cdot q}
e^{2\pi i \tau \left( L_0-\frac{c_L}{24}\right)}e^{-2\pi i \taubar
\left( \til{L}_0-\frac{c_R}{24}\right)} e^{2\pi i y\cdot q} \right]
\;,
\ee
where \(y\in \L \otimes \R\) can be thought of as being the ``potential" for the M2 charges \(q\), $F={J}^3_0$  and the \(F^2\) insertion is needed to absorb the four leftover right-moving fermionic zero modes as we explained in section \ref{The Universal Sigma Model}. 

The purpose of this subsection is to discuss the following three properties of the generalised elliptic genus that we will need in order to give the microstates contributing to this index a gravitational interpretation.

\subsection{The Modified Fermion Number}
First we will explain the necessity for the extra \(e^{\pi ip\cdot q}\) phase insertion. The presence of it is closely connected to the presence of the Freed-Witten anomaly we discussed before. In particular, both effects are absent if \([{\cal P}] = c_1({\cal L}) = p^A \a_A\) is even. We will understand it in terms of the consistency of the conformal theory OPE, while other explanations can be found in \cite{Denef:2007vg,Belov:2006jd}. Later we will also see explicitly that the inclusion of this factor is crucial for the modular properties of the generalised elliptic genus. 

Note that the shift (\ref{qshift}) only affects the right-moving part of the charges \(q_-\), we will therefore concentrate our analysis on them. 
Consider the spectral flow vertex operators $e^{i k_- \varphi}$ acting on the state
$| q_-\rangle$, from the OPE between $e^{i k_- \varphi(\bar{z}')}$ and $e^{i q_- \varphi(\bar{z})}$
one has
\begin{equation} \label{aux9}
e^{i k_- \varphi(\zbar)} \left.| q_- \right>=\zbar^{(k\cdot q)_-}\left.|q_-\!+\!k_-\right>\;.
\end{equation}
The OPE will pick up a phase $\exp\left(2\pi i (k\cdot q)_-
\right)=\exp(2\pi i k\cdot q
)=(-1)^{k\cdot p}$ when $z$ circles around the origin.
Locality of the OPE requires projection onto the states with even
$k\cdot p$, which explains why the elliptic genus needs to contain
a factor $(-1)^{p\cdot k}$ for it to be modular invariant. For
convenience we include however a factor $e^{\pi i p\cdot q}=e^{\pi
i p\cdot\mu}e^{\pi i p\cdot k}$ in our definition of the elliptic
genus, where \(\m\) and \(k\) are given by the unique decomposition of the M2 charge vector (\ref{decomposition_m2_charge}).  The term $(p\cdot\mu)$ could be interpreted as an additional
overall phase or as a fractional contribution to the fermion
number.

\subsection{The Modular Properties}

Eventually our aim is to show that the partition function has an
asymptotic expansion in terms of semi-classical saddle-points of
the three-dimensional supergravity theory. A crucial ingredient for this to work is the property of
$Z'_{{\cal P}}(\t,y)$ that it is a modular
form. To be more precise, it is a modular form of weight $(0,2)$. To
show this, let us introduce a generalised partition function \be
\label{partW} W_{{\cal P}}(\t,y, z)= \Tr_R\left[ e^{2\pi i z
F} e^{\pi i p \cdot q} e^{2\pi i \tau
    \left( L_0-\frac{c_L}{24}\right)}e^{-2\pi i \taubar \left(
    \til{L}_0-\frac{c_R}{24}\right)} e^{2\pi i y\cdot q} \right] \ee

\noindent such that
$$
Z'_{{\cal P}}(\t,y)=-{1\over 4\pi^2} \partial_z^2
W_{{\cal P}}(\t,y, z)|_{z=1/2}\;.
$$
The function $W_{{\cal P}}(\t,y, z)$ can be
thought of as a generalised partition function of the (0,4) CFT on a torus with  Wilson lines parametrized by $y$
and $z$. It should be independent of the choice of cycles
on the torus, and hence be of weight \((0,0)\) under the modular
transformations
$$
\tau\to {a\tau+b\over c\tau+d},\quad \taubar\to {a\taubar+b\over
c\taubar+d}, \quad y_+\to \frac{y_+}{c\tau+d}, \quad
y_-\to \frac{y_-}{c\taubar+d}, \quad
z\to\frac{z}{c\taubar+d}\,. 
$$
In the above formula \(y_+, y_-\) are the Wilson line parameters coupling to the left- and right-moving charges respectively and are again given by the projection of the vector \(y \in \L \otimes \R \) into the positive- and negative-definite part as in (\ref{m2_charge_left_right}).  This together with the fact that \(\pa_z\) has weight one 
proves that $Z'_{p^A}(\t,y^A)$ has weight $(0,2)$.

As mentioned above, the partition function
$Z'_{{\cal P}}(\t,y)$ contains a continuous degeneracy
in the BPS states due to the zero-modes in the $\mathcal{N}=(0,4)$ ``universal" multiplet, which can be thought of the momenta in 
the $\mathbb R^3$ part
of the $S^1 \times \mathbb R^3$ target space of the universal sigma model. Macroscopically they correspond to the center of the mass degrees of freedom of the M5 brane which decouple from the rest of the degrees of freedom. We wish to extract this degeneracy by defining
\be \label{mod5}
Z'_{{\cal P}}(\tau,y) = Z_{{\cal P}}(\tau,y)\,\int d^3 \vec{\p} \,(e^{2\pi i \tau} e^{
-2\pi i \bar{\tau}} )^{\frac{1}{2} \vec{\p}^2}\,
\ee
where $Z_{{\cal P}}(\t,y)$ can be thought of as the trace among the BPS states with charges \(\vec{\p}=0\) . The Gaussian integral is
proportional to ${\rm Im}(\tau)^{-3/2}$ and therefore has weight
$\left(\frac{3}{2},\frac{3}{2}\right)$. We can therefore conclude that
$Z_{{\cal P}}(\tau,\bar{\tau},y)$ has weight $\left(-\frac{3}{2},\frac{1}{2}\right)$.

\subsection{The Theta-Function Decomposition}

The spectral flow symmetry discussed in section \ref{Wrapped branes on Calabi-Yau and the spectral flow}, or equivalently the isomorphism of the superconformal algebra discussed in (\ref{spectral_flow_SCA_L}) and (\ref{spectral_flow_SCA_R}), implies the presence of extra structures in the generalised elliptic genus. As we shall see, similar to the case of the elliptic genus of a \((2,2)\) CFT (\ref{(2,2)_elliptic_genus_decomposition}), these structures are most manifest when we write the generalised elliptic genus in a decomposed form in terms of the theta functions. 

First of all, since we have argued that the elliptic genus \(Z_{{\cal P}}(\t,y)\) only receives contribution from BPS states with \(\vec{\p}=0\), using (\ref{L0_R})  we can rewrite it as
\be\label{elliptic_version2}
Z_{{\cal P}}(\t,y)=  \,\Tr_R\left[ F^2 (-1)^F \,
{\bf e} [{\t}  (L_0 -\frac{c_L}{24}) - \frac{\tbar}{2}  q_-^2+ (q|y+\frac{p}{2})] \,\right]
\ee
in the shorthand notation introduced in section \ref{Elliptic Genus and Vector-Valued Modular Forms}. Recall that the bilinear \((|): (\L\otimes \R) \times (\L\otimes \R) \to \R\) , first defined in (\ref{bilinear_lattice_def}), is given by \(-\int_{{\cal P}} *\wedge *\) and of Lorentzian signature.
A direct consequence of this is that
the partition function depends on $\taubar$ in a specific way, namely the anti-holomorphic part is entirely  captured by the
``heat equation''
\be\label{heateq}
\left\lbrack \partial_{\taubar} + {1\over 4\pi i  }
\pa_{y_-}^2\right\rbrack
Z_{{\cal P}}(\tau,\taubar,y) =0\;. 
\ee
In particular, this implies that the anti-holomorphic part of 
\(Z_{{\cal P}}(\t,y)\) contains redundant information which is already encoded in the Wilson line \((y)\)-dependence of it.
It will turn out that by decomposing the elliptic genus in terms of the theta-functions we can indeed isolate the holomorphic factor factor which is all we need to determine the degeneracies of BPS states.   

Secondly, we are interested in the BPS degeneracies \(d_{{\cal P}}(q_A,q_0)\), but on the other hand the spectral flow symmetry implies a relation among those degeneracies for different charges \((q_A,q_0)\). A generating function for BPS degeneracies in terms of spectral flow invariant combinations of charges is therefore desirable. It will turn out that such generating functions are exactly the coefficients \(h_\m(\t)\) of the generalised elliptic genus \(Z_{{\cal P}}(\t,y)\) in its theta-function decomposition.

To begin let us write the type IIA D0 brane charge, proportional to the five- (or three-)dimensional angular momentum along the M-theory circle, as
\ben
q_0 &=& (L_0 - \frac{c_L}{24}) - (\til{L}_0 - \frac{c_R}{24}) \\
&=& {Q}_0 + \frac{1}{2}q^2 -\frac{\chi}{24}\;,
\een
where we have used \(c_L = \chi({\cal P})\). From the CFT point of view, \(\frac{1}{2}q^2\) is the contribution of the universal factor \(U(1)\) currents to the \(S^1\) momentum, and central charge contribution corresponds to the ground state energy of \(AdS_3\), while \({Q}_0\) is the contribution from the bosonic zero-modes of the part of the CFT which is not ``universal", together with the contribution from the left-moving excitations,. On the other hand, from the type IIA point of view this is also a natural split. Considering the D2 brane charges as fluxes on the D4 brane we recognise the second term as induced by the \(ch({\cal F})\) factor in the anomalous brane coupling (\ref{anomalous_brane}), while the third one induced by the A-roof genus curvature factor, leaving \({Q}_0\) being the number of pointlike D0 branes together with the contribution from the part of the world-volume flux which does not correspond to conserved D2 brane charges.

In terms of the spectral flow invariant combination \(\qhat\) (\ref{qhat}), we have
\be
L_0 -\frac{c_L}{24} = \qhat + \frac{1}{2}q_+^2\;.
\ee

Using this and the decomposition of the M2 charges (\ref{decomposition_m2_charge}), we can again rewrite the generalised elliptic genus (\ref{elliptic_version2}) as
\begin{align}\notag
&Z_{{\cal P}}(\t,y)=  \\ \notag
&\Tr_R\left[ F^2 (-1)^F \,
{\bf e}[\t \qhat]\, {\bf e}[\frac{\t}{2}  (\m+k+\frac{p}{2})_+^2- \frac{\tbar}{2}  (\m+k+\frac{p}{2})_-^2+ (\m+k+\frac{p}{2}|y+\frac{p}{2})] \,\right]\\
&\qquad\m \in \L^*/\L\quad,\quad k \in \L\;.
\end{align}

The statement of spectral flow that each \(k\in \L\) gives the same contribution to \(Z_{{\cal P}}(\t,y)\), can be now translated into the following statement
\begin{align}\label{decomposetheta}
Z_{{\cal P}}(\t,y)&= \sum_{\m \in \L^*/\L} \,\Theta_\m(\t,y) \,h_\m(\t)\\ \notag
\intertext{in which} \notag
h_\m(\t) &= \Tr_{q=\m+p/2}\left[ F^2 (-1)^F \,
{\bf e}[\t \qhat]\,\right] = \sum_{\qhat+\frac{c_L}{24}\geq 0} \, d_\m(\qhat)\,e^{2\p i \t \qhat}\\ \notag
\Theta_\m(\t,y)&= \sum_{q\in \m+\L+\frac{p}{2}} \, 
{\bf e} [\frac{\t}{2}q_+^2-\frac{\tbar}{2}q_-^2+(q|y+\frac{p}{2})]\\
&= {\bf e}[\frac{p}{4}(y+\frac{p}{2})]\,\th_{\m} (\t;-y-\frac{p}{2},\frac{p}{2})\;,
\end{align}
where  \(\th_\m(\t;\a,\b)\) is the Siegel theta function we introduced in (\ref{Siegel theta function}).

As promised, now we see that the entire information about BPS degeneracies contained in the generalised elliptic genus is encapsulated in the holomorphic modular forms \(h_\m(\t)\), whose Fourier coefficient \(d_\m(\qhat)\) depends only on the spectral invariant combination \((\m,\qhat)\) of the charges \((q_0,q_A)\). 

In order to evaluate the saddle-point contribution to \(h_\m(\t)\) we will need to know its modular property. The modular transformation of \(\Theta_\m(\t,y)\) can be computed by Poisson resummation, which has been performed in (\ref{modular_property_Siegel_theta}) and gives
\bea
\Theta_\m(-\frac{1}{\t},\frac{y}{\t}) = 
\frac{1}{\sqrt{|\L^*/\L|}}\,(\sqrt{-i\t})^{h^{1,1}-1}\, (\sqrt{i\tbar})\,\times\\ 
{\bf e}[\frac{1}{2\t}y_+^2 -\frac{1}{2\tbar}y_-^2]\,{\bf e}[-\frac{p^2}{4} ]\,
\sum_{\n \in \L^*/\L}\,{\bf e}[-(\m|\n)]\, \Theta_\n(\t,y)
\;.\eea 

While the extra exponential factors involving \({\bf e}[\frac{y^2}{2\t}]\) are expected as given by the modular transformation of the generalised elliptic genus in analogy with the elliptic genus case (\ref{elliptic_3}), the \(\tau\)-prefactors together with the knowledge that \(Z_{\cal P}(\t,y)\) has weight \((-\frac{3}{2},\frac{1}{2})\) shows that \(\{h_\m(\t)\}\) transforms as a vector-valued modular form of weight \(-\frac{h^{1,1}+2}{2}\). See section \ref{Elliptic Genus and Vector-Valued Modular Forms} for a small introduction of these the vector-valued modular forms. More explicitly, it transforms as 
\be\label{S_transform_h}
h_\m(-\frac{1}{\t}) =\sqrt{|\L^*/\L|}\,(\sqrt{-i\t})^{-(h^{1,1}+2)}\,
 {\bf e}[\frac{p^2}{4} ]\,
\sum_{\n \in \L^*/\L}\,{\bf e}[(\m|\n)]\, h_\n(\t,y)\;,
\ee
where \(\sqrt{|\L^*/\L|} = |{\text Vol}(\L)|\) is the volume of a unit cell of the lattice \(\L\).

The T-transformation of the theta-functions can also be computed. Recall that the lattice \(\L\) is not necessarily even and this makes the computation a bit more involved than usual. Representing a lattice vector \(k\in \L\) as an integral two-cycle imbedded in the hypersurface \({\cal P}\), then the adjunction formula together with the Riemann-Roch theorem gives
\be
Q\cdot Q + Q\cdot {P} = 2g-2\;,
\ee
and it turn shows that \(k^2 - (p|k) = 0\;\;\text{mod}\;\;2\). A direct computation then shows
\be
\Theta_\m (\t+1,y)= {\bf e}[\frac{(\m+p/2)^2}{2}]
\Theta_\m (\t,y)\;, 
\ee
which implies the T-transformation for the vector-valued modular form \(h_\m(\t)\)
\be\label{T_transform_h}
h_\m(\t+1) =  {\bf e}[-\frac{(\m+p/2)^2}{2}]\,h_\m(\t)\;.
\ee

\subsection{The (Modern) Farey Tail Expansion}
\label{The (Modern) Farey Tail Expansion}
After decomposing the BPS-states-counting elliptic genus of the low-energy CFT into combinations of theta-functions, we are now ready to employ some important mathematical properties of our vector-valued modular functions \(h_\m(\t)\) in order to give the elliptic genus a spacetime interpretation. The treatment of this part of the story in the original paper \cite{deBoer:2006vg} of the present author is not completely correct in the most general cases and was later improved in the publication \cite{Manschot:2007ha}. We will refer to this paper and the PhD thesis of fellow student Jan Manschot for further details and simply summarise the results we need here. 

Suppose we have a weight \(w\) vector-valued modular form \(f_\m(\t)\), where \(\m\in \L^*/\L\) for some lattice \(\L\), which transforms under 
\(\G=PSL(2,\Z)\) modular transformation as
\be
f_\m(\frac{a\t+b}{c\t+d}) = (c\t+d)^w \,M(\g)_\m^{\;\;\n}f_\n(\t)\quad,\quad \g = \bem a&b\\ c&d \eem \in PSL(2,\Z)\;.
 \ee
Given the Fourier expansion 
\be
f_\m(\t) = \sum_{n\geq0} D_\m(n)\,q^{n-\D_\m}\;,
\ee
recall that the polar part of \(f_\m(\t)\) is given by (\ref{polar_part_intro})
\be
f_\m^-(\t) = \sum_{0\leq n\leq\D_\m} D_\m(n)\,q^{n-\D_\m}\;.
\ee
The special property which will be useful for us is that the full vector-valued modular form is determined by its polar part alone. 

This property will give us an expression for the BPS partition function which looks like, at the cartoon level and ignoring regularisation,
\be\label{farey_naive}
Z_{{\cal P}}(\t,y) \sim \sum_{\G_\inf \backslash \G}
Z^-_{{\cal P}}(\frac{a\t+b}{c\t+d},\frac{y}{c\t+d})\;,
\ee
where the coset \(\G_\inf \backslash \G\) denotes \(\t\sim \t+1\), 
with each term in the sum lending itself to a natural interpretation in terms of the dual gravitational theory.

The full expression including all the details, however, is substantially more involved. It reads
\begin{align}\notag
f_\m(\t) &= \frac{1}{2} D_\m(\D_\m) + \frac{1}{2} \sum_{0\leq n\leq\D_\m} \lim_{K\to \inf}\sum_{(\G_\inf \backslash \G)_K}  \Bigg(\\ \notag
&  (c\t+d)^{-w}\, 
{\bf e}[\frac{a\t+b}{c\t+d}(n-\D_\n)] \,R_w\bigg(\frac{2\p i |n-\D_\n|}{c(c\t+d)}\bigg)\\ \label{huge}
&\times M^{-1}(\g)_\m^{\;\;\n} \,D_\n(n)\,\Bigg)\;,\end{align}
where 
\be
R_w(x) = \frac{1}{\G(1-w)}\,\int^x_0\, e^{-z}\,z^{-w}\,dz\ee
and 
\be
\sum_{(\G_\inf \backslash \G)_K} = \sum_{\substack{|c|,|d| \leq K \\ (c,d)=1}}\;.
\ee
Please see \cite{Manschot:2007ha} and the PhD thesis of Jan Manschot for the derivation of the above formula. 

Now we can apply this formula on our character \(h_\m(\t)\) of the elliptic genus, where the transition matrix \(M(\g)\) can be read off from the T- and S-transformation of \(h_\m(\t)\) that we calculated in (\ref{T_transform_h}) and (\ref{S_transform_h}). From this procedure and combining again with the appropriate theta-functions we obtain an expression for the generalised elliptic genus \(Z_{{\cal P}}(\t,y)\) which is in its spirit given by (\ref{farey_naive}). In the following section we will give this expansion a physical interpretation in terms of geometries.

\section{Spacetime Interpretation of the Attractor Farey Tail}
\label{Spacetime Interpretation of the Attractor Farey Tail}
\setcounter{equation}{0}

So far the generalised  Rademacher formula appears to be just a mathematical
result. What makes it interesting is that it has a very natural
interpretation from the point of view of a dual gravitational
theory. In this section we discuss the interpretation of the Farey tail expansion first in
terms of the effective supergravity action, and
subsequently from an M-theory/string theory perspective.
We will first discuss the gravitational interpretation of the
general formula presented in section \ref{The (Modern) Farey Tail Expansion}, and then turn to the present
case of the \({\cal N}=2\) black holes.

\subsection{Gravitational Interpretation of the Generalised Rademacher formula}

Microscopic systems described by a 2d CFT have a dual description
in terms of a string- or M-theory on a space that contains $AdS_3$
as the non-compact directions.  This is because $AdS_3$ is the
unique space whose isometry group is identical to the 2d conformal
group. The miracle of AdS/CFT is that the dual theory contains
gravity, which suggests that the partition function of the 2d CFT
somehow must have an interpretation as a sum over geometries in the classical limit. The
full dual theory is defined on a space that is 10- or
11-dimensional, but only the three directions of $AdS_3$
are non-compact. Hence, by performing a
dimensional reduction along the compact directions we find that
the dual theory can be represented as a (super-)gravity theory on
$AdS_3$. The effective action therefore contains the  Einstein
action for the 3d metric
$$
S_E={1\over 16\pi G_3} \int_{AdS}\!\!\! \sqrt{g} (R-\frac{2}{\ell^2})+{1\over 8\pi
G_3} \int_{\del (AdS)}\!\!\! \sqrt{h} (K-\frac{1}{\ell})
$$
where we have included the Gibbons-Hawking boundary term. Here
$\ell$ represents the AdS-radius. According to the AdS/CFT
dictionary, the 3d Newton constant $G_3$ is related to the central
charge $c$ of the CFT by \cite{Henningson:1998ey} \be {3\ell\over 2G_3}= c\;. \label{Gc} \ee
The dictionary also states that the partition function $Z(\tau)$
of the CFT is equal to that of the dual gravitational theory on (a quotient of)
$AdS_3$, whose boundary geometry coincides with the 2d torus on
which the CFT is defined. The shape of the torus is kept fixed and
 parametrized by the modular parameter $\tau$.

The rules of quantum gravity tell us to sum over all possible
geometries with the same asymptotic boundary conditions. For the
case at hand, this means that we have to sum over all possible three dimensional
geometries with the torus as the asymptotic boundary. Semi-classically,
these geometries satisfy the equations of the motion of the
supergravity theory, and hence are locally $AdS_3$. There
indeed exists an Euclidean three geometry with constant curvature
which has $T^2$ with modular parameter \(\tau\) as its boundary. It is the BTZ black hole, which
is described by the Euclidean line element
$$
ds^2 = N^2(r) dt_E^2
+\ell^{-2}N^{-2}(r)dr^2+r^2(d\phi+N_\phi(r)dt_E)^2
$$
with
$$
N^2(r)={(r^2-\tau_2^2)(r^2+\tau_1^2)\over r^2},\qquad\quad
N_\phi(r)={\tau_1\tau_2\over r^2}\;.
$$
Here $\tau=\tau_1+i\tau_2$ is the modular parameter of the
boundary torus. Using (\ref{Gc}), one can compute the Euclidean action of this solution
and obtain \cite{Maldacena:1998bw}
$$
S=-{\pi c\over6 } {\rm Im}{1\over \tau}\;.
$$

For the present purpose of counting BPS states, one needs to
consider extremal BTZ black holes. With the Minkowski signature
this means that its mass and angular momentum are equal. After
analytic continuation to a Euclidean complexified geometry, one
finds that the action has become complex and equals $i\pi {c\over
12}\tau$.

Note that a torus with modular parameter $\tau$ is equivalent to a
torus with parameter $a\tau+b\over c\tau+d$, since they differ
only by a relabelling of the $A$- and $B$-cycles. But the
Euclidean BTZ solution labelled by $a\tau+b\over c\tau+d$ in
general differs from the one labelled by $\tau$, with the
difference being that these three-dimensional geometries fill up
the boundary torus in distinct ways. Namely, for the above BTZ
solution the torus is filled in such a way that its $A$-cycle is
contractible. After a modular transformation, this would become
the $\gamma(A)=cA+dB$ cycle. In fact, the BTZ black hole is
related to thermal $AdS_3$ with metric
\be
ds^2 =(r^2+\ell^2)dt_E^2+{dr^2\over r^2+\ell^2} +r^2d\phi^2
\ee
after interchanging the $A$- and $B$-cycles and with $t_E$ and \(\phi\) periodically identified as
$$
t_E\equiv t_E+ 2\pi n \tau_2\qquad, \qquad \phi\equiv \phi+2\pi n\tau_1\;.
$$
In this case the $B$-cycle is non-contractible, while the $A$-cycle is now contractible.
Notice that the metric is manifestly invariant under $\tau\to\tau+1$ gives
the same geometry. The Euclidean action for this geometry is $S=i\pi {c\over 12}\tau$.


\begin{figure}[htb!]
\centering%
\includegraphics[height=3.5cm]{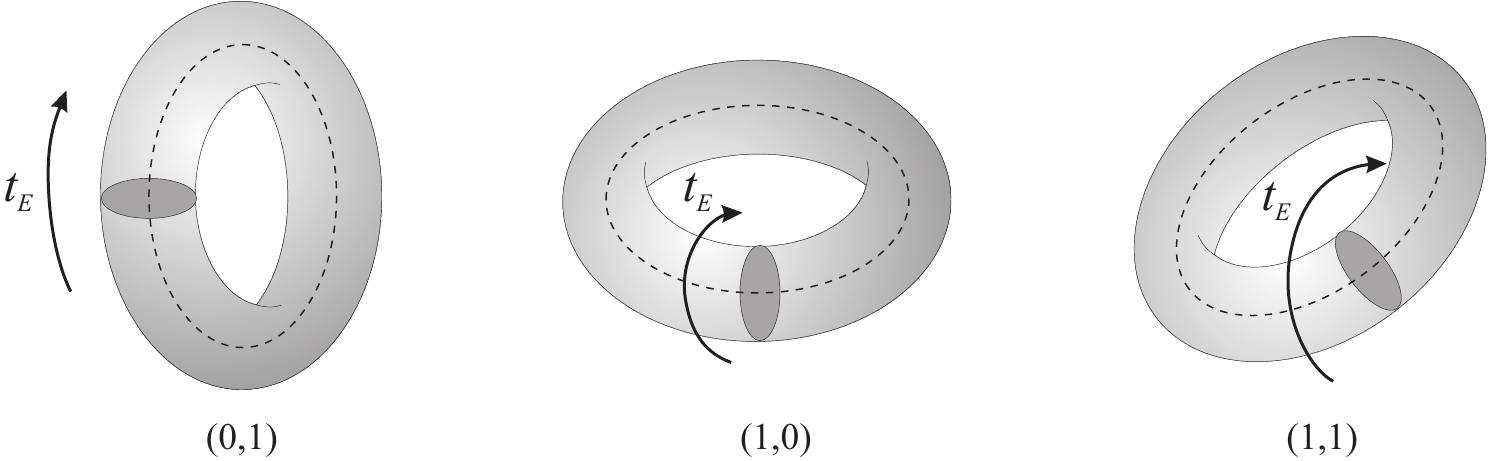}
\caption{\small{(i) Thermal \(AdS_3\), with the B-cycle being
contractible. (ii) BTZ black hole, with the A-cycle being
contractible. (iii) The geometry with the (A+B)-cycle being
contractible.}} \label{fig:Farey}
\end{figure}

The classical geometries with a given boundary torus with modular
parameter $\tau$ can now be obtained from either the thermal AdS
background or the BTZ background by modular transformations. For
definiteness let us take the thermal AdS as our reference point,
so that the classical action for the geometry obtained by acting
with an element $\gamma$ of the modular group is
$$
S=i\pi{c\over 12}\left({a\tau+b\over c\tau+d}\right)\;.
$$
One easily recognizes that these solutions precisely give all  the
leading contributions in the Farey tail expansion corresponding to
${\bf e}[-\frac{c_L}{24}\,\frac{a\t+b}{c\t+d}]$ factor in (\ref{huge}). In fact, these terms occur with
a multiplicity one since they represent the vacuum of the SCFT.
The other terms in the expansion should then be regarded as
dressing the Euclidean background with certain contributions that
change the energy of the vacuum. These contributions were already given an interpretation as coming from
virtual particles that circle around the
non-contractible cycle in the previous work on black hole Farey tails \cite{Dijkgraaf:2000fq}. In fact, in the next section we will give
a further justification of this interpretation for the case of the
attractor black holes. Specifically, by using the arguments of
Gaiotto et al.\cite{Gaiotto:2006ns}, we find that the subleading
contributions are due to a gas of wrapped M2-branes which carry
quantum numbers corresponding to the charges and the spin in the
$AdS_3$ geometry. The truncation to the polar terms can in turn be
interpreted as imposing the restriction that the gas of particles
are not heavy enough yet to form a black hole. The $AdS_3$
geometry carries a certain negative energy which allows a certain
amount of particles to be present without causing gravitational
collapse. However, when the energy surpasses a certain bound then
a black hole will form through a Hawking-Page transition. In the
case of the $AdS_3/CFT_2$ correspondence such an interpretation
was first proposed by Martinec \cite{Martinec:1998wm}. 

Recall that the way  we obtain the generalised Rademacher formula for the five-brane CFT is essentially to apply it (\ref{huge}) on the characters \(h_\m(\t)\), while the theta-functions are then combined with the resulting expression and are therefore present in each thermal or BTZ background.
The origin of the presence of the theta
functions lies in the chiral bosons on the world-sheet of the
string reduction of the M5-brane. From a spacetime perspective,
on the other hand, the theta functions arise due to the presence of gauge fields and in particular the presence of the Chern-Simons term in the action, as discussed
in section \ref{Wrapped branes on Calabi-Yau and the spectral flow}. In fact, the partition function of a spacetime
effective theory that includes precisely such Chern-Simons terms
in addition to the usual Yang-Mills action was analyzed in detail
by Gukov et al.  in \cite{Gukov:2004id}. These authors showed that
the partition function indeed decomposes in to a sum of
Siegel-Narain theta functions. There the  $\taubar$-dependence arises
because one of the gauge field components is treated differently
from the others, to ensure that the partition function indeed
converges.  In this part of the thesis we will
not give further details of this calculation. A more recent discussion
in which parts of this calculation were carefully worked out is
\cite{Kraus:2006nb}.

\subsection{Wrapped M2-branes}

In the previous subsection, we have interpreted the sum over modular orbits of the most polar term as the sum over gravitational background,  and the appearance of the theta functions as the effect of the spectral flow symmetry induced by the gauge Chern-Simons term in the action. 
In this subsection we would like to further give a spacetime interpretation to the rest of the polar terms 
\be\label{polar_terms_MSW}
\sum_{\qhat<0} \, d_\m(\qhat)\,{\bf e}[\qhat \frac{a\t+b}{c\t+d}]\,R_w\bigg(\frac{2\p i |n-\D_\n|}{c(c\t+d)}\bigg)
\ee
in (\ref{huge}) as a dilute gas of wrapped M2 branes. 
 In particular one would like to
give a more detailed accounting of those polar states from the point of
view of string theory on CY$\times S^2 \times\! AdS_3$.  In fact, a
nice physical picture of a large class of these states was given
in \cite{Gaiotto:2006ns} in terms of M2 and anti-M2 branes which fill up
Landau levels near the north and south pole of the $S^2$
respectively. In a dilute gas approximation,
their macroscopic computation gives rise to the following contribution to the elliptic genus $Z$:
\be
\label{sugra1} Z_{\rm gas}(\tau,y) = {\bf e}[-\frac{c_L}{24}]\,
Z_{\rm sugra}(\tau) Z_{GV}(\tau, {\textstyle{\frac{1}{2}}} p \tau
+ y) Z_{GV}(\tau, {\textstyle{\frac{1}{2}}} p \tau - y)\;.
\ee

Let's now explain the different factors in the above formula. First of all, the presence of the factor \({\bf e}[-\frac{c_L}{24}]\) is due to the fact that the supergravity partition function should be computed with the NS boundary condition, since the \(AdS_3\) circle is contractible. Second, \(Z_{\rm sugra}(\tau)\) denotes the contribution of the supergravity modes, which is basically given by the MacMahon function (\ref{macmahon}) as \cite{Gaiotto:2006ns,Kraus:2006nb,Kraus:2006wn}
\be
Z_{\rm sugra}(\tau)  = M(q)^{-\chi}
\ee
where \(\chi\) is the Euler number (\ref{euler_number}) of the Calabi-Yau manifold in question. Finally, the two (reduced) Gopakumar-Vafa partition functions (\ref{Z_GV}) account for the contribution from the wrapped M2 and anti-M2 charges, since it's known that the Gopakumar-Vafa invariants \(\a_g^q\) counts wrapped membranes with charge \(q\in H_2(X,\Z)\) when the non-compact directions are flat five-dimensional Minkowski space \cite{GV1,GV2}. Using the relationship between the Gopakumar-Vafa partition function and topological string partition function discussed in section \ref{Topological String Theory}, together with the modular invariance of the elliptic genus \(Z_{\cal P}(\t,y)\), the expression (\ref{sugra1}) provides a justification of the OSV conjecture (\ref{OSV1}) when the gas approximation is to be trusted \cite{Gaiotto:2006ns}.

As is clear from the form of the Gopakumar-Vafa partition function (\ref{Z_GV}), what it counts are non-interacting wrapped membranes. 
 From
the point of view of the world-sheet SCFT, the M2/anti-M2 brane
gas describes a collection of states that is freely generated by a
collection of chiral vertex operators. It is clear that
(\ref{sugra1}) suffers from various kinds of limitations. The dilute
gas approximation will eventually break down, there could be other
BPS configurations that contribute, the Landau levels can start to
fill out the entire $S^2$, the $SU(2)$ quantum numbers are bounded
by the level of the $SU(2)$ current algebra, etc. Furthermore, it
also does not exhibit the right behavior under spectral flow.

Therefore, the gas expression (\ref{sugra1}) is really only valid in the limit in which the five-branes charges are large and membranes charges are small. In this limit, the factor 
\be
d_\m(\qhat)\,{\bf e}[\qhat \frac{a\t+b}{c\t+d}]\quad,\quad \qhat <0 
\ee
can be thought of as counting the degeneracies of a gas of charged particles in the appropriate thermal \(AdS_3\) or BTZ background, which are made of membranes wrapping internal cycles and are not heavy enough to form a black hole. The latter statement can be seen directly from the expression for the black hole entropy (\ref{M5_entropy}).

Finally, the extra factor \(R_w\bigg(\frac{2\p i |n-\D_\n|}{c(c\t+d)}\bigg)
\) in (\ref{polar_terms_MSW}) should be thought of as a regularising  factor for the gravitational path integral. Again we refer to the PhD thesis of fellow student Jan Manschot for further details.

\section{Summary and Conclusion}
\label{Summary and Conclusion}
In this part of the thesis we present the Farey tail expansion for \({\cal N}=2\)
D4-D2-D0 black holes. The central idea of this expansion is to first truncate the partition function so that it
includes only particular low excitation states, and then sum over all
images of it under the modular group. Each term can be interpreted
as representing the contribution of a particular (semi-)classical
background. The formula can thus be regarded as partly microscopic
(as the states counted in the "tail") as well as macroscopic (as
the sum over classical backgrounds). We would like to emphasize
that in this expansion, there is no one to one correspondence
between microstates and gravitational backgrounds and the
major part of the entropy is carried by one particular black hole
background.

The supergravity interpretation of the Farey expansion involves a
natural complete collection of backgrounds of a given type. It is
natural to ask whether the expansion can be refined by including
more general macroscopic backgrounds. It is indeed likely that
such refinements exist, but one expects that their contribution will follow a
similar pattern: one has to truncate the microscopic spectrum even
further and replace the contribution of the omitted states by
certain classical backgrounds. Here one can think of various type
of backgrounds, such as multi-centered solutions, bubbling solutions that deform the horizon geometry,
black rings...etc.  A large class of such solutions is known, but
the list is presumably incomplete, and it remains an interesting
problem to use them in a systematic  manner.

\part{${\cal N}=4$ Dyons}
\setcounter{equation}{0}
In this last part of the thesis we focus on one single theory, namely the heterotic string compactified on six-torus, or equivalently the type II string theory compactified on \(K3\times T^2\). An introduction of the basic properties of K3 manifolds and the basic features of the low-energy effective theory of the IIA/\(K3\times T^2\) compactification can be found in chapter \ref{K3 Compactification}.

There are two chapters in this part of the thesis. In the first one we discuss the microscopic degeneracies of the BPS states of this theory. In section \ref{Microscopic Degeneracies} we review the counting of \(1/\!2\)-BPS and \(1/\!4\)-BPS states using various duality frames and in particular the derivation of the dyon-counting formula. This counting formula turns out to have various seemingly unrelated mathematical properties. In section \ref{The Counting Formula and a Borcherds-Kac-Moody Algebra} we will review them using the theme of a generalised (Borcherds-) Kac-Moody algebra as the connecting point of these different properties. 

After introducing the mathematical background we need, in chapter \ref{Counting the Dying Dyons} we will present our study of the BPS spectrum of the theory. This chapter is based on the publications \cite{Cheng:2007ch,to_appear}.
First we address the issue of the moduli dependence of the spectrum. The main tool used here is the \({\cal N}=4\), \(d=4\) supergravity effective theory and the stability condition of certain 
multi-centered solutions in this theory. Second we study the ambiguity of choosing a contour of integration, when one attempts to retrieve the actual BPS degeneracies from its generating function. In section \ref{The Contour Prescription and its Interpretation} we show that the contour-dependence of the dyon degeneracies is related to its moduli-dependence, and show how an appropriate choice of contour can incorporate the moduli-dependence of the spectrum into the counting formula. 

After that we turn to the role of the Borcherds-Kac-Moody algebra in the BPS spectrum of the theory. 
First we argue that  the Weyl group of the algebra plays the role of the group of a discretised version of attractor flows of the theory, with the walls of marginal stability identified with the walls of the Weyl chambers. We then comment on some arithmetic properties of this discrete attractor flow. Second we argue that, with an appropriate 
prescription for choosing the simple roots and the highest weight, 
 the counting formula is related to a certain character formula for the Verma module of the algebra. We then see how this correspondence between a choice of charge and moduli with a representation of the Borcherds-Kac-Moody algebra provides some microscopic understanding of the BPS states of the theory, in particular how the wall-crossing formula can be re-derived from this microscopic picture.

In this part of the thesis we will focus on the cases which are relatively well-understood, namely
 we will assume that the total charges of the states, given by two vectors \(P\) and \(Q\) in the 28-dimensional charge lattice 
(\ref{dyon_charge_lattice_P_Q}), satisfies the following ``co-prime condition" 
\be\label{co_prime_condition}
\text{g.c.d}(P^a Q^b - P^b Q^a) = 1 \quad,\quad a,b = 1,\dotsi,28\;,
\ee
which ensures that the degeneracies are completely determined by the set of three T-duality invariants (\(P^2/2, Q^2/2, P\cdot Q\)). Please see \cite{Maldacena:1999bp,Banerjee:2008ri,Banerjee:2008pv,Banerjee:2008pu,Dabholkar:2008zy} for discussions about the cases in which the above condition is not satisfied.

\chapter{Microscopic Degeneracies and a Counting Formula}
\label{Microscopic Degeneracies and a Counting Formula}

Our main reason for being interested in the \({\cal N}=4,\,d=4\) string theory compactification introduced in chapter \ref{K3 Compactification}
lies in the possibility of studying the supersymmetric states in great details. In particular, the presence of many supersymmetries and a long chain of dualities relating different corners of moduli space make possible a microscopic understanding of the supersymmetric spectrum of the theory, and this is something that cannot be said for a generic \({\cal N}=2,\,d=4\) string theory. 

In this chapter we will review the microscopic counting of BPS states in the present theory. In section \ref{Microscopic Degeneracies} we recall the microscopic origin of the \(1/\!2\)- and  \(1/\!4\)-BPS states, and in particular we will see how a microscopic counting formula for dyonic states can be derived using the known D1-D5-P degeneracies. In section \ref{The Counting Formula and a Borcherds-Kac-Moody Algebra} we review various mathematical properties of this counting formula, which are connected to each other by their relations to a certain Borcherds- (or generalised-) Kac-Moody algebra. These properties will be important for our physical discussion in the next chapter.

\section{Microscopic Degeneracies}
\label{Microscopic Degeneracies}

In this section we will discuss the microscopic counting of the \(1/\!2\)- and  \(1/\!4\)-BPS states of the theory, exploiting the chain of dualities introduced in section \ref{Heterotic String Dualities}.

\subsection{$1/\!2$- and $1/\!4$-BPS Solutions}

The central charge in the \({\mathcal{N}}=4\) supersymmetry algebra can be written as
\be \label{central_charge_N4}
\hat{Z}=\frac{1}{\sqrt{\l_2}}(P_L - \l Q_L)^m\Gamma_m\quad,\quad m=1,..,6\;,
\ee
where $\l=\l_1+i\l_2$ is the complex scalar which is a part of the \({\cal N}=4, d=4\) supergravity multiplet introduced in (\ref{subsec_N4d4_sugrav}). In the heterotic frame it is the 
usual axion-dilaton field while in the IIA/\(K3\times T^2\) frame the K\"ahler moduli of the torus, in the IIB/\(K3\times T^2\) frame the complex moduli of the torus. And  \(P_L\), \(Q_L\) denote the moduli-dependent left-moving charges given in (\ref{left_moving_charge_dyon}). Here and from now on all the moduli fields should be understood as being evaluated at spatial infinity.

As mentioned in (\ref{2_bps_N=4}), there are two BPS bounds in \({\cal N}=4, d=4 \) supersymmetry algebra. Indeed, from 
$$
\hat{Z}^\dagger \hat{Z}=\frac{1}{\l_2}| P_L- \l Q_L|^2\,{\mathds {1}} -2i  P_L^m Q_L^n \Gamma_{mn}\;
 $$
 and 
 the fact that the operator \( i P_L^m Q_L^n \Gamma_{mn}\) satisfies
 \be\label{wedge_expression}
 (i P_L^m Q_L^n \Gamma_{mn})^{2} =   |P_L\wedge Q_L|^2 \equiv Q_L^2 \,P_L^2 - (Q_L\cdot P_L)^2\;,
 \ee
one concludes that $\hat{Z}^\dagger \hat{Z}$ has the following two eigenvalues
\bea \label{mass_dyon_formula}
 |Z_{P,Q}|^2&=& {1\over \l_2}| P_L-\l Q_L|^2+2|P_L\wedge Q_L|\\  
 \nonumber\text{and            }\;\;\;\;\;\;\;
|Z'_{P,Q}|^2& =&{1\over \l_2}| P_L-\l Q_L|^2-2|P_L\wedge Q_L|\;.
\eea
Therefore the \(1/\!4\)-BPS states of the theory satisfy 
$$
M_{P,Q} =  |Z_{P,Q}| > |Z'_{P,Q}|\;,
$$
while states that preserve half of the supersymmetries must have
\be\label{half_BPS_condition}
|P_L\wedge Q_L|=0  \quad \Leftrightarrow \quad P\parallel Q \;.
\ee

\subsection{Microscopic Degeneracies of $1/\!2$-BPS States}

Let's begin with the microscopic counting of states which preserve half of the supersymmetries. From the supersymmetry algebra we have seen that the electric and magnetic charges have to be parallel to each other (\ref{half_BPS_condition}). Together with the co-prime condition (\ref{co_prime_condition}) this means that we can always find the S-duality transformation such that the charges are purely magnetic, namely now we can put \(Q=0\) without loss of generality. The microscopic degeneracy therefore becomes a function of only one T-duality invariant \(P^2/2\).

In other words, to count these \(1/\!2\)-BPS states it is enough to count the perturbative heterotic string states, for example the momentum and winding modes along the internal six-torus listed in Table \ref{N=4_brane_dualities}. Recall that the right-moving sector of the heterotic string theory, which is the same as the open bosonic string theory, has non-vanishing zero point energy level \(-1\). The mass shell condition and the level matching condition of the heterotic string therefore read
\index{level matching condition}
\be
m^2 = N_L + \frac{1}{2} P_L^2 = N_R-1 +\frac{1}{2} P_R^2 \;.
\ee
Furthermore, supersymmetry requires the supersymmetric (left-moving) side of the string to be at its ground state, namely \(N_L=0\). Combining these we conclude the right-moving oscillator number is given in terms of the charges as
\be\label{oscilation_level_het}
N_R = 1 + \frac{1}{2} \,(P_L^2 - P_R^2 )  =1+\frac{1}{2}P^2\;.
\ee
Recall that \(\frac{1}{2}P^2 \in \Z\) because the charge lattice \(\G^{6,22}\) is even and self-dual (unimodular).

There are 24 bosonic oscillators in the right-moving sector, which can be understood as the 24 bosonic oscillators in the light-cone quantisation of the bosonic string theory, which implies that the generating function of the degeneracies of the above \(1/\!2\)-BPS states is
$$
\sum_{\frac{P^2}{2}\in \Z} \,d(P)\,q^{1+\frac{P^2}{2}} =\prod_{n=1} ^{\inf}\left( \frac{1}{1-q^n} \right)^{24}\;,
$$
or equivalently
\be\label{half_BPS}
d(P)= \oint d\s \, \frac{e^{-\p i P^2 \s}}{\eta^{24}(\s)}\,,
\ee
where \(\eta(\s)\) is the Dedekind eta-function
$$
\eta(\s) = q^{1/24}\prod_{n=1} ^{\inf} (1-q^n)\quad,\quad q= e^{2\p i \s}\;.
$$

The \(1/\!2\)-BPS states in this context are sometimes called the Dabholkar-Harvey states \cite{Dabholkar:1989jt}. 
Notice that from the modular transformation of the eta-function one can see that the asymptotic growth of degeneracy is
\be
\log d(P) \sim \sqrt{P^2/2} 
\ee
and scales linearly with the charges, which is slower than the quadratic growth one expects for the Bekenstein-Hawking entropy of a four-dimensional black hole. In this sense we say that \(1/\!2\)-BPS states are ``small" and form ``small" black holes. 

\index{Dedekind eta-function} 
\index{Dabholkar-Harvey states}
\subsection{Microscopic Degeneracies of $1/\!4$-BPS States}
\label{section_quartBPS_counting}

More interesting and more complicated are the dyonic states, meaning states with both magnetic and electric charges non-vanishing in all duality frames and therefore must preserve only four of the sixteen supercharges, as can be seen from the supersymmetry algebra (\ref{half_BPS_condition}). In this part of the section, following \cite{David:2006yn,Shih:2005uc}, we will derive a microscopic formula counting these \(1/4\)-BPS states, which will be play an important role in the following chapters of the thesis.

Under the assumption (\ref{co_prime_condition}) that the degeneracies depend only on the three quadratic invariants (\(P^2/2, Q^2/2, P\cdot Q\)), it is enough to understand the degeneracies of the states with charges highlighted in Table \ref{N=4_brane_dualities}. Namely, let's now consider states with the following charges in the type IIB frame: \(Q_1\) D1 and \(Q_5\) D5 strings with \(k\) units of momenta along the first circle, together with a Taub-NUT along the third circle and \(\til{k}\) units of momenta along that direction, assuming that the size of the third circle is large compared to the rest of the internal directions\footnote{Notice that we use \(Q_1\) and \(Q_5\) to denote the components of the charge vector corresponding to D1 and D5 branes respectively, but not the actual number of branes wrapped. The relevant subtlety here is that there is also the geometrically induced D1 charge when a D5 brane is wrapped around the K3.}.

One can immediately  work out the three invariants for these charges 
\ben
P^2 &=& 2Q_1 Q_5 \\
Q^2 &=& 2 k\\
P\cdot Q &=& \til{k}\;,
\een
and see that for given \(P^2/2, Q^2/2, P\cdot Q\) we can always find the corresponding D1, D5 charges, and momenta \(Q_1,Q_5, k, \til{k}\).

The advantage of studying this relatively simple  system is that its microscopic description is relatively well understood, namely the D1-D5-P system in five dimensions. 
Going back to the Table \ref{N=4_brane_dualities}, let's first decompactify the circle \(S_{(3)}^1\), meaning that we take the limit that the circle is very large in the four-dimensional Planck unit and the theory becomes the five-dimensional theory obtained by compactifying type IIB string theory on \(S_{(1)}^1\times K3\). In the five-dimensional description, the KK monopole becomes a Taub-NUT space (\ref{TN}) with the \(S_{(3)}^1\)-direction being the direction of the circle fibration.
Secondly, we assume that the internal circle \(S_{(1)}^1\) is much larger than the size of the \(K3\) manifold, then it was proposed that the six-dimensional world-volume theory of the D5-D1 branes is reduced to a two-dimensional supersymmetric sigma model, whose target space is the symmetric product of \(\frac{P^2}{2}+1\) copies of \(K3\) \cite{Breckenridge:1996is,Strominger:1996sh,Vafa:1995zh}. Furthermore, there are decoupled modes which are present even when \(P^2=0\), corresponding to the closed string modes localised at the tip of the Taub-NUT that may also carry momenta along the internal circle, and the center of mass modes of the D1-D5 system, which may carry momenta along the internal and the Taub-NUT circle. In other words, we can break the theory into three separated parts \cite{David:2006yn,Dabholkar:2008zy}
\be
\S(\text{\small TN}_1) \times \S({\text{\small C.O.M.}}) \times \S(S^{Q_1Q_5+1} K3)\;.
\ee
From Table \ref{N=4_brane_dualities} we can see that the first part can be dualized to a perturbative heterotic string system and is therefore again counted by the partition function
\be
\frac{1}{\eta^{24}(\s)} = \frac{1}{q}\prod_{m\geq 1} \frac{1}{(1-q^m)^{24}}
\;.\ee
The contribution of the second part is computed to be \cite{David:2006yn,Dabholkar:2008zy}
\be
\frac{1}{(y^{1/2}-y^{-1/2})^2}\, \prod_{m\geq 1} \frac{(1-q^m)^4}{(1-q^m y)^2(1-q^m y^{-1})^2}\;.
\ee

Now let's look at the third factor of the CFT. 
Recall that the K3 elliptic genus has the following Fourier expansion
\begin{align}\nonumber
\chi(\t,\n) &= \Tr_{\text{\tiny{RR}}}\,(-1)^F \, e^{2\p i \n J_0}\, e^{2\p i \t (L_0-\frac{c}{24})}\, e^{-2\p i \tbar (\til{L}_0-\frac{c}{24})}\\
&= \sum_{n\in\Z_+, \ell \in \Z}
c(4n-\ell^2) q^n y^\ell\;,
\end{align}
with \(c(-1)=2, c(0)=20\). 

The elliptic genus of the symmetric products of \(K3\) has the following generating function given in terms of the Fourier coefficients of the K3 elliptic genus \cite{Dijkgraaf:1996xw}
\be
\sum_{N\geq0} p^N \chi(S^NK3;q,y) = \prod_{n>0,m\geq 0, \ell}\left(
\frac{1}{1-p^nq^my^\ell}\right)^{c(4nm-\ell^2)}\;,
\ee
and this is the contribution from the symmetric product part of the CFT. 
Identifying the CFT and the spacetime data as 
\begin{align}\notag
k &= \frac{Q^2}{2} = L_0 -\bar{L}_0 = L_0 -\frac{c}{24}
= \text{\small momenta along internal circle}\\ \notag
\til{k} &= P\cdot Q = J_0 =  \text{\small momenta along TN circle}
\end{align}
and combining the three factors, we conclude that the generating function for the degeneracies of the \(1/\!4\)-BPS states is
\begin{align} \nonumber
&\sum_{P,Q}\,(-1)^{P\cdot Q+1}\,D(P,Q) \,e^{\p i (P^2 \r + Q^2 \s + 2 P\cdot Q \n)}
= \frac{1}{pqy}  \prod_{(n,m, \ell)>0}\left(
\frac{1}{1-p^nq^my^\ell}\right)^{c(4nm-\ell^2)}\\ 
\label{DVV_1}
&p=e^{2\p i \r}\quad,\quad q=e^{2\p i \s}\quad,\quad y=e^{2\p i \n}\quad\end{align}
and \((n,m,\ell)>0\) means \(n,m\geq 0,\; \ell \in \Z\) but \(\ell<0 \) when \(n=m=0\). 

In particular, the above formula has been shown \cite{Dijkgraaf:1996it} to reproduce the asymptotic growth which agrees with the macroscopic black hole entropy \cite{Cvetic:1995uj,Cvetic:1995bj}
\be\label{dyon_bh_entropy}
S(P,Q) = \p \sqrt{P^2 Q^2 - (P\cdot Q)^2} \;.
\ee

This is the dyon counting formula, sometimes referred to as the Dijkgraaf-Verlinde-Verlinde formula, conjectured more than ten years ago \cite{Dijkgraaf:1996it}. 

\section{The Counting Formula and a Borcherds-Kac-Moody Algebra}
\label{The Counting Formula and a Borcherds-Kac-Moody Algebra}
\setcounter{equation}{0}
The above dyon counting formula (\ref{DVV_1}) turns out to have many seemingly unrelated mathematical properties, such as being an automorphic form, having an infinite product expansion, and being the ``lift" of a modular form related to the elliptic genus of K3. For later use we will now review the relevant mathematical properties of the following object appearing at the right-hand side of (\ref{DVV_1})
\begin{align}\label{F_10}
&\F(\O) = pqy \prod_{(n,m, \ell)>0}\left(
{1-p^nq^my^\ell}\right)^{c(4nm-\ell^2)}\\ \nonumber
&\O = \bem\s & -\n \\ -\n & \r \eem\quad,\quad p=e^{2\p i \r}\,,\, q=e^{2\p i \s}\,,\, y=e^{2\p i \n}\;,
\end{align}
using the presence of a Borcherds-Kac-Moody algebra as the theme connecting these various properties.

\subsection{Dyons and the Weyl Group}
\label{Dyons and the Weyl Group}

In this subsection we will introduce a vector space of Lorentzian signature which appears naturally in the dyon-counting problem. In particular we consider the vector of quadratic invariants in this vector space, and define a basis for these ``charge vectors". This basis defines a Lorentzian lattice of signature \((2,1)\) and generates a group of reflection with respect to them. We then briefly argue the physical relevance of this group while leaving the details for later sections.

In the above formula (\ref{F_10}) we have written the inverse partition function \(\F\) as a function of a \(2\times 2\) symmetric complex matrix \(\O\). Indeed, anticipating the important role played by the S-duality group \(PSL(2,\Z)\) (\ref{S-duality_dyons}), it will turn out to be convenient to introduce a space of \(2 \times 2\) symmetric matrices with real entries
\be\label{space_of_matrices}
 \left\{X\Big\lvert \;X= \bem x_{11} & x_{12} \\ x_{21} & x_{22} \eem,\; X=X^T\;, x_{ab} \in \R\right\}\;.
\ee

Besides the matrix \(\O\), the left-hand side of the counting formula (\ref{DVV_1}) involves another matrix \(\L_{P,Q}\) constituted of the three T-duality invariants \((P^2, Q^2, P\cdot Q)\)
\be\label{matrix_charge_vector}
\L_{P,Q} = \bem P\cdot P & P\cdot Q\\ P\cdot Q & Q\cdot Q \eem\;.
\ee
Since this is the vector of invariants of charges that determine the counting of states, in the following we will often refer to this vector \(\L_{P,Q}\)  also as the ``charge vector".

From the expression for the Bekenstein-Hawking entropy  (\ref{dyon_bh_entropy})
for a \(1/\!4\)-BPS dyonic black hole, which is manifestly invariant under the S-duality group (\ref{S-duality_dyons}), we see that the vector \((P^2,Q^2,\; P\cdot Q)\) naturally lives in a space of Lorentzian signature (\(+,+,-\))  on which the S-duality group acts as a Lorentz group \(PSL(2,\Z)\sim SO^+(2,1;\Z)\), where the ``\(+\)" denotes the time-orientation preserving component of the group.

This motivates us to 
equip the vector space of \(2 \times 2\) symmetric real matrices (\ref{space_of_matrices}) with the following metric
\be
(X,Y) =  - \, \e^{ac}\,\e^{bd}\, x_{ab}\, y_{cd}\;= -\text{\small det}Y\, \Tr(XY^{-1}),
\ee
where \(\e\) is the usual epsilon symbol \(\e^{12} = -\e^{21} =1\) .

Especially, the norm of a vector is given by\footnote{The ``-2" factor here and in many places later is due to the fact that we choose the normalisation of the metric to be consistent with the familiar convention for Kac-Moody algebras that the length squared of a real simple root is 2.}
\be\label{metric_M2_space}
(X,X)= -2\, \text{\small det} X\;.
\ee
For later notational convenience we will also define 
$|X|^2 = -\frac{1}{2} (X,X)= \text{\small det} X$.

One can immediately see that this is indeed a vector space of signature (2,1), in which the diagonal entries of the \(2\times 2\) matrix play the role of the light-cone directions. In other words, the space of all real $2\times 2$ symmetric matrices defined above is nothing but the Lorentzian space $\R^{2,1}$.
As mentioned earlier, an element of the S-duality group \(PSL(2,\Z)\) acts as a Lorentz transformation on this space:
for any real matrix \(\g\) with determinant one, one can check that the following action 
\be\label{action_PGL2Z}
X \to \g(X) := \g X \g^T
\ee
is indeed a Lorentz transformation satisfying \(|X|^2 =| \g(X) |^2\) .

Using this metric, the entropy of a dyonic black hole (\ref{dyon_bh_entropy}) becomes nothing but given by the length of the charge vector \(\L_{P,Q}\) as
\be
S(P,Q) =\p|\L_{P,Q}|\;.
\ee

Similarly, the counting formula (\ref{DVV_1}) can now (at least formally) be rewritten as the following contour integral
\be\label{DVV_integral_1}
D(P,Q) = (-1)^{P\cdot Q+1}\oint_{{\cal C}} \, d\O \,\frac{e^{\p i (\L_{P,Q},\O) }}{\F(\O)}\;.
\ee

Next we would like to consider a basis for the charge vectors \(\L_{P,Q}\). From the fact that \(P^2, Q^2\) are both even, it is easy to check that for any dyonic charge which permits a black hole solution, namely for all \((P, Q)\) with \(S(P,Q)>0\), the charge vector \(\L_{P,Q}\) is an integral positive semi-definite linear combination of the following basis vectors
\be\label{3_simple_roots}
\a_1 =\left(\!\!\begin{array}{rrr} 0 &-1 \\ -1& 0 \end{array}\!\!\right)\quad,\quad
\a_2 =\bem 2 &1 \\ 1& 0 \eem\quad,\quad
\a_3 =\bem 0 &1 \\ 1& 2 \eem\,.
\ee
In other words, for all black hole dyonic charges we have
\be
\L_{P,Q} \in \G_+ :=\{\Z_+\a_1+\Z_+\a_2+\Z_+\a_3 \}\;.
\ee

As a side remark we note that there is another place where the positive part \(\G_+\) of the lattice generated by the three vectors \(\a_{1,2,3}\) appears naturally. Consider the integral vector 
\be
\a= \bem2n & \ell \\ \ell & 2m \eem\quad,\quad n,m,\ell \in \Z\;,
\ee
from the fact that the Fourier coefficients of the K3 elliptic genus satisfies \(c(k)=0\) for \(k<-1\) (\ref{K3_elliptic}), one can easily show that the microscopic partition function can be rewritten in the following suggestive form
\be\label{DVVPF_2}
\F(\O) = e^{-2\p i(\varrho,\O)}\,\prod_{\a\in \G_+}\,\Big(1-e^{-\p i (\a,\O)}\Big)^{c(|\a|^2)}\;,
\ee
where 
\be\label{weyl_vector_for_W}
\varrho = \frac{1}{2} \sum_{i=1}^3 \a_i 
\ee
is the Weyl vector corresponding to the above basis vectors, a name that will be justified later in section \ref{The Borcherds-Kac-Moody Superalgebra and the Denominator Formula}.

The matrix of the inner products of the above basis is
\be\label{cartan_1}
(\a_i,\a_j) = \left( \begin{array}{rrr}2&-2&-2\\ -2&2&-2\\ -2&-2&2\end{array}\right)\;.
\ee

We can now define in the Lorentzian vector space \(\R^{2,1}\) the group \(W\) generated by the reflection with respect to the spacelike vectors \(\a_{1,2,3}\):
\be\label{generator_weyl}
s_i : X \to X - 2\frac{(X,\a_i)}{(\a_i,\a_i)}\a_i\quad,\;\;i=1,2,3\;.
\ee
\index{Coxeter group}
\index{Weyl group}
This group turns out to be a hyperbolic Coxeter group with the Coxeter graph shown in Fig \ref{Coxeter_graph}. The definition and basic properties of Coxeter groups can be found in the Appendix \ref{Appendix: Properties of Coxeter Groups}. We will from now on refer to this group as the Weyl group, anticipating the role it plays in the Borcherds-Kac-Moody algebra discussed in the following sections. In particular, we will denote as \(\D_+^{re} \) the set of all positive roots of the Weyl group (\ref{posi_negi_roots})
\be\label{positive_real_roots}
\D_+^{re} =  \{\a = w (\a_i),w\in W, i =1,2,3 \} \,\cap \,\G_+ \;,
\ee
as it will turn out to be the set of all positive {\it real} ({\it i.e.} spacelike) roots of the Borcherds-Kac-Moody algebra discussed in section \ref{The Borcherds-Kac-Moody Superalgebra and the Denominator Formula}.
\index{positive root}
\index{real roots}
\index{simple roots}

\begin{figure}
\centering
\includegraphics[width=3cm]{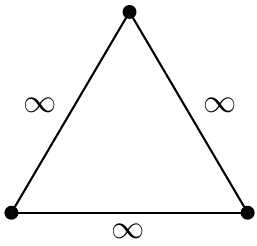} \setlength{\abovecaptionskip}{5pt}
\caption{\footnotesize{\label{Coxeter_graph}The Coxeter graph of the hyperbolic reflection group generated by (\ref{generator_weyl}). See (\ref{def_coxeter_grp}) for the definition of the Coxeter graph.}}
\setlength{\belowcaptionskip}{5pt}
\end{figure}


\index{extended S-duality group}
The physical relevance of this group can be intuitively understood in the following way. We have seen that the S-duality group \(PSL(2,\Z)\), which acts like (\ref{S-duality_dyons}), is a symmetry group of the theory. We can further extend this symmetry group with the spacetime parity reversal transformation
\be\label{parity_reversal}
\l \to - \bar{\l}\quad,\quad \bem P\\Q\eem \to \bem P\\-Q\eem
\ee
and thereby extend the group \(PSL(2,\Z)\) to \(PGL(2,\Z)\). 
From the point of view of the Lorentzian space \(\R^{2,1}\), the above element, 
when acting as (\ref{action_PGL2Z}),  
 augments the restricted (time-orientation preserving) Lorentz group with the spatial reflection.
Notice that the requirement that the inverse of an element \(\g \in PGL(2,\Z)\) is also an element implies \({\text{\small det}}\g=\pm 1\).
Explicitly, this group acts on the charges and the (heterotic) axion-dilaton as
\begin{align}\nonumber
&\bem P\\Q\eem \to \bem P_\g\\Q\g\eem:= \g \bem P\\Q\eem\quad,\quad \g = \bem a&b\\c&d \eem \in PGL(2,\Z)\\ 
\label{extended_S-dual}
&\l \to \l_\g := \frac{a\l+b}{c\l+d} \;\;\Big( \frac{a\bar{\l}+b}{c\bar{\l}+d}\Big)  \;\;\text{when}\;\; ad-bc=1 \;(-1)\;.
\end{align}

\index{Weyl chamber}
\index{fundamental Weyl chamber}

\index{dihedral group}
As we will prove now, this is nothing but the semi-direct product of the Weyl group \(W\) and the automorphism group of its fundamental domain (the fundamental Weyl chamber), which is in this case the dihedral group \(D_3\) that maps the regular triangle whose boundaries are orthogonal to \(\{\a_1,\a_2,\a_3\}\) to itself. Explicitly, the \(D_3\) is the group with six elements generated by the following two generators: the order two element corresponding to the reflection 
\be \label{dihedral_action_1}
\a_1 \to \a_1 \quad,\quad \a_2 \leftrightarrow \a_3
\ee
and order three element corresponding to the \(120^\circ\) rotation 
\be\label{dihedral_action_2}
\a_1 \to \a_2 \quad,\quad \a_2 \to \a_3 \quad,\quad \a_3 \to \a_1 
\ee
of the triangle. See Figure \ref{dihedral_grp}.

\begin{figure}
\centering
\includegraphics[width=10cm]{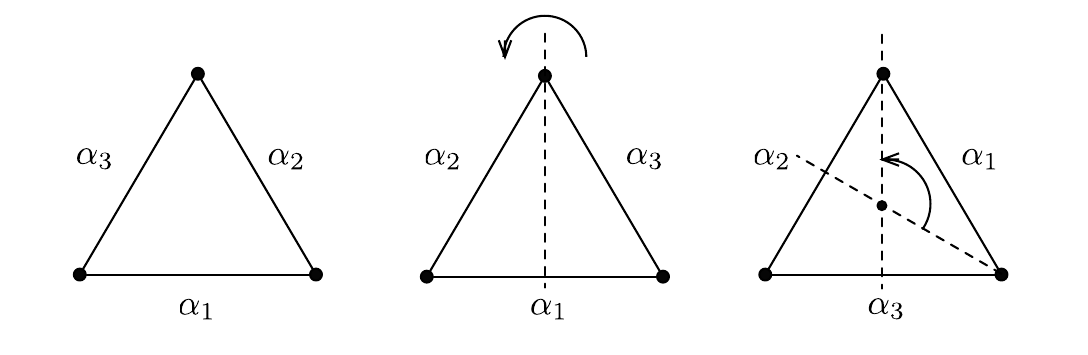} \setlength{\abovecaptionskip}{5pt}
\caption{\footnotesize{\label{dihedral_grp}The dihedral group \(D_3\), which is the symmetry group of an equilateral triangle, or the outer automorphism group of the real roots of the Borcherds-Kac-Moody algebra (the group of symmetry mod the Weyl group), is generated by an order two element corresponding to a reflection and an order three element corresponding to the \(120^\circ\) rotation.}}
\setlength{\belowcaptionskip}{5pt}
\end{figure}

Recall that the usual S-duality group \(PSL(2,\Z)\) is generated by the two elements \(S\) and \(T\), with the relation \(S^2=(ST)^3 =1\). In terms of \(2\times 2\) matrices, they are given by
$$
S =\bem0&-1\\ 1&0\eem\quad,\quad T=\bem1 & 1\\ 0&1\eem\;.
$$
The extended S-duality group \(PGL(2,\Z)\) is then generated by the above two generators, together with the other one corresponding to the parity reversal transformation (\ref{parity_reversal})
$$
R = \bem 1&0\\0&-1 \eem\;.
$$

On the other hand, in terms of these matrices and the \(PGL(2,\Z)\) action (\ref{action_PGL2Z}) on the vectors in the vector space \( \R^{2,1}\), one of the three generators of the Weyl group \(W\), corresponding to the reflection with respect to the simple root \(\a_1\), is given by
\be
s_1: X \to X - 2 \frac{(X,\a_1)}{(\a_1,\a_1)}\,\a_1 = R(X)\;.
\ee
For the dihedral group \(D_3\),
the reflection (\(\a_2\leftrightarrow \a_3\)) generator  is given by
\be
X\to RS(X)
\ee
and the order three \(120^\circ\) rotation generator is given by
\be
X\to TS(X)\;. 
\ee

From the expression for these three generators one can deduce the rest of the elements of \(W\). For example, the reflections \(s_2,s_3\) with respect to the other two simple roots \(\a_2\), \(\a_3\) are given by \(R\) conjugated by the appropriate power (1 and 2 respectively) of the rotation generator \(TS\).  

In particular, we have shown that the extended S-duality group can be written as
\be\label{sym_root_extended_S}
 PGL(2,\Z) \cong O^+(2,1;\Z)\cong W \rtimes D_3\;.
\ee
This means that the Weyl group is a normal subgroup of the group \( PGL(2,\Z)\), namely that the conjugation of a Weyl group element with any element of $PGL(2,\Z)$ is again a Weyl group element.

This relation between the symmetry of the root system of the present Weyl group and the physical phenomenon of crossing the walls of marginal stability will be further explored later in section \ref{Weyl Chambers and Discrete Attractor Flow Group}.

\subsection{K3 Elliptic Genus and the Siegel Modular Form}

In this subsection we will focus on the automorphic property of the microscopic partition function \(\F(\O)\). As we have discussed in the previous subsection, the theory has an extended S-duality group \(PGL(2,\Z)\), which acts naturally on the argument \(\O\) of the partition function as (\ref{action_PGL2Z}). We therefore expect \(\F(\O)\) to transform nicely under this group action. But it will turn out that this function has a much larger automorphic group \(Sp(2,\Z) \supset SL(2,\Z)\) under which it displays a nice transformation property. We will now motivate and explain the presence of this automorphic group from a mathematical point of view. The material covered here can be found in, for example, \cite{Borcherds1,Harvey:1995fq}
\index{Siegel upper-half plane}
\index{Siegel modular form}

As it stands in equation (\ref{F_10}), \(\F(\O)\) is a function of the \(2\times 2\) symmetric complex matrix \(\O\). But as \(c(n)\) grows with \(n\), it is clear that in order for the function to be convergent  \(\O\) should be restricted to lie in the Siegel upper-half plane, obtained by complexifying the vector space \(\R^{2,1}\) introduced before and taking only the future light-cone for the imaginary part
\begin{align}\nonumber
&\O \in \R^{2,1}+ iV^+\\ \label{future_light_cone_for_O}
&V^+ = \{X|\,(X,X) < 0 \,, \Tr X > 0\}\;.
\end{align}
See Figure \ref{lightcone}. In other words, \(\F(\O)\) should be considered as a function on the space \(\R^{2,1} + iV^+\). But there is another equivalent presentation of this space, namely the Grassmannian of a higher dimensional space
\be
\R^{2,1} + iV^+ = \frac{O(3,2)}{O(3)\times O(2)} = \{u\in \C^5, \langle u, u\rangle =0 , \langle u, \bar{u}\rangle < 0\}/(u\sim \C^* u)\;.
\ee
To see the second equivalence, simply observe that the real and imaginary part of \(u\) are indeed two mutually perpendicular timelike vectors which span a maximally timelike surface in the total space \(\R^{3,2}\), a phenomenon we have used in our discussion of the K3 moduli space in section \ref{Moduli Space of K3}.
 To see the first equivalence, separate the five-dimensional one \(\R^{3,2}\) into \(\R^{2,1}\oplus \R^{1,1}\) with the inner product
 $$
 \langle (X;z_+,z_-),   (X;z_+,z_-)\rangle = -2|X|^2 + 2 z_+ z_-\;,
 $$
then using the \(\C^*\) identification we obtain the following one-to-one mapping between a vector \( \O \in \R^{2,1}  + iV^+\)  and \(u \in \frac{O(3,2)}{O(3)\times O(2)}\): 
$$
\O\to u(\O) = (\O;1,|\O|^2)\;.
$$

From this point of view, \(\F(\O)\) is a function on the coset space \({O(3,2)}/{O(3)\!\times \!O(2)}\), and it is therefore not surprising that \(\F(\O)\) should have automorphic properties with respect to the automorphism group \(SO^+(3,2;\Z) \cong Sp(2,\Z)/\{\pm{\mathds   {1}}_4 \}\). For the explicit map between these two groups, see for example \cite{GN1}.
\index{Borcherds' lift}

To be more precise, what we have here is actually a special case of the following theorem of R. Borcherds (Theorem 10.1 of \cite{Borcherds1}).
\begin{theorem}
Let \(g(\t) = \sum \til{c}(4n) q^n \) be a meromorphic modular form with all poles at the cusps of weight \(-s/2\) for \(SL(2,\Z)\) with integer coefficients, with \(24\vert \til{c}(0)\) if \(s=0\). There is a unique vector \(\varrho\) in a Lorentzian lattice \(\G=\G^{s+1,1}\) such that 
\be\label{Borcherds_product}
F(\O) = e^{-\p i (\O,\varrho)} \prod_{\a\in\G_+} \big(1- e^{-\p i (\O,\a)} \big)^{\til{c}(|\a|^2)}\;
\ee 
is a meromorphic automorphic form of weight \(\til{c}(0)/2\) for \(O(s+2,2;\Z)\). 

Furthermore, define a rational quadratic divisor  to be the following zero locus
\be
\langle (\a;2r,2s), u(\O) \rangle = \langle (\a;2r,2s), (\O; 1, | \O |^2)\rangle =0 \quad,\quad r,s \in \Z
\ee
for 
$$
\langle (\a;2r,2s), (\a;2r,2s) \rangle > 0\;,
$$
then all the zeros and poles of \(F\) lie on the rational quadratic divisors with the multiplicities of the zeros being
$$
\sum_{n>0} \til{c}(-\frac{n^2}{2} \langle (\a;2r,2s), (\a;2r,2s) \rangle )\;.
$$
\end{theorem}

In some cases the above product formula is known to be the denominator formula of a certain Borcherds-Kac-Moody algebra. In this case the vector \(\varrho\) is the Weyl vector of the algebra.

\index{rational quadratic divisor }
To see how this theorem applies to our \(\F(\O)\), let's first recall a few facts about the elliptic genus of K3. As was discussed in section \ref{Elliptic Genus and Vector-Valued Modular Forms}, the elliptic genus has a theta-function decomposition given in (\ref{theta_decomposition1}). From the transformation properties of the theta-function we conclude that \(h_\m(\t)\)'s transform as modular forms of weight \(-1/2\).

For the K3 case, we know the full answer in terms of Eisenstein series 
(\ref{K3_elliptic}). Now consider the case when the modular form in the above theorem is given by \(h_\m/2\), where
\ben
 2\f_{0,1} (\t,z) &=& \chi_{K3} (\t,z)
 =\sum_{\m=0,1} h_{\m}(\s) \theta_\m(\s,z)\\
 &=& 2y^{-1} +20 + 2y + {\cal O}(q)
\;,\\\nonumber
h_{\m}(\s) &=& c(4n-\m^2) q^{n-\frac{\m^2}{4}} \\\nonumber
\theta_\m   &=& \sum_{\ell \in \Z} q^{(\ell+\frac{\m}{2})^2} \,y^{\m + 2\ell}\;.
\een
\index{Igusa cusp form}
By taking \(\til{c}(n)=\frac{1}{2}c(n)\) and comparing the result to (\ref{DVVPF_2}) we see that \(\F(\O) = (F_5(\O))^2\) is a weight 10 automorphic form for the group \(SO^+(3,2;\Z) \cong Sp(2,\Z)/\{\pm{\mathds   {1}}_4 \}\), which is also the modular group of a genus two Riemann surface. 
In other words, \(\F(\O)\) transforms as
\be \label{weight10_modular_form}
\F(\O) \to \big(\text{\small det}(C\O+D)\big)^{10}\,\F(\O) 
\ee
when 
$$
\O \to (A\O+B)(C\O+D)^{-1}\;,
$$
for \(2\times 2\) matrices of integral entries satisfying the symplectic condition
$$
AB^T = B^T A \quad,\quad CD^T = D^T C \quad,\quad AD^T -BC^T = {\mathds   {1}}_{2\times2}\;.
$$

In particular, this gives a product formula for the Igusa cusp form of weight 10, which is one of the five generators of the ring of genus two modular forms of all weights, which can also be written as 
\be\label{Igusa_cusp_product}
\F(\O)  = \prod_{(a,b) \text{ even}} \th_{a,b}^2(\O) = e^{-2\p i(\varrho,\O)}\,\prod_{\a\in \G_+}\,\Big(1-e^{-\p i (\a,\O)}\Big)^{c(|\a|^2)}
\ee
 where the product of the theta function is taken over all \(a,b \in (\Z/2\Z)^2\) with \(a^T b = 0 \text{  mod   }2\).

Furthermore, using the second part of the above theorem, we see that all the zeros of \(\F(\O)\) are of multiplicity two and lie on the rational quadratic divisor
$$
 \frac{1}{2}\langle (-\a;2r,2s), (\O; 1,|\O |^2)\rangle = 
  r(\r\s-\n^2) + n\r + m\s +\ell \n + s = 0  \;,
  $$
 with
 \be\label{RQdivisor}
\frac{1}{2}\langle (\a;2r,2s),(\a;2r,2s)\rangle = \ell^2 - 4nm + 4rs = 1\;\;,\;\; \ell,n,m,r,s \in \Z\;.
\ee
In the counting formula (\ref{DVV_integral_1}), these zeros lead to double poles in the integrand.

It is clear that all the above zero are related to each other by \(Sp(2,\Z)\) transformations. Indeed, when one identifies $\Omega$ with the period matrix of a genus two surface, the poles in $1/\Phi$ occur
precisely at those values of $\Omega$ at which the genus two surface degenerates into two disconnected genus one surfaces through the pinching of
a trivial homology cycle. These degenerations are labelled by elements of $Sp(2,\Z)/\{\pm{\mathds   {1}}_4 \}$ and
are characterized by the fact that the transformed period matrix is diagonal. From this consideration and from the knowledge that \((\O,\a_1)\) describes such a degeneration, 
we see that the location of the above rational quadratic divisor can also be written as
\be
\big( (A\O+B)(C\O+D)^{-1}, \a_1\big)=0\;.
\ee

For later use we will now review some properties of a class of  poles of the partition function $1/\F(\O)$. This property of the automorphic form $\F(\O)$ can be best understood as a special property of the Fourier coefficients of the K3 elliptic genus. Recall that when we put the parameter $z=0$, the elliptic genus reduces to the Witten index of the conformal field theory, which is given by the Euler number of the target manifold and equals $24$ in the present case of $K3$ manifold. In other words
\bea\notag
&&\chi(\t,z=0) = 24\\
\Rightarrow&&  \sum_{ \ell \in \Z}c(4n-\ell^2)  = 0 \quad\text{for all}\quad 
n>0
\eea

One can check that this property of the integers $c(4n-\ell^2)$ translates into the following expression for the pole of the automorphic form $\F(\O)$
$$
\frac{\F(\O)}{(y^{1/2}-y^{-1/2})^2}\Big\lvert_{y\to1} = \eta^{24}(\r) \, \eta^{24}(\s)\;. 
$$

Employing the invariance of $\F(\O)$ under the symmetry group $PGL(2,\Z)$ of the root system, this can be generalised into
\be\label{pole_look_like}
\frac{\F(\O)}{(e^{-\p i(\O,\a)/2} - e^{\p i(\O,\a)/2})^2}\Big\lvert_{(\O,\a)\to 0} =
\eta^{24}(-(\a^+,\O))\,\eta^{24}(-2(\a^-,\O))
\ee
which holds for arbitrary positive real root $\a$ and the associated $\a^\pm$ are defined later in (\ref{charge_split_new})\footnote{This equation might be thought of as a ``automorphized" version of the G\"ottsche's generating function of the Euler characteristic of the symmetric products \cite{Gottsche,Dijkgraaf:1996xw}
$$
\sum_{m=0}^\inf\, p^m \chi(\text{Sym}^mM) = \prod_{k=1}^\inf \left(\frac{1}{1-p^k}\right)^{\chi(M)}\;,
$$
where the manifold $M$ is the $K3$ manifold in this case.}.

As will be seen in section \ref{Contour Dependence of the Counting Formula}, this property makes possible 
that the wall-crossing phenomenon can be incorporated in the dyon-counting integral by a moduli-dependent contour prescription. 
In section \ref{Microscopic Derivation of the Wall-Crossing Formula} we show how it is possible to derive the wall-crossing formula microscopically using the above property of the denominator formula and the different representations of the BKM algebra at different point in the moduli space proposed in section \ref{Wall-Crossing and Representations of the Algebra}.

\subsection{The Borcherds-Kac-Moody Superalgebra and the Denominator Formula}
\label{The Borcherds-Kac-Moody Superalgebra and the Denominator Formula}

After the discussion in the last two sections about the related mathematical properties of the counting formula, now we are ready to see how it is associated with a Borcherds-Kac-Moody, or generalised Kac-Moody, superalgebra. A Borcherds-Kac-Moody superalgebra is a generalisation of the usual Lie algebra by the following facts: (i) the Cartan matrix is no longer positive-definite (``Kac-Moody"), (ii) there are also the so-called ``imaginary" simple roots with lightlike or timelike length (``Borcherds"), (iii) it is \(\Z_2\)-graded into the ``bosonic" and the ``fermionic" part (``super"). We will summarise the important properties of these algebras that we will use later. See \cite{Ray} or the appendix of \cite{GN1} for a more systematic treatment of the subject.

Consider a set \(I = \{1,2,\dotsi,n\} \) and a subset \(S\subset I\). A  generalised Cartan matrix is a real \(n\times n\) matrix \(A=(h_i,h_j)\) that satisfies the following properties
\begin{enumerate}
\item{either \(A_{ii}=2\) or \(A_{ii}\leq0\)  .}
\item{\(A_{ij}<0\) if \(i\neq j\)   , \(A_{ij}\in \Z\) if \(A_{ii}=2\)  .}
\index{Borcherds-Kac-Moody algebra}
\index{Cartan matrix (generalised)}
\index{imaginary roots}

Furthermore we will restrict our attention to the special case of BKM algebra without odd real simple roots, which means
\item{\(A_{ii}\leq 0\) if \(i\in S\)  .}
\end{enumerate}

Then the BKM superalgebra \({\mathfrak{g}}(A,S)\) is the Lie superalgebra with even generators \({\mathfrak{h}}\), \(e_i\), \(f_i\) with \(i \in I-S\) and odd generators  \(e_i\), \(f_i\) with \(i \in S\), satisfying the following defining relations
\begin{align}\nonumber
&[e_i,f_j] = \d_{ij} h_i\\ \nonumber
&[h,h']=0\\ \nonumber
&[h,e_i] = (h,h_i) \,e_i\quad,\quad [h,f_i] =- (h,h_i) \,f_i\\ \nonumber
&(\text{ad} e_i)^{1-A_{ij}}e_j = (\text{ad} f_i)^{1-A_{ij}}f_j = 0 \quad\text{if   } A_{ii}=2, i\neq j \\ \nonumber
& [e_i,e_j] =   [f_i,f_j] = 0\quad \text{if   } A_{ij}=0\;.
\end{align}
\index{root space}
\index{root}
\index{even root}
\index{odd root}
\index{Weyl group}
\index{Weyl vector}

Another important concept we need is that of the root space. For later use we have to introduce more terminologies. The {\it root lattice} \(\G\) is the lattice (the free Abelian group) generated by \(\a_i\), \(i \in I \), with a real bilinear form \((\a_i,\a_j) = A_{ij}\). The Lie superalgebra is graded by \(\G\) by letting \({\mathfrak{h}}\), \(e_i\), \(f_i\) have degree \(0\), \(\a_i\) and \(-\a_i\) respectively. Then a vector \(\a\in \G\) is called a {\it root} if there exists an element of \({\mathfrak{g}}\) with degree \(\a\). A root \(\a\) is called {\it simple} if \(\a=\a_i\), \(i \in I\), {\it real} if it's spacelike \( (\a,\a)  > 0\) and {\it imaginary} if otherwise. It is called {\it even (odd)} if the elements in  \({\mathfrak{g}}\) with degree \(\a\) are generated by the even (odd) generators, and {\it positive (negative)} if it is a positive- (negative-) semi-definite linear combinations of the simple roots. It can be shown that a root is either positive or negative, and either even or odd. Furthermore, the {\it Weyl group} \(W\) of  \({\mathfrak{g}}\) is the group generated by the reflection in \(\G\otimes \R\) with respect to all real simple roots. A Weyl vector \(\varrho\) is the vector with the property
$$
(\varrho,\a_i) = -\frac{1}{2} (\a_i,\a_i)
$$
for all simple real roots \(\a_i\). It is easy to see that, for the case discussed in section \ref{Dyons and the Weyl Group}, the vector (\ref{weyl_vector_for_W}) is indeed the Weyl vector satisfying the above condition.
\index{denominator formula}

Just as for ordinary Kac-Moody algebras, there is the following so-called denominator formula
\be\label{denominator_general}
e(-\varrho) \prod_{\a \in \D_+} \big( 1- e(-\a) \big)^{\text{mult} \a} = \sum_{w\in W}\,\e(w) w(e(-\varrho) \,S) \;,
\ee
where \(\D_+\) is the set of all positive roots, \(\e(w) = (-1)^{\ell(w)}\) where \(\ell(w)\) is the length of the word \(w\) in terms of the number of generators, defined in (\ref{length_function})\footnote{Warning: the conventions of the signs of the above formula, in particular the signs of the Weyl vector, do vary in the existing literature. }. There are differences between this product formula and the usual Weyl denominator formula due to, first of all, the fact that it's ``super". Concretely, we have used the following definition for the ``multiplicity" of roots {\small mult}\(\a\): 
$$
{\text{\small mult}}\a = {\text{\small dim}}{\mathfrak{g}}_\a  (-{\text{\small dim}}{\mathfrak{g}}_\a)\quad {\text{when}} \;\;\a \;\;\text{ is even (odd)}\;.
$$ 
Furthermore, there is a correction term \(S\) on the right-hand side of the formula due to the presence of the imaginary roots. The exact expression for \(S\) is rather complicated for generic BKM superalgebras and can be found in \cite{Ray,GN1}.

In the above formula, the \(e(\m)\)'s are formal exponentials satisfying the multiplication rule \(e(\m)e(\m')=e(\m+\m')\). Taking \(e(-\m) \to e^{-\p i (\m,\O)}\), the left-hand side becomes the product formula of (\ref{Borcherds_product}) with {\small mult}\(\a= \til{c}(|\a|^2)\).

Now we will concentrate on the case discussed in the last subsection. We have seen that (\ref{Igusa_cusp_product})
\be
\F(\O)^{1/2}  =  e^{-\p i(\varrho,\O)}\,\prod_{\a\in \D^+}\,\Big(1-e^{-\p i (\a,\O)}\Big)^{\frac{1}{2}c(|\a|^2)}\;.
\ee
From the transformation property of the above automorphic form under the Weyl group \(W\) introduced in (\ref{generator_weyl}), we can also rewrite it in a form as the right-hand side of (\ref{denominator_general}). From this equivalence one can therefore read out the set of even and odd imaginary simple roots and therefore construct a ``automorphic-form corrected" Borcherds-Kac-Moody superalgebra, whose denominator is the Siegel modular form \(\F(\O)^{1/2}\) of weight five and have 
\be
\text{{\small mult}}\a= \frac{1}{2}\,c(|\a|^2)\;,
\ee
where \(c(n)\) is the Fourier coefficient of the K3 elliptic genus.

By construction, the real simple roots can be chosen to be the basis of the vector of T-duality invariants appearing in the dyon counting formula (\ref{3_simple_roots}), while the Weyl group is the group generated by the three reflections with respect to them (\ref{generator_weyl}). In particular, the part of the generalised Cartan matrix corresponding to the real roots is given by (\ref{cartan_1}).
The above expression for the root multiplicity together with the property \(c(n)=0\) for \(n<-1\) is indeed consistent with the fact that all real roots have length \((\a,\a) = 2\). The set of all real positive roots is denoted as \(\D_+^{re}\) as was announced in (\ref{positive_real_roots}).

The fact that the dyons are counted with a generating function which is simply the square of the denominator formula of a generalized Kac-Moody algebra, and that the 
charge vectors naturally appear as elements of its root lattice, strongly suggests a physical relevance of this superalgebra in the BPS sector of the theory. Later we will see how features of this algebra appear in a physical context and elucidate (part of) the role of this algebra.

\chapter{Counting the Dying Dyons}
\label{Counting the Dying Dyons}
\section{Introduction}
\setcounter{equation}{0}

In the last chapter we have reviewed the microscopic counting of \(1/\!2\)- and \(1/\!4\)-BPS states, in particular the derivation of a dyon-counting formula and various mathematical properties of the dyon-counting partition function. 

Recently, various puzzles have been raised about this formula \cite{Sen:2007vb,Dabholkar:2007vk}.
First of all, a subtlety in checking their \(S\)-duality invariance has been observed.
Secondly, there is an ambiguity in choosing the integration contour arising from the complicated pole structure of
the modular forms that enter the formulas. Finally, it has been noted that the BPS spectrum in the
macroscopic supergravity theory is subjected to moduli dependence due to the presence of walls of marginal stability for some
multi-centered bound states. See section \ref{Walls of Marginal Stability} for a discussion of this phenomenon in the
\({\mathcal{N}}=2\) context.
Finally, either by using a duality argument \cite{Sen:2007vb}, or by studying a specific example in great details \cite{Dabholkar:2007vk},
there have been some hints that all the above issues might actually have something to do with each other.

The goal of the first part of the chapter is to address these issues and provide a resolution to some of these puzzles.
In particular, our aim is to present a precise contour prescription that will lead to a counting formula that is
manifestly \(S\)-duality invariant and suitable for all moduli. To arrive at this prescription for contours, an important role is played by the one-to-one correspondence between various
poles in the integrand of the counting formula, and the different decay channels in which a dyon can be split
into two \(1\!/2\)-BPS particles. This correspondence between poles and bound states was envisaged in \cite{Dijkgraaf:1996it}, and was
recently reiterated in \cite{Sen:2007vb,Dabholkar:2007vk}.
It turns out that the only poles that can be crossed when the choice of contour is
varied are precisely the ones that admit such a correspondence.
Moreover, we find that the contributions of the poles exactly match the
expected number of states corresponding to the two-centered configurations of
BPS dyons (See also the work by Sen around the same time 
\cite{Sen:2007pg}).
The key observation which allows us to identify the correct contour prescription is that
the resulting integration contour should render the counting formula explicitly \(S\)-duality invariant, and
should furthermore automatically take into account the (dis-)appearance of the two-centered bound states when a wall of
marginal stability is crossed. This leads to a moduli-dependent degeneracy (or index-) formula that counts all the living dyons
in every region of moduli space.
In particular, we note that the walls of marginal stability have the property that for large black hole charges
(as opposed to ``small black holes" with vanishing leading macroscopic entropy),
none of the two-centered bound states of \(1\!/2\)-BPS particles can exist when the background moduli are fixed at their attractor values.
Using this fact we also propose a second, moduli-independent contour prescription,
which has the property of counting only the ``immortal dyons" which exist everywhere in the moduli space.

The goal of the second part of this chapter is to explore the role of the Borcherds-Kac-Moody algebra in the BPS spectrum. Despite the progress mentioned above which has been made in understanding the dyonic spectrum of the theory, so far no concrete interpretation is given to the appearance of the Borcherds-Kac-Moody algebra. We will address the following two aspects of this issue. 

First of all, we note that the counting formula, now equipped with a moduli-dependent contour, can be identified with the character formula of a Verma module of the algebra with an appropriate choice of highest weight depending on the moduli. Specifically, the dictionary is such that the attractor moduli corresponds to a Verma module of dominant highest weight. This sheds light on the question of how the BPS states form a representation of the algebra. 

Secondly, using the realisation that the walls of moduli stability of the supergravity theory are given by the walls of Weyl chambers of the Borcherds-Kac-Moody algebra, we show that the discrete dependence of the BPS spectrum on the moduli is described by the Weyl group of the algebra. From this we conclude that, just from the low-energy supergravity theory we should be able to derive the existence of such a group as the group of a discretized version of the attractor flow. More concretely, given a beginning point in the moduli space, there is a unique sequence of Weyl reflections, each represents the crossing of a wall of marginal stability, which brings the moduli to their attractor values. This path of wall-crossing is exactly the one taken by the usual attractor flow. In particular this gives an order among possible decay channels, which has the structure of an order given by an RG-flow. 
We hope that this chapter elucidates the relationship between the wall crossing, the counting formula, and the Borcherds-Kac-Moody algebra, in the context of the \({\cal N}=4\) dyonic spectrum.

\section{Dying Dyons and Walls of Marginal Stability}
\label{Dying Dyons and Walls of Marginal Stability}
\setcounter{equation}{0}

\subsection{Determining the Walls}

Earlier in this thesis we have seen in section \ref{Properties of Multi-holes} the phenomenon of moduli dependence of solutions of \({\cal N}=2\), \(d=4\) supergravity. Here we are interested in the question whether there is a similar phenomenon in the present \({\cal N}=4\), \(d=4\) theory. Especially, we are interested in the bound states of two \(1/\!2\)-BPS objects since these are the only bound states whose degeneracies might jump in different regions in the moduli space, which can be understood either by zero-mode counting or from the fact that the corresponding walls of marginal stability are of higher co-dimension in the moduli space in all other cases \cite{Sen:2007nz}

As discussed in chapter \ref{Properties of Multi-holes}, a decay of a two-centered supergravity  solution is possible only when the background moduli at infinity is such that the mass satisfies the condition \(M=M_1+M_2\). 

We want to determine when  a dyonic bound state might decay in this theory.
First we concentrate on the
specific decay channel of a dyonic, \(1\!/4\)-BPS state with charges \((P,Q)\) splitting into two \(1\!/2\)-BPS
particles with charges \((P,0)\) and \((0,Q)\).
For this case, the condition for a wall of  marginal stability is
\be \label{mass}
M_{P,Q}=M_{P,0}+M_{0,Q}\;,
\ee
which can be rewritten as
$$
|Z_{P,Q}(\l)| = |Z_{P,0}(\l)| + |Z_{0,Q}(\l)| \;.
$$
Notice that we have temporarily suppressed the dependence on the Narain part of the moduli in our notation, which determines the left-moving part of the charges \(P_L = \m\!\cdot\! P\) (\ref{left_moving_charge_dyon}), since these fields \(\m\) do not transform under the extended S-duality group \(PGL(2,\Z)\).

Using the fact that the total central charge obeys
$$
Z_{P,Q}(\l)=Z_{P,0}(\l)+Z_{0,Q}(\l)\;,
$$
one finds that the condition of marginal stability can only be satisfied when
the phases of the central charges are aligned, a phenomenon that is already familiar to us from chapter \ref{Properties of Multi-holes}. 

Using the explicit expression for the central charge (\ref{central_charge_N4}), the above equation leads to the condition
\be \label{line1}
{\l_1\over \l_2}+{{P_L\cdot Q_L}\over |P_L\wedge Q_L|} =0\;, 
\ee
where 
$$
|P_L\wedge Q_L| = \sqrt{P_L^2 Q_L^2 - (P_L \cdot Q_L)^2} 
$$
is the quantity defined earlier in (\ref{wedge_expression}).

The next step will be to consider the other ways in which a dyon can split into two \(1\!/2\)-BPS particles, and determine the
corresponding walls of marginal stability. The decay channels are determined by the fact that that a \(1\!/2\)-BPS state must satisfy the condition on their charges that the magnetic and electric charges have to be parallel to each other (\ref{half_BPS_condition}). 
It is now easy to check that every element of \(\g\in PGL(2,\Z)\) gives the following split of charges
\be\label{split_1}
\bem P \\ Q \eem = \bem P_1 \\ Q_1 \eem +\bem P_2 \\ Q_2 \eem
=  P_\g \,\g^{-1}  \bem 1 \\ 0\eem + Q_\g\, \g^{-1} \! \bem 0 \\ 1\eem\;,
\ee
where 
$$
\bem P_\g\\ Q_\g\eem = \g \bem P \\ Q \eem
$$
as before. Furthermore, from the quantisation condition of the charges one can also show that the converse is also true \cite{Sen:2007vb}. Namely,  for every possible \(1/\!2\)-BPS split one can always find a (not unique) \( PGL(2,\Z)\) element \(\g\) such that the charges can be  written in the above form. 

To determine the location of walls of marginal stability in the moduli space for the two-centered solutions with the above charges, we can just plug in the above split of charges into the marginal stability equation \(M=M_1+M_2\) and solve for the solution. But it will turn out to be a much more economic way to study the transformation of the central charge matrix (\ref{central_charge_N4}) under the extended duality group \(PGL(2,\Z)\) (\ref{extended_S-dual}). It is easy to check that this transformation has the effect of shifting the phase of the central charges by
\be
\label{central_charge_transform}
Z_{P,Q}(\l) = 
\begin{cases}
e^{i\alpha_\g} Z_{P_\g,Q_\g}(\l_\g)  \quad&\text{for     \small{det}}\g = 1
\\
e^{i\alpha_\g} \bar{Z}_{P_\g,Q_\g}(\l_\g)  \quad&\text{for     \small{det}}\g = -1
\end{cases} \ee
with some charge-independent phase \(\alpha_\g\) which depends on the group element \(\g\) and the axion-dilaton \(\l\).
Due to the fact that the phase shift is independent of the charges, all magnitudes of the relative phases will be duality invariant.
In particular, we have
$$
|Z_{P,Q}(\l)| =|Z_{P_\g,0}(\l_\g)+Z_{0,Q_\g}(\l_\g)|=|Z_{P_\g,0}(\l_\g)|+|Z_{0,Q_\g}(\l_\g)|\;.
$$

This shows that the position of the walls of marginal stability corresponding to the charge splitting (\ref{split_1}) are simply the $PGL(2,\zz)$ image of the one for the bound state of the
purely electric and purely magnetic \(1\!/2\)-BPS states. Namely, the walls of marginal stability for all two-centered \(1\!/2\)-BPS splits are
\be
\label{wallgamma}
\frac{\l_{\g,1}}{\l_{\g,2}}+ \frac{(P_{L} \cdot Q_{L})_\g}{|P_{L} \wedge Q_{L} |_\g}
=0\;,
\ee
where \(P_{L,\g}, Q_{L,\g}, \l_\g\) are given by (\ref{extended_S-dual}) as before.

The above formula might not look too complicated, but we should keep in mind that these are really infinitely many equations since \(PGL(2,\Z)\) is not a finite group. 
Inspired by the fact that all the quantities involved in the above equation have simple transformation rules under the extended S-duality group, we would like to look for a way to organise the above equations of walls of marginal stability in a form that is directly an equation on the {un-transformed} fields \(P_L,Q_L\) and \(\l\).

First note that, from the fact that 
$$
\L_{P,Q}  = \bem P \\ Q \eem \cdot \bem P & Q\eem
$$
we see that  the three-dimensional charge vector \(\L_{P,Q} \) transforms as\\
$$
\L_{P,Q} \to \g(\L_{P,Q}) = \g \L_{P,Q} \g^T 
$$
under the S-duality transformation (\ref{extended_S-dual})
\be\label{transformation_LPQ}
\bem P \\ Q\eem \to\g \bem P \\ Q\eem := \bem P_\g \\ Q_\g \eem \quad,\quad \g= \bem a&b\\ c&d\eem \in PGL(2,\Z)\;.
\ee

Since the S-duality group leaves the Narain moduli \(\m\) invariant, we conclude that the same transformation rule holds for \\
$$
\bem P_L \cdot P_L &  P_L\cdot Q_L \\  P_L\cdot Q_L &  Q_L\cdot Q_L \eem \to \g 
\bem P_L \cdot P_L &  P_L\cdot Q_L \\  P_L\cdot Q_L &  Q_L\cdot Q_L \eem  \g^T\;.
$$
Especially, the norm of the this vector \(\sim |P_L\wedge Q_L |\) is invariant under the transformation.
For the axion-dilaton, from our experience with the S-duality of type IIB supergravity (\ref{axion-dilaton_matrix_transf1}), we have seen that the following matrix also transforms under \(PSL(2,\Z)\) in the same way as the charge vector
$$
\frac{1}{\l_2} \bem|\l|^2 & \l_1 \\ \l_1 & 1 \eem \to \g\;
\frac{1}{\l_2} \bem|\l|^2 & \l_1 \\ \l_1 & 1 \eem\,\g^T \;, 
$$ 
and it is trivial to check that the same transformation rule extends to the extended duality group \(PGL(2,\Z)\).

In particular, if we choose the following combination of these two vectors 
\be\label{def_Z_vec}
{\mathcal Z} = \frac{1}{\sqrt{|P_L \wedge Q_L|}} \bem P_L \cdot P_L &  P_L\cdot Q_L \\  P_L\cdot Q_L &  Q_L\cdot Q_L \eem + \frac{\sqrt{|P_L \wedge Q_L|}}{\l_2} \bem|\l|^2 & \l_1 \\ \l_1 & 1 \eem\;,
\ee
then it transforms as 
$$
{\mathcal Z} \to \g({\mathcal Z}) = \g {\mathcal Z} \g^T 
$$
under an extended S-duality transformation. 

As a side remark, let us note that the above somewhat awkward-looking normalisation has the advantage that now the mass of a dyon, which involves \(134\) moduli fields (\ref{mass_dyon_formula}),  is nothing but the norm of this single vector
$$
M_{P,Q}^2 = |Z_{P,Q}|^2 = |{\mathcal Z}|^2\;.
$$
But we are going to see in a moment that it is only the direction, not the length, of the vector \({\mathcal Z}\) which determines the existence or not of a certain two-centered solution. For convenience we will therefore define a ``unit vector" \(X\) by 
\be\label{def_X_vec}
X = \frac{{\mathcal Z} }{M_{P,Q}} =  \frac{{\mathcal Z} }{|{\mathcal Z}|}\;. 
\ee 
Apparently this vector also transforms in the same way as \({\mathcal Z}\) under the \(PGL(2,\Z)\) transformation.

With this notation, the walls of marginal stability for the \((P,0)\), \((0,Q)\) split of charges (\ref{line1}) lie on the co-dimensional one space characterised by the following equation
$$
\big(X,\a_1 \big) = 0 \;,
$$
and similarly for other splits (\ref{wallgamma})
$$
\big(\g(X) ,\a_1\big) =  \big( X,\a \big) =0\quad,\quad \a = \g^{-1}(\a_1)\;.
$$
 
Now we have achieved the goal of organising the equations of walls of marginal stability as equations directly constraining the untransformed quantities \(P_L,Q_L\) and \(\l\). 


To discuss in more details the possible relationship between the contour dependence of the integral and the physical walls of marginal stability, it is necessary to label these walls. As we mentioned before, the map (\ref{split_1}) between \(PGL(2,\Z)\) elements and the split of charges into two \(1/\!2\)-BPS charges is not one-to-one. For example, the element which simply exchanges what we call the ``1st" and the ``2nd" decay products will not give a different split of charges. On the other hand, from the expression for the walls of marginal stability (\ref{wall_2}), we see that the element \(\g\) which gives \(\g(\a_1)= \pm \a_1\) will not give a new physical wall. It can indeed be checked that these are also the elements which give the same split of charges when using the map (\ref{split_1}). 
Using the fact that \(PGL(2,\Z)\) is the group of symmetry for the root system of the Coxeter (Weyl) group \(W\) (\ref{sym_root_extended_S}) and by inspecting the action of the dihedral group (\ref{dihedral_action_1}),(\ref{dihedral_action_2}), it is not hard to convince oneself that the two-centered solutions with two \(1/\!2\)-BPS charges  discussed in this section are actually given by the positive real roots of the Borcherds-Kac-Moody algebra.
So we arrive at the conclusion that the relevant two-centered solutions are in one-to-one correspondence with the positive real roots of the Borcherds-Kac-Moody algebra, whose walls of marginal stability are given by
\be\label{wall_2}
(X,\a) = 0  \quad,\quad \a \in \D_+^{re}
\ee
and whose decay products can be represented by the split of the charge vector as
\bea\nonumber
\L_{P_1,Q_1} &=& P_\a^2\,\a^+ \quad,\quad\L_{P_2,Q_2} =  Q_\a^2\,\a^-\\
\label{charge_split_new}
\L_{P,Q} &=& P_\a^2\,\a^+ + Q_\a^2\,\a^-  - |(P\cdot Q)_\a|\,\a\;,
\eea
where for a given \(\a\), the set of the two vectors \(\a^\pm\) is given by the requirement that 
(i) they are both lightlike and future-pointing and perpendicular to the root \(\a\),
(ii) \(\a^\pm/2\) lie in the weight lattice, namely that $\a^\pm$ as matrices have integral entries  (iii) they have inner product \((\a^+,\a^-)=-1 \). 
See figure \ref{lightcone}.  In particular, the combination \(P_\a^2\) and \(Q_\a^2\), which is related to the ``oscillation level" of the heterotic string as  (\ref{oscilation_level_het}) and which determines the degeneracy of the \(1/\!2\)-BPS states, can be thought of as the ``affine length" of the lightlike charge vector \(\L_{P_1,Q_1}\) and \(\L_{P_2,Q_2}\) of the decay products. Conversely, given the charges of the two centers, the walls of marginal stability is given by the requirement that the moduli vector \(X\) is the linear combination of  the two charge vectors   
\(\L_{P_1,Q_1}\) and \(\L_{P_2,Q_2}\).

More concretely, what the above formula (\ref{charge_split_new}) means is the following: given a spacelike vector, any future-pointing timelike vector can be split into a component parallel to it and a future-pointing timelike vector perpendicular to it. And the latter can be further split into two future-pointing lightlike vectors lying on the plane perpendicular to the given spacelike vector. See figure \ref{lightcone}. When the timelike vector is taken to be the original charge vector \(\L_{P,Q}\), the two future-pointing lightlike vectors are then the charge vectors of the two decay products.

For instance, suppose \(\a = \g^{-1}(\a_1)\) for some \(\g \in PGL(2,\Z)\), then \(\{\a^+,\a^-\}\) is given by
$$
\g^{-1}(\text{\small diag}(0,1))\;,\;\;\;\g^{-1}(\text{\small diag}(1,0))\;.
$$
In this case one can show that \(P_\a^2 = P_\g^2\) and similar for the Q's, and the ambiguity of relating a \(PGL(2,\Z)\) element \(\g\) to a specific decay channel lies in the fact that there are different \(\g\)'s that give the same set \(\{\a^+,\a^-\}\).

\subsection{Stability Conditions from Supergravity Solutions}

The meaning of the presence of a wall of marginal stability is that a BPS bound state of two particles
exists on one side of the wall and disappears when crossing into the other side. After deriving the location
of the walls for these bound states, we would like to know on which side these states are stable and on which side unstable. As discussed in section \ref{Walls of Marginal Stability}, this can either be determined using the heuristic argument that the attractor moduli of the single-centered black hole solution with the given total charges should lie on the unstable side of the wall, or by analysing the integrability condition of the supergravity solution. Now we will perform the latter analysis as a check. 
For this purpose we need more information about the corresponding supergravity solutions.

Let us now consider the four-dimensional \({\cal{N}}= 4\) supergravity theory describing the low energy
limit of the heterotic string compactified on a six-torus.
The metric part of a stationary solution reads
\bea\nonumber
ds^2&=& -e^{-2U} (dt + \vec{\w}\cdot d\vec{x})^2 + e^{2U}d\vec{x}^2\\
\nonumber
e^{2U}&=&{ |\cal{P}\wedge \cal{Q}|} \,{\equiv}
\sqrt{  {\mathcal{P}}^{2} {\mathcal{Q}}^2 - ( {\mathcal{P}} \cdot {\mathcal{Q}})^2 } \\
\label{angular_momentum1}
\vec{\nabla} \times \vec{\w} &=& \cal{P}\cdot \vec{\nabla} \cal{Q}- \cal{Q}\cdot \vec{\nabla} \cal{P} \;,
\eea
where the indices are contracted using the standard \(SO(6,22)\)-invariant \(28\times 28\) matrix \(\eta_{AB}\), for
example \({\cal{P}}^2 \equiv {\cal{P}}^A {\cal{P}}^B \eta_{AB}\).

The 56 harmonic functions appearing in the above solution are
\ben
{\cal{P}}^A(\vec{x}) = C^A + \sum_i \frac{P^A_i}{|\vec{x}-\vec{x}_i|}&\\
\nonumber
{\cal{Q}}_A(\vec{x}) = D_A + \sum_i \frac{Q_{A,i}}{|\vec{x}-\vec{x}_i|}&\;,
\een
with the 56 constants given by the asymptotic value of 23 complex scalar fields (the axion-dilaton moduli
\(\l\) and the 22 complex moduli projected from the aforementioned \(6\times 22\) moduli) as\footnote
{By evaluating the \({\mathcal{N}}=4\) central charge operator \(\hat{Z}\) (\ref
{central_charge_N4}) in the eigen basis of  \(\hat{Z}^\dagger\hat{Z}\), one
can write the BPS equations in a way analogous to the \({\mathcal{N}}=2\) case as in section \ref{General Stationary Solutions}.
Only 22 complex moduli made out of the \(6\times 22\) real moduli fields play a role in the solution.
It is indeed known that the \({\mathcal{N}}=4\) moduli space locally decomposes as a product of 22 vector-, 44 hyper-,
and 1 tensor-multiplet scalars in the \({\mathcal{N}}=2\)  language (see, for example, \cite{Moore:1998pn}).}
\ben
C^A =-   \im\left(e^{-i\a_{P,Q}} \frac{\pa Z_{P,Q}}{\pa Q_A} \right) &\\
\nonumber
D_A =   \im\left(e^{-i\a_{P,Q}} \frac{\pa Z_{P,Q}}{\pa P^A} \right) &\;,
\een
where the \(P^A\)'s and the \(Q_A\)'s denote the total charges coming from all the centers.
From this expression one immediately sees that these coefficients satisfy \( Q_A C^A = P^A D_A \),
since the central charge is linear in all charges.

For the specific two-center bound state with charges \((P,0)\), \((0,Q)\) considered earlier, the corresponding supergravity solution
has harmonic functions given by
\ben
{\cal{P}}^A = C^A +  \frac{P^A}{|\vec{x}-\vec{x}_P|}&\\
\nonumber
{\cal{Q}}_A = D_A +  \frac{Q_{A}}{|\vec{x}-\vec{x}_Q|}&\;.
\een

In this case the coordinate distance between the two centers \(|\vec{x}_P-\vec{x}_Q|\) is fixed by the
integrability condition \cite{Denef:2000nb}, obtained by taking the divergence of the both sides of (\ref{angular_momentum1}), and reads
$$
\frac{P\cdot Q }{|\vec{x}_P-\vec{x}_Q| } = - C^A Q_A\;.
$$

After some algebra this becomes
$$
{|\vec{x}_P-\vec{x}_Q|} =  -\frac{1}{\sqrt{|P_L\wedge Q_L|}}\,\frac{(\L_{P,Q},\a_1)}{(X,\a_1)}\;.
$$

Since the distance between the two centers is always a positive number, one finds that, in order for the bound state to exist,
the expression on the r.h.s. should better be positive as well.
We therefore conclude that the bound state only exists when
\be\label{stability_1}
(\L_{P,Q},\a_1)(X,\a_1) < 0\;,
\ee
and decays when one dials the background moduli to hit the wall where \((X,\a_1)\) vanishes.
More precisely,  one finds that the distance between the two centers goes to infinity,
and the bound state no longer exists as a localisable state.

Using the \(PGL(2,\Z)\) transformation as before, it is now easy to write down for all the other two-centered solutions with \(1/\!2\)-BPS centers 
the expression for the coordinate distance between the centers
\be \label{distance_dyon_center}
{|\,\vec{x}_{P_\a}-\vec{x}_{Q_\a}|} =  -\frac{1}{\sqrt{|P_L\wedge Q_L|}}\,\frac{(\L_{P,Q},\a)}{(X,\a)}\;,
\ee
and hence the stability condition reads
\be\label{stability_2}
(\L_{P,Q},\a)(X,\a) < 0\quad,\quad \a \in \D_+^{re}\;.
\ee

We have therefore achieved the goal of studying the stability condition for all two-centered bound states of \(1/\!2\)-BPS objects in the present \({\cal N}=4\), \(d=4\) supergravity theory.

\section{Contour Dependence of the Counting Formula}
\label{Contour Dependence of the Counting Formula}
\setcounter{equation}{0}

In section \ref{section_quartBPS_counting} we have derived the microscopic counting formula (\ref{DVV_1}), expressed in terms of the generating function \(1/\F(\O)\). Formally, to extract the actual degeneracies \(D(P,Q)\) from the generating function we can invert the formula into a contour integral (\ref{DVV_integral_1}). 

But in this formula we have not specified how the contour of integration should be chosen. It would not be a problem if the integral were contour-independent, but as it turns out it is not the case here. As discussed in (\ref{RQdivisor}), the generating function \(1/\F(\O)\) has poles lying on the rational quadratic divisors which are related to each other by \(Sp(2,\Z)\) modular transformation. Due to the presence of these poles, one has to be careful with choosing the contour \(\mathcal{C}\): the counting formula will ``jump'' when the contour crosses
one of these poles. Therefore, strictly speaking the formula (\ref{DVV_integral_1}) for $D(P,Q)$ is not just a function of the
charges $P$ and $Q$ but also depends on the contour. 

To determine what the appropriate contour should be, the symmetry of the theory, in particular the extended S-duality \(PGL(2,\Z)\) symmetry in this case, will be provide us with important hints. 

First recall the transformation of property of of charge vector \(\L_{P,Q}\)  (\ref{transformation_LPQ}) under the extended S-duality group
$$
\bem P\cdot P &  P\cdot Q \\  P\cdot Q &  Q\cdot Q \eem \to \g 
\bem P\cdot P &  P\cdot Q \\  P\cdot Q &  Q\cdot Q \eem  \g^T\;. 
$$
Recall that \(PGL(2,\Z)\) acts as a Lorentz plus spatial reflection transformation in the space  of \(2\times 2\) symmetric matrices. In particular, the inner product of two vectors is invariant under such a transformation
$$
(\g(\O),\g(\L_{P,Q})) = (\O,\L_{P,Q}) \;.
$$
Together with the fact that \(\F(\O)\) is invariant under \(\O\to \g(\O)\) (\ref{weight10_modular_form}), we conclude that the integrand of the contour integral (\ref{DVV_integral_1}) is invariant under the following S-duality transformation 
$$
(-1)^{(P\cdot Q)_\g+1} \,\frac{e^{\p i (\g(\O),\g(\L_{P,Q}))}}{\F(\g(\O))} 
= (-1)^{P\cdot Q+1}\, \frac{e^{\p i (\O,\L_{P,Q})}}{\F(\O)}\;,
$$
where we have also used the fact that \(P^2,Q^2 = 0\) mod \(2\) and as a consequence \((P\cdot Q)_\g=P\cdot Q\) mod \(2\).
Therefore, if we ignore the ambiguity of the contour, the degeneracy formula (\ref{weight10_modular_form}) indeed satisfies the physical condition of being invariant under the duality group.

This fact is not yet sufficient, however,  to prove the invariance of the degeneracies. Namely, due to the presence of the poles,
the expression for $D(P,Q)$ fails to be \(S\)-duality invariant, unless the contour ${\mathcal C}$ is also transformed to a
new contour ${\mathcal C}_\g$. Explicitly, the equality
\be
\label{Sinvariance}
\oint_{\mathcal{C}} d\O \;
(-1)^{P\cdot Q+1}\, \frac{e^{\p i (\O,\L_{P,Q})}}{\F(\O)}
=\
\oint_{\mathcal{C_\g}} d\O\;
(-1)^{(P\cdot Q)_\g+1} \,\frac{e^{\p i (\g(\O),\g(\L_{P,Q}))}}{\F(\g(\O))} \ee
only holds when  the
new contour ${\mathcal C}_\g$ in the $\g(\O)$-plane is the same as ${\mathcal C}$ in the $\Omega$-plane. If this is not true, in general different contours cannot be deformed into one another without picking up any residue and the answer we get for the degeneracies will therefore not be S-duality invariant. 

A natural guess for a remedy for the present situation is to let the contour depend on the charges, and possibly also the moduli fields, since these quantities do
transform under \(S\)-duality. Indeed, there is an important reason to suspect that the dyon counting formula is moduli-dependent, since as we have seen in the last section, certain multi-centered BPS solutions only exist in some range of background moduli and decay when a wall of marginal stability is crossed. 

As a first step towards understanding the moduli dependence of the integration contour, we will now study the dependence of  \(D(P,Q)\) on the choice of the contour in the integral formula (\ref{DVV_integral_1}), by analysing the contribution from the poles of the partition function \(1/\F(\O)\) to the integral. 

Let us have a closer look at the possible choice of the contour, namely a choice of three-cycle on which we perform the integral over in the three-complex dimensional space parametrised by 
\be
\O = \bem \s & -\n \\ -\n & \r \eem \in \R^{2,1}  + i V^+\;.
\ee
Due to the fact that we are dealing with a modular form, the contour will have to be inside a fundamental domain of the
$Sp(2,\zz)$ modular group. A natural choice of contour is to perform the integral over the real parts of $\r$, $\s$ and $\n$,
while keeping the imaginary parts fixed.
Specifically, the range of integration of the real variables is
\be
\label{realdomain}
0\leq \re\r,\re\s,\re\n < 1\;.
\ee
The integration contour is thus a three-torus.
The location of the contour is determined by a choice of the imaginary parts.
To make sure that $1/\Phi$ has a well-defined expansion up to high order, we will choose these imaginary parts so that $\Omega$ lies well inside the Siegel
upper-half plane, that is
\be\label{contour_space}
| \im\O |^2 = \im \r\, \im\s-(\im \n)^2 \ = \varepsilon^{-2}\gg 1\,.
\ee
To visualize the location of the poles relative to the contours, we note that the above condition defines a
sheet of a hyperboloid high up inside the future light-cone.
This is shown in Figure \ref{lightcone}.
\begin{figure}
\centering
\includegraphics[width=12cm]{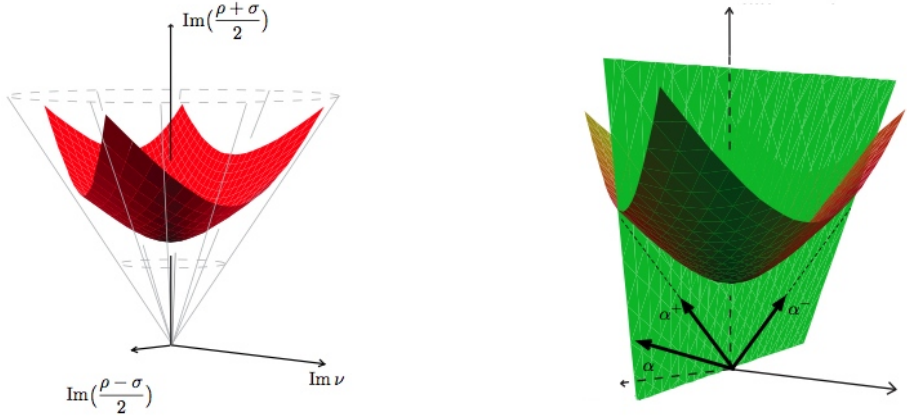} 
\setlength{\abovecaptionskip}{5pt}
\caption{\footnotesize{\label{lightcone} {\bf{(a)}} The imaginary part of the Siegel upper-half plane for the modular form \(\F\) is the future light-cone \(V^+\) in the
Minkowski space \(\rr^{2,1}\), and we consider the space of all contours to be a sheet of hyperboloid inside this light-cone, with all the points
on the hyperboloid having the same large distance from the origin.
{\bf{(b)}} A plane \((X,\a)=0\) given by a positive real root \(\a\) always intersects the hyperboloid, or equivalently the upper-half plane or the Poincar\'e disk. And the root \(\a\) defines two lightcone directions \(\{\a^+\), \(\a^-\}\) perpendicular to it, given by the intersection of the plane with the future light-cone. }}
\setlength{\belowcaptionskip}{5pt}
\end{figure}
As mentioned before, all the double poles of the generating function \(1/{\F}\) are located at divisors
given by the \(Sp(2,\zz)\)  modular images of the divisor \((\O,\a_1)= -2 \n=0\)  in the
\(\O\)-space, where \(\a_1\) coincides with one of the simple real roots given in (\ref{3_simple_roots}).
 As we have seen before (\ref{RQdivisor}), these poles take the form 
$$
 r(\r\s-\n^2) + n\r + m\s +\ell \n + s = 0\;.
$$

 The poles at divisors with \(r=1\) have exponentially dominant contribution to the degeneracy formula
(\ref{DVV_integral_1}) compared to the rest in the case of large charges, as explained in the appendix of
\cite{Dijkgraaf:1996it}.
In \cite{Sen:2007vb} it was observed
that the contour space (\ref{contour_space}) does not intersect any of the
poles having \(|r| \geq1 \). Indeed, a look at the real part of the above equation reveals that, since
all the entries of \(|\re\O|\) run between 0 and 1, there is nothing to compensate the large contribution from
$|\im \O|^2 \gg 1$ contained in the real part of the term $|\O|^2 = \r\s\!-\!\n^2$.
In other words, these poles will always contribute to the degeneracy formula no matter which contour we
choose, since they lie  lower in the light-cone. Therefore, we
never run into the danger of having a contour which crosses one of these poles.
For our purpose of studying the contour dependence of the integral, it is hence sufficient to concentrate on the poles with \(r=0\).

Since we are only interested in the poles inside the real domain of integration (\ref{realdomain}), we can restrict our attention to
the poles with $r\!=\!s\!=\!0$. It is easy to see that an image of 
the pole \((\O,\a_1)=0\) under \(PGL(2,\zz)\) transformation is another  rational quadratic divisor, since  the length condition (\ref{RQdivisor})
$$
-| \a |^2 = -| \g^{-1}(\a_1) |^2 =1
$$
follows directly from $| \a |^2 = | \g^{-1}(\a_1) |^2 $.

On the other hand, it can be shown, by classifying both entries of \(PGL(2,\Z)\) elements and (\(k,\ell,m\)) by their prime
factorizations for example, that one can always find a (not unique) \(PGL(2,\zz)\) element \(\g\) for each pole (\ref{RQdivisor}) with \(r=s=0\) such that it is the image of the pole \(( \O,\a_1)=0\) under the group transformation.

In other words, these poles can be written in the form
$$
\big(\g\O\g^T, \a_1\big)=\big(\g(\O), \a_1\big) = \big(\O, \g^{-1}(\a_1) \big) = 0 \quad\text{for some   }\g\in PGL(2,\Z)\;,
$$
and its imaginary part
$$
\big(\im\O,\a \big) = 0\quad,\quad \a = \g^{-1}(\a_1) \quad \text{for some  }\g \in PGL(2,\Z)\;.
$$
defines a plane inside the space \(\R^{2,1}\). 
 
The fact that \((\a, \a) > 0\) implies that the normal vector to the plane is spacelike, and hence
these planes always intersect the contour space hyperboloid (\ref{contour_space}) along a hyperbola. Therefore, each plane
divides the contours into two sub-classes \((\im\O,\a)>0\) and \((\im\O,\a)<0\).
See Figure \ref{lightcone}.
Whether the corresponding poles  contribute to the degeneracy formula for a given charge configuration will therefore
depend on the contour we choose.

Let us now determine the condition under which these poles contribute to the integral, and, if they do contribute,
what their contribution is. We first concentrate on  the double pole at $(\O,\a_1)=2\nu=0$.
Near the $\nu=0$ divisor the generating function has the limit
\be \label{pole1}
\frac{1}{\F(\r,\s,\n)} = -\frac{1}{4 \p^2} \frac{1}{\n^2} \frac{1}{\eta^{24}(\r)} \frac{1}{\eta^{24}(\s)} \;(1+{\cal{O}}
(\n^2))\;.
\ee
Notice that the last two factors in the limiting
expression (\ref{pole1}) are exactly the generating function for the \(1\!/2\)-BPS degeneracies (\ref{half_BPS}).
By plugging the above expression into the degeneracy formula (\ref{DVV_integral_1}) and performing the integration over the real part of $\r$ and $\s$,
one gets
\begin{displaymath}
\frac{(-1)^{P \cdot Q+1}}{4\p^2} d(P) d(Q)  \oint_{{\mathcal{C}}_\n}  d\n\,\frac{e^{-2 \p i (P\cdot Q) \n}}{\n^2}\;,
\end{displaymath}
where we have made use of (\ref{half_BPS}).
To evaluate the remaining integral over $\nu$, we first consider a contour with 
$$
\big(\im\O, \a_1\big) =2 {\rm Im}\n> 0\;.
$$
For this case the contour is shown in the Figure \ref{contour_graph}.
When the charges under consideration satisfy 
$$
 \big(\L_{P,Q},\a_1\big) = -2 P\!\cdot\! Q > 0\;,
 $$ 
one can deform the contour to the upper infinity
of the cylinder (\(\im\n  \rightarrow \inf \))
where the integrand is zero without crossing any pole. One thus concludes that the integral yields zero.
On the other hand, in the case \(P\cdot Q > 0\), the contour can be moved to the lower infinity (\(\im\n
\rightarrow -\inf \)) where the integrand is again zero, but now by doing so we pick up the contribution of
the pole
$$
-2\p i \pa_{\n} (e^{-2 \p i (P\!\cdot\!Q) \n})| _{\n=0} = -4\p^2 \; (P\cdot Q)\;,
$$
where the extra minus sign comes from the fact that we are enclosing the pole in a clockwise direction.
For the contours with \(\im\nu < 0 \), a similar argument shows that the pole only contributes when $(P\cdot Q)<0$, but now with the opposite sign as above due to the reverse orientation in which the pole is enclosed.  One therefore concludes that the contribution of
this specific pole to the degeneracy formula (\ref{DVV_integral_1}) is
\be\label{degeneracy_PQ1}
(-1)^{(P\cdot Q)+1}\, |P\cdot Q|\,d(P)\,d(Q)\;\qquad \mbox{when}\ \ 
 \big(\L_{P,Q},\a_1\big) \big(\im\O, \a_1\big) <0
\ee
and zero otherwise.
\begin{figure}
\centering
\includegraphics[width=12cm]{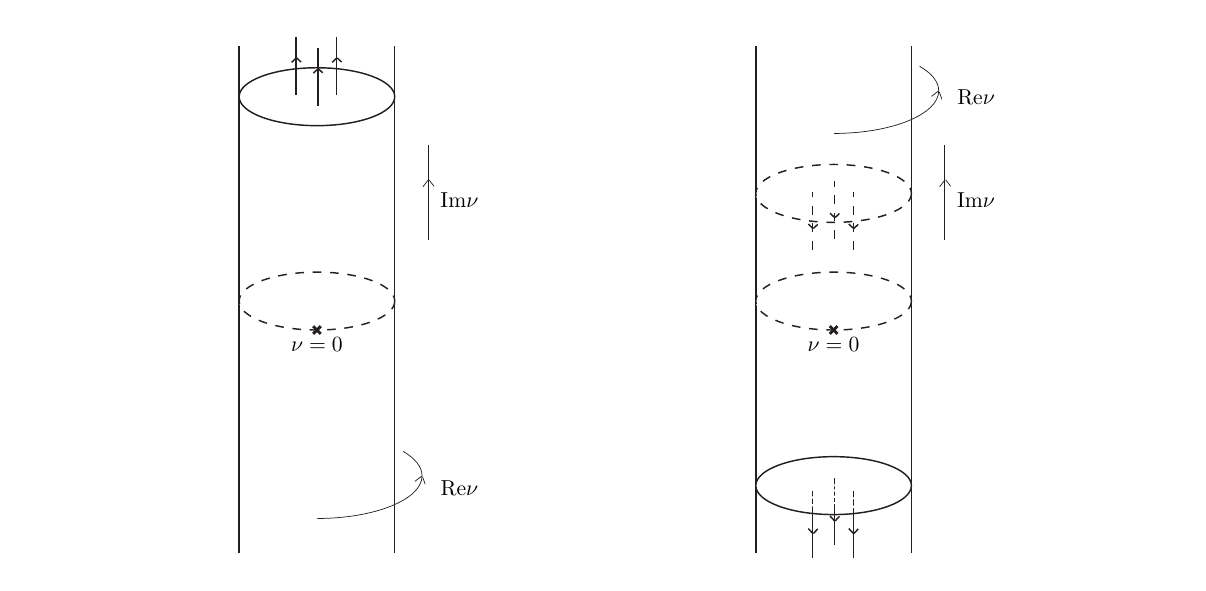} 
\setlength{\abovecaptionskip}{5pt}
\caption{\footnotesize{\label{contour_graph}In this figure we show how the pole located at \(\n=0\)
contributes to the degeneracy formula for contours with \(\im\nu>0\). {\bf{(a)}} For charges with \(P\cdot Q
< 0\), one can deform the contour to the upper infinity of the cylinder where the integrand goes to zero
without hitting the pole. {\bf{(b)}} For charges with \(P\cdot Q > 0\), one can deform the contour to the
lower infinity of the cylinder, and by doing so pick up the residue of the pole. }}
\setlength{\belowcaptionskip}{5pt}
\end{figure}
The contributions of the other poles can be determined directly in a similar fashion. However, they are more easily
obtained by making use of the fact that they are the $PGL(2,\zz)$ images of the $\n=0$ pole.
Together with the fact that the integrand is invariant under extended \(S\)-duality group (\ref{Sinvariance}),
it follows that the double pole of \(1/\F\)
located at
\be\label{pole_gamma}
(\O, \a) = 0 \quad,\quad\a = \g^{-1}(\a_1) \;\;\;\text{for some   }\g\in PGL(2,\Z)
\ee
gives the contribution
\be \label{jump_gamma}
(-1)^{P_\g\cdot Q_\g+1} \,|P_\g\cdot Q_\g| \,d(P_\g)\,d(Q_\g)\;\qquad \mbox{when}\ \ \big(\L_{P,Q},\a\big) \big(\im\O, \a\big) <0 
\ee
and zero otherwise.
The equation (\ref{jump_gamma}) summarizes all the contour dependence in the degeneracy formula (\ref{DVV_integral_1}).

As we will see, the jumps in the counting formula when a contour crosses one of the poles are related to the decay of marginally
bound \(1\!/2\)-BPS particles.
Specifically, we will argue that (\ref{degeneracy_PQ1}) precisely counts the number of states associated with the bound state of
a purely electric \(1\!/2\)-BPS object and a purely magnetic \(1\!/2\)-BPS object, while (\ref{jump_gamma}) is associated with more
general dyonic bound states that are obtained by electric-magnetic duality. This interpretation will be discussed in more
details in section \ref{The Contour Prescription and its Interpretation}.

In order to make the above correspondence more precise, we will now discuss the labelling of the poles susceptible to contour-dependence studied above. 
As we mentioned earlier, the group element \(\g \in PGL(2,\Z)\) associated with a given such pole might not be unique. 
Indeed, the elements \(\g\) which gives \(\g^{-1}(\a_1) = \pm \a_1\) will of course not give another equation. 
This redundancy is exactly the one we encountered in the last section when we were deriving the labelling of the decay channels (\ref{stability_2}). Hence we conclude that these poles are also labelled by the positive real roots of the Borcherds-Kac-Moody algebra.
 In other words, the full contour-dependence of the integral (\ref{DVV_integral_1}) can be summarised as follows: the hyperplanes 
\be\label{wall_of_contour}
(\im \O,\a ) = 0 \quad,\quad \a \in\D_+^{re}
\ee
cut the space of contours into different regions, and the integral changes its value by the amount (\ref{jump_gamma}) when one of the ``walls of contours" is crossed.

\section{The Contour Prescription and its Interpretation}
\label{The Contour Prescription and its Interpretation}
\setcounter{equation}{0}

\subsection{A Contour Prescription}
Let us now return to the problem of identifying the contour that should be used in the counting formula, such that it counts the right number of states for a given value of the moduli, and therefore by definition yields a duality-invariant answer. The key observation which will allow us to find the correct prescription is that the contour dependence due to the crossing of
 the pole labelled by the positive real roots $\a$ should exactly match the physical decay process of the corresponding dyonic bound state. For example, at the wall
of marginal stability of the bound state of an electric \(1\!/2\)-BPS particle with charge $(P,0)$ and a magnetic
\(1\!/2\)-BPS particle with charge $(0,Q)$, one expects the degeneracy $D(P,Q)$ to be adjusted by a certain amount
corresponding to the degeneracy of this \((P,0)\), \((0,Q)\) bound solution. This degeneracy can be found in the following way
\cite{Sen:2007vb,Dabholkar:2007vk,Denef:2007vg}.
Firstly, each of the two centers has its respective degeneracy \(d(P)\), \(d(Q)\), which is given by the \(1\!/2\)-BPS partition function of the theory as (\ref{half_BPS}).
Secondly, there is an extra interaction factor due to the fact that the spacetime is no longer static.
The conserved angular  momentum, after carefully quantizing the system \cite{Denef:2002ru},  turns out to be
\be\label{angular_momentum}
2J+1= |P\cdot Q|\;.
\ee
One therefore concludes that the jump in the counting formula when one crosses the wall of marginal stability from the stable to the unstable side
is given by
\be
D(P,Q)\to D(P,Q) + (-1)^{P\cdot Q}\, |P\cdot Q|\,d(P)\,d(Q)\;.
\ee
This jump in the degeneracy is precisely the contribution from the pole at $(\O,\a_1)=2\nu=0$ that we found in  (\ref{degeneracy_PQ1})!
Similar jumps occur when one crosses the walls of marginal stability for the other dyonic solutions labelled by positive real roots $\a$. 
As explained in (\ref{charge_split_new}) and (\ref{jump_gamma}), both the jump in degeneracies and the contribution of the pole are given by
$$
(-1)^{(P\cdot Q)_\a+1} \,|(P\cdot Q)_\a| \,d(P_\a)\,d(Q_\a)\;,
$$
where \((P\cdot Q)_\a\), \(P_\a^2\) and \(Q_\a^2\) are determined by \(\a\) as (\ref{charge_split_new}).

Since the amount of discontinuity matches between the contour side and the supergravity side, now the aim is to find the contour prescription such that the condition for contribution matches as well. Notice that it is a priori not clear whether this would be possible or not, since there are infinitely many potential two-centered solutions and infinitely many poles susceptible to contour dependence. However, from the condition for two-centered solution to exist (\ref{stability_2}) and for the poles to contribute (\ref{jump_gamma}), we see that this can be done simply by choosing the contour of the integral (\ref{DVV_integral_1}) to be the three-torus lying at 
\be\label{contour_summary}
\im\O = \varepsilon^{-1} X\;.
\ee

Here $\varepsilon \ll 1$ is taken to be small and positive to ensure that the series expansion of
$1/ \Phi$ converges rapidly. Moreover, as explained earlier, for sufficiently small $\varepsilon $ the contour avoids all other poles
except the ones given by the positive real roots as \((\O,\a)=0\).

To sum up, this prescription gives the location of the contour \({\cal C}\) in terms of the charges and moduli: 
\bea\nonumber
D(P,Q)\lvert_{\l,\m} = (-1)^{P\cdot Q+1}\oint_{{\cal C}(P,Q)\lvert_{\l,\m} } \, d\O \,\frac{e^{\p i (\L_{P,Q},\O) }}{\F(\O)}\\ \label{DVV_integral_2}
{\cal C}(P,Q)\lvert_{\l,\m}  = \{\re \O \in T^3 , \im\O =  \varepsilon^{-1} X\}\;,
\eea
where \(X\) is given in terms of total charges and moduli as (\ref{def_Z_vec}), (\ref{def_X_vec}) and \(\varepsilon \ll 1\) is some arbitrary small positive number. 

Furthermore, from the fact that the contour transforms in the same way as the charge vector under the \(PGL(2,\Z)\) S-duality group
$$
X \to \g(X) \quad,\quad \L_{P,Q}\to \g(\L_{P,Q})\;,
$$
we can now finish the argument in (\ref{Sinvariance}) and show that the counting formula (\ref{DVV_integral_2}), now coming with a contour prescription, is indeed consistent with the S-duality symmetry of the theory.

A similar result also holds for the so-called CHL models \cite{Chaudhuri:1995bf,Chaudhuri:1995fk}. A dyon-counting formula has been proposed for these appropriate \(\zz_N\) orbifolds of the above theory for \(N=2,3,5,7\) \cite{Dabholkar:2006bj,David:2006ud,David:2006ru,Dabholkar:2006xa,Jatkar:2005bh,LopesCardoso:2006bg}.
In these theories, the rank of the gauge group is reduced and the \(S\)-duality group is now the following subgroup of \(SL(2,\zz)\):
$$
\G_1(N) = \left\{\bem a& b\\ c&d\eem\in SL(2,\zz)\;\vert\; c= 0 \text{ mod } N\;, a,d =  1 \text{ mod } N\right\}\;.
$$
Moreover, the family of the  the contour-dependent poles of the proposed generating function \(\frac{1}{\til{\F}_k(\til{\O})}\),
which is now a modular form of a subgroup of \(Sp(2,\zz)\), and the ways in which a dyon can split into two \(1\!/2\)-BPS particles,
are both modified compared to the original theory. Nevertheless, we find that they can again both be given by the elements of the reduced
\(S\)-duality group \(\G_1(N)\),
and these poles again give the same jump of index as the decaying of these bound states.
In particular, following the same arguments we make exactly the same proposal (\ref{contour_summary})
for the integration contour for the dyon counting formula of this class of models.

\subsection{The Attractor Contour for Large Charges}

\noindent
For large charges corresponding to a macroscopic black hole, it is natural to ask what happens to our prescription when one takes the
moduli at infinity to be at the attractor value. Since the attractor values of the moduli are completely determined by the charges, this procedure
leads to a degeneracy formula that is independent of the moduli. At the attractor point in moduli space the following equations hold for the Narain moduli
\be
\label{attractor}
P_R|_{\text{\tiny{attr.}}}=0,\qquad Q_R|_{\text{\tiny{attr.}}}=0\;,
\ee
and the axion and dilaton are given by
\be
\label{attractor_1}
\l_1|_{\text{\tiny{attr.}}} = { P\cdot Q\over Q^2},\qquad \l_2|_{\text{\tiny{attr.}}}={|P\wedge Q|\over Q^2}\;.
\ee

In our favourite matrix notation, this reads
$$
\frac{1}{|P_L\wedge Q_L|}\!\bem P_L \cdot P_L &  P_L\cdot Q_L \\  P_L\cdot Q_L &  Q_L\cdot Q_L \eem\Bigr\rvert_{\text{\tiny{attr.}}} = 
\frac{1}{\l_2} \bem|\l|^2 & \l_1 \\ \l_1 & 1 \eem\Bigr\rvert_{\text{\tiny{attr.}}} = 
\frac{1}{|P\wedge Q|}\bem P \cdot P &  P\cdot Q \\  P\cdot Q &  Q\cdot Q \eem . 
$$

In this way, we find that at the attractor point our moduli-dependent contour reduces to the following moduli-independent expression
\be
\label{proposal_2}
\im\O= \varepsilon^{-1} X\vert_{\text{\tiny{attr.}}} = \varepsilon^{-1} \frac{\L_{P,Q}}{| \L_{P,Q} |} =\p\varepsilon^{-1}\, \frac{\L_{P,Q}}{S(P,Q)}
\;.
\ee

Again the $PGL(2,\zz)$ invariance is manifest, since both sides transform in the same way, and hence this prescription
also leads to a \(S\)-duality invariant counting formula. But what are the states that are being counted by this formula? 

In fact, we will now argue that these are precisely the \(1\!/4\)-BPS states that are not given by the bound states of two
\(1\!/2\)-BPS particles, and therefore cannot decay.
Namely, when one fixes the moduli to be at the attractor values, the stability condition (\ref{stability_2}) reduces to
$$
(\L_{P,Q},\a)^2 < 0\;,
$$
which can clearly never be satisfied.
In other words, none of the bound states of two \(1\!/2\)-BPS particles can exist at the attractor moduli,
which is a fact consistent with the general phenomenon that  an attractor flow always flows from the stable to the unstable side, a fact that we are already familiar with from our discussion of the walls of marginal stability of the \({\cal N}=2\) theory in section \ref{Walls of Marginal Stability}.

In this sense, our moduli-independent contour prescription leads to a counting formula which counts only the ``immortal" dyonic states that exist everywhere in the
moduli space.
Notice further that this class of contours is not defined for
charges with negative discriminant, since they lie outside of the Siegel domain. This is consistent with the fact that they do not have an attractor point, and there is no single-centered supergravity solution carrying these charges.

Finally we would like to briefly comment on the role of the number \(\varepsilon\) in our proposed contours
(\ref{contour_summary}), (\ref{proposal_2}).
It can be seen as playing the role of a regulator for the convergence of the generating function. To see
this, notice that when we take the contour according to our prescription (\ref{proposal_2}), the contribution
\be
\Bigl\lvert \, D(P,Q) \, e^{-i \p (\O,\L_{P,Q}) }  \Bigr\rvert =  |D(P,Q)| e^{- 2\varepsilon^{-1}  \p |P\wedge Q|}  \sim e^S e^{- 2\varepsilon^{-1}  S}\;
\ee
of certain large charges to the partition function is highly suppressed when  \(\varepsilon\ll 1\), and we are therefore left with  a rapidly converging generating function.

\section{Weyl Chambers and Discrete Attractor Flow Group}
\label{Weyl Chambers and Discrete Attractor Flow Group}
\setcounter{equation}{0}

Classically, a moduli space is a continuous space in which the vev's of the moduli fields of the theory can take their values. A distinct path in this space  for a given superselection sector, namely the total conserved charges, is the attractor flow of a single-centered black hole solution with a given starting point. 

As we have established in section \ref{Dying Dyons and Walls of Marginal Stability}, the asymptotic values of these scalar fields play a role in the spectrum of BPS states through the presence/absence of certain \(1/\!4\)-BPS bound states of two \(1/\!2\)-BPS objects. The walls of marginal stability for these bound states divide the moduli space into different regions in which the BPS spectrum is predicted to be constant by our supergravity analysis. This is because the spectrum only jumps when a wall of marginal stability is crossed, and for the purpose of studying the BPS spectrum of the theory we can identify the region bounded by a set of walls to be a point. 

From the above consideration, it is useful to consider a ``discrete attractor flow", which brings one region in the moduli space to another. It is not difficult to see that such an operation forms a group. Recall that in this specific theory we study, 
the walls of marginal stability are in one-to-one correspondence with the positive roots of a Weyl group \(W\), or the positive real roots of a Borcherds-Kac-Moody algebra introduced in section \ref{The Borcherds-Kac-Moody Superalgebra and the Denominator Formula}, as we have seen in section \ref{Dying Dyons and Walls of Marginal Stability}.  This implies that we should be able to identify the group of discrete attractor flow to be the same Weyl group \(W\). The aim of this section is to make this statement precise, and to address the implication of having such a group structure underlying the attractor flow.

\subsection{Weyl Chamber and Moduli Space}

To make the discussion more concrete let us visualise the situation. First recall that given a point in the 134-dimensional moduli space, whether a given two-centered solution exists only depends on the combination of the moduli field encapsulated in the following ``unit vector" (\ref{def_X_vec})
$$
X =  \frac{{\mathcal Z} }{ | {\mathcal Z}|}\;. 
$$
where 
$$
{\mathcal Z} = \frac{1}{\sqrt{|P_L \wedge Q_L|}} \bem P_L \cdot P_L &  P_L\cdot Q_L \\  P_L\cdot Q_L &  Q_L\cdot Q_L \eem + \frac{\sqrt{|P_L \wedge Q_L|}}{\l_2} \bem|\l|^2 & \l_1 \\ \l_1 & 1 \eem\;.
$$

This can be understood in the following way. Recall that the largest central charge is related to the above vector as
\be
|Z_{P,Q}|^2 = M_{P,Q}^2 = |{\mathcal Z} |^2
\ee
and the fact that, in the \({\cal N}=2\) language of chapter \ref{Black Holes and Multi-Holes}, the attractor flow is a gradient flow of the central charge \(|Z_{P,Q}|\) (see Figure \ref{central_charge_fig}), it is not surprising that the relevant part of the moduli is encoded in its direction \(X\).

Using the fact that the sheet of hyperbola of all future-pointing vectors of fixed norm \(| X|=1\) is equivalent to the upper-half plane \({\cal H}_1\) and the Poincar\'e disk (see Figure \ref{models_disk_fundamental_domain}), we can therefore map the relevant part (the \(X\)-space) of the moduli space onto \({\cal H}_1\) or the Poincar\'e disk. Concretely, we will use the map (\ref{tau_upper_half})
\bea\nonumber
X&=& \frac{1}{\t_2} \bem  |\t|^2 & \t_1 \\ \t_1 & 1  \eem\\
z&=& i \Big(\frac{\t+e^{-\frac{i\p}{3}}}{\t+e^{\frac{i\p}{3}}}\Big)\;.
\eea

Now we are ready to draw the walls of marginal stability (\ref{wall_2}). First we will begin with \(\a_1\), which corresponds to the two-centered solution with charges \((P,0)\), \((0,Q)\), for which the wall of marginal stability reads
$$
(X,\a_1) = 0\;.
$$
This gives an arc of a circle on the Poincar\'e disk, which is a geodesic with respect to the hyperbolic metric, and a straight line (a degenerate circle) in the upper-half plane. See Figure \ref{first_three_walls_graph}. 

As the next step, we will draw the walls given by other two roots \(\a_2\) and \(\a_3\) defined in (\ref{3_simple_roots}). Notice that the three walls \((X,\a_i) = 0\), \(i=1,2,3\) bound a triangle on the disk, and furthermore it is easy to show that the interior of the disk satisfies \((X,\a_i)< 0 \). For example, the center of the triangle is given by the normalised version \(\varrho/|\varrho|\)
 of the Weyl vector \(\varrho\) satisfying \((\varrho,\a_i)=-1\).

\begin{figure}
\centering
\includegraphics[width=10.5cm]{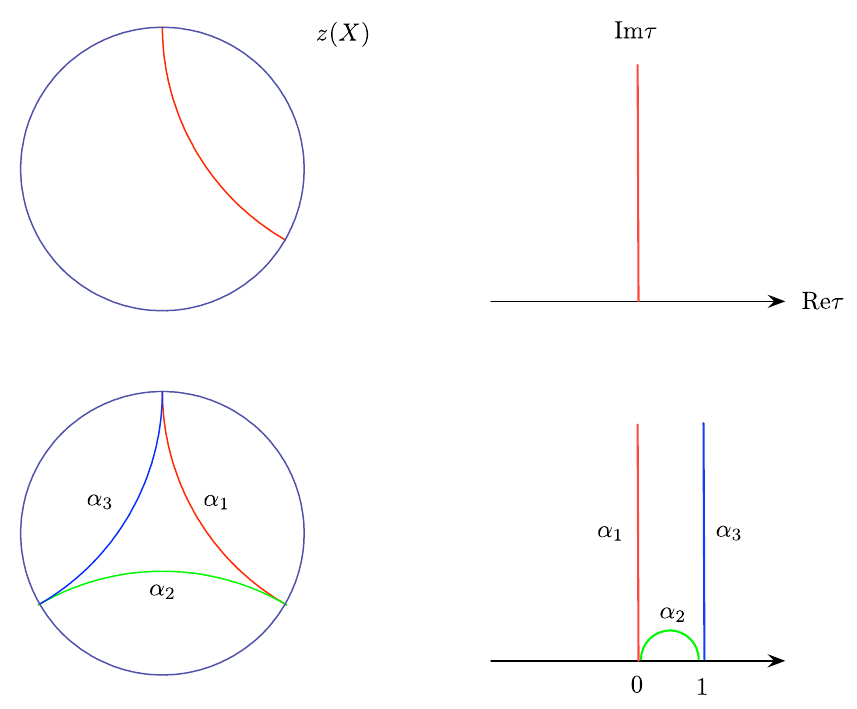} \setlength{\abovecaptionskip}{5pt}
\caption{\footnotesize{ \label{first_three_walls_graph} {\bf (i)} 
The wall of marginal stability for the two-centered solution with charges (P,0) and (0,Q),
projected onto the two-dimensional slice of moduli space equipped with a natural hyperbolic metric, and mapped to the Poincar\'e disk and the upper-half plane. {\bf (ii)} The basic three walls \((X,\a_i)=0\).
}}
\setlength{\belowcaptionskip}{5pt}
\end{figure}

As will be explained in the next section, given a charge vector one should choose the simple roots such that \(\L_{P,Q}\) lies inside the fundamental Weyl chamber. For concreteness of the discussion and without loss of generality, we will now assume that such a choice is given by the three simple roots we used in section \ref{Dyons and the Weyl Group}, namely that the charge vector satisfies 
\begin{align}\nonumber
& (\L_{P,Q},\a_i )  <   0  \quad,\quad i =1,2,3\\ \nonumber
&\a_1 =\left(\!\!\begin{array}{rrr} 0 &-1 \\ -1& 0 \end{array}\!\!\right)\quad,\quad
\a_2 =\bem 0 &1 \\ 1& 2 \eem\quad,\quad
\a_3 =\bem 2 &1 \\ 1& 0 \eem\,.
\end{align}
In other words, from now on we will consider charges such that the vector \(\L_{P,Q}\) lies  inside the central triangle in Figure \ref{first_three_walls_graph}. By the virtue of the relation between the Weyl group and the extended S-duality group \(W \subset  PGL(2,\Z)\), we can always use the duality group to map a set of charges \((P,Q)\) to another set of charges for which the above is true.

By definition, all the other real roots, in particular all the other positive real roots, are related to these three simple roots by a Weyl reflection \(\a= w(\a_i)\). We can thus draw the rest of the walls of marginal stability 
$$
(X,\a) = 0 \quad,\quad \a \in \D_+^{re}
$$
by reflections of the triangle in Figure \ref{first_three_walls_graph} with respect to the three sides. This gives a tessellation of the Poincar\'e disk as shown in Figure \ref{xxx}. Notice that the figure we draw can never be a faithful presentation of the real situation, since the group tessellates the disk with an infinite number of triangles.

By definition, the Weyl group divides the relevant part of the moduli space, namely the \(X\)-space, into different Weyl chambers bounded by the walls of orthogonalities with the positive roots. In other words, for any point in the moduli space, there exists a unique element of the Weyl group \(w\in W\) such that the corresponding moduli vector \(X\) lies in the Weyl chamber 
$$
X \in w({\cal W}) \Leftrightarrow (X,w(\a_i)) < 0 \;.
$$

Because these walls, or mirrors, of the Weyl chambers are exactly the physical walls of marginal stability \((X,\a) = 0 \) corresponding to the split into two centers (\ref{charge_split_new}),
we conclude that the BPS spectrum does not jump when the moduli move inside a given triangle. In other words,  
the Weyl chambers are exactly the region in the moduli space where the BPS spectrum is constant, and there is a different dyon degeneracy associated to every different Weyl chamber \(w({\cal W})\).

\begin{figure}
\centering
\includegraphics[width=10.5cm]{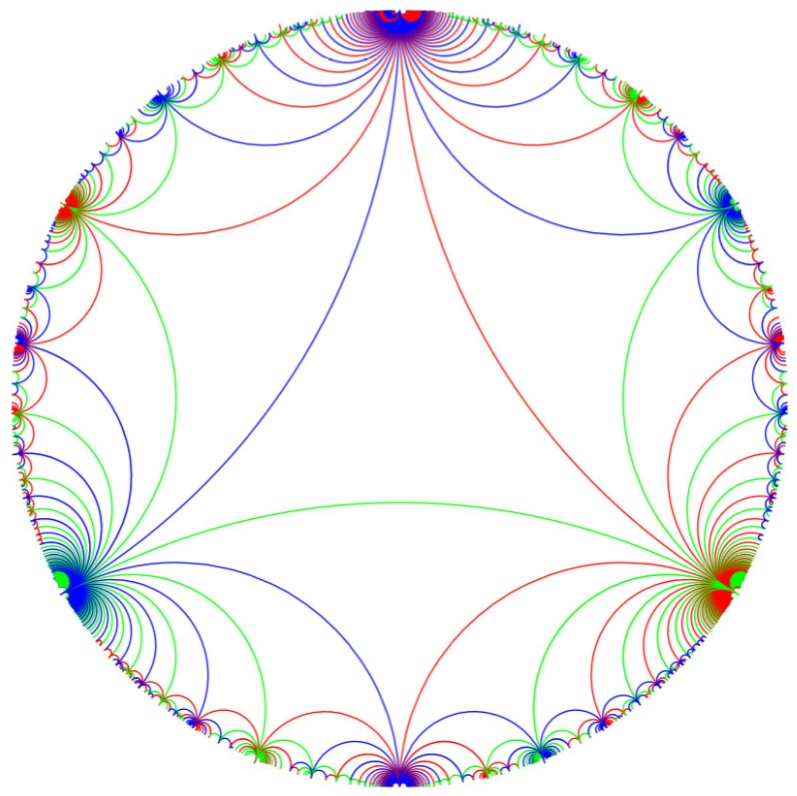} \setlength{\abovecaptionskip}{5pt}
\caption{\footnotesize{\label{xxx}Tessellation of the Poincar\'e disk using the group \(W\) generated by the reflection with respect to the three ``mirrors", namely the three sides of the regular triangle in the middle. Walls of the same color are mirror images of each other. Notice that each triangle has the same volume. The slight inhomogeneity of colours on the edge is an artifact of the computer algorithm we use.}}
\setlength{\belowcaptionskip}{5pt}
\end{figure}

\subsection{A Hierarchy of Decay}

Now we would like to discuss what the group structure of the moduli space discussed above implies for the dyon spectrum at a given moduli. In principle, given a point in the moduli space we know exactly which two-centered solutions exist by using the stability condition we worked out in section \ref{Dying Dyons and Walls of Marginal Stability}, namely that the solution given by the above split of charges exists if and only if
\be\label{stability_herhaling}
(X,\a)(\L_{P,Q},\a) < 0\;.
\ee
But in fact we know less than it might seem. It is because there are infinitely many decay channels, or to say that there are infinitely many positive real roots in the Borcherds-Kac-Moody algebra, so given a moduli vector \(X\), even equipped with the stability condition we will never be able to give a list of two-centered solutions within finite computation time. In this subsection we will see how the group structure changes this grim outlook.

First note that, in the central triangle, namely the fundamental Weyl chamber, there is no two-centered bound states. This can be seen as follows. Recall that we have chosen the simple roots such that 
$$
(\L_{P,Q},\a_i) < 0  \quad,\quad \a_i \in \{\text{simple real roots}\}\;,
$$
which implies
$$
(\L_{P,Q},\a) < 0  \quad,\quad \text{for all}\; \a \in \D_+^{re}\;
$$
simply from the definition of the positive real roots. Now the same thing holds for a point \(X\in {\cal W}\) inside the fundamental Weyl chamber, namely
$$
(X,\a) < 0  \quad,\quad \text{for all}\; \a \in \D_+^{re}\;,
$$
hence we see that the stability condition (\ref{stability_herhaling}) will never be satisfied for any split of charges. 

We therefore conclude that the fundamental Weyl chamber represents the ``attractor region" in the moduli space, namely the same region (chamber) where the attractor point lies and in which none of the two-centered solutions exists.

This is of course not true anymore once we move out of the fundamental Weyl chamber. Consider for example the neighbouring Weyl chamber \(s_1({\cal W})\), obtained by reflecting the fundamental chamber with respect to one of the simple roots \(\a_1\). Because this reflection takes \(\a_1 \to -\a_1\) and permutes the rest of the positive real roots (\ref{length_posi_negi}), we conclude that 
$$
(X,\a_1) >0\;,\;(X,\a)< 0 \quad\text{for all}\;\; X\in s_1({\cal W}), \a\neq \a_1\;, \a \in \D_+^{re}\;,
$$
which is also obvious from the picture. 
This means that there is now one two-centered solution corresponding to the split into charges \((P,0)\), \((0,Q)\) (\ref{stability_1}) and no others. 

We can now go on with this process to every Weyl chamber in the (reduced) moduli space: go to the next-neighbouring chamber, and the next-next-neighbouring, and so on, with the condition that the path doesn't walk ``backwards", or more precisely that the length function (see \ref{length_function}) of the corresponding group element always increases.  

In general, considering an arbitrary point on the disk
$$
X \in w({\cal W})
$$
for some \(w\in W\). For any group element, we can always decompose it as the following ``shortest word" in terms of the three group generators
\begin{align}\nonumber
&w = s_{i_1}s_{i_{2}}\dotsi s_{i_n}\\
\label{decomposition_1}
&i_m \in \{1,2,3\}\;\;,\; i_m \neq i_{m\pm 1} \;\text{   for all    } m = 2,\dotsi,n-1\;.
\end{align}
From the fact that a reflection with respect to the root \(w(\a_i)\) is given by the group element \(w s_i w^{-1}\):
$$
w s_i w^{-1}\big( w(\a_i) \big) = w s_i(\a_i) = -w(\a_i)\;,
$$
 we can describe the Weyl chamber \(w({\cal W})\) as given by the following successive reflection of the fundamental Weyl chamber: 
\be\label{bruhat_decay}
w_0 = {\boldsymbol 1}  \xrightarrow[\a_{i_1} ]{}  \,w_{1} \,\xrightarrow[w_{1}(\a_{i_{2}}) ]{}\, w_{2} \;\dotsi \,\xrightarrow[w_{n-1} (\a_{i_n})]{}  \quad w_n=w\\ 
\ee

where `` \(\xrightarrow[\;\;\a\;\; ]{}\)" means ``reflecting with respect to the wall \((\a,X)\)=0 given by the root \(\a\)"
and the intermediate group elements \(w_m\) are given by
$$
w_m = s_{i_1}\dotsi s_{i_m}\quad,\quad m\leq n\;.
$$

In other words, for a given point in the moduli space and its Weyl chamber
$$
X\in w({\cal W})\;,
$$
by following the above path from the attractor region \({\cal W}\), or the fundamental Weyl chamber, to \(w({\cal W})\), we can read out the two-centered solutions which exist at the point \(X \in {\cal W}\). These are given by the charge splitting
\ben
\L_{P_1,Q_1} &=& P_\a^2\,\a^+ \quad,\quad\L_{P_2,Q_2} =  Q_\a^2\,\a^-\\
\L_{P,Q} &=& P_\a^2\,\a^+ + Q_\a^2\,\a^-  - |(P\cdot Q)_\a|\,\a\;,
\een
with now
$$
\a \in \{ w_{m-1}(\a_{i_m})\;,\quad m=1,\dotsi,n\} \subset \D_{+}^{re} \;.
$$
 
 In other words, when we follow the journey from the attractor region \({\cal W}\) to the Weyl chamber \(w({\cal W})\) where the moduli is, namely when we follow the inverse attractor flow to the point under consideration, we will successively cross the walls of marginal stability corresponding to the roots \(\a_{i_{1}}\), \(w_{1}(\a_{i_{2}})\), ... , and finally \(w_{n-1}(\a_{i_{n}})\).
 
This gives a simple dictionary to read out the list of dying dyons for any given point in the moduli space, provided that we know the shortest decomposition of the group element \(w\) in terms of a string of generators (``letters"). 

To complete the algorithm, we also give a very simple algorithm to determine such a decomposition given an arbitrary lightlike, future-pointing vector \(X\). Given a point \(X\), we would like to determine the shortest-length string 
$$
w = s_{i_1}\dotsi s_{i_n}
$$
such that its corresponding chain of reflection induces the following successive mapping of the vector \(X\) into the fundamental Weyl chamber
\begin{align}\nonumber
w_0 = {\boldsymbol 1}   \xrightarrow[\a_{i_1} ]{}  \,w_{1} \,\xrightarrow[w_{1}(\a_{i_{2}}) ]{}\,   \dotsi \,&\xrightarrow[w_{n-1} (\a_{i_n})]{}  w_n=w\\ \nonumber
 X_0 \in {\cal W}  \xrightarrow[\a_{i_1} ]{}  \,X_1\,
 \xrightarrow[w_{1}(\a_{i_{2}}) ]{} \, \dotsi&\xrightarrow[w_{n-1} (\a_{i_n})]{}  X_n=X \in w({\cal W})\;.
\end{align}
One can show that the string is determined as follows: suppose
\be
X_m = s_{i_1}\dotsi s_{i_m} X_0\quad,\quad m\leq n\;,
\ee
then  \(i_m\in\{1,2,3\}\) is given by the condition
\be
(\a_{i_m},X_m) > 0\;,
\ee
which has at most one solution for \(i_m\). We can go on with this process for \(X_{m-1}\) until the above equation has no solution anymore, corresponding to when the three expansion coefficients of \(X_{0} = \sum_{i=1}^3 \a_i X_0^{(i)} \) satisfy the triangle inequality. This is when the decay ends and when the moduli flow to the attractor region given by the fundamental Weyl chamber.

Notice that there is a hierarchy of decay (or a ``death row") in this process. Namely, considering another point in the moduli space which is in the Weyl chamber \(w_m({\cal W})\) with \(m< n\), applying the same argument as above shows that the two-centered solutions existing in that Weyl chamber are given by the first \(m\) positive roots in the above list. Specifically, this argument shows that there is nowhere in the moduli space where the bound states given by the root \(w_{n-1}(\a_{i_{n}})\) exists without all the other \(n-1\) bound states given by \(w_{m-1}(\a_{i_{m}}), m<n \) in front of it in the row. 

 This hierarchy among two-centered solutions clearly stems from the hierarchy among elements of the group \(W\). Indeed, the ordering in (\ref{bruhat_decay}) is an example of what is called the ``weak Bruhat order" \(w_m < w_{m+1}\) among elements of a Coxeter group. See (\ref{weak_order}) for the definition. 
 
In our case this weak Bruhat ordering has an interpretation reminiscent of the RG-flow of the system. From the integrability condition (\ref{distance_dyon_center})
$$
\sqrt{|P_L\wedge Q_L|}\, {|\,\vec{x}_{P_\a}-\vec{x}_{Q_\a}|} = \Big\rvert
\frac{(\L_{P,Q},\a)}{(X,\a)}\Big\lvert\;,
$$
now with \(\a = \{ w_{m-1}(\a_{i_m}), m \leq n\}\), 
we see that the ordering of the decay is exactly the ordering of the coordinate size of the bound state. In other words, roughly speaking, the ordering we discussed above can be summarised as the principle that the bound state which is the bigger in size are more prone to decay than the smaller ones, which is a fact parallel to the usual RG-flow phenomenon. See Fig \ref{messy_disk} for a simple example of the flow beginning from a point in \(s_1s_3({\cal W})\).

\begin{figure}[h]
\centering
\includegraphics[width=9cm]{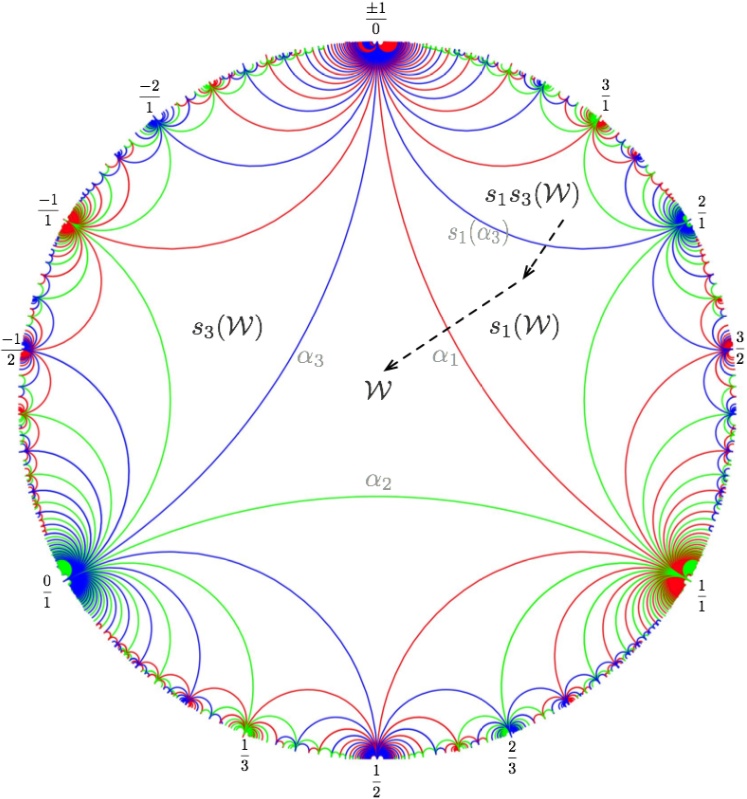} \setlength{\abovecaptionskip}{5pt}
\caption{\footnotesize{\label{messy_disk}(i) An example of a discrete attractor flow from \(X \in s_1s_3({\cal W}) \) to the attractor region 
\({\cal W}\), passing through two walls of marginal stability \((s_1(\a_3),X)=0\) and \((\a_1,X)=0\).  (ii) The boundary of the disk can be identified with the boundary of the light-cone, and in turn be identified with a compactified real line using the map (\ref{map_lightcone_number}). In this way there is a pair of rational numbers associated to each positive real roots, and a Weyl chamber is given by such a pair with its mediant. Furthermore, the discrete attractor flow can be thought of as a process of ``coarse-graining" the rational numbers.}}
\setlength{\belowcaptionskip}{5pt}
\end{figure}

As was alluded to earlier, we can identify the path taken by such a string of reflections as the path taken by a discrete version of attractor flows. In other words, if we identify all points in a given Weyl chamber, justified by the fact that all points there have the same BPS spectrum, the usual continuous attractor flow reduces to the successive reflections discussed above. To argue this, note that the (single-centered) attractor flow also gives such a structure of hierarchy among multi-centered solutions, because an attractor flow can only cross a given wall at most once. Using the \({\cal N}=2\) language of chapter \ref{Black Holes and Multi-Holes}, this is because for a single-centered solution, the quantity \(\im(\bar{Z} Z')/|Z|\) is linear in \({1}/{r}\), the inverse of the coordinate distance from the black hole, where \(Z'\) is the central charge of another arbitrary charge at the moduli where the flow is, and can therefore only pass zero at most once \cite{Denef:2001xn}.  On the other hand, such a hierarchy among solutions can only be a property of the structure of the moduli space, since the stability condition is a local condition on the moduli and is in particular path-independent. Hence we have to conclude that the group and the flow must cross the walls in the same order. This justifies our claim that the group is simply the discretised group of attraction.

\subsection{Arithmetic Attractor Flows}

The title of this subsection is very similar the title ``Arithmetic and Attractors" of the classic paper by Moore \cite{Moore:1998pn,Moore:1998zu}. As suggested in the title, in this section we will discuss the arithmetic aspects of our newly defined discrete attractor ``flow"  group.

First let us again review the equation for the walls of marginal stability (\ref{charge_split_new}), which states that for any positive real root \(\a\), the charge vector \(\L_{P_1,Q_1}\), \(\L_{P_2,Q_2}\) of the associated two-centered solution are given by the component of the total charge vector \(\L_{P,Q}\) along the two lightcone direction \(\a^\pm\) perpendicular to \(\a\). 

From the above conditions, it is not difficult to see that the two-centered solutions can equivalently be given by a pair of rational numbers \(\{b/a,d/c\}\) satisfying \(ad-bc=1\),
such that the lightlike vectors are given by
\bea\label{map_lightcone_number}
\{\a^+,\a^-\} = \Bigg\{ \bem b^2 & ab \\ ab & a^2\eem, \bem d^2 & cd \\ cd & c^2\eem \Bigg\}
\eea
Notice that exchanging the two rational numbers amounts to exchanging \((P_1,Q_1)\) and \((P_2,Q_2)\), which obviously does not give a new solution.
Without loss of generality, we now impose that \(a\geq 0, c\geq 0\), 
while \(b,d\) can take any sign. 

To be more precise, given such a pair of rational numbers, the corresponding positive roots is
\be
\a = \bem 2bd & ad+bc \\ ad+bc & 2ac \eem\;, 
\ee
and the corresponding charge splitting is 
\be\label{charge_split_rational_number}
\bem P_1 \\ Q_1 \eem = (-c P + dQ) \,\bem b \\ a \eem \quad,\quad
\bem P_2 \\ Q_2 \eem = (a P - bQ) \,\bem d \\ c \eem\;.
\ee

In other words, the above formula gives an alternative labelling of the two-centered solutions of the theory by a pair of rational numbers 
\be\label{pair_rational_number}
\Big\{\frac{b}{a},\frac{d}{c}\Big\}\quad,\quad ad-bc=1\quad,\quad a,c \geq 0\;.
\ee

In particular, the three simple roots (\ref{3_simple_roots}) correspond to the three sets of rational numbers \(\{\frac{-1}{0},\frac{0}{1}\}\),  \(\{\frac{0}{1},\frac{1}{1}\}\),  \(\{\frac{1}{1},\frac{1}{0}\}\) respectively. 

Now we would like to know what the discrete attractor flow, defined in the last subsection, looks like in terms of the presentation in terms of rational numbers. From the figure \ref{messy_disk} it is obvious that, for a positive root bounding the Weyl chamber \(w({\cal W})\), one of the following two roots \(wOw^{-1}(\a)\) and \(wO^2w^{-1}(\a)\) must be negative, where \(O\) is the order three generators of the dihedral group corresponding to a rotation of \(120^\circ\) (\ref{dihedral_action_2}), and this must be the root with respect to which the last reflection in the sequence  (\ref{bruhat_decay}) is. Therefore, from the computation which gives the following expression for \(-wOw^{-1}(\a)\) and \(-wO^2w^{-1}(\a)\)
$$
\bem 2b(d-b) & bc+ad-2ab \\ bc+ad-2ab & 2c(c-a) \eem \;,\;\bem 2d(b-d) & bc+ad-2cd \\ bc+ad-2cd & 2a(a-c) \eem\;,
$$
we conclude that given a two-centered solution corresponding  to the pair rational numbers \(\{b/a\), \(d/c\}\) with \(a,c \geq 0\) and \(ad-bc=1\), we can read out the next two-centered solution on the list of decadence (or the death row) as another pair of rational numbers
$$
\Big\{\frac{b}{a},\frac{d-b}{c-a}\Big\}\;\;\text{if}\;\;c\geq a, bd \geq b^2\quad,\quad
\Big\{\frac{d}{c},\frac{b-d}{a-c}\Big\}\;\;\text{if}\;\;a \geq c, bd \geq d^2\;.
$$

This rule is actually much simpler than it might seem. Consider the boundary of the Poincar\'e disk as a compactified real line, namely with \(\frac{\pm1}{0}\) identified, then the map (\ref{charge_split_rational_number}), (\ref{pair_rational_number}) associates to each wall of Weyl chamber, or equivalently wall of marginal stability, a pair of rational numbers satisfying the above conditions. See Figure \ref{messy_disk}. Now what we saw above simply means: firstly, any triangle is bounded by a set of three rational numbers of the form
\be
\Big\{\frac{b}{a},\frac{b+d}{a+c},\frac{d}{c}\Big\}\quad,\quad ad-bc=1\quad,\quad a,c \geq 0\;.
\ee
The middle element \((b+d)/(a+c)\) is called the ``mediant" of the other two, and can be easily shown to always lie between the two
$$
\frac{b}{a}<\frac{b+d}{a+c}<\frac{d}{c}\;.
$$ 
Secondly, the direction of the attractor flow is that it always flows to the numbers with smaller \(|a|+|b|\), \(|c|+|d|\), and can therefore be seen as a flow of coarse-graining the rational numbers. 

\begin{figure}
\centering
\includegraphics[width=14cm]{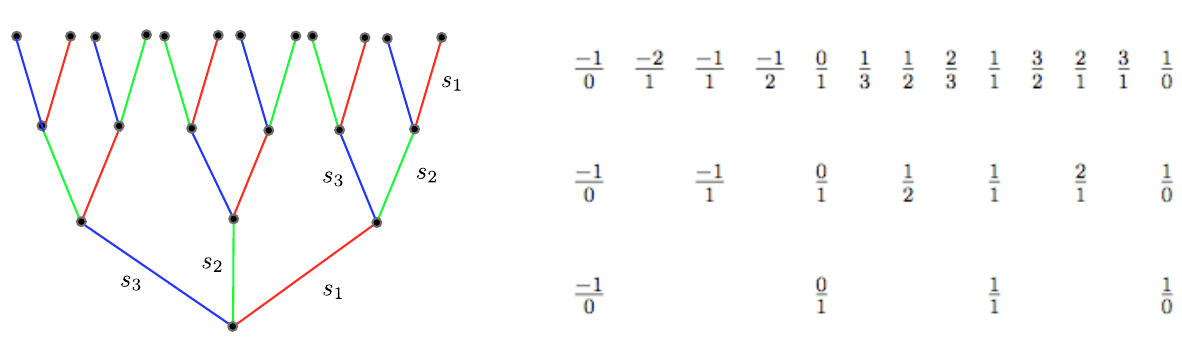} \setlength{\abovecaptionskip}{5pt}
\caption{\footnotesize{\label{stern_brocot_bruhat} ({\bf}l) The (first three levels of the) weak Bruhat ordering of the group \(W\), which corresponds to the hierarchy of the wall-crossing of the theory. ({\bf}r) The corresponding coarse-graining of the rational numbers, which is a part of the Stern-Brocot tree generalised to the whole real line. }}
\setlength{\belowcaptionskip}{5pt}
\end{figure}
\index{Farey series}
\index{Stern-Brocot tree}
As an illustration of the above statement, in Figure \ref{stern_brocot_bruhat} we show the rational numbers corresponding to the first three levels of inverse attractor reflections from the attractor region, corresponding to the first three levels of the weak Bruhat tree.

Note that this is simply the generalisation of the so-called Stern-Brocot tree to the negative part of the real line, and notice that {\it all} the rational numbers are contained in this tree. In particular, the Farey series is contained in the middle part of the tree. 

\section{The Algebra and the BPS States}
\label{The Algebra and the BPS States}
\setcounter{equation}{0}

\subsection{Wall-Crossing and Representations of the Algebra}
\label{Wall-Crossing and Representations of the Algebra}

Now we would like to pause and ask ourselves the following question: what does it mean that a different choice of contour gives a different answer for the BPS degeneracy? After all, the counting formula (\ref{DVV_1}) we derived using the D1-D5 CFT does not seem to have any ambiguity. And since we know that the difference between different answers are exactly accounted for by the two-centered solutions in supergravity, what can we say about the states corresponding to these solutions? 

First we begin with the first question. Although the formula (\ref{DVV_1}) might look unambiguous, the ambiguity really lies in how we expand the right-hand side of the equation. For example \cite{Dabholkar:2007vk}, the two possible ways of expanding the following factor of the partition function
\ben
\frac{1}{(y^{1/2}-y^{-1/2})^2} &=& \frac{1}{y(1-y^{-1})^2} = y^{-1} + 2y^{-2} + \dotsi \\ \nonumber
&=& \frac{1}{y^{-1}(1-y^{1})^2} = y^{1} + 2y^{2} + \dotsi \;,
\een
corresponding to two possible ranges for the parameter \(y>1\) and \(y<1\), will give different answers for the degeneracies. It is not hard to convince oneself that this ambiguity of choosing expansion parameters is exactly the same ambiguity as that of choosing integration contours when we invert the equation. To be more precise, rewrite the equation (\ref{DVV_1}) in the following form as in section \ref{The Borcherds-Kac-Moody Superalgebra and the Denominator Formula}
$$
\sum_{P,Q} (-1)^{P\cdot Q+1} D(P,Q) e^{-\p i (\L_{P,Q},\O)} = \left( 
 \frac{1}{e^{-\p i(\varrho,\O)}\,\prod_{\a\in \D^+}\,\Big(1-e^{-\p i (\a,\O)}\Big)^{{\text {mult}}\a}}\right)^2\;.
$$
Apparently, we should expand the product factor in powers of \(e^{-\p i (\a,\O)}\) when (\(\im\O, \a) < 0\) and in powers of \(e^{\p i (\a,\O)}\) when \(({\mathrm{Im}}\O, \a) > 0\).
In other words, the prescription for contours is equivalent to the prescription for a way to expand the generating function.

We can give the above prescription another interpretation which makes the role of the Borcherds-Kac-Moody algebra more manifest. Let's consider the Verma module \(M(\L)\) of this algebra  with highest weight \(\L\) and its super-character. Besides the denominator (\ref{denominator_general}), the character formula also contains a numerator. Using the formal exponential introduced in section \ref{The Borcherds-Kac-Moody Superalgebra and the Denominator Formula}, 
the super-character reads
\bea\label{verma_character}
\text{sch} M(\L) &=& \sum_{\m\leq \L}\,\text{sdim}( M(\L)_\m)\,e(\m)\\
\nonumber
&=&
\frac{e(-\varrho+\L)}{e(-\varrho){\prod_{\a\in \D^+}\big( 1-e(-\a)\big)^{\text{mult}\a}}}  
=\frac{e(\L)}{\prod_{\a\in \D^+}\big( 1-e(-\a)\big)^{\text{mult}\a}}\;,
\eea
where \(M(\L)_\m\) denotes the weight-\(\m\) sub-module of the Verma module \(M(\L)\), and ``\(\m\leq \L\)" means that \(\L - \m\) is a sum of simple roots. The ``s" of ``sdim" denotes the fact that we are dealing with the graded characters counting the graded degeneracies, taking the plus or minus sign depending on whether the root involved is even or odd. Indeed, recall that in section \ref{The Borcherds-Kac-Moody Superalgebra and the Denominator Formula} we have defined the number {\small mult}\(\a\) to be the graded multiplicities of the root \(\a\). 

Let's now compare this character formula with the integral 
\begin{align}\nonumber
&(-1)^{P\cdot Q+1} D(P,Q)\vert_{\l,\m} = \oint_{{\cal C}(P,Q)\vert_{\l,\m}} \, d\O \,\left( \frac{e^{i\frac{\p}{2}  (\L_{P,Q},\O) }}{e^{-\p i(\varrho,\O)}\prod_{\a\in \D^+}\big( 1-e^{-\p i(\a,\O)}\big)^{\text{mult}\a}} \right)^2\\ 
\label{DVV_integral2_herhaling}
&{\cal C}(P,Q)\vert_{\l,\m} = \{\re \O \in T^3 , \im\O =  \varepsilon^{-1} X\}\;,
\end{align}
we note that the integrand is exactly the square of sch\(M(\L)\),
namely the square of the super-character of the Verma module with highest weight
\be
\L= \varrho + \frac{1}{2} \L_{P,Q}\;,
\ee
and the contour integral has the function of picking up the zero-weight sub-module. Hence the dyon degeneracy \((-1)^{P\cdot Q+1} D(P,Q)\) has the interpretation of counting the (graded) number of ways the weight \(2\L\) can be written as a sum of two copies of positive roots. Equivalently, one can interpret the dyon degeneracy as the ``second-quantized multiplicity" of the weight $2\L$, generated by a gas of freely-acting bosonic and fermionic generators given by two copies of the positive roots of the algebra.

What we just saw is that the dyon degeneracies have a nice interpretation in terms of positive roots of the algebra which seems to be free of ambiguities, so we might wonder where the contour/moduli ambiguities we have seen from the integral/supergravity viewpoint goes in this picture.  The subtlety lies in the fact that, the character formula (\ref{verma_character}) written in terms of the formal exponentials \(e(\m)\) satisfying \(e(\m)e(\m')=e(\m+\m')\) contains the same information as the integral formula in terms of functions \(e^{\p i (\O,\m)}\) of \(\O\) only  if we expand all the expressions in the latter formula in powers of \(e^{\p i (\O,\m)}\). From our discussion above, this means that for this interpretation to be correct, one must make the (unique) choice of simple roots, and thereby the choice of positive roots, in the character formula such that the moduli vector \(X\) lies in the fundamental Weyl chamber of the root system.
A manifest but crucial fact to keep in mind here is that the character formula (\ref{verma_character}) for Verma modules in {\it not} invariant under a change of simple roots. 

What we have concluded from the above reasoning is that, for a super-selection sector with given total charges, a different choice of moduli corresponds to a different choice of positive roots in the algebra. This is nevertheless not very convenient. Instead we will use the equivalent description of letting the highest weight of the module be moduli-dependent while keeping the simple roots fixed. 

No matter whether we choose to keep the highest weight fixed and vary the simple roots when we vary the moduli, or keep the simple roots fixed and vary the highest weight, from our contour condition 
 it's clear that only in the attractor region, characterised by 
$$
(X,\a) (\L_{P,Q},\a) > 0 \quad,\quad \text{for all}\quad\a\in \D^{re}\;,
$$ 
does the counting formula corresponds to a Verma module of dominant highest weight. Recall that a dominant weight is a vector lying in the fundamental Weyl chamber and having integral inner product with all roots. 

For a given set of total charges, a natural choice for the simple roots is therefore such that the charge vector lies in of the fundamental domain\footnote{We ignore the special cases where the charge vector lies on the boundary of some fundamental domain, corresponding to the situation of having two-centered scaling solutions.}
$$
\L_{P,Q} \in{\cal W}\;. 
$$

When the moduli do not lie in the attractor region, the corresponding Verma module will not have dominant highest weight.  
Indeed, when considering a point in the moduli space corresponding to
another Weyl chamber
$$
X \in w({\cal W})\quad,\quad w\in W\;,
$$
either from the contour integral (\ref{DVV_integral2_herhaling}) or from the character formula (\ref{verma_character}) we see that the dyon degeneracy is encoded in the Verma module of the following highest weight  
$$
\L_w =  \varrho + w^{-1}(\L-\varrho) \;.
$$
Notice that the ambiguity we discuss above does not involve the imaginary positive roots, defined as those positive roots that are timelike or lightlike, which give the majority of factors in the product formula, and are therefore responsible for the asymptotic growth of the degeneracies. This is because, because our \(X\) is in the future lightcone by construction, the convergence criterion
$$
(X,\b) < 0 \quad,\quad \b \in \D_+^{im}
$$
is guaranteed to be met, independent of the choice of \(X\). This also justifies the fact that the Weyl group of a Borcherds-Kac-Moody algebra is the reflection group with respect to the real roots only. 

Therefore, we arrive at the conclusion that crossing a wall of marginal stability corresponds to a change of representation of the BPS algebra microscopically. More precisely, the change is such that the highest weight of the Verma module is changed by a Weyl reflection, and away from the attractor region, the highest weight of the representation will no longer be dominant.

\subsection{Microscopic Derivation of the Wall-Crossing Formula}
\label{Microscopic Derivation of the Wall-Crossing Formula}

In the last subsection we have seen how the dyon-counting formula together with the prescription for its contour of integration leads us to
a moduli-dependent prescription of a highest weight, whose Verma module encodes the degeneracies of BPS dyons at the given moduli. In this subsection we will see what the group structure of wall-crossing implies for this microscopic description of BPS states. In particular, we will see how the wall-crossing formula predicted using the supergravity analysis can be derived from this microscopic point of view. 

To begin let us consider the sequence of reflections (\ref{bruhat_decay}) as described in section \ref{Weyl Chambers and Discrete Attractor Flow Group}. 

From the above discussion about the correspondence between the moduli and the highest weights appearing in the dyon-counting formula, we see that the above weak Bruhat ordering gives the following sequence of the highest weights: 
\be
\L =\L_{w_0}\to \L_{w_1}\to \dotsi \to  \L_{w_n} \;.
\ee
From the fact that the charge vector \(\L_{P,Q}\) lies inside the fundamental chamber in the future light-cone, it is not difficult to show that 
$ w^{-1}_{m}(\L_{P,Q})> w^{-1}_{m-1}(\L_{P,Q}) $
 at all steps, namely that the difference of the two and therefore the difference $\L_{w_m} -\L_{w_{m-1}} $
 is always some sum of simple real roots. 

To see this explicitly, 
consider two points in the moduli space inside the two neighbouring Weyl chamber $w_{m-1}({\cal W})$ and $w_{m}({\cal W})$ respectively, separated by the wall $(X,\a)=0$  with
$$
\a= w_{m-1}(\a_{i})\;
$$
where $\a_i$ can be any of the three simple roots. In other words, we have
$$ 
w_m^{-1} = s_{i} w_{m-1}^{-1}\;.
$$

It will be convenient to use the root $\a$ and the corresponding light-like vectors $\a^\pm$ as a basis and 
decompose the charge vector $\L_{P,Q}$ in the way introduced in (\ref{charge_split_new}).
 
This gives the image of the charge vector under the Weyl group element $w_{m-1}^{-1}$ as
\be
w^{-1}_{m-1} (\L_{P,Q})= P_\a^2 \a_{i}^+ + Q_\a^2 \a_{i}^- - |(P\cdot Q)_\a| \,\a_i \;.
\ee

By construction we have $(\a^{\pm},\a ) = 0$, which implies that the light-like vectors $\a_i^\pm$ are invariant under the reflection $s_i$ and therefore
\be
w^{-1}_{m}(\L_{P,Q}) = P_\a^2 \a_{i}^++ Q_\a^2 \a_{i}^- + |(P\cdot Q)_\a|\, \a_i \;.
\ee

We have therefore shown that 
$$
\L_{w_{m}} - \L_{w_{m-1}}  =\,(\L_{w_{m}}-\varrho,\a_i)\, \,\a_i
$$

Therefore, the hierarchy (the weak Bruhat ordering) of the group elements, which have been identified with the direction given by the attractor flows, induces the following hierarchy of Verma modules of the BKM algebra
\be\label{sequence_module}
M(\L)= M(\L_{w_0})\subset M(\L_{w_1})\subset \dotsi \subset M(\L_{w_n})\;.
\ee

In this sequence, every module on the left in a sub-module on the right. Explicitly, the highest weight vector of the $(m-1)$-th module is the highest weight vector of the $m$-th module acted on by the element
$$
f_{i_m}^{\,(\L_{w_{m}}-\varrho,\,\a_{i_m})\,}
$$
of the algebra. Indeed, one can show that the combination $e_j f_{i_m}^{\,(\L_{w_{m-1}}-\varrho,\,\a_{i_m})\, }
$  annihilates the highest weight vector for all $j$ and this guarantees that the next module $M(\L_{w_{m-1}})$ along the attractor flow is again a highest weight module.

Next we are interested in the following question: what does this hierarchy of representations of the algebra imply for the counting of BPS states when the background moduli are dialled along the attractor flow?

Let's again concentrate on the neighbouring regions $w_{m-1}({\cal W})$ and $w_{m}({\cal W})$. We have just seen that the difference between the corresponding highest weights of the relevant Verma modules is a multiple of the real simple roots $\a_i$. For counting the difference between the number of dyonic states we want to exploit the fact that all but one factor in the product $\prod_{\b\in \D_+} (1- e(-\b))^{\text{mult}\b}$ is invariant under the Weyl reflection $s_i$, 
which can be either concluded from the fact that the Weyl reflection $s_i$ reflects the positive root $\a_i$ and permutes all the others (Theorem \ref{length_posi_negi}), or by inspection of the explicit expression for the present algebra (\ref{DVVPF_2}). This difference in  transformation properties of the positive roots under the Weyl reflection with respect to the simple root $\a_i$ naturally divides the product formula into two factors
$$
\prod_{\a\in \D_+} \left(\frac{1}{1-e(-\a)}\right)^{2{\text {mult}}\a} =  \frac{1}{(1-e(-\a_i))^2} \times\prod_{\substack{\b\in \D_+\\ \b\neq \a_i}} \Big(\frac{1}{ 1- e(-\b)}\Big)^{c(|\b|^2)} 
$$

We will thus from now on use the simple root $\a_i$ and the corresponding light-like vectors $\a_i^\pm$ as our basis for the space $\R^{2,1}$. In particular, it's easy to check that the Weyl vector can be written in this basis as
$$
\varrho = \a_{i}^+ + \a_{i}^- -\frac{1}{2} \a_i \;.
$$

Now consider the highest weight  
$$
2 \L_{w_{m}} 
= (P_\a^2 +2 )\a_i^+ + (Q_\a^2 +2 )\a_i^- + \Big( |(P\cdot Q)_\a| -1\Big)\a_i
$$
as the sum of two vectors $n \a_i$ and $2 \L_{w_{m}} -n\a_i$, where the first vector comes from the first factor of the product formula, corresponding to the generator $\a_i$. We can then write the dyon degeneracy at moduli space $X\in w_{m}({\cal W})$ as
\be
(-1)^{P\cdot Q+1}D(P,Q)\lvert_{w_{m}}
= \sum_{ n\geq0} (n+1)\,f(n)\;.
\ee
where we have defined the integer $f(n)$ to be the degeneracy of the state corresponding to the weight $2 \L_{w_{m}} -n\a_i$, generated by
generators given by all positive roots but with $\a_i$ excluded. The pre-factor is simply the degeneracy of a state of oscillation level $n$ when there are two identical harmonic oscillators. From the same consideration we have for the other side of the wall
\be
(-1)^{P\cdot Q+1}D(P,Q)\lvert_{w_{m-1}}
= \sum_{ n\geq0} (n+1)\,f(n+2 |(P\cdot Q)_\a|)\;.
\ee

Now, the fact that the Weyl reflection with respect to the real simple root $\a_i$ only takes the simple root to minus itself while permuting the rest of the positive roots simply means the property of the degeneracy $f(n)$ that the two integers $n$ and $2|(P\cdot Q)_\a|-2-n$ have the same image under the map $f:\Z \to \Z$. Furthermore, the limiting expression (\ref{pole_look_like}) at the pole $(\O,\a_i)\to 0$ of the partition function is equivalent to the statement that
\be
\sum_{n=-\inf}^{\inf} f(n)  = d(P_\a) \,d(Q_\a)\;.
\ee

Using these two properties of the degeneracy $f(n)$ and after some manipulation of the formula, we obtain
\be
D(P,Q)\lvert_{w_{m}}
=D(P,Q)\lvert_{w_{m-1}} + (-1)^{(P\cdot Q)_\a+1}\, |(P\cdot Q)_\a|\,d(P_\a) \,d(Q_\a)\;,
\ee
where we have also used  $(P\cdot Q)_\a = P\cdot Q$ mod 2, which can be shown by the fact that the charge lattice $\G^{6,22}$ is even. Here we see that what we derived above from considering the different representations corresponding to different part of the moduli space, is exactly the wall-crossing formula derived using the macroscopic consideration. In other words, the difference between the microscopic degeneracies in two neighbouring Weyl chambers separated by the wall $(X,\a)=0$
 is precisely the entropy of carried by the two-centered supergravity solution given by $\a$. Therefore, the wall-crossing formula which follows from our proposal for counting dyon as the second-quantized multiplicity of a moduli-dependent highest weight coincides with the two-centered wall-crossing formula derived using supergravity analysis, and in fact can be 
thought of  providing a microscopic derivation of the two-centered wall-crossing formula.

\section{Summary and Conclusion}
\setcounter{equation}{0}
In this chapter we study the moduli dependence of the BPS dyon degeneracies of a \({\cal N}=4\), \(d=4\) string theory, and its relationship to a counting formula and a Borcherds-Kac-Moody algebra. In section \ref{Dying Dyons and Walls of Marginal Stability} we study the stability condition for a solution in the low-energy supergravity theory which describes the bound state of two \(1/\!2\)-BPS objects, to exist. We show that these are in one-to-one correspondence with the positive real roots of the Borcherds-Kac-Moody algebra. In section \ref{Contour Dependence of the Counting Formula} we study the dependence of the dyon-counting formula on the possible choices of integration contours, and chart the difference between different BPS degeneracies predicted using different contours of integration. In the following section we show how a moduli-dependent contour prescription can relate the two ambiguities: that of choosing a contour and that of choosing the moduli of the theory, such that the dyon-counting formula correctly accounts for the moduli-dependence of the BPS spectrum. 

After that  we turn to the question of how a certain Borcherds-Kac-Moody algebra is related to the above phenomenon, since we have seen in the previous chapter that the counting formula is mathematically related to the denominator formula of this  algebra. 

First we show that the Weyl group of the algebra has the physical interpretation as the group of wall-crossing, or the group of discrete attractor flow of the theory. This is done by relying on the analysis of the low-energy effective supergravity theory which shows how the walls of the Weyl chambers can be identified with the physical walls of marginal stability in the moduli space. Especially, the ordering of the group elements induces an ordering of the two-centered solutions which gives the structure of an RG flow. 

Second we propose that the moduli-dependent dyon degeneracy is given by enumerating the ways a certain charge- and moduli-dependent highest weight can be generated by a set of freely acting bosonic and fermionic generators, which are given by two copies of positive roots of the Borcherds-Kac-Moody superalgebra. In other words we propose an equivalence between the dyon degeneracy and the second-quantized multiplicity of a moduli-dependent highest weight in the algebra.
 In particular, using our identification of the Weyl group as the discrete attractor group, we show that the attractor flow generates a sequence of Verma sub-modules (\ref{sequence_module}), which terminates at the smallest module when all moduli have flown to their attractor values. By comparing the neighbouring module and using the properties of the multiplicities of the roots, one can furthermore derive a wall-crossing formula from our counting proposal, which exactly reproduces the wall-crossing formula obtained from supergravity consideration.

These results strongly suggest the following picture: the BPS states are generated by the generators corresponding to the positive roots of the algebra; crossing a wall of marginal stability corresponds to a change of vacuum state, or a change of representation of the algebra, or alternatively and equivalently, exchanging some pairs of creation and annihilation operators. We believe that this sheds light on both the role of the Borcherds-Kac-Moody algebra in the BPS sector of the theory and the nature of wall-crossing in this theory. 
What we have not answered in the following question: can we have an interpretation of this Borcherds-Kac-Moody superalgebra from the low-energy effective action point of view or microscopic point of view? In particular, the positive real roots of the algebra have clear supergravitational interpretation as corresponding to the stationary solutions with two centers carrying $1/\!2$-BPS charges, but how about the positive imaginary roots which are responsible for the bulk of the black hole entropy? Alternatively, we might also ponder about the microscopic interpretation of the present algebra. While the $1/\!2$-BPS degeneracies have an interpretation of counting curves in the $K3$ manifold \cite{Yau:1995mv}, our result suggests that any such interpretation for $1/\!4$-BPS states should have an underlying BKM algebra. 
Similarly, one expects an interpretation of this algebra in terms of the three string junctions as discussed in \cite{Gaiotto:2005hc,Sen:2007ri}.
Finally, on the gauge theory side, a similar Weyl-group-like structure also appears in Seiberg-Witten theory and in Seiberg-like dualities \cite{Cachazo:2001sg}. It would be very interesting if any connection can be found between the appearances of the Weyl group in theories with and without gravity. 

There are numerous immediate generalizations one might explore. The first one is how the algebra and the Weyl group are modified when the charges have other T-duality invariants being non-trivial, namely when the condition (\ref{co_prime_condition}) is no longer satisfied. In this case there are extra two-centered solutions and the wall-crossing formula is modified. The second one is the generalization to other $\cn=4$ theories, in particular the CHL models \cite{w_in_progress}. Finally, the most challenging and tantalizing idea to explore is how much of the group and algebraic structure unveiled in the present paper can survive in the theories with less supersymmetries. In particular whether there is any concrete relation between the general wall-crossing formulas \cite{Kontsevich_Soibelman} and such Borcherds-Kac-Moody superalgebras.

\section{Appendix: Properties of Coxeter Groups}
\label{Appendix: Properties of Coxeter Groups}
\setcounter{equation}{0}

In this appendix we collect various definitions and facts about Coxeter groups. The proofs of them can be found in \cite{coxeter_book1,coxeter_book2,coxeter_book3}. Our presentation is very similar to that of \cite{Henneaux:2007ej}.
\index{Coxeter group}
\index{Coxeter system}
\begin{definition}{\bf (Coxeter System)}
\label{def_coxeter_grp}
A Coxeter system \((W,S)\) consists of a Coxeter group \(W\) and a set of generators \(S=\{s_i, i = 1,\dotsi,n\}\), subjected to the relations
\begin{align} \label{def_coxeter}
s_i^2 &= 1 \qquad;\qquad (s_is_j)^{m_{ij}}=1\\
\intertext{with}
m_{ij} &= m_{ji} \geq 2 \;\;\;\text{for    } i\neq j\;.
\end{align}

A Coxeter graph has \(n\) dots connected by single lines if \(m_{ij} > 2\), with \(m_{ij}\) written on the lines if \(m_{ij} > 3\). 

\end{definition}
\begin{theorem}[Geometric Realization of Coxeter Groups]
\label{def_coxeter_geometric}
Define a basis \(\{\a_1,\dotsi,\a_n\}\) of an \(n\)-dimensional vector space \(M\). Define a metric by
\ben
(\a_i,\a_i) &=& (\a_j,\a_j) \qquad \forall \;\;i,j\\
(\a_i,\a_j) &=&  - (\a_i,\a_i) \cos(\frac{\p}{m_{ij}})\;,
\een
then the reflection
$$
s_i: x \longmapsto x -2 \a_i\, \frac{(\a_i,x) }{(\a_i,\a_i)} 
$$
satisfies the definition of Coxeter group (\ref{def_coxeter}).

Notice that there are null directions of the metric \(\a_i+\a_j\) if \(m_{ij}=\inf\) and the metric is not decomposable iff the Coxeter graph is connected. 

\end{theorem}

\begin{definition}{\bf (Length Function)}
Define the length function
\be\label{length_function}
\ell: W \rightarrow \Z_+
\ee
such that an element has length \(\ell\) if there is no way to write the element in terms of a product of less than \(\ell\) generators. For \(w\in G\), from 
\ben
\ell(ws_i) &\leq &\ell(w) + 1\\
\ell(w) = \ell(ws_i^2 ) & \leq &\ell(ws_i) + 1
\een
we see that 
$$
\ell(w) - 1 \leq \ell(ws_i) \leq \ell(w) + 1\;.
$$

Furthermore, the length function defines a distance function on the group \(d: W\times W \rightarrow \Z_+\) as 
$$
d(w,w') = \ell(w^{-1}w') = d(w',w) \;.
$$ 
One can easily check that this is a metric, especially that the triangle inequality is satisfied.

\end{definition}

\begin{definition}{\bf (Roots of the Coxeter Group)}
\label{root1}
Define the set of roots
$$
\D^{re} = \bigl\{ w(\a_i)\,, w\in W,\; i = \{1,\dotsi,n\} \bigr\}
$$
especially \(\a_i\)'s are called the {\it simple roots}\footnote{The notation \(\D^{re}\) is adapted to the fact that the real roots of the set of Borcherds-Kac-Moody algebra is the set of roots of the Weyl Coxeter group.}. A root
$$
\a = \sum_i a^{(i)} \a_i 
$$
is called a {\it positive root} if all \(a^{(i)}>0\) and a {\it negative root} if all \(a^{(i)}<0\). We will denote them as \(\a>0\) and \(\a<0\) respectively.
\begin{theorem}\label{posi_negi_roots}
 A root of a Coxeter group is either positive or negative, namely
 $$
 \D^{re} = \D_+^{re} \cup \D_-^{re}
 $$
where 
$$
 \D_+^{re}  = \{\a \in  \D^{re} \lvert \, \a >0 \}\quad,\quad
 \D_-^{re}  = \{\a \in \D^{re} \lvert \, \a <0 \}\;.
$$
Furthermore
$$
 \D_-^{re}  = -  \D_+^{re}  \;.
$$

\end{theorem}
\index{root}
\index{positive root}

\begin{theorem}\label{length_posi_negi}
$$
\ell(ws_i) = \ell(w) +1
$$
iff \(w(\a_i)>0\).

\end{theorem}
\begin{corollary}
A root is either positive or negative.
\end{corollary}

\end{definition}

\begin{theorem}
For hyperbolic Coxeter group, the Tits cone, namely the image of a connected fundamental domain under the group action, is the future light-cone. Projected onto a constant length surface, this gives a tessellation of the hyperbolic space. 

\end{theorem}
\index{Tits cone}
\index{Bruhat order}
\index{weak Bruhat order}
\begin{definition}{\bf (Bruhat Order)}
\label{bruhat_order}

Given the Coxeter system (\(W,S\)) and its set of reflections 
\({\cal R} = \{w s_i w^{-1}\lvert\, w\in W \;,\; s_i \in S\}\), and let \(u,u' \in W\), then
\begin{enumerate}
\item{\(u\rightarrow u'\) means that \(u^{-1}u' \in {\cal R}\) and \(\ell(u)<\ell(u')\).}
\item{\(u \leq_B u'\) means that there exists \(u_k \in W\) such that}
\be
u \rightarrow u_1 \rightarrow \dotsi \rightarrow u_m \rightarrow u'
\ee
\end{enumerate}
\end{definition}

\begin{definition}{\bf (weak Bruhat order)}
\label{weak_order}
Given two elements \(u,u' \in W\), repeating the above definition but now restrict further to \(u^{-1}u' \in S\), we obtain the ``weak Bruhat order" \(u\leq u'\).

Two elements are said to be {\it comparable} if \(u\leq u'\) or \(u\geq u'\) and {\it incomparable} otherwise. See \cite{coxeter_book2} for more details.

\end{definition}

\appendix
\chapter{Mathematical Preliminaries}
\label{Mathematical Preliminaries}

In this appendix we will collect various definitions and mathematical results for the purpose of self-containedness. Nevertheless, we do not seek to prove nor to explain them, since they can easily be found in various places in a compact and readable form. See for example  \cite{BBS,Greene:1996cy,Candelas:1987is,Hori:2003ic,Eguchi:1980jx}  for some physics literature. 


\begin{definition}{\bf (complex manifold)} A complex manifold is a topological space \(M\) together with a holomorphic atlas. Equivalently, define the  almost complex structure \({\cal J}\) to be a map between the tangent bundle 
\({\cal J}: TM \rightarrow TM\) satisfying \({\cal J}^2 = -\mathds{1} \) and its Nijenhuis tensor
\(N:TM \times TM \rightarrow TM\)
$$
N[X,Y] = [X,Y] + {\cal J}[{\cal J}X,Y] + {\cal J}[X,{\cal J}Y] - [{\cal J}X,{\cal J}Y]\;.
$$
Then 
$$
\text{N=0} \Leftrightarrow\; {\cal J}\; \text{integrable} \Leftrightarrow M\text{  Complex Manifold}
$$
\end{definition}
In local coordinates, we can write 
\({\cal J} =- i dz^i \otimes \frac{\pa}{\pa z^i}+ i d\bar{z}^{\bar{\imath}} \otimes \frac{\pa}{\pa \bar{z}^{\bar{\imath}} }\).

\index{almost complex structure}
\index{complex manifold}
\index{Nijenhuis tensor}

\begin{definition}{\bf (hermitian metric)} A  {hermitian metric} \(g: TM\times TM \rightarrow \R \) of a complex manifold is a metric which satisfies
$$
g({\cal J}X,{\cal J}Y) = g(X,Y)\;.
$$
In local coordinates it can be written in the form
$$
g= g_{i\jbar} \,dz^i \otimes dz^{\overline{\jmath}}
+ g_{\ibar j } \,d\zbar^{\ibar} \otimes dz^j\;
$$
where the reality of the metric implies \(g_{i\jbar}=\overline{g_{ j \ibar}}\).
\end{definition}

\begin{theorem} Every complex manifold admits a hermitian metric. 
\end{theorem}
With the hermitian metric one can turn the complex structure \({\cal J}\) into a (1,1)-form, called the  K\"ahler form
$$
 J =i g_{i \jbar} \,dz^i\otimes d\zbar^{\jbar}  - i g_{\jbar i} \,d\zbar^{\jbar} \otimes dz^i  = i g_{i \jbar} \,dz^i\wedge d\zbar^{\jbar}  \,.
$$
\index{hermitian metric}
For a complex manifold with \(n\) complex dimension, the \((n,n)\)-form
$$
\underbrace{J\wedge \dotsi\wedge J}_n
$$
is nowhere vanishing can therefore serve as a volume element.
\index{K\"ahler form}

\begin{definition}{\bf (K\"ahler manifold)} A { K\"ahler manifold} is a hermitian manifold with closed K\"ahler form \(dJ=0\).
\index{K\"ahler manifold}
In local coordinates, this implies that the metric can be written as
$$
g_{i \jbar} = \pa_i \pa_{\jbar} {\cal K} 
$$
for some  K\"ahler potential \({\cal K}\). In the overlap of different coordinate charts, the K\"ahler potentials are related by \({\cal K} \rightarrow {\cal K} + f(z) + \bar{g}(\zbar)\) for some (anti-)holomorphic function \(f\) (\(\bar{g}\)) and therefore have the same metric.
\end{definition}

\begin{definition}{\bf(fibre bundle)}
A {fibre bundle} consists of data (\(E,B,\p,F\)), often denoted by \( E \xrightarrow{\p} B\), where \(E\) (total space)
, \(B\) (base space) and \(F\) (fibre) are differential manifolds, and \(\p\) (projection) is a surjection \(\p: E \rightarrow B\) such that the ``fibre at " \(p\in B\) satisfies  \(\p^{-1}(p) = F_p \simeq F\). A local trivialization \(\f_i\) in an open neighbourhood \(U_i\) in \(B\) is a map \(\f_i : U_i \times F \rightarrow \p^{-1}(U_i)\). A transition function \(t_{ij}: U_i \cap U_j \rightarrow G\), where G is a Lie group called the structure group, satisfies \(\f_j (p,f) = \f_i(p,t_{ij}(p) f)\). A section \(v\) of a fibre bundle is a map \(v: B \rightarrow E\) such that \(\p(v(p)) = p\) for all \(p \in B\).
\end{definition}
\begin{figure}
\centering
\includegraphics[width=12cm]{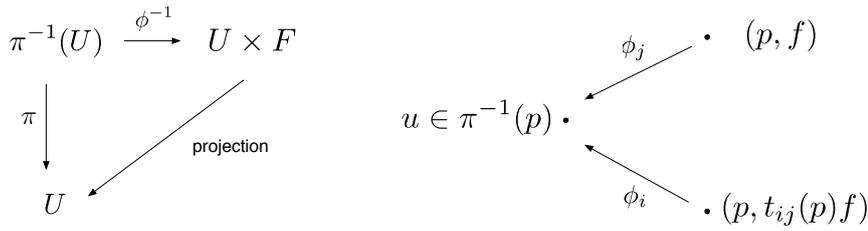}\setlength{\abovecaptionskip}{5pt}
\caption{\footnotesize{The definition of (a) a fibre bundle (b) its structure group.}}
\setlength{\belowcaptionskip}{5pt}
\end{figure}
\index{fibre bundle}
\index{structure group}

\begin{definition}{\bf (vector bundle)}
 A {vector bundle} \( E \xrightarrow{\p} B\) is a fibre bundle whose fibre is a vector space. 
 \end{definition}
 \index{vector bundle}

 In analogy to the concept of parallel transport on a Riemannian manifold, we can also define such a connection on a vector bundle.
Formally, suppose \(\G(E)\) is the space of smooth sections of \(E\), then a connection is a map \({\cal D}: \G(E) \rightarrow \G(E\otimes T^*B)\) such that the Leibnitz' rule
$$
{\cal D} (f \z) = f  {\cal D}\z + \z \otimes df
$$
for any \(\z\in \G(E) \) and smooth function \(f\) on  \(B\). 
For a local coordinate \(e^a\) for \(E\) such that any \(\z\in \G(E)\) can be locally written as \(\z_a e^a\), the connection one-form and the curvature two-form are given by the  Cartan's structure equation
\ben
de^a + \w^a_{\;\;b} \wedge e^b &=& 0 \\ \notag
d \w^a_{\;\;b} + \w^a_{\;\;c}\wedge \w^c_{\;\;b} &=& R^a_{\;\;b}
\een
\begin{definition}{\bf(Chern class)}

\index{connection one-form}
\index{curvature two-form}

Given a complex vector bundle \(E\xrightarrow{\p} B\), let \(R\) be its curvature two-form, then the total {\it Chern class} is defined as
$$
c(R) = \text{det}\left( \mathds{1} + \frac{i}{2\p} R t \right) = \sum_{i=0}^n 
t^i c_i(E)\;,
$$
where \(c_i(R)\) is called the \(i\)-th Chern class. 
It would be useful to diagonalise the curvature two-form \(  R\) into a diagonal \(n\times n\) matrix of two-forms with eigen-two-forms \(-2\p ix_{i=1,\dotsi,n}\). Then the Chern class can be written as 
$$
c(R) = \text{det}\left( \mathds{1} + \frac{i}{2\p} R t \right) = \prod_{i=1}^n (1+x_i t)
$$
For example, for a \(n\)-(complex) dimensional fibre we have
\ben
c_0(R) &=& 1\\ \notag
c_1(R) &=&  \frac{i}{2\p} \Tr R = \sum_{i=1}^n x_i\\ \notag
\vdots\\ \notag
c_n (R) &=& (\frac{i}{2\p})^n \text{det}R =\prod_{i=1}^n x_i\;.
\een 
\end{definition}
Chern classes are topological invariants, meaning that all the curvatures in the same cohomology class have the same Chern classes, which capture many topological properties of a bundle. For example, when \(E\) is a line bundle the first Chern class is the only non-trivial one. In this case the first Chern class completely fixes the topology of the line bundle. 

Chern classes satisfies the following property for the Whitney sum of two bundles \(E\) and \(F\), namely
\be\label{whitney_sum_formula}
c(E \oplus F) = c(E) \wedge c(F)\;.
\ee

\vspace{1cm}
\begin{tabular}{p{12cm}}
\hline\\
\end{tabular}
\vspace{1cm}
\index{Chern class}
\begin{digression}ÊMore Characteristic Classes and Some Index Theorems

For future use we list here more relevant characteristic classes and some index theorems related to them. 
\begin{itemize}

\item{The Chern character \(\text{ch}(R)=\Tr\exp (\frac{iR}{2\p})  \), is related to the Chern classes as}
\be\label{chern_character}
\text{ch}(R) =\Tr\exp (\frac{iR}{2\p}) = 2n + c_1(R) + \left( \frac{1}{2} c_1^2(R) - c_2(R)  \right) +\dotsi\,.
\ee

\item{For a real \(2n\)-dimensional vector bundle with structure group \(O(2n)\), the field strength \(R'\) is anti-symmetric and can be diagonalise with \(2n\) imaginary eigenvalues (\({i}{2\p}x_1, -{i}{2\p}x_1,\dotsi, {i}{2\p}x_n, -{i}{2\p}x_n\))}. Define the Pontrjagin class to be
\be
p(R) = \text{det}( \mathds{1} + \frac{R'}{2\p}t) = \text{det}( \mathds{1} - \frac{R'}{2\p}t) = \sum_{i=0}^n 
t^i p_i(R)\;,
\ee

By expressing both \(p(R) \) and \(c(R)\) in terms of the eigenvalues \(x_i\), we obtain the following relations
\ben
p_1 &=& c_1^2 -2 c_2 \\
p_2 &=& c_2^2 - 2 c_1 c_3 + 2 c_4 \qquad \text{etc.}
\een

The special case in which the bundle \(E\) is the tangent bundle \(TM\) of a manifold \(M\) will be especially useful for us. We will often omit writing out the tangent bundle explicitly and just denote the characteristic classes by the manifold \(M\), for example, it should be understood that \(c_1(M)= c_1(TM)\) etc.

\item{ The Dirac genus (A-roof genus) is defined as
\bea\notag
\hat{\cal A}(M) = \prod \frac{x_i/2}{\sinh(x_i/2)} &=& 1- \frac{1}{24} p_1 +\dotsi\;\\\label{A_roof}
&=& 1 -\frac{1}{24} (c_1^2 -2 c_2) +  \dotsi\;. 
\eea
}
\item{ The Hirzebruch \({\hat{\cal L}}-\)polynomial is defined as
\bea\notag
{\hat{\cal L}}(M) = \prod \frac{x_i}{\tanh(x_i)} &=& 1+ \frac{1}{3} p_1 +\dotsi\;\\\label{Hirzebruch}
&= &1+ \frac{1}{3} (c_1^2 -2 c_2) +  \dotsi\;.
\eea
}
\item{The Todd class is defined as
\be
{\text Td}(M) = \prod \frac{x_i}{1-e^{-x_i}}  = 1 +\frac{1}{2}c_1 + \frac{1}{12} (c_1^2 + c_2) +\dotsi\;.
\ee}

\item{The Euler class is defined for an oriented real (\(2n\))-dimensional fiber with SO(2n) structure group as}
\be
e(M)= x_1 \,\dotsi x_n =( \,p_n(M)\,)^{1/2}=c_n(M)\;.
\ee

\item{Gauss-Bonnet Theorem}

\index{Gauss-Bonnet theorem}
For an even dimensional manifold \(M\)
\be \label{gauss_bonnet}
\chi(M) = \int_M e(T(M)) = \sum (-1)^i \, b^i\;,
\ee
where \(b^i\) denotes the \(i\)-th Betti number.

\item{Signature Index Theorem}

\index{signature index theorem}
For a real (\(4n\))-dimensional manifold \(M\), the signature is given by the dimension of the self-dual and anti-self-dual (under the Hodge star operation) part of the middle cohomology as
\be
\s(M) = \text{dim}H^{2n}_+(M;\R)- \text{dim}H^{2n}_-(M;\R)\;,
\ee
and the index theorem states
\be\label{signature_index_theorem}
\s(M) =  \int_M {\hat{\cal L}}(M)\;.
\ee
\item{Hirzebruch-Riemann-Roch Theorem}
\index{Hirzebruch-Riemann-Roch Theorem}

The betti numbers of bundle-valued forms satisfy the following relation
\be\label{Hirzebruch_Riemann_Roch}
w(M;V) = \sum_{k=0}^{\text{dim} M} \, (-1)^k\, \text{\small dim} H^{k}(M;V)
= \int_M \, ch(V) Td(M)\;.
\ee

\end{itemize}
\end{digression}
\index{chern character}
\index{Pontrjagin class}
\index{A-roof (Dirac) genus}
\index{Hirzebruch \({\hat{\cal L}}-\)polynomial}

\vspace{1cm}
\begin{tabular}{p{12cm}}
\hline\\
\end{tabular}
\vspace{1cm}

Finally we are ready to introduce the kind of manifold which will play an important role in the discussion about the moduli space of Calabi-Yau three-folds, namely the special K\"ahler manifold. Historically a special K\"ahler manifold is first defined as the scalar manifold of the \({\cal N}=2, d=4\) supergravity theory. But since this kind of manifold appears also in other context, we will proceed by first define the manifold abstractly, and later show that the \({\cal N}=2, d=4\) supergravity scalar manifold is an example of them.

\begin{definition}{\bf(Special K\"ahler manifold)}
A {(local) special K\"ahler manifold} \(M\) is a \(n\)-(complex) dimensional manifold with a holomorphic Sp(2n+2,\(\R\)) vector bundle \({\cal E}\) over  \(M\), a line bundle \({\cal L}\) over  \(M\) with \((2\p\) times) the first Chern class equal to the K\"ahler form, and a 
holomorphic section \(\O\) in \({\cal L} \otimes {\cal E}\) such that
the K\"ahler form is given by 
\be \label{special_K_def}
J = i \pa \bar{\pa} \log[-i\langle \O, \bar{\O}\rangle]
\ee
and such that \(\langle \pa_a \O,\O\rangle\) = 0 \;, 
where \(\langle\,,\rangle\) denotes the symplectic product \cite{Strominger:1990pd,Craps:1997gp}.
\end{definition}
\index{special K\"ahler manifold}

In order to define the Calabi-Yau manifolds themselves, we still have to introduce a last element, meaning the holonomy group of a manifold. With the Levi-Civita connection we have a concept of parallel transport on a \(n\)-(complex) dimensional Riemannian manifold. When parallel transporting an orthonormal frame around a closed loop, it doesn't come back to itself generically, but is rather related by a SO(\(2n\)) transformation to the original frame. By combining different closed loops it's not hard to see that holonomy forms a group. There are special kinds of manifolds whose holonomy group does not cover the the whole SO(\(2n\)) but only a subgroup of it. They will be called manifolds with special holonomy. As we will see later, a smaller holonomy group is crucial for having unbroken supersymmetry after compactification.

\begin{definition}{\bf(Calabi-Yau manifold)}
A {Calabi-Yau} manifold is a compact K\"ahler manifold with vanishing first Chern class.
\end{definition}
\index{Calabi-Yau manifold}

\begin{theorem}(Yau's Theorem)
If \(X\) is a complex K\"ahler manifold with vanishing first Chern class and with K\"ahler form \(J\), then their exists a unique Ricci-flat metric on \(X\) whose K\"ahler form is in the same cohomology class as \(J\).
\end{theorem}
\index{Yau's theorem}

From the above theorem and from the fact that Ricci-flatness is a sufficient condition for a vanishing first Chern class, we can conclude that an equivalent definition of a Calabi-Yau manifold is a compact 
K\"ahler manifold admitting a Ricci-flat metric. This is a very powerful theorem since it is in general very hard to find the explicit metric but relatively easy to compute the Chern class. 

The Ricci-flatness also implies special holonomy properties. Consider the integrability condition for parallel transporting a spinor 
$$
[\nabla_k,\nabla_m]\eta = -\frac{1}{4} R_{kmpq} \g^{pq} \eta\;,
$$ 
where \(\g\)'s are the gamma-matrices satisfying the Clifford algebra. From the index structure of a K\"ahler manifold we can already see that a \(n-\)(complex) dimensional K\"ahler manifold has holonomy group 
\(U(n) \subset SO(2n)\). But this is not enough to ensure the existence of a constant spinor when \(n=3\). For that we need a holonomy group \(SU(n) \subset SO(2n)\). The vanishing of the \(U(1)\) part of holonomy is given exactly by Ricci-flatness. 
Furthermore, employing the covariantly constant spinors one can actually show the presence of another equivalent relation, namely the existence of a nowhere vanishing, holomorphic \((n,0)\) form. 

We can now therefore gather the above facts and give four equivalent definitions of a Calabi-Yau manifold. 

\begin{definition}{\bf(Calabi-Yau manifold)}
A {Calabi-Yau manifold} is a compact, complex, 
\(n\) (complex)-dimensional K\"ahler manifold \(X\) which 
\begin{align}\label{def_calabi_yau}
&\text{has  } c_1(TX) = 0 \\ \notag
&\Leftrightarrow \text{admits a Ricci-flat metric}\\ \notag
&\Leftrightarrow \text{has holonomy group  } SU(n)\\ \notag
&\Leftrightarrow \text{admits a nowhere vanishing holomorphic  } (n,0) \text{  form}\;.
\end{align}
\end{definition}

 Furthermore this holomorphic (n,0) form 
 \be\label{complex_structure}
 \O  =  \frac{1}{3!} \O_{i_1 \dotsi i_n} \,dz^{i_1} \wedge \dotsi \wedge dz^{i_n}
 \ee
is harmonic and covariantly constant when the Ricci-flat metric is chosen.

Before going into the details of Calabi-Yau manifolds of different dimensions, we would like to end this appendix by discussing the cohomology structure of Calabi-Yau manifolds. 

Let's denote by \(h^{r,s}\) the dimension of the vector space of harmonic \((r,s)\)-form on X. From the Hodge  star operation we see that these Hodge numbers have the symmetry \(h^{r,s}=h^{n-r,n-s}\). Furthermore, from the complex conjugation and K\"ahlerity \(\D = 2\D_\pa = 2\D_{\bar{\pa}}\) we have \(h^{r,s}=h^{s,r}\). What is special about Calabi-Yau manifolds compared to the usual K\"ahler manifolds are, first of all, \(h^{n,0}= h^{0,n} = 1\) as mentioned before. Moreover,  Using the Ricci-flatness it's not hard to show that \(h^{r,0}=0\) for \(0<r<n\), at least when the Euler characteristic is non-zero. In the following we sum up the above properties in the so-called Hodge diamond for Calabi-Yau two- and three-folds. 
 
\index{Hodge diamond}
\begin{align}\label{hodge_diamond_K3}
\begin{array}{ccccc}
&&h^{0,0}&&\\
&h^{1,0}&&h^{0,1}&\\
h^{2,0}&&h^{1,1}&&h^{0,2}\\
&h^{2,1}&&h^{1,2}&\\
&&h^{2,2}&&
\end{array}
&=\begin{array}{ccccc}
&&1&&\\
&0&&0&\\
1&&20&&1\\
&0&&0&\\
&&1&&
\end{array}
\\ \notag
&\\ \notag
\begin{array}{ccccccc}
&&&h^{0,0}&&&\\
&&h^{1,0}&&h^{0,1}&&\\
&h^{2,0}&&h^{1,1}&&h^{0,2}&\\
h^{3,0}&&h^{2,1}&&h^{1,2}&&h^{0,3}\\
&h^{3,1}&&h^{2,2}&&h^{1,3}&\\
&&h^{3,2}&&h^{2,3}&&\\
&&&h^{3,3}&&&
\end{array}
&=\begin{array}{ccccccc}
&&&1&&&\\
&&0&&0&&\\
&0&&h^{1,1}&&0&\\
1&&h^{1,2}&&h^{1,2}&&1\\
&0&&h^{1,1}&&0&\\
&&0&&0&&\\
&&&1&&&
\end{array}
\end{align}

We immediately see that the Euler numbers 
$$
\chi(X) = \sum_{r,s} (-1)^{r+s} h^{r,s} 
$$
of these Calabi-Yau manifolds are
\be\label{euler_number}
\chi(\text{CY}_2) = 24\;\;;\;\;\chi(\text{CY}_3) = 2 (h^{1,1}-h^{1,2})\;.
\ee

Notice that apart from the above discussion about the cohomology properties of Calabi-Yau n-folds in general, we have used an extra piece of information for the two-folds. Namely we have put in \(h^{1,1} = 20\). This is because there is only one Calabi-Yau two-folds with \(SU(2) = Sp(1)\) holonomy group (so they are also hyper-K\"ahler) but not a subgroup of it, in the sense that all of them are diffeomorphic to each other. From now on we will refer to this unique two-fold as K3 manifold and reserve the name ``Calabi-Yau" for the three-folds.

A construction of K3 manifolds of special interests is given by orbifolding a \(T^4\). Let's for example consider an orbifold by \(\Z_2\): \((z_1,z_2) \rightarrow (-z_1,-z_2)\) where \(z_{1,2}\) are the coordinates for the square torus. There are clearly \(2^4=16\) fixed points, each can be ``blown up" by replacing the singularity with an Eguchi-Hanson metric. See (\ref{E_H}) for an description of the Eguchi-Hanson space.
This is a gravitational instanton and therefore the first Chern class indeed vanishes. From the 16 harmonic anti-self-dual \((1,1)\)-form of the  Eguchi-Hanson metric, combined with the 4 \((1,1)\)- and one \((0,2)\)-, one \((2,0)\)-form from the original torus, we indeed obtain the above Hodge diamond. Furthermore, from the above analysis we also know the self-dual and anti-self-dual splitting of the two-forms 
\be\label{K3_signature_1}
b_2^+ (K3) = 3 \;\;;\;\; b_2^- (K3) = 19\;.
\ee
In the eigenbasis of the Hodge star operator on a four dimensional manifold \(S\), which satisfies \(\star^2 =1\), the symmetric bilinear on the space of even-forms satisfies
\be\label{symmetric_bilinear_four-fold}
(\a,\a) := \int_{S}  \a\wedge \a = \pm   \int_{S}  \a\wedge \star\a \;.
\ee
Using the information about the dimensions of the eigenspace of the Hodge star operator (\ref{K3_signature_1}), we conclude that the signature of the space of middle cohomology classes of the K3 surface \(S\) is (3,19), namely 
\be\label{K3_lattice_1}
H^2(S,\Z) \cong \G^{3,19}\,.
\ee
Furthermore, we can incorporate the whole Hodge diamond (\ref{hodge_diamond_K3}) by introducing also the basis \(\a_0\) and \(\a^0\) for \(H^0(S,\Z)\) and \(H^4(S,\Z)\) respectively, which is dual to each other in the sense that  \( \int_{S}  \a_0\wedge \a^0 = 1\), and therefore enlarge the lattice with an extra piece
\be\label{lattice_1_1_U}
U = \G^{1,1} = \bem 0&1\\1&0\eem\;.
\ee
We therefore conclude that the space of integral cohomology classes of a K3 manifold is the following lattice 
\be\label{K3_lattice_2}
H^{2*}(S,\Z) \cong \G^{3,19} \oplus \G^{1,1} \cong \G^{4,20}\;.
\ee
Finally let us remark that we have used the fixed point counting of the \(\Z_2\) orbifold limit of the K3 manifold, and the fact that all K3 manifolds are diffeomorphic with each other and have therefore the same topological invariants, to obtain the signature (\ref{K3_signature_1}). But one can also obtain it from using the signature index theorem (\ref{signature_index_theorem}). See, for example, \cite{Eguchi:1980jx}.

\index{Eguchi-Hanson space}
This situation is in stark contrast with that of Calabi-Yau three-folds, in which case we don't even know whether the possibilities of different Hodge numbers are finite. Or, in another (not incompatible) extreme, whether all these possible Calabi-Yau's with different Hodge numbers are actually all connected by non-perturbative conifold transitions and in this sense there is only one Calabi-Yau. For the purpose of our thesis we can just stick to thinking of the space of all CY's as... rather complicated.




\cleardoublepage
\bibliography{bib_arxiv}

\cleardoublepage
\printindex

\end{document}